\pdfoutput=1
\documentclass[%
aps, physrev,
showkeys,
 amsmath,amssymb,
reprint,%
floatfix,
]{revtex4-2}

\usepackage{graphicx}
\usepackage{dcolumn}
\usepackage{bm}
\usepackage[mathlines]{lineno}
\usepackage{fix-cm}

\usepackage[
scale=0.7, marginratio={1:1, 2:3}, 
text={7in,10in},centering,
]{geometry}

\usepackage[colorlinks=true,linkcolor=blue,citecolor=blue]{hyperref}

\usepackage{commath}
\usepackage{mathtools}
\usepackage{chemformula}
\usepackage{booktabs}
\usepackage{multirow}

\usepackage[hyphenation, lastparline, rivers]{impnattypo}
\usepackage[activate={true, nocompatibility}, kerning=true, tracking=true,final,nopatch=footnote]{microtype}
\microtypecontext{spacing=nonfrench}

\begin{document}

\preprint{Submitted to Phys Rev E} 

\title{\textbf{Optimizing reaction and transport fluxes in temperature gradient-driven chemical
  reaction-diffusion systems}}

\author{Mohammed Loukili} 
\affiliation{Institut de Recherche de l'École Navale, EA 3634, IRENav, Brest, France}

\author{Ludovic Jullien} \affiliation{CPCV, Département de chimie, École normale supérieure, PSL
  University, Sorbonne Université, CNRS, 24, rue Lhomond, 75005 Paris, France}

\author{Guillaume Baffou}
\affiliation{Institut Fresnel, CNRS, Aix Marseille University, Centrale Med, 13013 Marseille, France}

\author{Raphaël Plasson}
\email{raphael.plasson@univ-avignon.fr}
\affiliation{Avignon University, INRAE, UMR408 SQPOV, 84000 Avignon, France}

\date{\today}

\begin{abstract}  
  Temperature gradients represent energy sources that can be harvested to generate steady reaction or transport fluxes. Technological developments could lead to the transfer of free energy from heat sources and sinks to chemical systems for the purpose of extraction, thermal batteries, or
  nonequilibrium synthesis.

  We present a theoretical study of 1D chemical systems subjected to temperature gradients, for sustaining nonequilibrium chemical fluxes. A complete theoretical framework describes the behavior of the system induced by various temperature profiles.  An exact mathematical derivation was established for a simple two-compartment model and was generalized to arbitrary reaction-diffusion systems based on numerical models. An experimental system was eventually scaled and tuned to optimize either nonequilibrium chemical transport or reaction.

  The relevant parameters for this description were identified; they focused on the system symmetry for chemical reaction and transport. Nonequilibrium thermodynamic approaches lead to a description analogous to electric circuits. Temperature gradients lead to the onset of a steady chemical
force, which maintains steady reaction-diffusion fluxes moderated by chemical resistance. The system activity was then assessed using the entropy production rate as a measure of its dissipated power.

  The chemical characteristics of the system can be tuned for general optimization of the
  nonequilibrium state or for the specific optimization of either transport or reaction processes. The shape of the temperature gradient can be tailored to precisely control the spatial localization of active processes, targeting either precise spatial localization or propagation over large areas. The resulting temperature-driven chemical system can in turn be used to drive secondary processes into either nonequilibrium reaction fluxes or concentration gradients.
\end{abstract}


\maketitle

\section{Introduction}
\label{sec:introduction}

The logic of preparative chemistry is in sharp contrast with the one of biochemical systems. The former focuses primarily on multistep synthesis, with the objective of producing pure compounds in optimal yield through individually optimized reactions. In living cells, the reactions are also grouped in sets. However, they occur in cycles, like the Krebs and Calvin cycles, where reactants and products are not distinguished from each other, but simultaneously coexist in non-equilibrium steady states. Hence, although sharing a same common canon of reactivity, a cell operates differently than a chemical reactor seeking maximal yields. What advantages does this mode of operation afford?

Thermodynamics fixes the final composition in which a reactive system reaches equilibrium under
chemical and physical constraints. In practice, it governs the realization of multiple chemical
reactions of academic and industrial interest, as well as the elementary steps of most separation
processes, such as distillation or chromatography. In contrast, kinetics drives the trajectory of
the evolution of a reactive system. It also governs the dynamic behavior of the latter when
sustained in an nonequilibrium steady state, which can exhibit much richer phenomena
(e.g. oscillations, waves, patterns, and chaos) than in the equilibrium state. In particular, such
nonequilibrium steady states are encountered in living biological matter, where they control its
dynamics \cite{beard2008,lane.allen.ea-10} and lead to the propagation of a sustained energy flux
throughout metabolism by exploiting free energy transduction
\cite{hill2012,morowitz.smith-07,pascal.boiteau-11}.

The dynamic behavior of the nonequilibrium steady states is theoretically understood
\cite{nicolis1977,vidal1988,epstein1998}. In contrast, the issue of steadily sustaining an
nonequilibrium reactive state has received much less attention and remains a major challenge
\cite{grzybowski2016}. In practice, its realization requires either introducing constraints to
prevent the effective relaxation of the reactive processes towards equilibrium
\cite{jullien2003,lemarchand2004,ragazzon2018}, or steadily applying a periodic excitation on the
reactive system. \cite{astumian2011} We previously demonstrated the relevance of the first approach
by frustrating the relaxation of reactive processes by the interference of molecular motion
\cite{emond_energy_2012, saux_energy_2014}, a strategy to generate reaction-diffusion cycles that
are thought to establish spatial gradients of signaling activities in living cells
\cite{kholodenko-06,soh.byrska.ea-10}. However, the efficiency of this first approach has not been
theoretically analyzed, and is the object of the present manuscript.

\begin{figure}[t]
  \includegraphics[width=\linewidth]{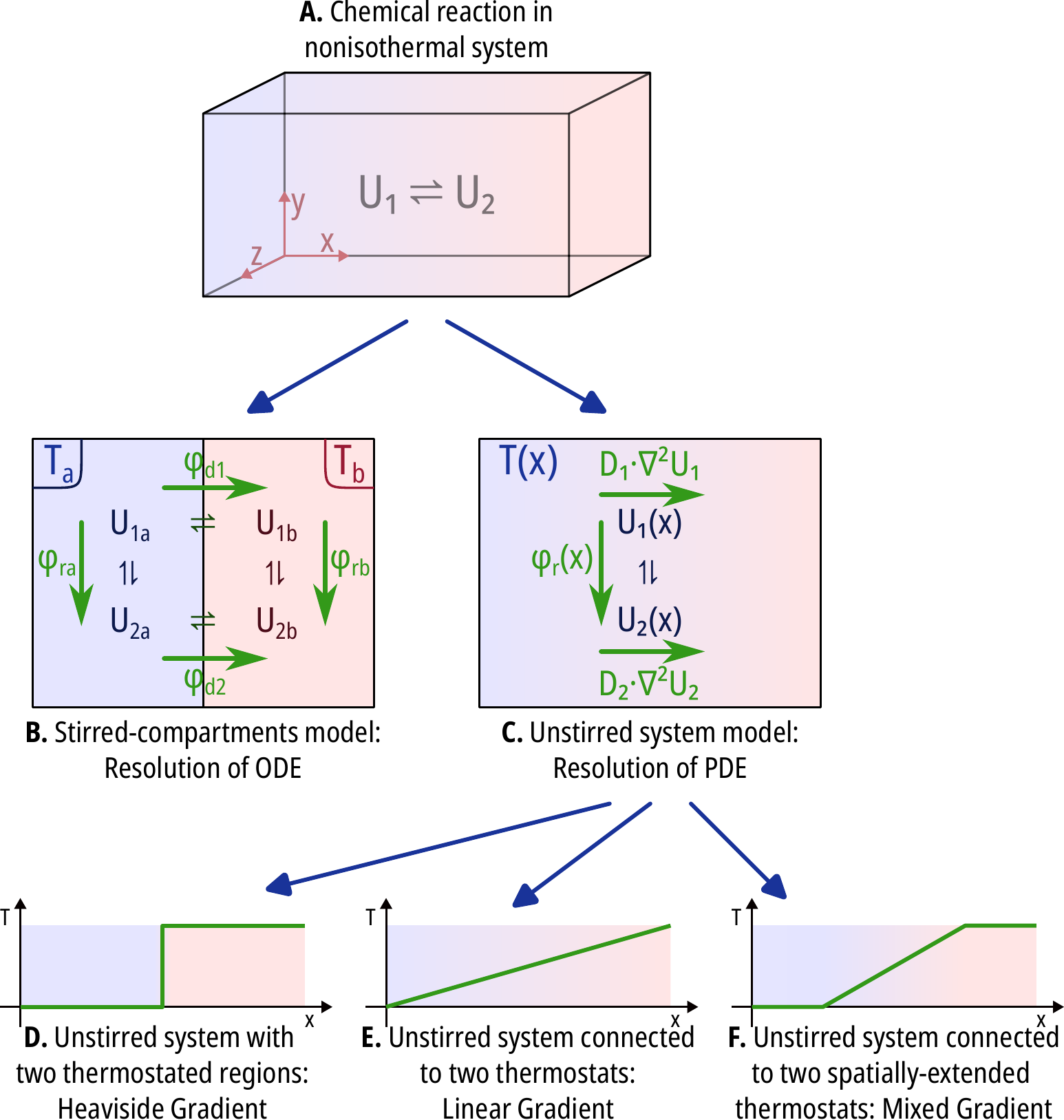}
  \caption{Description of the chemical system. \textbf{A}, general representation;
    \textbf{B}, two-compartment model; \textbf{C}, unstirred model; \textbf{D-F}, applied
    temperature gradient profiles.}
  \label{fig:Fig1}
\end{figure}

In our previous work \cite{emond_energy_2012, saux_energy_2014}, we used light as an energy source
to maintain a reactive system in a nonequilibrium state. In the present study, we favored a more
general approach and adopted the application of a temperature gradient.  Despite its chemical
relevance, the behavior of reactive media in contact with two thermostats at different temperatures
has included glorious names (including Nernst \cite{nernst1904}, Dirac \cite{dirac1924}, Prigogine
\cite{prigogine1952a,prigogine1952b}).  Several papers have been published since then, but have been
mainly concerned with the gas phase and the question of how a chemical reaction that occurs in a
system modifies its thermal conductivity
\cite{meixner1952,haase1953,butler_thermal_1957,brown1958,kehlen1981,xu2006,xu2007,deZarate2011,skorpa_diffusion_2015}.
Recently, heat flow has been applied to generate proton gradients and pH oscillations in microscale
aqueous solutions \cite{keil2017,matreux2023}. In the latter works, the reactive system was often
assumed to be at chemical equilibrium within the temperature gradient.

In this paper, we level off this assumption, which has been predicted to generate attractive
behavior such as dissipation-driven selection of states in nonequilibrium chemical networks
\cite{busiello2021}, or emergent thermophoretic behavior in chemical reaction systems
\cite{liang2022}.  We investigate the behavior of an active chemical system submitted to a
  nonequilibrium force, induced by an externally imposed temperature gradient. The primary objective
  is to determine the optimal characteristics and conditions that will lead to maximizing an
  internal nonequilibrium reaction-diffusion flux within the system. This consists in a simple
  chemical reaction subjected to an inhomogeneous field of temperature.  The associated temperature
  gradient directly generates a large entropy production by heat diffusion; the aim is to redirect a
  maximal part of this source of nonequilibrium to chemical processes.

  In this context, the temperature gradient-induced steady chemical force and the resulting steady
  chemical flux must both be maximized, with the ultimate objective of maximizing the entropy rate
  produced specifically by this internally sustained cyclic process. This objective was further
  refined by recognizing that the focus might entail optimizing either the transport process or the
  reaction process on their own, or balancing both processes. To achieve this optimization, we
  benefited from a fruitful analogy between entropy production by a chemical reaction and the ohm
  dissipated power by electrical resistors.

  We then identified the conditions for the spatial localization or delocalization of nonequilibrium
  activity, emphasizing the important role of the system scaling and of the shape of the temperature
  gradient. Finally, we assessed the possibility of propagating this primary energy transfer from
  the heat sources and sinks to secondary athermic reaction or transport processes.

\section{Results and discussion}
\label{sec:results-discussion}

\subsection{Chemical model}
\label{sec:chemical-model}

The studied chemical system was reduced to experiencing a single isomerization reaction
\begin{align}
    & \ch{U_1  <=>[ $k_{+}$ ][ $k_{-}$ ] U_2} \label{eq:1}\\
  \shortintertext{where the rate constants $k_{+}$ and $k_{-}$ are linked to the thermodynamic constant $K$ with}
    & K = \frac{k_{+}}{k_{-}} \label{eq:2}
\end{align}

This represents a generic reaction, in its simplest form. It can be easily extended to more complex
mechanisms, as detailed further, namely for the description of second-order reactions or more
complex mechanisms.

This chemical reaction was driven in a steady nonequilibrium state by an imposed temperature
gradient (see Fig.~\ref{fig:Fig1}). We studied different configurations embodied by various
temperature profiles associated with different experimental setups, leading to the study of
theoretical models with increasing complexities.

Using thermodynamic and Arrhenius relationships, the variation in the constants with temperature
can be expressed as:
\begin{subequations}
  \begin{align}
    K & = K^{\infty}   e^{-\frac{\Delta_{r}H_0}{RT}} \label{eq:3}\\
    k_{+} & = k_{+}^{\infty}   e^{-\frac{E_{a,+}}{RT}} \label{eq:4}\\
    k_{-} & = k_{-}^{\infty}   e^{-\frac{E_{a,-}}{RT}} \label{eq:5}\\
    \shortintertext{with}
    \Delta_{r}H_0 & = E_{a,+} - E_{a,-} \label{eq:6}
  \end{align}
\end{subequations}

\subsection{Chemical response to a temperature gradient between two homogeneous compartments}
\label{sec:case-two-homogeneous}

The general behavior of a reaction driven by temperature gradients was first characterized in the limits
of two compartments, each of which was homogeneous in temperature and concentration. This
corresponds to two stirred systems; each compartment $A/B$ is connected to the other by chemical
exchange with the reaction rate constant $k_{d,i}$ for each compound of concentration $c_{i}$:
\begin{subequations}
  \begin{align}
    \shortintertext{Compartment $A$       :}
    \ch{U_{1a} &<=>[ ${k}_{+,a}$ ][ ${k}_{-,a}$ ] U_{2a}} \label{eq:7}\\
    \shortintertext{Compartment $B$       :}
    \ch{U_{1b} &<=>[ ${k}_{+,b}$ ][ ${k}_{-,b}$ ] U_{2b}} \label{eq:8}\\
    \shortintertext{Diffusion of \ch{U1}:}
    \ch{U_{1a} &<=>[ $k_{d,1}$ ][ $k_{d,2}$ ] U_{1b}}          \label{eq:9}\\
    \shortintertext{Diffusion of \ch{U2}:}
    \ch{U_{2a} &<=>[ $k_{d,2}$ ][ $k_{d,2}$ ] U_{2b}}                   \label{eq:10}
  \end{align}
\end{subequations}

\begin{table*}
  \begin{center}
    \begin{tabular}{llclrll}
      \toprule
      & Description                & Dimensionalized & Unit                         &               & Nondimenzionalized                                               \\
      \midrule
      \multirow{6}{2.6cm}{Characteristic dimensions}     & Concentration              & $c_i$           & mol.m$^{-3}$                 & $u_{i}$       & $=\sfrac{c_{i}}{c_{0}}$                                          \\
      & Length                     & $x$             & m                            & $\bar{x}$     & $=\sfrac{x}{l_{0}}$                                              \\
      & Temperature                & $T$             & K                            & $\theta$           & $=\sfrac{(T-T_{0})}{T_{0}}$                                      \\
      & Diffusion constant         & $D_{i}$         & m$^{2}$.s$^{-1}$             & $d_{i}$       & $=\sfrac{D_{i}}{D_{0}}$                                          \\
      & Time                       & $t$             & s                            & $\bar{t}$     & $=\sfrac{t}{t_{0}}$  (with $t_{0}=\sfrac{l_{0}^{2}}{D_{0}}$)     \\
      \midrule
      \multirow{4}{2.6cm}{Chemical characteristics}     & Reaction rate constant     & $k_{\pm}$         & s$^{-1}$                     & $\bar{k}_{\pm}$ & $=k_{\pm}  t_{0}$                                                  \\
      & Exchange rate constant     & $k_{d,i}$       & s$^{-1}$                     & $\delta_{i}$       & $=k_{d,i}  t_{0}$                                                \\
      & Activation energy          & $E_{a,\pm}$       & J.mol$^{-1}$                 & $\varepsilon_{\pm}$       & $=\sfrac{E_{a,\pm}}{RT_{0}}$                                       \\
      & Reaction enthalpy          & $\Delta_{r}H_{0}$    & J.mol$^{-1}$                 & $\eta$           & $=\sfrac{\Delta_{r}H_{0}}{RT_{0}}$                                    \\ 
      \midrule
      \multirow{8}{2.6cm}{Primary parameters}           & Equilibrium constant       &                 &                              & $\kappa$           & $ = \sqrt{\sfrac{\bar{k}_{+,0}}{\bar{k}_{-,0}}} $            \\
      & Exchange ratio             &                 &                              & $\lambda$           & $ = \sqrt{\sfrac{\delta_{2}}{\delta_{1}}}  = \sqrt{\sfrac{d_{2}}{d_{1}}} $ \\
      & Mean kinetic rate          &                 &                              & $\rho$           & $ =\sqrt{\bar{k}_{+,0} \bar{k}_{-,0}} $                      \\
      & Mean exchange rate         &                 &                              & $\delta$           & $ =\sqrt{\delta_{1} \delta_{2}} $                                          \\
      & Mean diffusion constant    &                 &                              & $d$           & $ =\sqrt{d_{1}  d_{2}} $                                         \\
      & Average activation energy  &                 &                              & $\beta$           & $= \sfrac{(\varepsilon_{+} + \varepsilon_{-})}{2} $                                  \\
      & Energy profile asymmetry   &                 &                              & $\gamma$           & $=\sfrac{(\varepsilon_{+}-\varepsilon_{-})}{(\varepsilon_{+}+\varepsilon_{-})}$                          \\
      & Gradient intensity         &                 &                              & $\theta_{g}$       & $\: :\: \theta \in[-\theta_{g}, + \theta_{g}]$                                    \\
      \midrule
      \multirow{2}{2.6cm}{Secondary parameters}         & Reaction asymmetry         &                 &                              & $\kappa'$          & $= \sfrac{2 \kappa}{(1 + \kappa^{2})}$                                     \\
      & Exchange asymmetry         &                 &                              & $\lambda'$          & $= \sfrac{2\kappa\lambda}{(1+\kappa^{2}\lambda^{2})}$                                  \\
      \midrule
      \multirow{4}{2.6cm}{Steady state characteristics} & Chemical force             & $\sfrac{\mathcal{A}}{T}$  & J.mol$^{-1}$.K$^{-1}$        & $\alpha$           & $=\sfrac{\mathcal{A}}{RT}=\ln\sfrac{K}{Q}$                                 \\
      & Chemical flux              & $\phi$             & mol.m$^{-3}$.s$^{-1}$        & $\varphi$           & $=\sfrac{\phi   t_{0}}{c_{0}}$                                      \\
      & Diffusion flux             & $J_{i}$         & mol.m$^{-2}$.s$^{-1}$        & $j_{i}$       & $=d_{i}\sfrac{\partial u_{i}}{\partial \bar{x}}$                                  \\
      & Rate of entropy production & $\dot{S}$       & J.m$^{-3}$.s$^{-1}$.K$^{-1}$ & $\sigma$           & $=\sfrac{\dot{S}   t_{0}}{Rc_{0}}$                               \\
      \bottomrule
    \end{tabular}
  \end{center}
    \caption{Dimensionalized and nondimensionalized parameters, classified as characteristic
      dimensions, chemical characteristics, primary parameters (control parameters for the
      description of a given system), secondary parameters (used for the simplification of the
      final equations), and steady state characteristics (describing the resulting steady dynamic
      properties of the system). $R=8.314$~J.mol$^{-1}$.K$^{-1}$. }
  \label{tab:table1}
\end{table*}

\subsubsection{Parameters}
\label{sec:parameters}

Identifying the relevant parameters for the description of the system is a critical step. We
focused on describing the system in general terms by assessing the global symmetry or asymmetry of
the parameters, as this leads to a description of the system behavior in the simplest mathematical
terms.

The general properties of the system are described based on its characteristic dimensions:
concentration $c_{0}$ (corresponding to the total concentration), temperature $T_{0}$ (corresponding
to the median system temperature) and time $t_{0}$. All subsequent system parameters were
nondimensionalized using these characteristic parameters (see Table~\ref{tab:table1}). This  leads to:
\begin{subequations}
  \begin{align}
    K & = K_{0}   e^{
        \frac{\Delta_{r}H_0}{RT_{0}} 
        \left(\frac{T-T_{0}}{T}\right)
        } \label{eq:11}\\
    k_{+} & = k_{+,0}   e^{
            \frac{E_{a,+}}{RT_{0}}
            \left(\frac{T-T_{0}}{T}\right)
            } \label{eq:12}\\
    k_{-} & = k_{-,0}   e^{
            \frac{E_{a,-}}{RT_{0}}
            \left(\frac{T-T_{0}}{T}\right)
            } \label{eq:13}
  \end{align}
\end{subequations}
with $K_{0}=K(T_{0})$, $k_{+,0} = k_{+}(T_{0})$, $k_{-,0} = k_{-}(T_{0})$.

This is then nondimensionalized as:
\begin{subequations}
  \begin{align}
    \label{eq:14}
    \bar{k}_+ & = \kappa\rho e^{\beta(1-\gamma)\frac{\theta}{\theta+1}} \\
    \label{eq:15}
    \bar{k}_- & = \kappa^{-1}\rho e^{\beta(1+\gamma)\frac{\theta}{\theta+1}} \\
    \label{eq:16}
    K & = \kappa^{2} e^{2\beta\gamma\frac{\theta}{\theta+1}}\\
    \label{eq:17}
    \eta & = 2 \beta\gamma \\
    \varepsilon_{+} & = \beta(1-\gamma) \label{eq:18} \\
    \varepsilon_{-} & = \beta(1+\gamma) \label{eq:19}
  \end{align}
\end{subequations}

The diffusion is modeled as a first-order exchange reaction between two
compartments as a function of the geometric mean of the exchange kinetic constant $\delta$:
\begin{subequations}
  \label{eq:20}
  \begin{align}
    \label{eq:21}
    \delta_{1} & = \lambda \delta \\
    \label{eq:22}
    \delta_{2} & = \lambda^{-1} \delta
  \end{align}
\end{subequations}

The system can then be reduced to a set of ordinary differential equations with a set of
  nondimensionalized parameters:
  \begin{subequations}
  \label{eq:23}
  \begin{align}
    \label{eq:24}
    \frac{du_{1a}}{d\bar{t}} & = \lambda^{-1} \delta \left(- u_{1a} + u_{1b}\right) - \bar{k}_{+,a} u_{1a} +
                               \bar{k}_{-,a}  u_{2a} & \\
                             &=  -\varphi_{d1} - \varphi_{ra} \label{eq:25}     \\
    \label{eq:26}
    \frac{du_{1b}}{d\bar{t}} & = \lambda^{-1}\delta \left(u_{1a} - u_{1b}\right) - \bar{k}_{+,b} u_{1b} +
                               \bar{k}_{-,b} u_{2b}     & \\
                             &=  +\varphi_{d1}  -\varphi_{rb}     \label{eq:27} \\
    \label{eq:28}
    \frac{du_{2a}}{d\bar{t}} & =  \lambda \delta \left(- u_{2a} + u_{2b}\right) + \bar{k}_{+,a} u_{1a} -
                               \bar{k}_{-,a} u_{2a}      & \\
                             &=  -\varphi_{d2} + \varphi_{ra}  \label{eq:29}    \\
    \label{eq:30}
    \frac{du_{2b}}{d\bar{t}} & =  \lambda \delta \left(u_{2a} - u_{2b}\right) + \bar{k}_{+,b} u_{1b} -
                               \bar{k}_{-,b} u_{2b}        & \\
                             &=  +\varphi_{d2} + \varphi_{rb} \label{eq:31}
  \end{align}
\end{subequations}
 The system is thus described in the nondimensionalized concentrations $u_{i,j}$ of compound $i$ in
compartment $j$, relative temperature deviation $\theta$, and time $\bar{t}$.  Each compartment is
thermostated at temperature $\theta_{a} = -\theta_{g}$ and $\theta_{b} = +\theta_{g}$, with
$\theta_{g} \in [0,1]$; the limit cases correspond, respectively, to a uniform temperature $T_{0}$ and a
temperature gradient from $0$~K to $2T_{0}$; the median temperature is $T_{0}$ in all
situations. The initial concentration is $1$ in each compartment, i.e.:
\begin{align}
  \label{eq:32}
  u_{1a}(0)+u_{2a}(0) & =u_{1b}(0)+u_{2b}(0)=1
\end{align}

Analytical solutions for the resulting steady state were obtained.  A second-order
Taylor development as a function of the temperature gradient $\theta_{g}$ leads to simpler but more
informative solutions (see Appendix \ref{sec:math-deriv} for details).

The resulting system behavior in the steady state was then characterized in terms of three
dimensionless parameters: chemical force $\alpha$, chemical flux $\varphi$, and entropic production
$\sigma$, which can be expressed as:
\begin{align}
  \label{eq:33}
 \sigma & = \alpha\varphi 
\end{align}

\subsubsection{Parameter description}
\label{sec:param-interpr}

The chemical reaction characteristics can be described using the following parameters:
\begin{itemize}
\item $\rho$ is the geometric mean of the nondimensionalized kinetic constants at $\theta=0$ (i.e.
  $T_{0}$), which reports the global reaction rate,
\item $\kappa$ is the square root of the thermodynamic constant at $T_{0}$, which reports the reaction
  asymmetry between the forward and backward reactions, with $\kappa=1$ for $u_{1}=u_{2}$ at equilibrium
    at $T_{0}$,
\item $\beta$ is the arithmetic mean of activation energies, which reports the average activation
  energy,
\item $\gamma$, which reports  the asymmetry of the activation energy profile (see Fig.~S3 in SI for a
  graphical representation).
\end{itemize}
Similarly, the characteristics of the chemical exchanges can be described in terms of:
\begin{itemize}
\item $\delta$ is the geometric mean of the exchange rate constants, which reports  the global exchange
  rate between the two compartments,
\item $\lambda$ is the squared root of the exchange rate constant, which reports the asymmetry between the
  exchange rate of the two compounds (see Eqs.~\eqref{eq:20}).
\item $\kappa$ and $\lambda$ can be combined into two secondary parameters $\kappa'$ and $\lambda'$, leading to simpler
mathematical representations. They represent the reaction and exchange asymmetry, respectively, with
a maximal value of $1$ for a perfectly symmetric process and $0$ for a process totally
displaced in a given direction.
\end{itemize}

\subsubsection{Establishment of a nonequilibrium steady state}
\label{sec:orig-noneq-steady}

The difference in temperature implies a difference in the equilibrium constant in each
compartment. Thus, a temperature gradient necessarily leads to a frustrated state, which is
characterized by the impossibility of simultaneously reaching chemical equilibrium in each
compartment (which would imply that each compound coexists at different concentrations in both
compartments) and the transport equilibrium between each compartment (which would imply identical
concentrations for each compound in both compartments).

This results in a nonequilibrium state, whose distance from equilibrium can be quantified by the
chemical force $\alpha$, as the sum of the chemical forces of each chemical reaction in each compartment
and the chemical forces of each chemical exchange between the two compartments. It can then be
expressed for low values of $\theta_{g}$ (see Eqs.~\eqref{eq:124}-\eqref{eq:130}) as:
\begin{align}
  \alpha & = 2\eta\theta_{g} \label{eq:34}
\end{align}
In all cases, a steady chemical force is thus sustained by the temperature gradient. It is
proportional to the intensity of the gradient $\theta_{g}$ and the enthalpy of the reaction $\eta$ and does not depend on
the other parameters.

\begin{figure*}
    \includegraphics[width=\linewidth]{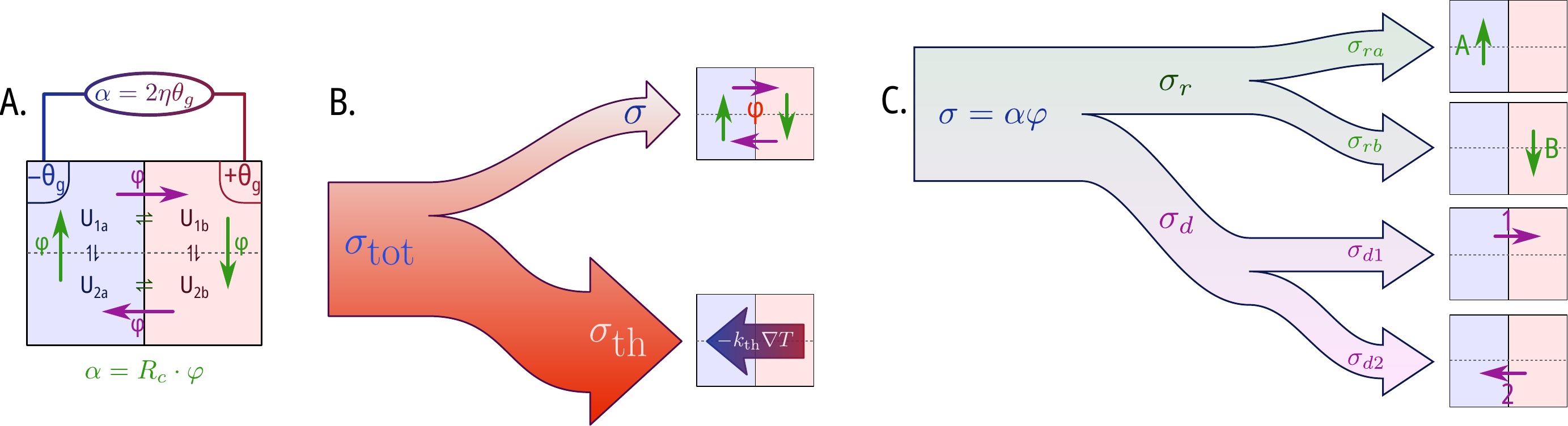}
    \caption{Establishment of a nonequilibrium steady reaction-diffusion cycle driven by a
      temperature gradient. Cyclic flux $\varphi$ in response to nonzero chemical force $\alpha$ (\textbf{A});
      full entropy production $\sigma_{\mathrm{tot}}$ as a sum of thermal
        ($\sigma_{\mathrm{th}}$, see appendix~\ref{sec:entr-prod-due}) and chemical ($\sigma$) contributions
        (\textbf{B}); distribution of the entropy production in each chemical reaction or exchange
      process (\textbf{C}).}
  \label{fig:Fig2}
\end{figure*}

In response to this non-zero chemical force, a circular reaction-diffusion flux is established.
Chemical reaction fluxes are processed in opposite directions in each compartment, compensated by the
continuous exchange of each compound in the opposite direction (see Fig.~\ref{fig:Fig2}A). By
analogy to electric circuits, $\alpha$ can be interpreted as the potential difference and
$\varphi$ as the intensity. These are expressed as (see Eqs.~\eqref{eq:154} and \eqref{eq:169}):
\begin{align}
  \label{eq:35}
  \varphi & = \frac{\alpha}{R_{\text{tot}}}
\end{align}

In the case of the linear regime, characterized by low values of $\theta_{g}$, this chemical resistance
can be expressed as:
\begin{align}
  \label{eq:36}
    R_{\text{tot}} &  = \frac{4}{\kappa'} 
                          \left(
                          \frac{1}{\rho} + \frac{1}{\lambda'\delta}
                          \right)
\end{align}
It is decomposed as the sum of two terms: the first one is directly linked to the chemical reaction
(term in $\sfrac{1}{\rho}$), and the second one to the chemical exchanges (term in
$\sfrac{1}{\lambda'\delta}$). Outside the linear regime, for large values of $\theta_{g}$, this chemical resistance
will not be constant and may depend on the chemical force.

\subsubsection{Process Distribution}

This steady nonequilibrium reaction-diffusion flux leads to a continuous dissipative process
characterized by $\sigma$, as an analog of electric power dissipation \cite{plasson2009} according to
Eq.~\eqref{eq:33}. The corresponding chemical entropy production taps on the important thermal
  entropy production by heat diffusion generated by the temperature gradient (Fig.~\ref{fig:Fig2}B
  and appendix~\ref{sec:entr-prod-due}). The energy extracted from the temperature gradient is
dissipated differently in each process (see Fig.~\ref{fig:Fig2}C), and can be decomposed as a sum of
terms specific to each process, according to each chemical resistance as (see
Eqs.~\eqref{eq:100}-\eqref{eq:104}):
\begin{align}
  \label{eq:37}
  \sigma & = \sum_{i} \sigma_{i} = \sum_{i}R_{i} \varphi^{2}\\
  \label{eq:38}
   R_{\text{tot}} & = \sum_{i} R_{i}
\end{align}

\paragraph{Reactions and exchanges}

The full entropy production $\sigma$ is first dispatched between the reaction term $\sigma_{r}$ and the
exchange term $\sigma_{d}$:
\begin{align}
  \label{eq:39}
  \sigma_{r}  = \frac{\lambda'\delta}{\lambda'\delta+\rho} \sigma \qquad &; \quad  \sigma_{d}   = \frac{\rho}{\lambda'\delta+\rho} \sigma\\
  \shortintertext{with}
  \label{eq:40}
  R_{r}  = \frac{4}{\kappa'\rho} \qquad &; \quad   R_{d}  = \frac{4}{\kappa'\lambda'\delta} \\
  \shortintertext{and}
  \label{eq:41}
    \sigma    = & \frac{\lambda'\delta\rho}{\lambda'\delta+\rho}\kappa'\cdot \eta^{2}\theta_{g}^{2}
\end{align}

Two limit regimes are observed:
\begin{itemize}
\item If $\lambda'\delta \gg \rho$, then $\sigma \approx \sigma_{r}$; this is the case for a fast exchange system. Both compartments are in equilibrium with each other, and most of the dissipative process is directed to the chemical reactions within each compartment.
\item If $\lambda'\delta \ll \rho$, then $\sigma \approx \sigma_{d}$; this is the case for a fast-reaction system. Chemical equilibrium is observed in each separate compartment, and most of the dissipative process is directed to the chemical exchanges between compartments.
\end{itemize}
In the intermediate regime $\rho \sim \lambda'\delta $, both the chemical reaction and the transport are
actively processed.

\paragraph{Reaction balance between the compartments}

The reaction entropy production is decomposed in each compartment as:
\begin{align} 
  \label{eq:42}
  \sigma_{ra} & = \frac{1+\varepsilon\theta_{g}}{2} \sigma_{r} \\
  \label{eq:43}
  \sigma_{rb} & = \frac{1-\varepsilon\theta_{g}}{2} \sigma_{r} 
\end{align}
The parameter $\varepsilon$ (see Eqs.~\eqref{eq:141}) comes down to an apparent activation
energy for the full system, varying from $\varepsilon_{+}$ to $\varepsilon_{-}$ through the average
$\beta$ value, depending on the global symmetry of the system: it tends to $\beta$ when
$\kappa \lambda \sim 1$, to $\beta(1-\gamma)=\varepsilon_{+}$ when $\kappa\lambda \gg 1$, and to $\beta(1+\gamma)=\varepsilon_{-}$ when $\kappa\lambda \ll 1$.

Qualitative changes are observed as functions of $\varepsilon\theta_{g}$. When
$\varepsilon\theta_{g} \ll 1$, the dissipation due to the chemical reaction is equally shared between both
compartments as $\sigma_{ra}\approx\sigma_{rb}$. This dissipation is asymmetric for higher
values of $\varepsilon\theta_{g}$, increasing the dissipation in the colder compartment and reducing it by the same
value in the warmer compartment.

This asymmetry originates from the slowing of the chemical reaction in the cold compartment.  The
chemical flux $\varphi$ is necessarily of the same intensity in each compartment to guarantee a
steady state; this implies an increase in the chemical force $\alpha$ in the cold compartment due to a
less efficient chemical relaxation toward equilibrium and its decrease in the warm compartment
owing to a more efficient relaxation.  Consequently, the more dissipative compartment is necessarily
the colder.

\paragraph{Exchange balance between compartments}

The exchange entropy production can be decomposed by the exchange of each compound as:
\begin{align}
  \label{eq:44}
  \sigma_{d1} & = \frac{\lambda^{2}\kappa^{2}}{1+\lambda^{2}\kappa^{2}} \sigma_{d} \\
  \label{eq:45}
  \sigma_{d2} & = \frac{1}{1+\lambda^{2}\kappa^{2}} \sigma_{d}           
\end{align}

It is symmetrically distributed between the two compounds when $\lambda\kappa=1$. The asymmetrization of this
exchange is quantified by $\lambda\kappa$, with a $\sigma_{d} \approx \sigma_{d1}$ for
$\lambda\kappa \gg 1$ (the exchange entropy production is then governed by $U_{1}$), and with a
$\sigma_{d} \approx \sigma_{d2}$ for $\lambda\kappa \ll 1$ (the exchange entropy production is then governed by $U_{2}$).

\subsubsection{Optimal conditions}
\label{sec:syst-optim-cond}

Considering that the temperature gradient is fixed and cannot be optimized, several criteria can
  be used to evaluate the efficiency of the system to generate a steady nonequilibrium
  reaction-diffusion process. The chemical force $\alpha$ reflects how far the system is pushed away from
  the equilibrium state ($\alpha=0$ corresponding to the equilibrium state). The chemical flux
  $\varphi$ quantifies the system response as the intensity of the resulting reaction-diffusion flux from
  the chemical force, as the quantity of matter that is converted at each moment in each
  compartment, and the quantity of matter that is transferred at each moment from one compartment to
  another ($\varphi=0$ corresponding to a static, kinetically locked system). The rate of entropy
  production $\sigma$ combines both thermodynamic and kinetic aspects, and quantifies the nonequilibrium
  activity of the system.

At first, according to Eq.~\eqref{eq:34}, the chemical force $\alpha$ is simply proportional to both
$\theta_{g}$ and $\eta$; this entails that the temperature gradient shall be applied to a high-enthalpy chemical reaction as a first requirement from driving the system in a non-equilibrium state.

Secondly, from Eqs.~\eqref{eq:35}-\eqref{eq:36}, maximizing the steady nonequilibrium flux
$\varphi=\sfrac{\alpha}{R_{\text{tot}}}$ implies the minimization of its chemical resistance
$R_{\text{tot}}$. This implies:
\begin{itemize}
\item to maximize parameter $\kappa'$; its largest possible value is $\kappa'=1$, corresponding to a chemical
  reaction perfectly balanced between the backward and forward reactions at $T_{0}$ (i.e.\ $K_{0}=1$),
\item to maximize parameter $\lambda'$ to $1$, corresponding to symmetric matter exchange (see
  Eqs.~\eqref{eq:44}-\eqref{eq:45}); this implies that an imbalance in the chemical equilibrium
  at $T_0$ (characterized by $\kappa$) can be compensated for by an imbalance in the chemical
  exchange of each compound (characterized by $\lambda$),
\item to maximize both the global chemical rate $\rho$ and the chemical exchange rate $\delta$, thus
  minimizing the chemical resistance of each process.
\end{itemize}

Overall, this implies that each process should be fast and symmetrical at mean temperature
$T_{0}$. The corresponding simultaneous optimization of $\alpha$ and $\rho$ implies the maximization of the
entropy production $\sigma$ (Eq.~\eqref{eq:41}), which represents the part of entropy production generated by the targeted chemical reaction-diffusion process. In practice, this represents only a fraction of the thermal entropy production (see Appendix~\ref{sec:entr-prod-due}).

Moreover, it is possible to focus on a specific process to be optimized, i.e.\ either the
  reaction or the transport process, via the $\sigma_{r}$ or the $\sigma_{d}$ values.  The optimization of
the reaction process via $\sigma_{r}$ (see Eqs.~\eqref{eq:183} and Fig.~S1 in the SI) implies that:
\begin{itemize}
\item for a given value of $\lambda'\delta$, a maximal $\sigma_{r}$ value is obtained for $\rho=\lambda'\delta$,
\item for a given value of $\rho$, high values of $\lambda'\delta$ are required; however, increasing
  $\lambda'\delta$ to much higher values than $\rho$ is useless, as $\sigma_{r}$ reaches a plateau.
\end{itemize}
A similar reasoning can be used to optimize the transport process via $\sigma_{d}$ (see
Eqs.~\eqref{eq:186}):
\begin{itemize}
\item for a given value of $\rho$, a maximal $\sigma_{d}$ value is obtained for $\lambda'\delta=\rho$,
\item for a given value of $\lambda'\delta$, high values of $\rho$ are required; however, increasing
  $\rho$ to much higher values than $\lambda'\delta$ is useless, as $\sigma_{d}$ reaches a plateau.
\end{itemize}

Consequently, general optimization involves maximizing $\theta_{g}$, $\eta$, $\lambda'$, and
$\kappa'$. Furthermore, $\lambda'\delta$ should be of the same order of magnitude as $\rho$, with
$\lambda'\delta > \rho$ to sustain chemical reactions, $\lambda'\delta < \rho$ for sustaining chemical transport, and
$\lambda'\delta \simeq \rho$ for sustaining reaction-diffusion cycles.

\subsection{Spatial localization of chemical processes}
\label{sec:unstirred-system}

\subsubsection{Model description}
\label{sec:model-description}

The spatial extension of the system was considered by introducing the chemical diffusion of all
compounds.  We limited our study to closed one-dimensional systems. A stationary temperature
gradient $\theta(\bar{x})$ was imposed over the entire system.  This corresponds to an unstirred system,
with free internal diffusion of chemical compounds, without matter exchange with the surroundings.
The system was described using a set of partial differential equations (see Eqs. \eqref{eq:189}), which
was numerically solved for a wide range of parameters, with a focus on determining its steady state
(see Appendix \ref{sec:steady state-determ}).  The characteristic parameter $l_{0}$ must be
introduced to account for the spatial extension, describing a one dimensional system:
\begin{align}
  \label{eq:46}
  \bar{x}=\frac{x}{l_{0}} \in [0,1]
\end{align}

Chemical diffusion is described based on the diffusion parameter $d_{i}$ for each compound
  $U_i$, which is nondimensionalized by the characteristic parameter $D_0$ corresponding to the
average diffusion constant of the chemical reactants. Consistently with the description of the chemical exchange
kinetics of the previous model, chemical diffusion is described as a function of the geometric mean
of the diffusion constant $d$ and the exchange ratio $\lambda$:
\begin{subequations}
  \label{eq:47}
  \begin{align}
    \label{eq:48}
    d_{1} &= \lambda d \\
    \label{eq:49}
    d_{2} & = \lambda^{-1}d
  \end{align}
\end{subequations}

In each point of the system, the reaction  processes is still characterized as (see Eq.~\eqref{eq:194}):
\begin{align}
  \label{eq:50}
  \sigma_{r} & = \varphi_{r} \cdot \alpha_{r}  
\end{align}
where the chemical flux $\varphi_{r}$ and force $\alpha_{r}$ are defined as in the two compartment system
(Eqs.~\eqref{eq:88}-\eqref{eq:89}), on each point of the system.

The transport process of compound $i$ is now expressed
as  \cite{groot_mazur_1969,mahara_yamaguchi_2010} (see Eq.~\eqref{eq:194}):
\begin{subequations}
  \label{eq:51}
  \begin{align}
    \label{eq:52}
    \sigma_{d} & = j_{i} \cdot \alpha_{d}\\
    \label{eq:53}
    \textrm{with} \qquad j_{i} & = d_{i}\left( \frac{\partial u_{i}}{\partial \bar{x}}
                            \right) \\
    \label{eq:54}
    \textrm{and} \qquad \alpha_{d,i} & = \frac{1}{u_{i}} \left( \frac{\partial u_{i}}{\partial \bar{x}} \right)
  \end{align}
\end{subequations}
$j_{i}$ is the transport flux of compound $i$, and $\alpha_{d,i}$ the corresponding transport force.

The characteristic time $t_{0}$ is additionally fixed by $l_{0}$ and $D_{0}$ (see
Table.~\ref{tab:table1}). All parameters corresponding to chemical reactions are defined as in the
stirred compartment model; however, they are now functions of position $\bar{x}$.

Taking into account both the reaction and diffusion parameters, each chemical system can be additionally
characterized by its nondimensionalized reaction-diffusion length $w$ defined as:
\begin{align}
  \label{eq:55}
  w & =\sqrt{\frac{\lambda' d}{2\rho}}
\end{align}
$w$ represents the scale at which both reaction and diffusion processes operate at similar rates.

For simplicity, and to clearly identify and quantify the specific role of a temperature gradient on
the generation of reaction diffusion fluxes, we made the following assumptions:
\begin{itemize}
\item We neglected the influence of temperature on the diffusion coefficient. Several simulations were performed, indicating that considering the variability of $d_{i}$ with $\theta$ did not qualitatively
  change the system behavior; this essentially led to an increase in the asymmetry between the cold and warm regions (see Fig.~S2 in SI).
\item The Soret effect was neglected, which is expected to be satisfactory for systems based on
  small molecular reactants \cite{wurger_2010}. However, emergent thermophoretic behaviors that
    originate from the coupling of chemical reactions and temperature gradients
     \cite{groot_mazur_1969,liang2022} may occur within the framework of this model, and were
    specifically studied.
\item Convection was not considered in this study. This assumption is anticipated to be satisfactory for sufficiently small systems in the direction of gravity or sufficiently highly viscous systems, so that the major transport mode is processed by molecular diffusion. This condition is
  compatible with a large system in which the temperature gradient is applied on the horizontal $x$
  axis, as long as the vertical spatial extension is sufficiently small (e.g., for horizontal
  capillary systems).
\item The endothermic or exothermic effects of the chemical reactions were neglected. This case is
  relevant to chemical systems connected to sufficiently efficient heat sources and sinks to
  reliably fix temperature profiles.
\end{itemize}

\begin{figure}
  \includegraphics[width=\linewidth]{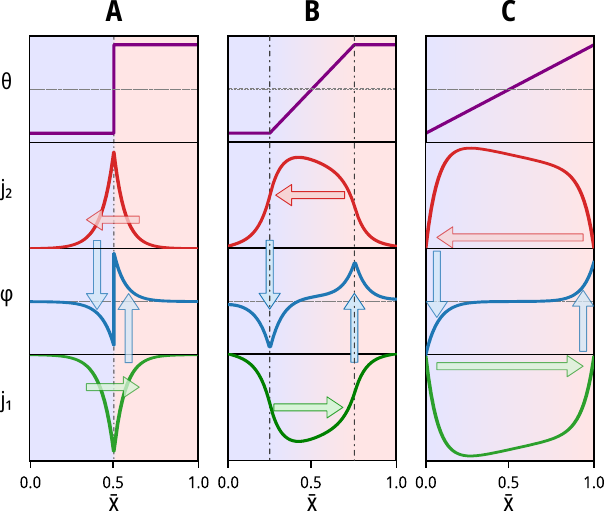}
  \caption{%
      Schematic representation of the steady reaction-diffusion cycle in unstirred systems driven by
      Heaviside (\textbf{A}), mixed (\textbf{B}) and linear (\textbf{C}) temperature profiles.  The
      horizontal wide arrows represent the direction of \ch{U1} and \ch{U2} diffusion and the vertical
      wide arrows represent the global reaction fluxes \ch{U1 -> U2} (up arrows) and \ch{U2 -> U1}
      (down arrows). The diffusion and reaction processes are processed in the same area for the Heaviside profile, close to the cold/warm interface (\textbf{A}). In contrast, they are physically separated with
     the chemical reaction occurring at the linear gradient boundaries, and the chemical diffusion taking place between these two
      points (\textbf{B,C}). Model parameters: $\kappa=1$, $\lambda=1$, $\gamma=0.2$, $d=1$,
      $\beta=1$, $\theta_{g}=10^{-1}$, and $\rho=10^{2}$.%
    .}
  \label{fig:Fig3}
\end{figure}

\begin{figure*}
  \includegraphics[width=\linewidth]{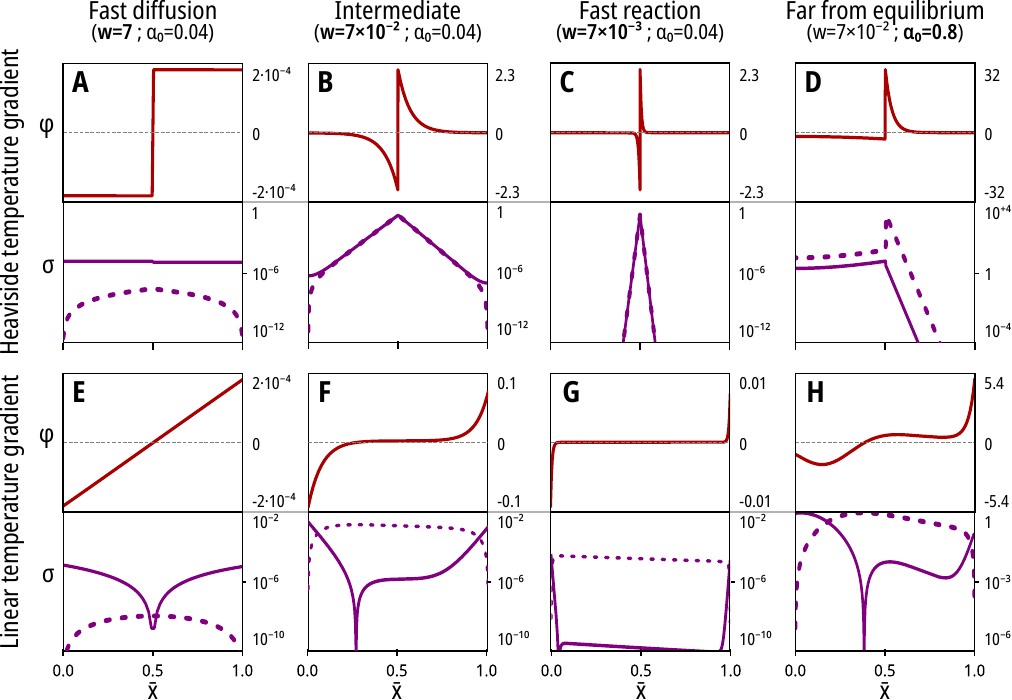}
  \caption{%
      Steady state spatial profiles of chemical flux of reaction $\varphi$ (red) and entropy production
      (solid purple, $\sigma_{r}$ and dotted purple, $\sigma_{d}$) as a function of the reaction-diffusion
      length $w$ as defined in Eq.~\eqref{eq:55}, and the characteristic chemical force $\alpha_{0}$ as defined in 
      Eq.~\eqref{eq:59}. Model parameters: $\kappa=1$, $\lambda=1$, and $\gamma=0.2$; $d=1$,
      $\rho=10^{-2}$ ($w=7$, fast diffusion: \textbf{A, E}), $d=1$, $\rho=10^{2}$ ($w=0.07$, intermediate:
      \textbf{B, D, F, H}) and $d=10^{-2}$, $\rho=10^{2}$ ($w=0.007$, fast reaction: \textbf{C, G});
      $\beta=1$, $\theta_{g}=10^{-1}$ ($\alpha_{0}=0.04$, linear regime: \textbf{A-C, E-G}) and
      $\beta=4$, $\theta_{g}=0.5$ ($\alpha_{0}=0.8$, far from equilibrium: \textbf{D, H}).%
    }
  \label{fig:Fig4}
\end{figure*}

\subsubsection{Steady state flux profiles}
\label{sec:steady state-flux}

\paragraph{System with two  homogeneously thermostated regions}
\label{sec:case-an-unstirred}

The first temperature gradient is defined as a Heaviside profile, defined by $\theta=-\theta_{g}$ for
$\bar{x} \in [0,0.5]$ and $\theta=+\theta_{g}$ for $\bar{x} \in ]0.5,1]$. This corresponds to a system in which
two contiguous spatial regions are maintained by an external thermostat at two different homogeneous
temperatures $-\theta_{g}$ and $+\theta_{g}$, while free chemical diffusion can occur throughout the chemical
system.

In the spatial region that is sufficiently close to the $\bar{x}=0.5$ cold / warm boundary---that is, at
a scale smaller than the reaction-diffusion length $w$---chemical diffusion is much faster than
chemical reactions. The concentrations are consequently essentially constant on each side of the
temperature boundary; the system is then equivalent to the ``two compartments'' model in the
vicinity of the boundary.  The values of the flux $\varphi$ and force $\alpha$ determined for the
two-compartment model correspond to the values observed at the boundaries $x=0.5^{+}$ and
$x=0.5^{-}$.

When $w>0.5$, the reaction-diffusion length is larger than the sizes of the cold and warm regions
$\bar{x} \in [0, 0.5]$ and $\bar{x} \in [0.5, 1]$. Both are homogeneous; the full system is then
equivalent to the stirred-compartment model. This system is characterized by fast diffusion
throughout the system, whereas the chemical reaction flux and dissipation are evenly distributed
throughout the entire system (see Fig.~\ref{fig:Fig4}A).

When $w<0.5$, an exponential decrease is observed in the values of $\varphi$ and $\alpha$, from their boundary values
at $\bar{x}=0.5$ to $x=0$ or $x=1$, with a half-length equal to $w$. If $w \ll 1$, the activity of the system is restricted to a short region close to the boundary, most of the system being in
equilibrium and characterized by $\varphi\approx0$, $\alpha\approx0$, $\sigma\approx0$ (see Fig.~\ref{fig:Fig4}C).

In both cases, a reaction-diffusion cycle is established, with chemical reaction and diffusion
occurring simultaneously in the same spatial region, at the system boundary (see
Fig.~\ref{fig:Fig3}A, more details are available in Fig.~S4).

The spatial extension of chemical activity in each region becomes asymmetric when the intensity of the temperature gradient is sufficiently high.  This can be described by considering the difference in the
kinetic rate in each region based on Eq.~\eqref{eq:14}-\eqref{eq:15}, as:
\begin{align}
  \label{eq:56}
  \rho_{\textrm{cold}} & = \rho e^{-\beta\frac{\theta_{g}}{1-\theta_{g}}} & \text{for}~\bar{x}\in[0,0.5]\\
  \label{eq:57}
  \rho_{\textrm{warm}} & = \rho e^{\beta\frac{\theta_{g}}{1+\theta_{g}}} & \text{for}~\bar{x}\in[0.5,1]  
\end{align}

Thus, the half-length of each exponential relaxation is different in each region; it is equal to the
corrected value $w_{i} = \sqrt{\sfrac{\lambda' d}{2\rho_{i}}}$. Consequently, chemical fluxes spread to a
greater extent in the cold zone than in the warm zone.

When $\beta\theta_{g} > 1$, the system becomes highly nonlinear, with a large behavioral asymmetry between
each region. Typically, situations with $w_{\text{cold}}>1$ and $w_{\text{warm}} \ll 1$ can be
obtained, with a homogeneous system in the cold zone and restricted to a short region in the warm
zone (see Fig.~\ref{fig:Fig4}D).

\paragraph{System connected to two thermostats at its extremities}
\label{sec:case-an-unstirred-1}

An imposed linear temperature profile corresponds to a chemical system in contact with two external
thermostats at its extremities, fixing the temperatures $-\theta_{g}$ and $+\theta_{g}$ at each boundary
$\bar{x}=0$ and $\bar{x}=1$. Similarly as in the previous model, two limit regimes are observed, but with
distinctive differences:
\begin{itemize}
\item Fast-diffusion systems (i.e.\ $w \gg 1$, see Fig.~\ref{fig:Fig4}E) are characterized by
  a linear chemical flux profile. The system is more active at each extremity of the system, and
    cancels close to $\bar{x}=\sfrac{1}{2}$, and the entropy production is dominated by the chemical
    reaction.
\item In fast reaction systems (i.e. $w \ll 1$, see Fig.~\ref{fig:Fig4}G), the central inactive
  region becomes much larger. The chemical reaction flux and entropy production by the chemical
    reaction is restricted at each extremity of the system, in a region of length $w_{\text{cold}}$
  at $\bar{x} \approx 0$ and $w_{\text{warm}}$ at $\bar{x} \approx 1$. In turn, the entropy production is
    dominated by the diffusion, and the chemical transport spreads, linking these two extremities
  see Fig.~\ref{fig:Fig3}C.
\end{itemize}

This general behavior implies a spatial separation \emph{de facto} of the different chemical
processes.  Three regions can be identified: cold and warm regions at each extremity, of
half-lengths equal to $w_{\text{cold}}$ and $w_{\text{warm}}$, separated by a central region of
length approximately equal to $1-2w$. The cold and warm regions are dominated by chemical reactions
($\sigma \approx \sigma_{r}$), and the central region is dominated by chemical transport
($\sigma \approx \sigma_{d}$). Thus, a spatially extended reaction-diffusion cycle is set in motion; the compounds
are transported across the entire system between each chemically active region, the central
transport region shrinking with $w$ (see Fig.~\ref{fig:Fig4}E-G, and Fig.~S5 for more
  details). This is in sharp contrast with the Heaviside gradient, where both chemical reactions
and transport are effective at the same spatial locations, close to the temperature boundary (see
Fig.~\ref{fig:Fig4}A-C).

Finally, the high asymmetrization of the profile induced by high $\beta\theta_{g}$ values leads to nontrivial
and complex flux profiles (see Fig.~\ref{fig:Fig4}H). The central region around
$\bar{x}=\sfrac{1}{2}$ is characterized by a non-negligible positive flux, corresponding to a
low-activity extension of the warm region. Consequently, the spatial position of the zero flux is
localized at $\bar{x} < \sfrac{1}{2}$; the cold region of the negative flux shrinks and the warm
region of the positive flux stretches.

\paragraph{System connected to two spatially extended thermostats}
\label{sec:mixed-profile}

A last temperature gradient was defined as a mix between a linear and a Heaviside gradient. This
is defined as $\theta=-\theta_{g}$ for $\bar{x} < \bar{x}_{1}=\sfrac{1}{4}$, $\theta=+\theta_{g}$ for
$\bar{x} > \bar{x}_{2}=\sfrac{3}{4}$, and a linear variation of $\theta$ from $-\theta_{g}$ to
$+\theta_{g}$ for $\bar{x} \in [\sfrac{1}{4}, \sfrac{3}{4}]$. This comes down to a system in contact with
two spatially extended external thermostats.

The behaviors of both the linear and Heaviside profiles can be recognized (see
Fig.~\ref{fig:Fig3}B, and Fig.~S6 for more details). The reaction-diffusion cycles are now in
motion, with two chemical reaction zones located at $\bar{x}_{1}$ and $\bar{x}_{2}$, connected by an
active chemical transport region across $\bar{x} \in [\bar{x}_{1}, \bar{x}_{2}]$.

It appears that controlling the precise shape of an imposed temperature gradient enables precise
control of the shape of the induced reaction-diffusion cycle. Steady dissipative chemical reactions
can be localized in regions characterized by temperature inflection points (i.e., high values of
$\sfrac{\partial^{2} T }{\partial x^{2}}$), whereas chemical transport can be sustained between these active regions
along temperature gradients (i.e., following high values of $\sfrac{\partial T }{\partial x}$).

\subsubsection{Steady state characteristics}
\label{sec:behavior-scaling}

The steady state can be described by a set of parameters $\alpha$, $\varphi$, $\sigma_{r}$,
$\sigma_{d}$ and $\sigma_{t}=\sigma_{r}+\sigma_{d}$. Their maximal value assesses the optimal activity in a specific
location, and their integral value assesses the optimal activity of the system as a whole (see the appendix
\ref{sec:steady state-char}).

The values of the steady chemical flux $\varphi$, chemical force $\alpha$ and entropy production
$\sigma$ can be expressed as functions of the following characteristic values:
\begin{subequations}
  \label{eq:58}
  \begin{align}
    \label{eq:59}
    \alpha_{0} & = 2\beta\gamma\theta_{g} \\
    \label{eq:60}
    \varphi_{0} &= \kappa'\beta\gamma\theta_{g}\rho \\
    \label{eq:61}
    \sigma_{0} &= 2 \kappa'\beta^{2}\gamma^{2}\theta_{g}^{2}\rho 
  \end{align}
\end{subequations}

\begin{table}
  \begin{tabular}{lcccccc} 
    \toprule
                                   & \multicolumn{2}{c}{Heaviside} & \multicolumn{2}{c}{Linear} & \multicolumn{2}{c}{Mixed} \\
                                   \cmidrule{2-7}
                                   & $w \ll 1$        & $w \gg 1$            & $w \ll 1$  & $w \gg 1$            & $w \ll 1$         & $w \gg 1$            \\
    \midrule
   $\sfrac{\varphi_{\text{max}}}{\varphi_{0}}$ & $1$            & $1$                & $2w$     & $1$                & $\sfrac{3w}{2}$ & $1$                  \\
   $\sfrac{\varphi_{\text{int}}}{\varphi_{0}}$ & $w$            & $\sfrac{1}{2}$     & $2w^2$   & $\sfrac{1}{4}$     & $4w^2$          & $\sfrac{3}{8}$     \\
    \midrule
   $\sfrac{\alpha_{\text{max}}}{\alpha_{0}}$ & $1$            & $1$                & $2w$     & $1$                & $\sfrac{4w}{3}$ & $1$                  \\
   $\sfrac{\alpha_{\text{int}}}{\alpha_{0}}$ & $w$            & $\sfrac{1}{2}$     & $2w^2$   & $\sfrac{1}{4}$     & $4w^2$          & $\sfrac{3}{8}$     \\
    \midrule
 $\sfrac{\sigma_{r,\text{max}}}{\sigma_{0}}$ & $1$            & $1$                & $4w^{2}$ & $1$                & $4w^{2}$        & $1$                  \\
 $\sfrac{\sigma_{r,\text{int}}}{\sigma_{0}}$ & $w$            & $1$                & $4w^3$   & $\sfrac{1}{3}$     & $8w^3$          & $\sfrac{2}{3}$     \\
    \midrule
 $\sfrac{\sigma_{d,\text{max}}}{\sigma_{0}}$ & $1$            & $\sfrac{1}{2w^2}$  & $4w^2$   & $\sfrac{1}{15w^2}$ & $20w^2$         & $\sfrac{2}{15w^2}$ \\
 $\sfrac{\sigma_{d,\text{int}}}{\sigma_{0}}$ & $w$            & $\sfrac{1}{12w^2}$ & $4w^2$   & $\sfrac{1}{30w^2}$ & $8w^2$          & $\sfrac{1}{15w^2}$ \\
    \midrule
 $\sfrac{\sigma_{t,\text{max}}}{\sigma_{0}}$ & $2$            & $1$                & $4w^2$   & $1$                & $20w^2$         & $1$                \\
 $\sfrac{\sigma_{t,\text{int}}}{\sigma_{0}}$ & $2w$           & $1$                & $4w^2$   & $\sfrac{1}{3}$     & $8w^2$          & $\sfrac{2}{3}$     \\
    \midrule
                           $y_{r}$ & $\sfrac{1}{2}$ & $1$                & $w$      & $1$                & $w$             & $1$                \\
                           $y_{d}$ & $\sfrac{1}{2}$ & $\sfrac{1}{12w^2}$ & $1$      & $\sfrac{1}{10w^2}$ & $1$             & $\sfrac{1}{10w^2}$ \\
    \bottomrule
  \end{tabular}
  \caption{First order values for $\varphi$, $\alpha$ and $\sigma$ for low values of $\beta\theta$, as identified from
    numerical models (see Fig.~S4-S6 in SI). The max index correspond to maximal values located at the
    cold/warm interface at $\bar{x}=\sfrac{1}{2}$ for the Heaviside profile, at the system
    extremities $\bar{x}=0$ and $\bar{x}=1$ for the linear profile, and the inflection point of the
    temperature profile at $\bar{x}=\sfrac{1}{4}$ and $\bar{x}=\sfrac{3}{4}$. The integral values
    correspond to the integration of the corresponding variable on the full system
    $\bar{x} \in [0,1]$ for the $\sigma$ variables, or on the region restricted to the positive values of
    $\varphi$ and $\alpha$ (i.e.\ $\bar{x} \in [\bar{x}_{0},1]$, where $\varphi >0$ or $\alpha > 0$ for $x > \bar{x}_{0}$).}
  \label{tab:table2}
\end{table}

These characteristic values are related with each other as:
\begin{align}
  \label{eq:62}
  \alpha_{0} & = \frac{2}{\kappa'\rho}   \varphi_{0} \\
  \label{eq:63}
  \sigma_{0} & = \alpha_{0}  \varphi_{0}
\end{align}
In the limit cases $w \ll 1$ and $w \gg 1$, and when the system is close to equilibrium (i.e.
$\beta\theta_{g} < 1$), the ratio of each variable to its reference value is only dependent on the
reaction-diffusion length $w$ as defined in Eq.~\eqref{eq:55}. The corresponding values are
summarized in Table~\ref{tab:table2} (see Fig.~S7 in SI for a graphical representation).

The situation $w \gg 1$ (that is, small systems) is essentially independent of scale. This implies
that, as long as the dimension of the system is smaller than the length of the reaction-diffusion, the
steady-state behavior is essentially described by the characteristic values of Eqs.~\eqref{eq:58},
modulated by a constant factor that depends on the shape of the temperature gradient profile.
Diffusion entropy production is the only varying parameter; it is proportional to $w^{-2}$, and thus is
essentially negligible for small systems. This situation is the optimal case for extracting energy
from the temperature gradient (by maximizing $\sigma_{t}$, as well as driving nonequilibrium chemical
reaction (by maximizing both $\sigma_{r}$ and $y_{r}$).

When $w \ll 1$ (that is, large systems), the system is effective only on a small fraction of the full
spatial extension, except for transport. This implies that all steady-state parameters are
proportional to $w^{n}$, with $n \in [1, 3]$, except for $\sigma_{d}$, which remains constant.  This is an
optimal situation for transporting chemical compounds over large distances.

The intermediate case $w \sim 1$ is characterized by the maximal value of $\sigma_{r}$, that is, by optimal
nonequilibrium transport. In this regime, the chemical reaction is also important, with a
production of entropy equally shared in both processes for $w=1$.  This situation is thus optimal
for observing nonequilibrium cycles of reaction-diffusion, as both diffusion and reaction can be
simultaneously efficient.

\begin{figure*}
  \includegraphics[width=\linewidth]{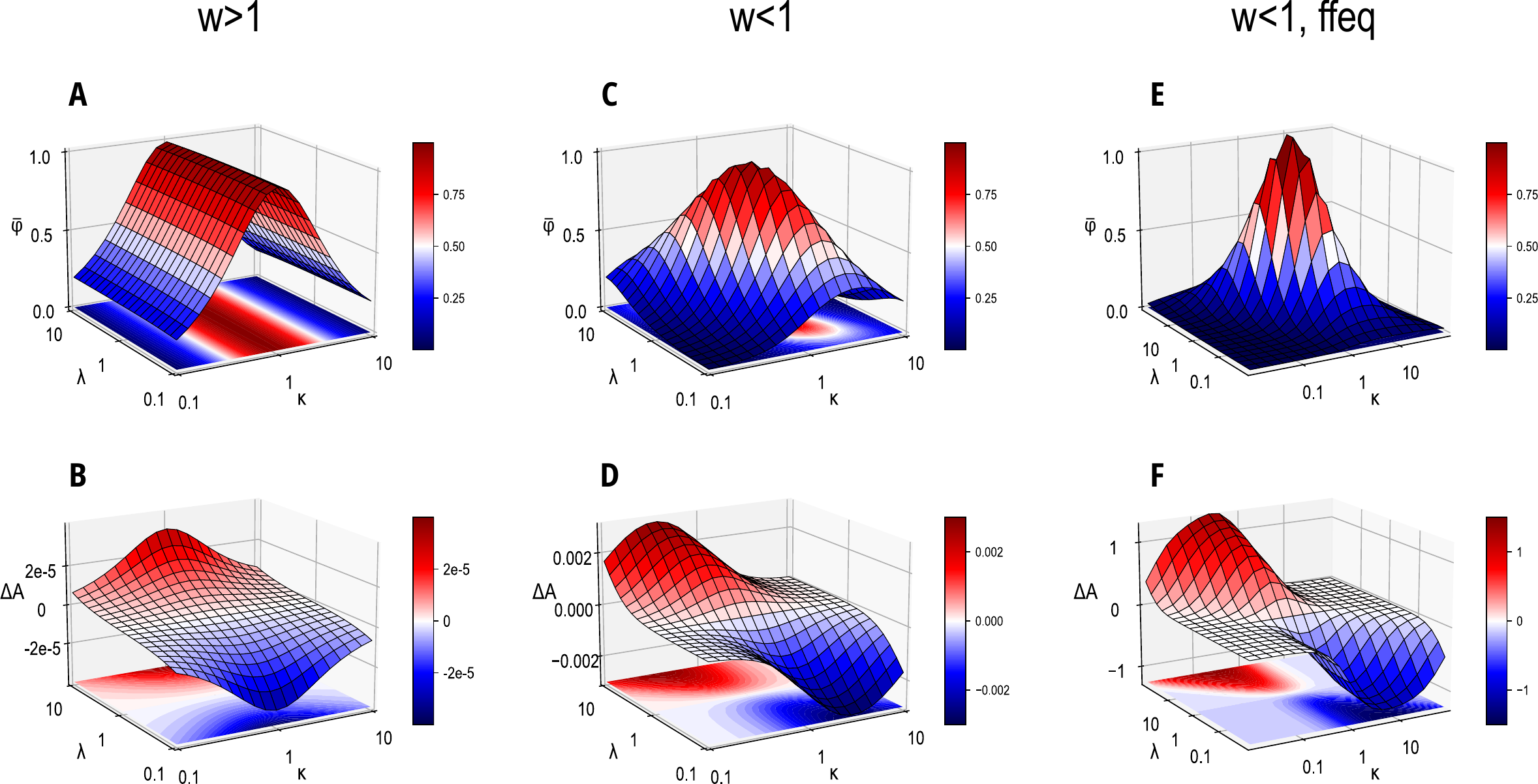} 
  \caption{Influence of the system asymmetry $\lambda$ and $\kappa$ on the steady normalized flux
    $\bar{\varphi}=\varphi_{\text{int}}/\varphi_{\text{int},0}$ (\textbf{A,C,E}) and the steady maximal difference of
    concentration across the system $\Delta A$(\textbf{B,D,F}), with $\varphi_{\text{int},0}$ the integral flux
    value observed in the same conditions, except for $\kappa=\lambda=1$. For all models, a linear temperature
    gradient was applied, with common model parameters $\gamma=0.2$, $d=1$, $\kappa\in[0.1; 10]$,
    $\lambda\in[0.1; 10]$. Additional parameters: $\theta_{g}=0.01$, $\beta=1$, $\rho=0.01$ (\textbf{A-B}),
    $\theta_{g}=0.01$, $\beta=1$, $\rho=100$ (\textbf{C-D}), and $\theta_{g}=0.04$, $\beta=4$, $\rho=100$ (\textbf{E-F}).}
  \label{fig:Fig5}
\end{figure*}

This behavior is also influenced by the symmetry of the system defined by the parameters $\lambda$ and $\kappa$.
Optimal nonequilibrium reaction fluxes are obtained for symmetrical systems characterized by
$\kappa=1$ when $w>1$ (Fig.~\ref{fig:Fig5}A), and $\kappa=1$ and $\lambda=1$ when $w<1$ (Fig.~\ref{fig:Fig5}C and
E). As observed by Liang \emph{et al} \cite{liang2022}, an asymmetry in the diffusion coefficient,
characterized by $\lambda \neq 1$, can lead to so-called ``emergent thermophoretic effect'', implying that
the coupling of the chemical reaction to the temperature gradient can lead to concentrating the
reactants in either the colder or the warmer region.

This effect is inefficient in small systems (Fig.~\ref{fig:Fig5}B). Thus, a sufficiently large
system is required (Fig.~\ref{fig:Fig5}D and F) such that the dissipative processes are dominated by
chemical transports. This process is especially efficient for values of $\lambda \neq 1$, and needs to be
driven away from the equilibrium by high intensity gradients (Fig.~\ref{fig:Fig5}F). Optimal
$\kappa$ values are also observed, with $\kappa > 1$ values associated with $\lambda < 1$ (and consequently
$\kappa < 1$ values associated with $\lambda > 1$ values).

\subsubsection{System scaling}
\label{sec:system-scaling}

\paragraph{Critical parameters}

From a general perspective, the efficiency of the system to be driven far from the equilibrium state
is directly linked to the maximization of $\beta$, $\gamma$, $\theta_{g}$ and $\rho$. This trivially implies that the
best chemical reaction candidates are characterized by high activation energies, high reaction
enthalpies, and high reaction rates and should be placed in high-temperature gradients.

More importantly, the results specifically emphasize the necessity of adapting the spatial scale of
the chemical setup to the characteristic scale of the reaction-diffusion system. The temperature
gradient should also be chosen carefully, so that the system is as symmetric as possible at the mean
temperature, in terms of chemical diffusion (that is, $D_{1}(T_{0}) \approx D_{2}(T_{0})$) and balance
between the reactants (i.e.\ $K(T_{0})\approx1$).

The reaction-diffusion length $w$ can be redimensionalized to the critical spatial dimension
$\mathcal{L}_{0}$ as:
\begin{align}
  \label{eq:64}
  \mathcal{L}_{0} & = \sqrt{\frac{D_{1}D_{2}}{k_{+,0}D_{2}+k_{-,0}D_{1}}}
\end{align}
If $D_{1}=D_{2}=D_{0}$ and $k_{+,0}=k_{-,0}=k_{0}$ (i.e.\ $\lambda=\kappa=1$), this expression comes down
to:
\begin{align}
  \label{eq:65}
  \mathcal{L}_{0} & = \sqrt{\frac{D_{0}}{2k_{0}}}
\end{align}
The system spatial dimension $l_{0}$ should be adjusted accordingly.  Nonequilibrium chemical
  reactions fluxes are obtained for $l_{0} < \mathcal{L}_{0}$, nonequilibrium chemical transport over
large distances for $l_{0} > \mathcal{L}_{0} $, and nonequilibrium reaction-diffusion cycle for
$l_{0} \sim \mathcal{L}_{0}$,

The second critical factor is the temperature gradient imposed on the system.  $\mathcal{T}_{0}$ indicates the
critical temperature variation above which nonlinear far-from-equilibrium effects are observed:
\begin{align}
  \label{eq:66}
  \mathcal{T}_{0} & = \frac{2RT_{0}^{2}}{E_{a,+}+E_{a,-}}
\end{align}
Symmetrical cold/warm behaviors will be obtained for $\Delta T \ll \mathcal{T}_{0}$, and an asymmetry with increased
activity in the cold region  for $\Delta T  > \mathcal{T}_{0}$.

Finally, the shape of the temperature gradient can be tailored to control the localization of
chemical activities. Typically, the chemical reaction will be concentrated in the spatial areas
characterized by large local variation of the temperature gradient slope (i.e. large values of
$\nabla^{2}T$) on space scale smaller than $\mathcal{L}_{0}$. In contrast, chemical transport is established across
temperature gradients (i.e following $\nabla T$).

\paragraph{Example of peptide exchange reaction}

Reversible peptide exchange reaction mediated by N-methylcysteine is a good candidate for being
driven by temperature gradients \cite{ruff_garavini_giuseppone_2014}. It possesses high activation
energies, and its equilibrium constant is $K=1$ close to the ambient temperature; its
  thermokinetic characteristics were experimentally obtained \cite{capit_2021}:
\begin{subequations}
  \label{eq:68}
  \begin{align}
  \label{eq:67}
  \ch{P1 + P2 & <=> P1P2 + mC}\\
    \label{eq:69}
    E_{a}^{+} & = 55\times 10^{3}~\text{J}.\text{mol}^{-1} \\
    \label{eq:70}
    E_{a}^{-} & = 43\times 10^{3}~\text{J}.\text{mol}^{-1} \\
    \label{eq:71}
    \Delta_{r}H_{0} & = 12\times 10^{3}~\text{J}.\text{mol}^{-1} \\
    \label{eq:72}
    k_{0}^{+}  = k_{0}^{-} & = 1.4  \times 10^{-3}~\text{M}^{-1}.\text{s}^{-1} \\
    \label{eq:73}
    \text{at} \qquad T_{0} & = 314~\text{K}
  \end{align}
\end{subequations}
The setup conditions for efficiently sustaining nonequilibrium chemical reaction fluxes for this
  reaction---leading to steady fluxes of polymerization-depolymerization---can be directly
determined from these characteristics.

This is a second-order reaction. An apparent first-order constant (that depends on the global
reactant concentrations) can be evaluated from $k_{0}^{+}$ to
$k_{0,\text{app}}=1.4\times 10^{-6}$~s$^{-1}$ assuming a concentration $c_{0}=10^{-3}$~M.  The diffusion
coefficients of short oligopeptides \cite{plasson-03,plasson_cottet_2005} are approximately equal to
$D_{0}=5\times10^{-10}$~m$^{2}$.s$^{-1}$, and will be assumed that they are identical (that is,$\lambda=1$).  Under such conditions, the critical parameters of this reaction-diffusion system are
$\mathcal{L}_{0}=1.3$~cm and $\mathcal{T}_{0}=17$~\textdegree{C}.

We consider three systems of characteristic lengths $l_{0}=10^{-2}$~m, $l_{1}=3\times 10^{-2}$~m (i.e.\
$\sim2 \mathcal{L}_{0}$) and $l_{2}=10^{-1}$~m, each of which is subject to a Heaviside temperature gradient
$T \in [9~\text{\textdegree{C}}, 71~\text{\textdegree{C}}]$. This scaling corresponds to a nondimensionalized system with
$\beta=18.8$, $\gamma=0.122$, $\theta_{g}=0.1$, $d=1$, $\kappa=1$, $\lambda=1$, and respectively
$\rho=0.28$, $\rho=2.5$, and $\rho=28$ for $l_{0}$, $l_{1}$ and $l_{2}$. These systems are characterized by
the following characteristic lengths:
\begin{itemize}
\item $l_{0}=1$~cm: $w_{\text{cold}}=3.8$, $w_{\text{warm}} = 0.57$. Both lengths are larger than
  the length of each cold/warm region; this corresponds to a small system in which the activity is
  fully extended to the entire system, dominated by dissipative reactions and small concentration
  gradients. Setups below this scale thus correspond to regimes for performing
    simultaneously a steady polymerization process in the warmer region, and a steady
    depolymerization process in the colder region.
\item $l_{2}=10$~cm: $w_{\text{cold}}=0.38$, $w_{\text{warm}} = 0.057$. Both lengths are smaller
  than the length of each cold/warm region; this corresponds to a large system where the activity is
  restricted in the region close to the $\bar{x}=0.5$ boundary, dominated by dissipative transport
  and large concentration gradients. Setups above this scale thus correspond to regimes for
    transporting peptides across the system, the chemical reaction being locally at equilibrium.
\item $l_{1}=3$~cm: $w_{\text{cold}}=1.27$, $w_{\text{warm}} = 0.19$.  This corresponds to an
  intermediate system in which the activity is fully extended in the cold region and restricted in
  the region close to the warm boundary, dominated by a dissipative reaction-diffusion cycle across
  both regions.
\end{itemize}
The numerical models of these three systems confirm this scaling analysis (see Fig.~S8 in SI).

From this analysis, the optimal setup conditions for sustaining steady
  poly\-meri\-za\-tion/de\-poly\-meri\-za\-tion process can thus be identified from the
  thermodynamic and kinetic characteristics of the system. The chemical system must be subjected to
  a temperature gradient centered on $T_{0}=314$~K (temperature for which $K=1$). Setting this
  gradient from $297$~K to $331$~K will drive the system in a nonlinear far-non-equilibrium state
  (as $T = T_{0} \pm \mathcal{T}_{0}$). Enforcing this gradient over a system length of $l_{0}=1$~cm (as
  $l_{0} < \mathcal{L}_{0}$) will then ensure to sustain continuous depolymerization-polymerization cycles.

\subsection{Coupling to a secondary reaction}
\label{sec:coupl-second-react}

Thus, energy can be directly extracted from a temperature gradient using these reaction-diffusion
systems. We further extended the minimalist network in Eq.~\eqref{eq:1} by adding additional
  chemical reactions. The purpose is to investigate the ability of the resulting chemical reaction
  network to efficiently perform free energy transduction
  \cite{wachtel_Rao_Esposito_2022,hill2012,emond_energy_2012,saux_energy_2014} from the temperature
gradient to athermal reactions, mediated by the previously studied \ch{U1/U2}-based
reaction-diffusion cycle.

\subsubsection{Maintaining a secondary chemical reaction flux}
\label{sec:maint-second-chem}

The \ch{U1/U2} reaction-diffusion cycle can be used to drive an athermal reaction in a
  steady reaction flux. In that purpose, an additional set of reactions involving a \ch{V1/V2}
  interconversion was introduced:
\begin{subequations}
  \label{eq:74}
  \begin{align}
    \label{eq:75}
    \ch{V1 & <=> V2}\\
    \label{eq:76}
    \ch{U1 + V1 & <=> U2 + V2 }
  \end{align}
\end{subequations}
A direct uncoupled reaction \eqref{eq:75} is first introduced; it is assumed to be slow compared to the
driving reaction \eqref{eq:1}, with a nondimensionalized reaction rate constant fixed at
$10^{-3}$. It can be driven by a coupled reaction \eqref{eq:76}, whose reaction rate constant is
equal to the kinetic rates of the driving reaction \eqref{eq:1}, multiplied by a factor
$\rho'$. Moreover, it is supposed to be athermic so that any steady nonequilibrium state would result
from a transfer from the \ch{U1/U2} couple.

\begin{figure}
  \includegraphics[width=\linewidth]{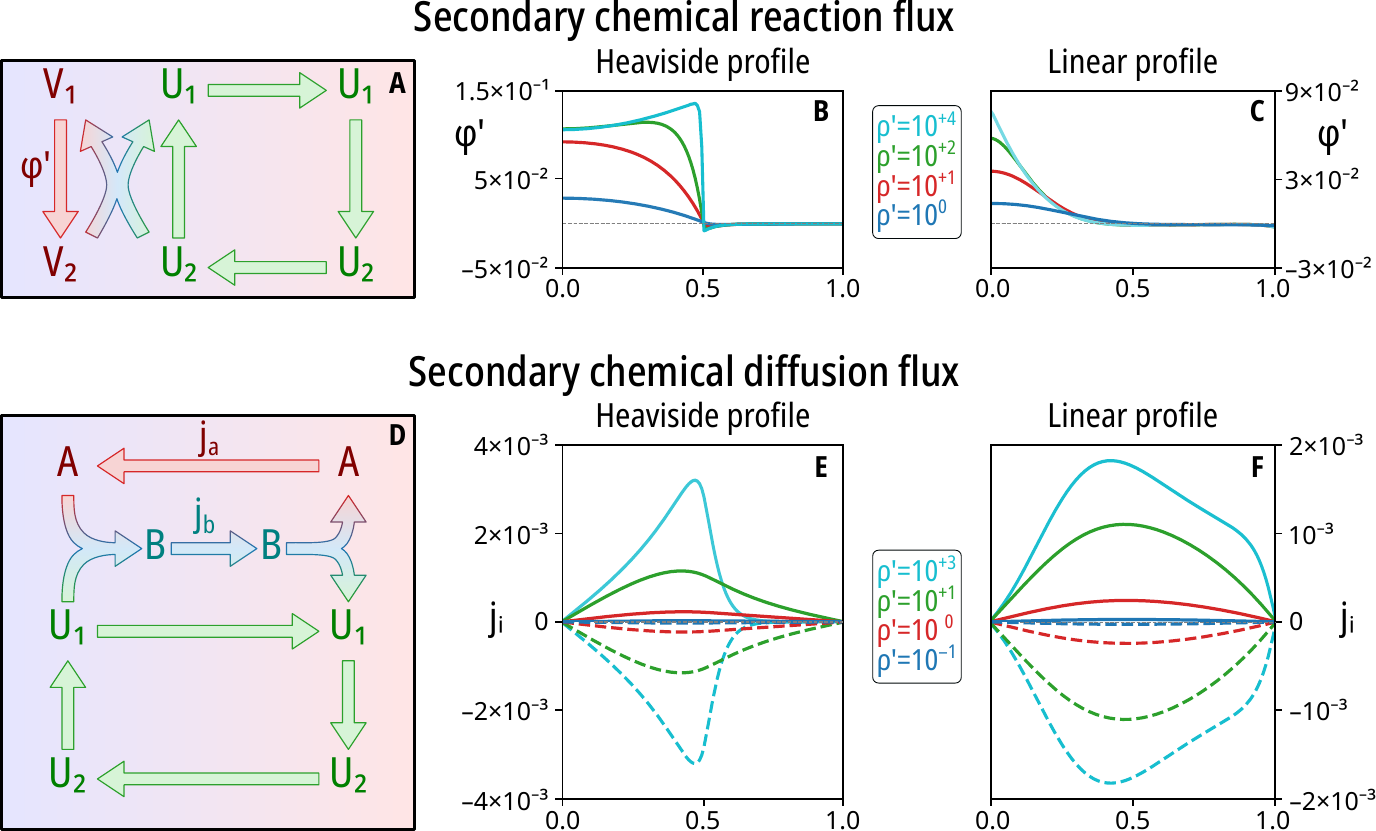}
  \caption{Coupling to secondary processes. Model parameters: $\kappa=1$, $\lambda=1$, $\gamma=0.2$,
    $d=1$, $\beta=4$, $\theta_{g}=0.5$, $\rho = 10^{2}$, $u'=0.1$, and \textbf{B,C}:
    $\rho' \in [10^{0}, 10^{4}]$, $\rho_{0}=10^{-3}$; \textbf{E, F}: $\rho' \in [10^{-1}, 10^{3}]$}.
  \label{fig:Fig6}
\end{figure}

Numerical models (see Fig.~\ref{fig:Fig6}A-C) show that the \ch{V1/V2} couple can be efficiently
driven to an active nonequilibrium state with respect to reaction \eqref{eq:75}, which is
characterized by a sustained chemical flux $\varphi'$. A necessary condition is $\rho' \gg 1$, implying that
the secondary driven reaction in Eq.~\eqref{eq:76} must be faster than the primary driving reaction
in Eq.~\eqref{eq:1}.  Furthermore, the total concentration of the compounds \ch{V_i} must be lower than
that of \ch{U_1}, to avoid perturbation of the primary reaction.

\subsubsection{Maintaining a secondary chemical diffusion flux}
\label{sec:maint-second-chem-1}

The \ch{U1/U2} reaction-diffusion cycle can also be used to drive the active transport of an
additional compound \ch{A} throughout the system. This was performed by adding the following
reaction for the initial system:
\begin{align}
  \label{eq:77}
  \ch{A + U1 & <=> B}
\end{align}
This reaction is assumed to be athermal, with a nondimensionalized kinetic rate $\rho'$ in both
directions, which does not depend on temperature. This implies that the compound \ch{A} can be bound
to \ch{U1} as a complex \ch{B}, then diffuses from the cold to the warm region, where it is released
back as \ch{A}; then it can diffuse back to the cold region.

Once again (see Fig.~\ref{fig:Fig6}D-F), large gradients can be obtained as long as the coupling
reaction \eqref{eq:77} is faster than the driving reaction \eqref{eq:1}.  Transport is also
essentially efficient for lower concentrations of \ch{A} compounds than for \ch{U1/U2} compounds.

\section{Conclusions}
\label{sec:conclusions}

A general model was established to describe how an nonequilibrium steady reaction-diffusion
process can be sustained by a temperature gradient. A difference in temperature across the system
induces spatial differences in the chemical equilibrium states. This results in a frustrated state
because it is impossible to simultaneously sustain the chemical equilibrium (which would induce a
concentration gradient) and the global transport equilibrium (which would imply homogeneous
concentrations). Consequently, reaction-diffusion cycles are necessarily established, with chemical
reactions counterbalancing chemical transports, as long as a temperature gradient is coupled to an
endothermic or exothermic chemical process. We determined the parameters that enable the system to
be dominated at will by either transport or reaction. Moreover, we demonstrated that the temperature
gradient shape can be tuned so that both the reaction and transport processes can be
spatially disconnected from each other or concentrated in the same spatial area.

We showed that this nonequilibrium steady state can be characterized in terms of:
\begin{itemize}
\item chemical force, describing the local distance from equilibrium. This essentially originates from
the reaction enthalpy and the intensity of the temperature gradient.
\item chemical flux, describing the resulting steady reaction rate.  It is proportional to the chemical force, but additionally depends on the reaction rate constants and diffusion constants,
  as well as the system symmetry at the mean temperature.
\item entropy production rate, which describes the dissipative efficiency. This results from the
combination of chemical force and flux.
\end{itemize}
These characteristics can be understood as analogous to the electromotive force, current intensity,
and dissipated power in an electric circuit, respectively. This yields a fruitful quantitative
description of the nonequilibrium state. The expression of these characteristics was established
analytically in the simple case of two homogeneous compartments that exchange matter. This result
was then extended to interpret the complete reaction-diffusion systems that were solved
numerically.

From the model established in this work, it is possible to identify the best experimental
  conditions (i.e in terms of the values of the minimal, mean and maximal temperatures, the shape of
  the temperature gradient, and the system dimension) for sustaining efficiently a given a chemical
  system in a nonequilibrium steady state (characterized by a maximal values of entropy production
  by the targeted chemical process).  Moreover, the entropy produced by the reaction-diffusion
system is distributed unequally among each process.  A net difference can be observed between the
diffusion-dominated and reaction-dominated systems. The desired regime can be tailored by scaling
the system to a critical system length $\mathcal{L}_{0}$. It should be noted that the traditional ``local
equilibrium'' approximation reduces to the limit case of fast reactions, that is, to large systems
whose physical length is well above $\mathcal{L}_{0}$. This regime is suboptimal for generating nonequilibrium
reaction fluxes; it is the worst case for extracting chemical energy from the temperature gradient,
as it leads to minimization of the chemical flux.

This framework also demonstrates the possibility of concentrating the chemical reactants in either the
cold or warm region under the effect of the temperature gradient. This global effect is only
effective in large systems compared to the critical length $\mathcal{L}_{0}$, when the reactants possess
sufficiently different diffusion coefficients, with large temperature gradients centered on an
optimal temperature.

This study is based on an imposed steady temperature gradient.  However, steady chemical reaction
fluxes can act as secondary heat sources that can influence back the temperature gradient from
which they originate. Having a steady temperature gradient comes down to the assumption of the presence of
infinitely efficient external heat sources that can seamlessly absorb this chemical heat. Further
work should describe the full coupling between the external and internal heat sources in these
reaction-diffusion systems, to evaluate their impact. In addition, we should be able to explain how
optimal reaction-mediated heat transport \cite{butler_thermal_1957}, leading to an increase in
thermal conductivity \cite{skorpa_diffusion_2015}, can be obtained.

\acknowledgments This work was supported by the ANR grant "Sacerdotal" (ANR-19-CE6-0010-02). The authors thank Alois Würger for the fruitful discussions on thermophoresis. The authors thank Nicolas Giuseppone, Nicolas
Capit, and Émilie Moulin for discussions on peptide exchange reactions.

\section*{Data Availability Statement}

Additional information, as referred to in the main text, is available in a supplementary PDF
file. Simulation files (XMDS files, raw simulation data, and processed data results) are available
upon request.


\appendix


\section{Numerical tools }
\label{sec:steady state-determ}

Four geometries of increasing complexity were used:
\begin{enumerate}
\item Two homogeneous stirred compartments, each thermostated at a different temperature, were
  connected to each other, allowing chemical exchange between them.
\item A single unstirred system was connected to two thermostats, each of which maintained a homogeneous temperature in two contiguous regions, with free chemical diffusion throughout the compartment.
\item A single unstirred system was connected to two thermostats, each of which maintained a fixed temperature at each extremity of the system, leading to a steady linear gradient throughout the system, with free chemical diffusion throughout the compartment.
\item A single unstirred system was connected to two spatially extended thermostats, each of which maintained a fixed temperature at each extremity over a length \sfrac{1}{4}, leading to a steady linear gradient over the central part of the system $\bar{x} \in [\sfrac{1}{4},\sfrac{3}{4}]$, with free chemical diffusion throughout the compartment.
\end{enumerate}

The first model was solved by calculating the steady state of the resulting ordinary differential
equation (ODE) system. An analytical solution was obtained for a low gradient intensity. The full
derivation of the steady state behavior was performed using the Python symbolic computation package
\texttt{Sympy} \cite{meurer_sympy_2017, millman_python_2011}. It is available as a \texttt{Jupyter}
notebook \cite{granger2021} in the SI, and the major results are detailed in Section
\ref{sec:math-deriv}.

The other models were solved numerically by integrating the resulting partial differential equation
(PDE) system (see Section \ref{sec:expression-pde-model}).  The steady states of the
reaction-diffusion systems were numerically calculated for a wide set of values of
$\beta \in [0.5, 5]$, $\theta_{g} \in [0.01, 0.5]$, $\rho \in [10^{-3}, 10^{+3}]$,
$d \in [10^{-3}, 10^{+3}]$, $\gamma \in [0.2, 1]$, $\kappa \in [0.1, 10]$, and
$\lambda \in [0.1, 10]$. The steady state was considered to be reached when the relative residual flux was
sufficiently small, following:
\begin{align}
  \label{eq:78}
  \frac{\int_{0}^{1} \varphi \dif \bar{x}}{|\int_{0}^{\bar{x}_{0}} \varphi \dif \bar{x}| + |\int_{\bar{x}_{0}}^{1} \varphi \dif \bar{x}|} < \omega
\end{align}
where $\bar{x}_{0}$ is the position at which $\varphi = 0$ (see \ref{sec:steady state-char}). In all
simulations, $\omega=10^{-6}$ was used. When necessary, the total simulation time was increased until the
condition of Eq.~\eqref{eq:78} was met.

The calculations were performed using XMDS2, an open source software for solving PDEs using
  spectral methods for computing spatial derivatives \cite{dennis_xmds2_2013}. In each simulation,
a steady temperature gradient was imposed with either a Heaviside, linear, or mixed profile. In all
models, the influence of temperature on the chemical reaction was introduced via the Arrhenius
relationship for the reaction rate constants.  Integration was performed using the
8\textsuperscript{th} (embedded 9\textsuperscript{th}) order adaptive Runge-Kutta method.  Spatial
derivation was performed on the basis of a discrete cosine transform for the implementation of zero
Neumann boundaries, thus describing a closed system.

\section{Entropy production due to thermal diffusion}
\label{sec:entr-prod-due}

The system is submitted to a temperature gradient. It is thus subjected to a steady heat flux,
generating a production of entropy $\dot{S}_{\mathrm{th}}$, nondimensionalized as
$\sigma_{\mathrm{th}}$. Assuming a linear gradient from $T_{0}(1-\theta_{g})$ to
$T_{0}(1+\theta_{g})$ over a length $l_{g}$, inside a system of length $l_{0}$, this term can be
expressed at each position $x$ as  \cite{groot_mazur_1969}:
\begin{subequations}
  \begin{align}
    \label{eq:79}
    T & = T_{0} \cdot 
        \left(
        1+ 
        \left(
        \frac{2 \theta_{g}}{l_{g}}x - \theta_{g}
        \right)
        \right)
    \\
    \label{eq:80}
    \dot{S}_{\mathrm{th}} (x) & = k_{\mathrm{th}} 
                                \left(
                                \frac{1}{T} \cdot \frac{\partial T}{\partial x}
                                \right)^{2} \\
    \label{eq:81}
    & =  k_{\mathrm{th}} \frac{4 \theta_{g}^{2}}{(l_{g}(1-\theta_{g}) + 2 \theta_{g} x)^{2}}
  \end{align}
\end{subequations}
$k_{\mathrm{th}}$ being the thermal conductivity of the medium (in J.s$^{-1}$.m$^{-1}$.K$^{-1}$).

On average, the entropy production in the full system is thus:
\begin{subequations}
  \begin{align}
    \label{eq:82}
    \dot{S}_{\mathrm{th}} & = \frac{1}{l_{0}} \int_{0}^{l_{g}} \dot{S}_{\mathrm{th}}(x) \dif x \\
    \label{eq:83}
    & =  \frac{4 k_{\mathrm{th}}}{l_{0} l_{g}} \cdot \frac{\theta_{g}^{2}}{1-\theta_{g}^{2}}    
  \end{align}
\end{subequations}

This can be nondimensionalized as:
\begin{subequations}
  \begin{align}
    \label{eq:84}
    \sigma_{\mathrm{th}} & = \dot{S}_{\mathrm{th}} \cdot \frac{l_{0}^{2}}{R c_{0} D_{0}} \\
    \label{eq:85}
    & =  \frac{4 \bar{k}_{\mathrm{th}}}{\bar{l}_{g}} \cdot \frac{\theta_{g}^{2}}{1-\theta_{g}^{2}} \\
    \label{eq:86}
     & \approx  \frac{4 \bar{k}_{\mathrm{th}}}{\bar{l}_{g}}  \theta_{g}^{2} \quad \textrm{for $\theta_{g} \ll 1$}\\
    \label{eq:87}
    \textrm{with} \qquad    \bar{k}_{\mathrm{th}} & = \frac{k_{\mathrm{th}}}{Rc_{0}D_{0}} 
  \end{align}
\end{subequations}

$\bar{k}_{\mathrm{th}}$ is the nondimensionalized thermal conductivity. For example, in the case of
water as the system solvent, $k_{\mathrm{th}} = 0.61$~J.s$^{-1}$.K$^{-1}$ at $300$~K
\cite{lide_1992}; with a characteristic reactant concentration $c_{0}=10^{3}$~mol.m$^{-3}$ and
chemical diffusion $D_{0}=10^{-9}$~m$^{2}$.s$^{-1}$, $\bar{k}_{\mathrm{th}} = 7.3 \times 10^{4}$. This
value can be compared with $\eta^{2}$, in order to compare the intensity of the thermal entropy
production with the chemical entropy production (Eq.~\eqref{eq:181}. An example system, with
$\eta^{2}=1600$ for a reaction enthalpy $\Delta_{r}H=100$~kJ.mol$^{-1}$ at $300$~K, would imply a chemical
entropy production that represents at most 1\% of the thermal entropy production.

$\bar{l}_{g}$ represents the fraction on the system where the gradient is established ($1$ for the
linear gradient, $\sfrac{1}{2}$ for the mixed gradient, and $0$ for the Heaviside gradient and a
two-compartment system). It shall be noted that the entropy production thus tends to infinity in the
case of the Heaviside gradient; this is due to the non-physicality of sustaining an infinitely sharp
temperature gradient.

\section{Mathematical derivation in the stirred compartments model}
\label{sec:math-deriv}

\subsection{Steady state}

The steady state $u_{ij}^{\infty}$ (with $i \in [1,2]$ and $j \in [a,b]$) can be obtained directly from the
resolution of the set of equations \eqref{eq:23}, for $\dfrac{du_{ij}}{d\bar{t}} = 0$. Each chemical
force and flux of any reaction or exchange process $k$ can be expressed as:
\begin{subequations}
  \begin{align}
    \label{eq:88}
    \varphi_{k} & = r_{k}^{+} - r_{k}^{-} \\
    \label{eq:89}
    \alpha_{k} & = \ln\frac{r_{k}^{+}}{r_{k}^{-}} 
  \end{align}
\end{subequations}
where $r_{k}^{+}$ and $r_{k}^{-}$ are the process rates of direct and indirect reactions,
respectively.

It is then possible to express the on-equilibrium steady state of each reaction and compound
exchange in terms of $[\alpha,\varphi]$ as:
\begin{subequations}
  \label{eq:90}
  \begin{align}
    \label{eq:91} 
    \varphi_{ra} & = \bar{k}_{+,a}u^{\infty}_{1a} - \bar{k}_{-,a}u^{\infty}_{2a} \\
    \label{eq:92}
    \alpha_{ra}  & =  \ln
              \left(
              K_{a} \frac{u^{\infty}_{1a}}{u^{\infty}_{2a}}
              \right) \\
    \shortintertext{for the reaction in compartment $A$,}
    \label{eq:93}
    \varphi_{rb} & = \bar{k}_{+,b}u^{\infty}_{1b} - \bar{k}_{-,b}u^{\infty}_{2b} \\
    \label{eq:94}
    \alpha_{rb}  & =  \ln
              \left(
              K_{b} \frac{u^{\infty}_{1b}}{u^{\infty}_{2b}}
              \right) \\
    \shortintertext{for the reaction in compartment $B$,}
    \label{eq:95}
    \varphi_{d1} & = \lambda^{-1}  \delta ( u^{\infty}_{1a} - u^{\infty}_{1b}) \\
    \label{eq:96}
    \alpha_{d1}  & =  \ln
              \left(
              \frac{u^{\infty}_{1a}}{u^{\infty}_{1b}}
              \right) \\
    \shortintertext{for the diffusion of $U_{1}$, and}
    \label{eq:97}
    \varphi_{d2} & = \lambda \delta (u^{\infty}_{2a} - u^{\infty}_{2b}) \\
    \label{eq:98}
    \alpha_{d2}  & =  \ln
              \left(
              \frac{u^{\infty}_{2a}}{u^{\infty}_{2b}}
              \right)
  \end{align}
\end{subequations}
for the diffusion of $U_{2}$.

At the steady state, all the chemical fluxes compensate for each other, following Eqs.~\eqref{eq:23}
such that:
\begin{align}
  \label{eq:99}
  \varphi & = - \varphi_{ra} =  \varphi_{rb} = \varphi_{d1} = - \varphi_{d2}  
\end{align}

This implies that each process $k$ is characterized by $\alpha_{k}$ and $\varphi_{k} = \pm \varphi$. Thus, each process
is characterized by its chemical resistance $R_{k}$:
\begin{align}
  \label{eq:100}
  R_{k}  & = \frac{\alpha_{k}}{\varphi_{k}}
\end{align}
Each process produces entropy as:
\begin{subequations}
  \label{eq:101}
  \begin{align}
    \label{eq:102}
    \sigma_{k} & = \alpha_{k}\varphi_{k} \\
    \label{eq:103}
          & = R_{k} \varphi_{k}^{2} = R_{k} \varphi^{2} 
  \end{align}
\end{subequations}
Each term $\sigma_{k}$ is positive, so $R_{k} \geq 0$; $\alpha_{k}$ and $\varphi_{k}$ are thus necessarily of the
same sign.

The full entropy production is then:
\begin{subequations}
  \label{eq:104}
  \begin{align}
    \label{eq:105}
    \sigma & = \sum_{k} \sigma_{k} \\
    \label{eq:106}
      & = \sum_{k}R_{k} \varphi^{2}
  \end{align}
\end{subequations}

This yields the total production of entropy in the system as:
\begin{subequations}
  \begin{align}
    \label{eq:107}
    \sigma_{ra}    &=  \alpha_{ra} \varphi_{ra}  = - \alpha_{ra} \varphi \\
    \label{eq:108}
              &= R_{ra} \varphi^{2} \\
    \label{eq:109}
    \sigma_{rb}    &= \alpha_{rb} \varphi_{rb}   = + \alpha_{rb} \varphi \\
    \label{eq:110}
              &= R_{rb} \varphi^{b}           \\
    \label{eq:111}                
    \sigma_{d1}  &=  \alpha_{d1} \varphi_{d1}  = + \alpha_{d1} \varphi \\
    \label{eq:112}
              &= R_{d1} \varphi^{2}         \\
    \label{eq:113}                
    \sigma_{d2}  &=  \alpha_{d2} \varphi_{d2}  = - \alpha_{d2} \varphi \\
    \label{eq:114}
              &= R_{d2} \varphi^{2}       \\
    \label{eq:115}                
    \sigma_{r}     &= \sigma_{ra} + \sigma_{rb} = (\alpha_{rb}-\alpha_{ra}) \varphi \\
    \label{eq:116}
              &= R_{r} \varphi^{2}\\
    \label{eq:117}                
    \sigma_{d}   &= \sigma_{d1} + \sigma_{d2} = (\alpha_{d1}-\alpha_{d2}) \varphi \\
    \label{eq:118}
              &= R_{d} \varphi^{2}\\
    \label{eq:119}                
    \sigma  &= \sigma_{r} + \sigma_{d} =  (\alpha_{r2}-\alpha_{r1}+\alpha_{d1}-\alpha_{d2}) \varphi \\
    \label{eq:120}
              & = \alpha   \varphi = R_{\text{tot}} \varphi^{2}\\
    \shortintertext{with}
    \label{eq:121}
    R_{\text{tot}} & = R_{r} + R_{d} \\
    \label{eq:122}
    R_{r} & = R_{ra} + R_{rb} \\
    \label{eq:123}
    R_{d} & = R_{d1} + R_{d2}
  \end{align}
\end{subequations}

According to Eqs.~\eqref{eq:90}, \eqref{eq:99} and \eqref{eq:16}, the total chemical force $\alpha$ can
be expressed as:
\begin{subequations}
  \label{eq:124}
  \begin{align}
    \label{eq:125}
    \alpha & =- \alpha_{ra}+\alpha_{rb}+\alpha_{d1}-\alpha_{d2} \\
    \label{eq:126}
      & = \ln
        \left(
        K_{a}^{-1} \frac{u^{\infty}_{2a}}{u^{\infty}_{1a}} \cdot
        K_{b} \frac{u^{\infty}_{1b}}{u^{\infty}_{2b}} \cdot
        \frac{u^{\infty}_{1a}}{u^{\infty}_{1b}} \cdot
        \frac{u^{\infty}_{2b}}{u^{\infty}_{2a}}
        \right)
    \\
    \label{eq:127}
      & = \ln
        \left(
        \frac{K_{b}}{K_{a}}
        \right) \\
    \label{eq:128}
      & = 
        \left(
        \frac{\theta_{g}}{\theta_{g}+1}-\frac{\theta_{g}}{\theta_{g}-1}
        \right) \eta \\
    \label{eq:129}
      & = \frac{2\eta\theta_{g}}{1-\theta_{g}^{2}}
  \end{align}
\end{subequations}

Solving the corresponding set of equations exactly leads to complex expressions of $\alpha_{i}$,
$\varphi_{i}$, and $\sigma_{i}$ parameters. Thus, we simplified the system by assuming relatively small
variations in temperature, using a second order Taylor expansion; $\alpha$ is further simplified to:
\begin{align}
  \label{eq:130}
  \alpha & = 2\eta\theta_{g} + O(\theta_{g}^{3}) \approx 2\eta\theta_{g}
\end{align}

\subsection{Expression of the steady chemical flux }

The expression of the steady chemical flux can also be evaluated with a second order Taylor
expansion as a function of $\theta_{g}$, leading to:
\begin{subequations}
  \label{eq:131}
  \begin{align}
    \varphi & = \frac{  \rho  \lambda' \delta}{\lambda' \delta + \rho}  \frac{ \kappa'  \eta\theta_{g}}{2}  \label{eq:132} \\
    \shortintertext{with}
    \label{eq:133}
    \kappa' & = \frac{2 \kappa}{1 + \kappa^{2}}\\
    \shortintertext{and:}
    \label{eq:134}
    \lambda' & = \frac{2\kappa\lambda}{1+\kappa^{2}\lambda^{2}}
  \end{align}
\end{subequations}
$\kappa'$ and $\lambda'$ are  parameters that characterize the asymmetry of the chemical reaction and
chemical exchange, respectively.  The chemical flux in Eq.~\eqref{eq:132} indicates that the
behavior of the system depends on the relative values of $\rho$ and $\lambda'\delta$, with the intensity proportional
to the reaction asymmetry $\kappa'$.

\paragraph{Distribution between chemical reaction and exchange}

The total entropy production is distributed between the chemical reactions and exchanges:
\begin{subequations}
  \label{eq:135}
  \begin{align}
    \label{eq:136}
    \alpha_{r} & = \frac{\lambda'\delta}{\lambda'\delta+\rho} \alpha \\
    \label{eq:137}
    \alpha_{d} & = \frac{\rho}{\lambda'\delta+\rho} \alpha 
  \end{align}
\end{subequations}
This implies the following distribution of entropy distribution:
\begin{subequations}
  \label{eq:138}
  \begin{align}
    \label{eq:139}  \sigma_{r} & = \frac{\lambda'\delta}{\lambda'\delta+\rho} \sigma \\
    \label{eq:140}  \sigma_{d} & = \frac{\rho}{\lambda'\delta+\rho} \sigma 
  \end{align}
\end{subequations}

\paragraph{Distribution of the reaction  entropy production in each compartment}

The  chemical forces for each chemical reaction can be evaluated as:
\begin{subequations}
  \label{eq:141}
  \begin{align}
    \label{eq:142}
    \alpha_{ra} & = -\frac{1+\varepsilon}{2} \alpha_{r} \\
    \label{eq:143}
    \alpha_{rb} & = + \frac{1-\varepsilon}{2} \alpha_{r} \\
    \shortintertext{with}
    \label{eq:144}
    \varepsilon & = \beta + \frac{\eta\rho}{2(\lambda'\delta+\rho)} \frac{1-\kappa^{2}\lambda^{2}}{1+\kappa^{2}\lambda^{2}}
  \end{align}
\end{subequations}
implying the following distribution of entropy production:
\begin{subequations}
  \label{eq:145}
  \begin{align}
    \label{eq:146}
    \sigma_{ra} & = \frac{1+\varepsilon\theta_{g}}{2} \sigma_{r} \\
    \label{eq:147}
    \sigma_{rb} & = \frac{1-\varepsilon\theta_{g}}{2} \sigma_{r}
  \end{align}
\end{subequations}
The parameter $\varepsilon$ tends to $\beta$ when $\kappa \approx \lambda^{-1}$, to
$\beta(1-\gamma)=\varepsilon_{+}$ when $\kappa$ or $\lambda$ tends to large values and to
$\beta(1+\gamma)=\varepsilon_{-}$ when $\kappa$ or $\lambda$ tends to zero. $\varepsilon$ can be interpreted as the global activation energy
for the whole system, varying from $\varepsilon_{+}$ to $\varepsilon_{-}$ through the average value of $\beta$, depending on the
global symmetry of the system.

\paragraph{Distribution of the exchange entropy production for each compound}

The  chemical forces for each chemical reaction can be evaluated as:
\begin{subequations}
  \label{eq:148}
  \begin{align}
    \label{eq:149}  \alpha_{d1} & = +\frac{\lambda^{2}\kappa^{2}}{1+\lambda^{2}\kappa^{2}} \alpha_{d} \\
    \label{eq:150}    \alpha_{d2} & = - \frac{1}{1+\lambda^{2}\kappa^{2}} \alpha_{d} 
  \end{align}
\end{subequations}
implying the following distribution of entropy production:
\begin{subequations}
  \label{eq:151}
  \begin{align}
    \label{eq:152}  \sigma_{d1} & = \frac{\lambda^{2}\kappa^{2}}{1+\lambda^{2}\kappa^{2}} \sigma_{d} \\
    \label{eq:153}  \sigma_{d2} & = \frac{1}{1+\lambda^{2}\kappa^{2}} \sigma_{d} 
  \end{align}
\end{subequations}
The dissipation due to the exchange is equally shared by the exchange of each compound \ch{U1} and \ch{U2} when
the system is symmetrical ($\kappa=\lambda^{-1}$), but is diverted towards \ch{U1} for large values of $\kappa\lambda$,
and towards \ch{U2} for small values of $\kappa\lambda$.

\subsection{Chemical resistance}
\label{sec:chemical-resistance}

By analogy with electric circuits, the relationship between the chemical fluxes $\varphi_{i}$ and the
chemical forces $\alpha_{i}$ leads to the chemical resistance of process $R_{i}$ as:
\begin{subequations}
  \label{eq:154}
  \begin{align}
    \label{eq:155}
    R_{i} & = \frac{\alpha_{i}}{\varphi_{i}} \\
    \label{eq:156}
    R_{ra}  &= \frac{2(1+\varepsilon\theta_{g})}{\kappa'\rho} \\
    \label{eq:157}
    R_{rb}  &= \frac{2(1-\varepsilon\theta_{g})}{\kappa'\rho} \\
    \label{eq:158}
    R_{d1}   &= \frac{2\kappa\lambda}{\kappa'\delta} \\
    \label{eq:159}
    R_{d2} &= \frac{2}{\kappa'\delta\kappa\lambda} \\
    \label{eq:160}
    R_{r} & = \frac{4}{\kappa'\rho} \\
    \label{eq:161}
    R_{d} & = \frac{4}{\kappa'\lambda'\delta} \\
    \label{eq:162}
    R_{\text{tot}} & = \frac{4}{\kappa'}
        \left(
        \frac{1}{\rho} + \frac{1}{\lambda'\delta}
        \right)
  \end{align}
\end{subequations}

\subsection{Entropy yield}
\label{sec:entropy-yield}

The ratio $y_{k} = \sfrac{\sigma_{k}}{\sigma}$ provides the fraction of entropy produced by the process
$k$. This can be linked to the energy yield, which is the fraction of the energy produced by
a specific process of interest. This can be calculated as:
\begin{subequations}
  \label{eq:163}
  \begin{align}
    \label{eq:164}
    y_{k} & = \frac{\sigma_{k}}{\sigma} \\
    \label{eq:165}
          & = \frac{R_{k} \varphi^{2}}{R_{\text{tot}}\varphi^{2}} = \frac{R_{k}}{R_{\text{tot}}}
  \end{align}
\end{subequations}

The efficiency of a given system to be used for chemical reactions is $y_{r}$, and that for chemical
transport is $y_{d}$:
\begin{subequations}
  \label{eq:166}
  \begin{align}
    \label{eq:167}
    y_{r} & = \frac{\lambda'\delta}{\lambda'\delta+\rho} \\
    \label{eq:168}
    y_{d} & = \frac{\rho}{\lambda'\delta+\rho}
  \end{align}
\end{subequations}

\subsection{System characteristics}

All $\varphi$, $\alpha_{k}$ and $\sigma_{k}$ values can be directly derived from the expression of
$R_{k}$ in Eqs.~\eqref{eq:154} and $\alpha$ from Eq.~\eqref{eq:130}:
\begin{subequations}
  \label{eq:169}
  \begin{align}
    \label{eq:170}
    \varphi & = \frac{\alpha}{R_{\text{tot}}} \\
    \label{eq:171}
    \alpha_{k} & = \frac{R_{k}}{R_{\text{tot}}} \alpha \\
    \label{eq:172}
    \sigma_{k} & = \frac{R_{k}}{R_{\text{tot}}^{2}}\alpha^{2}
  \end{align}
\end{subequations}

In all situations, the steady state can thus be described as:
\begin{subequations}
  \label{eq:173}
  \begin{align}
    \label{eq:174}
    \varphi & = \frac{\lambda'\delta\rho}{\lambda'\delta+\rho} \varphi_{0}\\
    \shortintertext{with}
    \label{eq:175}
    \varphi_{0} & = \frac{1}{2}\kappa'\eta\theta_{g} \\
    \shortintertext{for the chemical flux,}
    \label{eq:176}
    \alpha_{r} & = \frac{\lambda'\delta}{\lambda'\delta+\rho} \alpha_{0} \\
    \label{eq:177}
    \alpha_{d} & = \frac{\rho}{\lambda'\delta+\rho} \alpha_{0} \\
    \shortintertext{with}
    \label{eq:178}
    \alpha_{0} & = 2\eta\theta_{g} \\
    \shortintertext{for the chemical force, and}
    \label{eq:179}
    \sigma_{r}&=\frac{(\lambda'\delta)^{2}\rho}{(\lambda'\delta+\rho)^{2}}\sigma_{0} \\
    \label{eq:180}
    \sigma_{d}&=\frac{\lambda'\delta \rho^{2}}{(\lambda'\delta+\rho)^{2}}\sigma_{0}\\ \\
    \label{eq:181}
    \sigma  = \sigma_{r} + \sigma_{d} & = \frac{\lambda'\delta\rho}{\lambda'\delta+\rho}\kappa'\cdot \eta^{2}\theta_{g}^{2}
    \shortintertext{with}
    \label{eq:182}
    \sigma_{0} & = \kappa'(\eta\theta_{g})^{2}
  \end{align}
\end{subequations}
for the entropy production.

This results in a reference steady state ($[\alpha_{0}, \varphi_{0}, \sigma_{0}]$), which depends on the
thermokinetic characteristics of the chemical reaction. The state of a given system can be retrieved
from this reference state based on Eqs.~\eqref{eq:173}.

\subsection{Entropy production variations}
\label{sec:optim-entr-diss}

The variation of $\sigma_{r}$ can be expressed as:
\begin{subequations}
  \label{eq:183}
  \begin{align}
    \label{eq:184}
    \left(
    \frac{\partial \sigma_{r}}{\partial \rho}
    \right)_{\delta, \sigma_{0}}
    & = \frac{( \lambda'\delta)^{2} }{\left(\delta \lambda' + \rho\right)^{3}}  \sigma_{0}   \left( \lambda'\delta - \rho\right) \\
    \label{eq:185}
    \left(
    \frac{\partial \sigma_{r}}{\partial \delta}
    \right)_{\rho, \sigma_{0}}
    & = \frac{2 \lambda'\delta \rho^{2} }{\left(\delta \lambda' + \rho\right)^{3}}  \sigma_{0}
  \end{align}
\end{subequations}
For a given value of $\lambda'\delta$, $\sigma_{r}$ is thus maximal for $\rho=\lambda\delta'$; it varies proportionally to
$\rho$ for $\rho \ll \lambda' \delta$ and to $\rho^{-1}$ for $\rho \gg \lambda'\delta$. For a given value of
$\rho$, $\sigma_{r}$ increases monotonically with $\lambda'\delta$; it varies proportionally to
$(\lambda'\delta)^{2}$ for $\lambda'\delta \ll \rho$, and reaches a plateau for $\lambda'\delta \gg \rho$.

Similarly, the variation of $\sigma_{d}$ can be expressed as:
\begin{subequations}
  \label{eq:186}
  \begin{align}
    \label{eq:187}
    \left(
    \frac{\partial \sigma_{d}}{\partial \rho}
    \right)_{\delta, \sigma_{0}}
    & = \frac{2 (\lambda'\delta)^{2} \rho }{\left(\delta \lambda' + \rho\right)^{3}}  \sigma_{0} \\
    \label{eq:188}
    \left(
    \frac{\partial \sigma_{d}}{\partial \delta}
    \right)_{\rho, \sigma_{0}}
    & = \frac{\rho^{2} }{\left(\delta \lambda' + \rho\right)^{3}}  \sigma_{0}   \left( \rho - \lambda'\delta \right) 
  \end{align}
\end{subequations}
For a given value of $\rho$, $\sigma_{d}$ is thus maximal for $\lambda\delta'=\rho$; it varies proportionally to
$\lambda'\delta$ for $\lambda' \delta \ll \rho$ and to $(\lambda'\delta)^{-1}$ for
$ \lambda'\delta \gg \rho$. For a given value of $\lambda'\delta$, $\sigma_{d}$ increases monotonically with
$\rho$, varying proportionally to $\rho^{2}$ for $\rho \ll \lambda'\delta$, but reaches a plateau for
$\rho \gg \lambda'\delta$ (see fig.~S1 in SI.).

\section{PDE system}
\label{sec:pde}

\subsection{Expression of the PDE model}
\label{sec:expression-pde-model}

Chemical reaction-diffusion is described based on the following set of partial differential
equations (PDEs):
\begin{subequations}
  \label{eq:189}
  \begin{align}
    \label{eq:190}
    \frac{\partial u_{1}}{\partial \bar{t}} &= d_{1}  \frac{\partial^{2}u_{1}}{\partial\bar{x}^{2}}  - \varphi_{r}(\bar{x}),\\
    \label{eq:191}
    \frac{\partial u_{2}}{\partial \bar{t}} &= d_{2}   \frac{\partial^{2}u_{2}}{\partial\bar{x}^{2}} + \varphi_{r}(\bar{x}), \\
    \shortintertext{with}
    \label{eq:192}
    \varphi_{r}(\bar{x}) & = \bar{k}_{+}(\bar{x})  u_{1}(\bar{x}) -  \bar{k}_{-}(\bar{x})   u_{2}(\bar{x})
  \end{align}
\end{subequations}
In contrast to the two-compartment model, the variables are expressed as a function of the
nondimensionalized spatial position $\bar{x}$, following an imposed temperature gradient
$\theta(\bar{x})$, with $\theta(\bar{x}) \in [-\theta_{g}, +\theta_{g}])$.

The general properties of the system are described based on its characteristic concentration $c_{0}$
and temperature $T_{0}$. However, the introduction of the spatial dimension introduces two new
characteristics: length $l_{0}$ (corresponding to the system dimension), and diffusion
constant $D_{0}$ (corresponding to the average diffusion constant of the reactants). The
characteristic time is then directly obtained as $t_{0}=\sfrac{l_{0}^{2}}{D_{0}}$. The system is
thus described in terms of the nondimensionalized concentrations $u_{i}$, position $\bar{x}$,
temperature deviation $\theta$, diffusion constants $d_{i}$, and time $\bar{t}$.

Chemical transport is now defined as molecular diffusion; it cannot be treated anymore as a chemical
reaction. The rates of entropy production of the chemical reaction ($\sigma_{r}$) and chemical diffusion
of compound $U_{i}$ ($\sigma_{d,i}$) are expressed as \cite{groot_mazur_1969, mahara_yamaguchi_2010}:
\begin{subequations}
  \label{eq:193}
  \begin{align}
    \label{eq:194}
    \sigma_{r}  = \varphi_{r} \cdot \alpha_{r} & = (\bar{k}_{+}u_{1} - \bar{k}_{-}u_
                             {2}) \cdot \ln\frac{\bar{k}_{+}u_{1}}{ \bar{k}_{-}u_{2}}\\
    \label{eq:195}
    \sigma_{d,i}  = \varphi_{d} \cdot \alpha_{d} & = j_{i} \cdot \frac{1}{u_{i}} \left(
              \frac{\partial u_{i}}{\partial \bar{x}}
              \right) \\
    & = \frac{d_{i}}{u_{i}}
              \left(
              \frac{\partial u_{i}}{\partial \bar{x}}
              \right)^{2}
  \end{align}
\end{subequations}

\subsection{Steady state characteristics}
\label{sec:steady state-char}

The steady-state conditions must consider the spatial extension of the system. In steady state,
the total concentrations of each compound $U_{1}$ and $U_{2}$ are constant. This implies that the
global reaction flux is zero, i.e.:
\begin{align}
  \label{eq:196}
  \int_{0}^{1}\varphi \dif \bar{x} = 0
\end{align}

In all situations, $\varphi<0$ in the cold region (for $\bar{x} \in [0, \bar{x}_{0}[$), and
$\varphi>0$ in the warm region (for $\bar{x} \in ]\bar{x}_{0}, 1]$), where $\bar{x}_{0}$ is the position at
which $\varphi=0$. This implies that:
\begin{align}
  \label{eq:197}
  \int_{0}^{\bar{x}_{0}}\varphi\dif\bar{x} = -\int_{\bar{x}_{0}}^{1}\varphi\dif\bar{x}
\end{align}
For quantities whose sign changes across the system, this integral value is thus calculated as:
\begin{subequations}
  \label{eq:198}
  \begin{align}
    \label{eq:199}
    \varphi_{\textrm{int}} & = 
                       \left|
                       \int_{0}^{\bar{x}_{0}}\varphi\dif\bar{x}
                       \right| \\
    \label{eq:200}
    \alpha_{\textrm{int}} & = 
                       \left|
                       \int_{0}^{\bar{x}_{0}}\alpha\dif\bar{x}
                       \right|
  \end{align}
\end{subequations}

A positive rate of entropy production $\sigma$ is present at each system location. The full entropy production comes down to the integral value is calculated over the full system
extension:
\begin{align}
  \label{eq:201}
  \sigma_{\textrm{int}} & = \int_{0}^{1}\sigma\dif\bar{x}  
\end{align}


\bibliography{biblio}

\end{document}




\title{Optimizing reaction and transport fluxes in temperature gradient-driven chemical
  reaction-diffusion systems: Supplementary Information}

\author{Mohammed Loukili} 
\affiliation{Institut de Recherche de l'École Navale, EA 3634, IRENav, Brest, France}

\author{Ludovic Jullien}
\affiliation{CPCV, Département de chimie, École normale supérieure, PSL University, Sorbonne Université, CNRS, 24, rue Lhomond, 75005 Paris, France}

\author{Guillaume Baffou}
\affiliation{Institut Fresnel, CNRS, Aix Marseille University, Centrale Med, 13013 Marseille, France}

\author{Raphaël Plasson}
\email{raphael.plasson@univ-avignon.fr}
\affiliation{Avignon University, INRAE, UMR408 SQPOV, 84000 Avignon, France}

\maketitle{}


\tableofcontents
\newpage{}


\section*{Notebook: Analytical development of the compartment model}
\label{nb:comp-mod}

    \begin{tcolorbox}[breakable, size=fbox, boxrule=1pt, pad at break*=1mm,colback=cellbackground, colframe=cellborder]
\prompt{In}{incolor}{1}{\boxspacing}
\begin{Verbatim}[commandchars=\\\{\}]
\PY{k+kn}{import} \PY{n+nn}{sympy} \PY{k}{as} \PY{n+nn}{sp}
\end{Verbatim}
\end{tcolorbox}

    \hypertarget{solving-the-steady-state-from-the-equation-system}{%
\subsection{Solving the steady state from the equation
system}\label{solving-the-steady-state-from-the-equation-system}}

Chemical reactions:
\begin{itemize}
\item Compartment A: \(U_{1a} \rightleftharpoons U_{2a}\), kinetic rates \(\bar{k}_{+a}\) and
  \(\bar{k}_{-a}\) 
\item Compartment B: \(U_{1b} \rightleftharpoons U_{2b}\), kinetic rates \(\bar{k}_{+b}\) and
  \(\bar{k}_{-b}\) 
\item Exchange of \(U_1\): \(U_{1a} \rightleftharpoons U_{1b}\), kinetic rate
  \(\lambda^{-1}\delta\) 
\item Exchange of \(U_2\): \(U_{2a} \rightleftharpoons U_{2b}\), kinetic rate \(\lambda\delta\)
\end{itemize}
All following calculations are performed from nondimensionalized
parameters.

    \begin{tcolorbox}[breakable, size=fbox, boxrule=1pt, pad at break*=1mm,colback=cellbackground, colframe=cellborder]
\prompt{In}{incolor}{2}{\boxspacing}
\begin{Verbatim}[commandchars=\\\{\}]
\PY{n}{u1a}\PY{p}{,} \PY{n}{u1b}\PY{p}{,} \PY{n}{u2a}\PY{p}{,} \PY{n}{u2b}\PY{p}{,} \PY{n}{delta}\PY{p}{,} \PY{n}{kpa}\PY{p}{,} \PY{n}{kma}\PY{p}{,} \PY{n}{kpb}\PY{p}{,} \PY{n}{kmb} \PY{p}{,} \PY{n}{theta}\PY{p}{,} \PY{n}{lamb} \PY{o}{=} \PY{n}{sp}\PY{o}{.}\PY{n}{symbols}\PY{p}{(}\PY{l+s+s2}{\PYZdq{}}\PY{l+s+s2}{u1a, u1b, u2a, u2b, delta, }\PY{l+s+se}{\PYZbs{}\PYZbs{}}\PY{l+s+s2}{bar}\PY{l+s+si}{\PYZob{}k\PYZcb{}}\PY{l+s+s2}{\PYZus{}}\PY{l+s+s2}{\PYZob{}}\PY{l+s+s2}{+a\PYZcb{}, }\PY{l+s+se}{\PYZbs{}\PYZbs{}}\PY{l+s+s2}{bar}\PY{l+s+si}{\PYZob{}k\PYZcb{}}\PY{l+s+s2}{\PYZus{}}\PY{l+s+s2}{\PYZob{}}\PY{l+s+s2}{\PYZhy{}a\PYZcb{}, }\PY{l+s+se}{\PYZbs{}\PYZbs{}}\PY{l+s+s2}{bar}\PY{l+s+si}{\PYZob{}k\PYZcb{}}\PY{l+s+s2}{\PYZus{}}\PY{l+s+s2}{\PYZob{}}\PY{l+s+s2}{+b\PYZcb{}, }\PY{l+s+se}{\PYZbs{}\PYZbs{}}\PY{l+s+s2}{bar}\PY{l+s+si}{\PYZob{}k\PYZcb{}}\PY{l+s+s2}{\PYZus{}}\PY{l+s+s2}{\PYZob{}}\PY{l+s+s2}{\PYZhy{}b\PYZcb{}, theta, lambda}\PY{l+s+s2}{\PYZdq{}}\PY{p}{,} \PY{n}{real}\PY{o}{=}\PY{k+kc}{True}\PY{p}{)}
\end{Verbatim}
\end{tcolorbox}

    \hypertarget{solution}{%
\subsubsection{Solution}\label{solution}}

    \(\dfrac{d U_{1a}}{d t}\) :

    \begin{tcolorbox}[breakable, size=fbox, boxrule=1pt, pad at break*=1mm,colback=cellbackground, colframe=cellborder]
\prompt{In}{incolor}{3}{\boxspacing}
\begin{Verbatim}[commandchars=\\\{\}]
\PY{n}{eq1} \PY{o}{=} \PY{n}{delta} \PY{o}{/} \PY{n}{lamb}  \PY{o}{*} \PY{p}{(}\PY{n}{u1b}\PY{o}{\PYZhy{}}\PY{n}{u1a}\PY{p}{)} \PY{o}{\PYZhy{}} \PY{n}{kpa}\PY{o}{*} \PY{n}{u1a} \PY{o}{+} \PY{n}{kma} \PY{o}{*} \PY{n}{u2a}
\end{Verbatim}
\end{tcolorbox}

    \(\dfrac{d U_{1b}}{d t}\) :

    \begin{tcolorbox}[breakable, size=fbox, boxrule=1pt, pad at break*=1mm,colback=cellbackground, colframe=cellborder]
\prompt{In}{incolor}{4}{\boxspacing}
\begin{Verbatim}[commandchars=\\\{\}]
\PY{n}{eq2} \PY{o}{=} \PY{n}{delta} \PY{o}{/} \PY{n}{lamb}\PY{o}{*} \PY{p}{(}\PY{n}{u1a}\PY{o}{\PYZhy{}}\PY{n}{u1b}\PY{p}{)} \PY{o}{\PYZhy{}} \PY{n}{kpb}\PY{o}{*} \PY{n}{u1b} \PY{o}{+} \PY{n}{kmb} \PY{o}{*} \PY{n}{u2b}
\end{Verbatim}
\end{tcolorbox}

    \(\dfrac{d U_{2a}}{d t}\) :

    \begin{tcolorbox}[breakable, size=fbox, boxrule=1pt, pad at break*=1mm,colback=cellbackground, colframe=cellborder]
\prompt{In}{incolor}{5}{\boxspacing}
\begin{Verbatim}[commandchars=\\\{\}]
\PY{n}{eq3} \PY{o}{=} \PY{n}{lamb} \PY{o}{*} \PY{n}{delta} \PY{o}{*} \PY{p}{(}\PY{n}{u2b}\PY{o}{\PYZhy{}}\PY{n}{u2a}\PY{p}{)} \PY{o}{+} \PY{n}{kpa}\PY{o}{*} \PY{n}{u1a} \PY{o}{\PYZhy{}} \PY{n}{kma} \PY{o}{*} \PY{n}{u2a}
\end{Verbatim}
\end{tcolorbox}

    \(\dfrac{d U_{2b}}{d t}\) :

    \begin{tcolorbox}[breakable, size=fbox, boxrule=1pt, pad at break*=1mm,colback=cellbackground, colframe=cellborder]
\prompt{In}{incolor}{6}{\boxspacing}
\begin{Verbatim}[commandchars=\\\{\}]
\PY{n}{eq4} \PY{o}{=} \PY{n}{lamb} \PY{o}{*} \PY{n}{delta} \PY{o}{*} \PY{p}{(}\PY{n}{u2a}\PY{o}{\PYZhy{}}\PY{n}{u2b}\PY{p}{)} \PY{o}{+} \PY{n}{kpb}\PY{o}{*} \PY{n}{u1b} \PY{o}{\PYZhy{}} \PY{n}{kmb} \PY{o}{*} \PY{n}{u2b}
\end{Verbatim}
\end{tcolorbox}

    Matter conservation: total initial concentration is 1 in each
compartment

    \begin{tcolorbox}[breakable, size=fbox, boxrule=1pt, pad at break*=1mm,colback=cellbackground, colframe=cellborder]
\prompt{In}{incolor}{7}{\boxspacing}
\begin{Verbatim}[commandchars=\\\{\}]
\PY{n}{eq5} \PY{o}{=} \PY{n}{u1a} \PY{o}{+} \PY{n}{u1b} \PY{o}{+} \PY{n}{u2a} \PY{o}{+} \PY{n}{u2b} \PY{o}{\PYZhy{}}\PY{l+m+mi}{2}
\end{Verbatim}
\end{tcolorbox}

    The steady state comes thus down to solve \{eq1=0; eq2=0; eq3=0; eq4=0;
eq5=0\}

    \begin{tcolorbox}[breakable, size=fbox, boxrule=1pt, pad at break*=1mm,colback=cellbackground, colframe=cellborder]
\prompt{In}{incolor}{8}{\boxspacing}
\begin{Verbatim}[commandchars=\\\{\}]
\PY{n}{res} \PY{o}{=} \PY{n}{sp}\PY{o}{.}\PY{n}{solve}\PY{p}{(}\PY{p}{[}\PY{n}{eq1}\PY{p}{,} \PY{n}{eq2}\PY{p}{,} \PY{n}{eq3}\PY{p}{,} \PY{n}{eq4}\PY{p}{,} \PY{n}{eq5}\PY{p}{]}\PY{p}{,}\PY{p}{[}\PY{n}{u1a}\PY{p}{,} \PY{n}{u1b}\PY{p}{,} \PY{n}{u2a}\PY{p}{,} \PY{n}{u2b}\PY{p}{]}\PY{p}{)}
\PY{n}{u1as} \PY{o}{=} \PY{n}{res}\PY{p}{[}\PY{n}{u1a}\PY{p}{]}\PY{o}{.}\PY{n}{simplify}\PY{p}{(}\PY{p}{)}\PY{o}{.}\PY{n}{factor}\PY{p}{(}\PY{n}{delta}\PY{p}{)}
\PY{n}{u1bs} \PY{o}{=} \PY{n}{res}\PY{p}{[}\PY{n}{u1b}\PY{p}{]}\PY{o}{.}\PY{n}{simplify}\PY{p}{(}\PY{p}{)}\PY{o}{.}\PY{n}{factor}\PY{p}{(}\PY{n}{delta}\PY{p}{)}
\PY{n}{u2as} \PY{o}{=} \PY{n}{res}\PY{p}{[}\PY{n}{u2a}\PY{p}{]}\PY{o}{.}\PY{n}{simplify}\PY{p}{(}\PY{p}{)}\PY{o}{.}\PY{n}{factor}\PY{p}{(}\PY{n}{delta}\PY{p}{)}
\PY{n}{u2bs} \PY{o}{=} \PY{n}{res}\PY{p}{[}\PY{n}{u2b}\PY{p}{]}\PY{o}{.}\PY{n}{simplify}\PY{p}{(}\PY{p}{)}\PY{o}{.}\PY{n}{factor}\PY{p}{(}\PY{n}{delta}\PY{p}{)}
\end{Verbatim}
\end{tcolorbox}

    \hypertarget{check-equations}{%
\subsubsection{Check equations}\label{check-equations}}

    Checking that everything is correct:

    \begin{tcolorbox}[breakable, size=fbox, boxrule=1pt, pad at break*=1mm,colback=cellbackground, colframe=cellborder]
\prompt{In}{incolor}{9}{\boxspacing}
\begin{Verbatim}[commandchars=\\\{\}]
\PY{k}{def} \PY{n+nf}{replace}\PY{p}{(}\PY{n}{x}\PY{p}{)}\PY{p}{:}
    \PY{n}{res} \PY{o}{=} \PY{n}{x}\PY{o}{.}\PY{n}{replace}\PY{p}{(}\PY{n}{u1a}\PY{p}{,} \PY{n}{u1as}\PY{p}{)}\PY{o}{.}\PY{n}{simplify}\PY{p}{(}\PY{p}{)}
    \PY{n}{res} \PY{o}{=} \PY{n}{res}\PY{o}{.}\PY{n}{replace}\PY{p}{(}\PY{n}{u1b}\PY{p}{,} \PY{n}{u1bs}\PY{p}{)}\PY{o}{.}\PY{n}{simplify}\PY{p}{(}\PY{p}{)}
    \PY{n}{res} \PY{o}{=} \PY{n}{res}\PY{o}{.}\PY{n}{replace}\PY{p}{(}\PY{n}{u2a}\PY{p}{,} \PY{n}{u2as}\PY{p}{)}\PY{o}{.}\PY{n}{simplify}\PY{p}{(}\PY{p}{)}
    \PY{n}{res} \PY{o}{=} \PY{n}{res}\PY{o}{.}\PY{n}{replace}\PY{p}{(}\PY{n}{u2b}\PY{p}{,} \PY{n}{u2bs}\PY{p}{)}\PY{o}{.}\PY{n}{simplify}\PY{p}{(}\PY{p}{)}
    \PY{k}{return} \PY{n}{res}
\end{Verbatim}
\end{tcolorbox}

    \begin{tcolorbox}[breakable, size=fbox, boxrule=1pt, pad at break*=1mm,colback=cellbackground, colframe=cellborder]
\prompt{In}{incolor}{10}{\boxspacing}
\begin{Verbatim}[commandchars=\\\{\}]
\PY{n}{replace}\PY{p}{(}\PY{n}{eq5}\PY{p}{)}
\end{Verbatim}
\end{tcolorbox}

\prompt{Out}{outcolor}{10}{}
    
    \begin{equation*} 0\end{equation*}

    \begin{tcolorbox}[breakable, size=fbox, boxrule=1pt, pad at break*=1mm,colback=cellbackground, colframe=cellborder]
\prompt{In}{incolor}{11}{\boxspacing}
\begin{Verbatim}[commandchars=\\\{\}]
\PY{n}{replace}\PY{p}{(}\PY{n}{eq4}\PY{p}{)}
\end{Verbatim}
\end{tcolorbox}

\prompt{Out}{outcolor}{11}{}
    
    \begin{equation*} 0\end{equation*}

    \begin{tcolorbox}[breakable, size=fbox, boxrule=1pt, pad at break*=1mm,colback=cellbackground, colframe=cellborder]
\prompt{In}{incolor}{12}{\boxspacing}
\begin{Verbatim}[commandchars=\\\{\}]
\PY{n}{replace}\PY{p}{(}\PY{n}{eq3}\PY{p}{)}
\end{Verbatim}
\end{tcolorbox}

\prompt{Out}{outcolor}{12}{}
    
    \begin{equation*} 0\end{equation*}

    \begin{tcolorbox}[breakable, size=fbox, boxrule=1pt, pad at break*=1mm,colback=cellbackground, colframe=cellborder]
\prompt{In}{incolor}{13}{\boxspacing}
\begin{Verbatim}[commandchars=\\\{\}]
\PY{n}{replace}\PY{p}{(}\PY{n}{eq2}\PY{p}{)}
\end{Verbatim}
\end{tcolorbox}

\prompt{Out}{outcolor}{13}{}
    
    \begin{equation*} 0\end{equation*}

    \begin{tcolorbox}[breakable, size=fbox, boxrule=1pt, pad at break*=1mm,colback=cellbackground, colframe=cellborder]
\prompt{In}{incolor}{14}{\boxspacing}
\begin{Verbatim}[commandchars=\\\{\}]
\PY{n}{replace}\PY{p}{(}\PY{n}{eq1}\PY{p}{)}
\end{Verbatim}
\end{tcolorbox}

\prompt{Out}{outcolor}{14}{}
    
    \begin{equation*} 0\end{equation*}

    Total concentration in both compartments shall be 2 (matter conservation):

    \begin{tcolorbox}[breakable, size=fbox, boxrule=1pt, pad at break*=1mm,colback=cellbackground, colframe=cellborder]
\prompt{In}{incolor}{15}{\boxspacing}
\begin{Verbatim}[commandchars=\\\{\}]
\PY{p}{(}\PY{n}{u1as}\PY{o}{+}\PY{n}{u2as}\PY{o}{+}\PY{n}{u1bs}\PY{o}{+}\PY{n}{u2bs}\PY{p}{)}\PY{o}{.}\PY{n}{simplify}\PY{p}{(}\PY{p}{)}
\end{Verbatim}
\end{tcolorbox}

\prompt{Out}{outcolor}{15}{}
    
    \begin{equation*} 2\end{equation*}

    \(u_{1a} + u_{2a}\) should not be 1 (differences due to \(\lambda\))

    \begin{tcolorbox}[breakable, size=fbox, boxrule=1pt, pad at break*=1mm,colback=cellbackground, colframe=cellborder]
\prompt{In}{incolor}{16}{\boxspacing}
\begin{Verbatim}[commandchars=\\\{\}]
\PY{p}{(}\PY{n}{u1as}\PY{o}{+}\PY{n}{u2as}\PY{p}{)}\PY{o}{.}\PY{n}{simplify}\PY{p}{(}\PY{p}{)}
\end{Verbatim}
\end{tcolorbox}

\prompt{Out}{outcolor}{16}{}
    
\begin{equation*}
  \frac{2 \left(\bar{k}_{+a} \bar{k}_{+b} \lambda^{2} + \bar{k}_{+a} \bar{k}_{-b} + \bar{k}_{+b} \bar{k}_{-a} \lambda^{2} + \bar{k}_{-a} \bar{k}_{-b} + \delta \lambda \left(\bar{k}_{+a} + \bar{k}_{+b}\right) + \delta \lambda \left(\bar{k}_{-a} + \bar{k}_{-b}\right)\right)}{2 \bar{k}_{+a} \bar{k}_{+b} \lambda^{2} + \bar{k}_{+a} \bar{k}_{-b} \lambda^{2} + \bar{k}_{+a} \bar{k}_{-b} + \bar{k}_{+b} \bar{k}_{-a} \lambda^{2} + \bar{k}_{+b} \bar{k}_{-a} + 2 \bar{k}_{-a} \bar{k}_{-b} + 2 \delta \lambda \left(\bar{k}_{+a} + \bar{k}_{+b} + \bar{k}_{-a} + \bar{k}_{-b}\right)}
\end{equation*}

    \(u_{1b} + u_{2b}\) should not be 1 (differences due to \(\lambda\))

    \begin{tcolorbox}[breakable, size=fbox, boxrule=1pt, pad at break*=1mm,colback=cellbackground, colframe=cellborder]
\prompt{In}{incolor}{17}{\boxspacing}
\begin{Verbatim}[commandchars=\\\{\}]
\PY{p}{(}\PY{n}{u1bs}\PY{o}{+}\PY{n}{u2bs}\PY{p}{)}\PY{o}{.}\PY{n}{simplify}\PY{p}{(}\PY{p}{)}
\end{Verbatim}
\end{tcolorbox}

\prompt{Out}{outcolor}{17}{}
    
    \begin{equation*} \frac{2 \left(\bar{k}_{+a} \bar{k}_{+b} \lambda^{2} + \bar{k}_{+a} \bar{k}_{-b} \lambda^{2} + \bar{k}_{+b} \bar{k}_{-a} + \bar{k}_{-a} \bar{k}_{-b} + \delta \lambda \left(\bar{k}_{+a} + \bar{k}_{+b}\right) + \delta \lambda \left(\bar{k}_{-a} + \bar{k}_{-b}\right)\right)}{2 \bar{k}_{+a} \bar{k}_{+b} \lambda^{2} + \bar{k}_{+a} \bar{k}_{-b} \lambda^{2} + \bar{k}_{+a} \bar{k}_{-b} + \bar{k}_{+b} \bar{k}_{-a} \lambda^{2} + \bar{k}_{+b} \bar{k}_{-a} + 2 \bar{k}_{-a} \bar{k}_{-b} + 2 \delta \lambda \left(\bar{k}_{+a} + \bar{k}_{+b} + \bar{k}_{-a} + \bar{k}_{-b}\right)}\end{equation*}

    \hypertarget{data-derivation}{%
\subsection{Data derivation}\label{data-derivation}}

    \hypertarget{fluxes}{%
\subsubsection{Fluxes}\label{fluxes}}

    \(\varphi_a\) (Signs chosen according to the directions indicated in Fig.
1 of the main text)

    \begin{tcolorbox}[breakable, size=fbox, boxrule=1pt, pad at break*=1mm,colback=cellbackground, colframe=cellborder]
\prompt{In}{incolor}{18}{\boxspacing}
\begin{Verbatim}[commandchars=\\\{\}]
\PY{n}{phia} \PY{o}{=} \PY{p}{(}\PY{n}{kpa}\PY{o}{*} \PY{n}{u1as} \PY{o}{\PYZhy{}} \PY{n}{kma} \PY{o}{*} \PY{n}{u2as}\PY{p}{)}\PY{o}{.}\PY{n}{simplify}\PY{p}{(}\PY{p}{)}\PY{o}{.}\PY{n}{factor}\PY{p}{(}\PY{n}{delta}\PY{p}{)}
\end{Verbatim}
\end{tcolorbox}

    \begin{tcolorbox}[breakable, size=fbox, boxrule=1pt, pad at break*=1mm,colback=cellbackground, colframe=cellborder]
\prompt{In}{incolor}{19}{\boxspacing}
\begin{Verbatim}[commandchars=\\\{\}]
\PY{n}{phib} \PY{o}{=} \PY{p}{(}\PY{n}{kpb}\PY{o}{*} \PY{n}{u1bs} \PY{o}{\PYZhy{}} \PY{n}{kmb} \PY{o}{*} \PY{n}{u2bs}\PY{p}{)}\PY{o}{.}\PY{n}{simplify}\PY{p}{(}\PY{p}{)}\PY{o}{.}\PY{n}{factor}\PY{p}{(}\PY{n}{delta}\PY{p}{)}
\end{Verbatim}
\end{tcolorbox}

    \begin{tcolorbox}[breakable, size=fbox, boxrule=1pt, pad at break*=1mm,colback=cellbackground, colframe=cellborder]
\prompt{In}{incolor}{20}{\boxspacing}
\begin{Verbatim}[commandchars=\\\{\}]
\PY{n}{phidiff1} \PY{o}{=} \PY{p}{(}\PY{n}{delta}\PY{o}{/}\PY{n}{lamb} \PY{o}{*} \PY{p}{(}\PY{n}{u1as}\PY{o}{\PYZhy{}}\PY{n}{u1bs}\PY{p}{)}\PY{p}{)}\PY{o}{.}\PY{n}{simplify}\PY{p}{(}\PY{p}{)}\PY{o}{.}\PY{n}{factor}\PY{p}{(}\PY{n}{delta}\PY{p}{)}
\end{Verbatim}
\end{tcolorbox}

    \begin{tcolorbox}[breakable, size=fbox, boxrule=1pt, pad at break*=1mm,colback=cellbackground, colframe=cellborder]
\prompt{In}{incolor}{21}{\boxspacing}
\begin{Verbatim}[commandchars=\\\{\}]
\PY{n}{phidiff2} \PY{o}{=} \PY{p}{(}\PY{n}{lamb}\PY{o}{*}\PY{n}{delta} \PY{o}{*} \PY{p}{(}\PY{n}{u2as}\PY{o}{\PYZhy{}}\PY{n}{u2bs}\PY{p}{)}\PY{p}{)}\PY{o}{.}\PY{n}{simplify}\PY{p}{(}\PY{p}{)}\PY{o}{.}\PY{n}{factor}\PY{p}{(}\PY{n}{delta}\PY{p}{)}
\end{Verbatim}
\end{tcolorbox}

    Check that \(\varphi_a=-\varphi_b\)

    \begin{tcolorbox}[breakable, size=fbox, boxrule=1pt, pad at break*=1mm,colback=cellbackground, colframe=cellborder]
\prompt{In}{incolor}{22}{\boxspacing}
\begin{Verbatim}[commandchars=\\\{\}]
\PY{p}{(}\PY{n}{phia}\PY{o}{+}\PY{n}{phib}\PY{p}{)}\PY{o}{.}\PY{n}{simplify}\PY{p}{(}\PY{p}{)}
\end{Verbatim}
\end{tcolorbox}

\prompt{Out}{outcolor}{22}{}
    
    \begin{equation*} 0\end{equation*}

    Check that \(\varphi_{d,1}=-\varphi_{d,2}\)

    \begin{tcolorbox}[breakable, size=fbox, boxrule=1pt, pad at break*=1mm,colback=cellbackground, colframe=cellborder]
\prompt{In}{incolor}{23}{\boxspacing}
\begin{Verbatim}[commandchars=\\\{\}]
\PY{p}{(}\PY{n}{phidiff1}\PY{o}{+}\PY{n}{phidiff2}\PY{p}{)}\PY{o}{.}\PY{n}{simplify}\PY{p}{(}\PY{p}{)}
\end{Verbatim}
\end{tcolorbox}

\prompt{Out}{outcolor}{23}{}
    
    \begin{equation*} 0\end{equation*}

    Check that \(\varphi_a = -\varphi_{d,1}\)

    \begin{tcolorbox}[breakable, size=fbox, boxrule=1pt, pad at break*=1mm,colback=cellbackground, colframe=cellborder]
\prompt{In}{incolor}{24}{\boxspacing}
\begin{Verbatim}[commandchars=\\\{\}]
\PY{p}{(}\PY{n}{phidiff1}\PY{o}{+}\PY{n}{phia}\PY{p}{)}\PY{o}{.}\PY{n}{simplify}\PY{p}{(}\PY{p}{)}
\end{Verbatim}
\end{tcolorbox}

\prompt{Out}{outcolor}{24}{}
    
    \begin{equation*} 0\end{equation*}

    Define \(\varphi\) as the reaction-diffusion flux, defined as \(\varphi = -\varphi_a = \varphi_b = \varphi_{d,1} = -\varphi_{d,2}\)

    \begin{tcolorbox}[breakable, size=fbox, boxrule=1pt, pad at break*=1mm,colback=cellbackground, colframe=cellborder]
\prompt{In}{incolor}{25}{\boxspacing}
\begin{Verbatim}[commandchars=\\\{\}]
\PY{n}{phi} \PY{o}{=} \PY{o}{\PYZhy{}}\PY{n}{phia}
\PY{n}{phi}
\end{Verbatim}
\end{tcolorbox}

\prompt{Out}{outcolor}{25}{}
    
\begin{equation*}
- \frac{2 \delta \lambda \left(\bar{k}_{+a} \bar{k}_{-b} - \bar{k}_{+b} \bar{k}_{-a}\right)}{
  \begin{lgathered}
    2 \bar{k}_{+a} \bar{k}_{+b} \lambda^{2} + \bar{k}_{+a} \bar{k}_{-b} \lambda^{2} + \bar{k}_{+a} \bar{k}_{-b}
    + \bar{k}_{+b} \bar{k}_{-a} \lambda^{2} \\
    \quad + \bar{k}_{+b} \bar{k}_{-a} + 2 \bar{k}_{-a} \bar{k}_{-b} + \delta \left(2 \bar{k}_{+a} \lambda + 2
      \bar{k}_{+b} \lambda + 2 \bar{k}_{-a} \lambda + 2 \bar{k}_{-b} \lambda\right)
  \end{lgathered}
}
\end{equation*}

    \hypertarget{alpha-and-sigma}{%
\subsubsection{\texorpdfstring{\(\alpha\) and
\(\sigma\)}{\textbackslash alpha and \textbackslash sigma}}\label{alpha-and-sigma}}

    \hypertarget{chemical-reaction}{%
\paragraph{Chemical reaction}\label{chemical-reaction}}

    \begin{tcolorbox}[breakable, size=fbox, boxrule=1pt, pad at break*=1mm,colback=cellbackground, colframe=cellborder]
\prompt{In}{incolor}{26}{\boxspacing}
\begin{Verbatim}[commandchars=\\\{\}]
\PY{n}{Ka} \PY{o}{=} \PY{n}{kpa}\PY{o}{/}\PY{n}{kma}
\PY{n}{Kb} \PY{o}{=} \PY{n}{kpb}\PY{o}{/}\PY{n}{kmb}
\end{Verbatim}
\end{tcolorbox}

    \begin{tcolorbox}[breakable, size=fbox, boxrule=1pt, pad at break*=1mm,colback=cellbackground, colframe=cellborder]
\prompt{In}{incolor}{27}{\boxspacing}
\begin{Verbatim}[commandchars=\\\{\}]
\PY{n}{Qa} \PY{o}{=} \PY{n}{u2as}\PY{o}{/}\PY{n}{u1as}
\PY{n}{Qb} \PY{o}{=} \PY{n}{u2bs}\PY{o}{/}\PY{n}{u1bs}
\end{Verbatim}
\end{tcolorbox}

    \begin{tcolorbox}[breakable, size=fbox, boxrule=1pt, pad at break*=1mm,colback=cellbackground, colframe=cellborder]
\prompt{In}{incolor}{28}{\boxspacing}
\begin{Verbatim}[commandchars=\\\{\}]
\PY{n}{KQa}\PY{o}{=}\PY{p}{(}\PY{n}{Ka}\PY{o}{/}\PY{n}{Qa}\PY{p}{)}\PY{o}{.}\PY{n}{simplify}\PY{p}{(}\PY{p}{)}
\end{Verbatim}
\end{tcolorbox}

    \begin{tcolorbox}[breakable, size=fbox, boxrule=1pt, pad at break*=1mm,colback=cellbackground, colframe=cellborder]
\prompt{In}{incolor}{29}{\boxspacing}
\begin{Verbatim}[commandchars=\\\{\}]
\PY{n}{KQb}\PY{o}{=}\PY{p}{(}\PY{n}{Kb}\PY{o}{/}\PY{n}{Qb}\PY{p}{)}\PY{o}{.}\PY{n}{simplify}\PY{p}{(}\PY{p}{)}
\end{Verbatim}
\end{tcolorbox}

    \begin{tcolorbox}[breakable, size=fbox, boxrule=1pt, pad at break*=1mm,colback=cellbackground, colframe=cellborder]
\prompt{In}{incolor}{30}{\boxspacing}
\begin{Verbatim}[commandchars=\\\{\}]
\PY{n}{alpha_a} \PY{o}{=} \PY{n}{sp}\PY{o}{.}\PY{n}{log}\PY{p}{(}\PY{n}{KQa}\PY{p}{)}\PY{o}{.}\PY{n}{simplify}\PY{p}{(}\PY{p}{)}
\end{Verbatim}
\end{tcolorbox}

    \begin{tcolorbox}[breakable, size=fbox, boxrule=1pt, pad at break*=1mm,colback=cellbackground, colframe=cellborder]
\prompt{In}{incolor}{31}{\boxspacing}
\begin{Verbatim}[commandchars=\\\{\}]
\PY{n}{alpha_b} \PY{o}{=} \PY{n}{sp}\PY{o}{.}\PY{n}{log}\PY{p}{(}\PY{n}{KQb}\PY{p}{)}\PY{o}{.}\PY{n}{simplify}\PY{p}{(}\PY{p}{)}
\end{Verbatim}
\end{tcolorbox}

    \begin{tcolorbox}[breakable, size=fbox, boxrule=1pt, pad at break*=1mm,colback=cellbackground, colframe=cellborder]
\prompt{In}{incolor}{32}{\boxspacing}
\begin{Verbatim}[commandchars=\\\{\}]
\PY{n}{sigma_a} \PY{o}{=} \PY{p}{(}\PY{n}{phia}\PY{o}{*}\PY{n}{alpha_a}\PY{p}{)}\PY{o}{.}\PY{n}{simplify}\PY{p}{(}\PY{p}{)}
\end{Verbatim}
\end{tcolorbox}

    \begin{tcolorbox}[breakable, size=fbox, boxrule=1pt, pad at break*=1mm,colback=cellbackground, colframe=cellborder]
\prompt{In}{incolor}{33}{\boxspacing}
\begin{Verbatim}[commandchars=\\\{\}]
\PY{n}{sigma_b} \PY{o}{=} \PY{p}{(}\PY{n}{phib}\PY{o}{*}\PY{n}{alpha_b}\PY{p}{)}\PY{o}{.}\PY{n}{simplify}\PY{p}{(}\PY{p}{)}
\end{Verbatim}
\end{tcolorbox}

    \begin{tcolorbox}[breakable, size=fbox, boxrule=1pt, pad at break*=1mm,colback=cellbackground, colframe=cellborder]
\prompt{In}{incolor}{34}{\boxspacing}
\begin{Verbatim}[commandchars=\\\{\}]
\PY{n}{sigmar}\PY{o}{=}\PY{p}{(}\PY{n}{sigma_a}\PY{o}{+}\PY{n}{sigma_b}\PY{p}{)}\PY{o}{.}\PY{n}{simplify}\PY{p}{(}\PY{p}{)}
\PY{n}{sigmar}
\end{Verbatim}
\end{tcolorbox}

\prompt{Out}{outcolor}{34}{}
    
\begin{equation*}
  \frac{
    \begin{lgathered}
      2 \delta \lambda \left(\bar{k}_{+a} \bar{k}_{-b} - \bar{k}_{+b} \bar{k}_{-a}\right) \\
      \quad \cdot \Bigg(\log{\left(\frac{\bar{k}_{+a} \left(\bar{k}_{+b} \bar{k}_{-a} \lambda^{2} + \bar{k}_{-a}
              \bar{k}_{-b} + \delta \lambda \left(\bar{k}_{-a} + \bar{k}_{-b}\right)\right)}{\bar{k}_{-a}
            \left(\bar{k}_{+a} \bar{k}_{+b} \lambda^{2} + \bar{k}_{+a} \bar{k}_{-b} + \delta \lambda
              \left(\bar{k}_{+a} + \bar{k}_{+b}\right)\right)} \right)}  \\
      \qquad - \log{\left(\frac{\bar{k}_{+b} \left(\bar{k}_{+a} \bar{k}_{-b} \lambda^{2} + \bar{k}_{-a}
              \bar{k}_{-b} + \delta \lambda \left(\bar{k}_{-a} + \bar{k}_{-b}\right)\right)}{\bar{k}_{-b}
            \left(\bar{k}_{+a} \bar{k}_{+b} \lambda^{2} + \bar{k}_{+b} \bar{k}_{-a} + \delta \lambda
              \left(\bar{k}_{+a} + \bar{k}_{+b}\right)\right)} \right)}\Bigg)
    \end{lgathered}
  }{
    \begin{lgathered}
      2 \bar{k}_{+a} \bar{k}_{+b} \lambda^{2} + \bar{k}_{+a} \bar{k}_{-b} \lambda^{2} + \bar{k}_{+a}
      \bar{k}_{-b}  + \bar{k}_{+b} \bar{k}_{-a} \lambda^{2} \\
      \quad + \bar{k}_{+b} \bar{k}_{-a} + 2 \bar{k}_{-a} \bar{k}_{-b} + 2 \delta \lambda \left(\bar{k}_{+a} +
        \bar{k}_{+b} + \bar{k}_{-a} + \bar{k}_{-b}\right)
    \end{lgathered}
}
  \end{equation*}

    \hypertarget{chemical-diffusion}{%
\paragraph{Chemical diffusion}\label{chemical-diffusion}}

    \begin{tcolorbox}[breakable, size=fbox, boxrule=1pt, pad at break*=1mm,colback=cellbackground, colframe=cellborder]
\prompt{In}{incolor}{35}{\boxspacing}
\begin{Verbatim}[commandchars=\\\{\}]
\PY{n}{KQ\PYZus{}diff1} \PY{o}{=} \PY{p}{(}\PY{n}{u1as}\PY{o}{/}\PY{n}{u1bs}\PY{p}{)}\PY{o}{.}\PY{n}{simplify}\PY{p}{(}\PY{p}{)}
\end{Verbatim}
\end{tcolorbox}

    \begin{tcolorbox}[breakable, size=fbox, boxrule=1pt, pad at break*=1mm,colback=cellbackground, colframe=cellborder]
\prompt{In}{incolor}{36}{\boxspacing}
\begin{Verbatim}[commandchars=\\\{\}]
\PY{n}{KQ\PYZus{}diff2} \PY{o}{=} \PY{p}{(}\PY{n}{u2as}\PY{o}{/}\PY{n}{u2bs}\PY{p}{)}\PY{o}{.}\PY{n}{simplify}\PY{p}{(}\PY{p}{)}
\end{Verbatim}
\end{tcolorbox}

    \begin{tcolorbox}[breakable, size=fbox, boxrule=1pt, pad at break*=1mm,colback=cellbackground, colframe=cellborder]
\prompt{In}{incolor}{37}{\boxspacing}
\begin{Verbatim}[commandchars=\\\{\}]
\PY{n}{alpha\PYZus{}diff1} \PY{o}{=} \PY{n}{sp}\PY{o}{.}\PY{n}{log}\PY{p}{(}\PY{n}{KQ\PYZus{}diff1}\PY{p}{)}
\end{Verbatim}
\end{tcolorbox}

    \begin{tcolorbox}[breakable, size=fbox, boxrule=1pt, pad at break*=1mm,colback=cellbackground, colframe=cellborder]
\prompt{In}{incolor}{38}{\boxspacing}
\begin{Verbatim}[commandchars=\\\{\}]
\PY{n}{alpha\PYZus{}diff2} \PY{o}{=} \PY{n}{sp}\PY{o}{.}\PY{n}{log}\PY{p}{(}\PY{n}{KQ\PYZus{}diff2}\PY{p}{)}
\end{Verbatim}
\end{tcolorbox}

    \begin{tcolorbox}[breakable, size=fbox, boxrule=1pt, pad at break*=1mm,colback=cellbackground, colframe=cellborder]
\prompt{In}{incolor}{39}{\boxspacing}
\begin{Verbatim}[commandchars=\\\{\}]
\PY{n}{sigma\PYZus{}diff1} \PY{o}{=} \PY{p}{(}\PY{n}{phidiff1} \PY{o}{*} \PY{n}{alpha\PYZus{}diff1}\PY{p}{)}\PY{o}{.}\PY{n}{simplify}\PY{p}{(}\PY{p}{)}
\end{Verbatim}
\end{tcolorbox}

    \begin{tcolorbox}[breakable, size=fbox, boxrule=1pt, pad at break*=1mm,colback=cellbackground, colframe=cellborder]
\prompt{In}{incolor}{40}{\boxspacing}
\begin{Verbatim}[commandchars=\\\{\}]
\PY{n}{sigma\PYZus{}diff2} \PY{o}{=} \PY{p}{(}\PY{n}{phidiff2} \PY{o}{*} \PY{n}{alpha\PYZus{}diff2}\PY{p}{)}\PY{o}{.}\PY{n}{simplify}\PY{p}{(}\PY{p}{)}
\end{Verbatim}
\end{tcolorbox}

    \begin{tcolorbox}[breakable, size=fbox, boxrule=1pt, pad at break*=1mm,colback=cellbackground, colframe=cellborder]
\prompt{In}{incolor}{41}{\boxspacing}
\begin{Verbatim}[commandchars=\\\{\}]
\PY{n}{sigma\PYZus{}diff} \PY{o}{=} \PY{p}{(}\PY{n}{sigma\PYZus{}diff1}\PY{o}{+}\PY{n}{sigma\PYZus{}diff2}\PY{p}{)}\PY{o}{.}\PY{n}{simplify}\PY{p}{(}\PY{p}{)}
\end{Verbatim}
\end{tcolorbox}

    \begin{tcolorbox}[breakable, size=fbox, boxrule=1pt, pad at break*=1mm,colback=cellbackground, colframe=cellborder]
\prompt{In}{incolor}{42}{\boxspacing}
\begin{Verbatim}[commandchars=\\\{\}]
\PY{n}{alphar} \PY{o}{=} \PY{n}{alpha_b}\PY{o}{\PYZhy{}}\PY{n}{alpha_a}
\end{Verbatim}
\end{tcolorbox}

    \begin{tcolorbox}[breakable, size=fbox, boxrule=1pt, pad at break*=1mm,colback=cellbackground, colframe=cellborder]
\prompt{In}{incolor}{43}{\boxspacing}
\begin{Verbatim}[commandchars=\\\{\}]
\PY{n}{alpha\PYZus{}diff}\PY{o}{=}\PY{n}{alpha\PYZus{}diff1}\PY{o}{\PYZhy{}}\PY{n}{alpha\PYZus{}diff2}
\end{Verbatim}
\end{tcolorbox}

    \begin{tcolorbox}[breakable, size=fbox, boxrule=1pt, pad at break*=1mm,colback=cellbackground, colframe=cellborder]
\prompt{In}{incolor}{44}{\boxspacing}
\begin{Verbatim}[commandchars=\\\{\}]
\PY{n}{alpha\PYZus{}tot} \PY{o}{=} \PY{p}{(}\PY{n}{alphar}\PY{o}{+}\PY{n}{alpha\PYZus{}diff}\PY{p}{)}\PY{o}{.}\PY{n}{simplify}\PY{p}{(}\PY{p}{)}
\PY{n}{alpha\PYZus{}tot}\PY{o}{.}\PY{n}{simplify}\PY{p}{(}\PY{p}{)}
\end{Verbatim}
\end{tcolorbox}

\prompt{Out}{outcolor}{44}{}
    
\begin{align*}
  &- \log{\left(\frac{\bar{k}_{+a} \bar{k}_{+b} \lambda^{2} + \bar{k}_{+a} \bar{k}_{-b} + \delta \lambda
  \left(\bar{k}_{+a} + \bar{k}_{+b}\right)}{\bar{k}_{+a} \bar{k}_{+b} \lambda^{2} + \bar{k}_{+b}
  \bar{k}_{-a} + \delta \lambda \left(\bar{k}_{+a} + \bar{k}_{+b}\right)} \right)} \\
  &+ \log{\left(\frac{\bar{k}_{+b} \bar{k}_{-a} \lambda^{2} + \bar{k}_{-a} \bar{k}_{-b} + \delta \lambda
  \left(\bar{k}_{-a} + \bar{k}_{-b}\right)}{\bar{k}_{+a} \bar{k}_{-b} \lambda^{2} + \bar{k}_{-a}
  \bar{k}_{-b} + \delta \lambda \left(\bar{k}_{-a} + \bar{k}_{-b}\right)} \right)} \\
  &- \log{\left(\frac{\bar{k}_{+a} \left(\bar{k}_{+b} \bar{k}_{-a} \lambda^{2} + \bar{k}_{-a} \bar{k}_{-b}
  + \delta \lambda \left(\bar{k}_{-a} + \bar{k}_{-b}\right)\right)}{\bar{k}_{-a} \left(\bar{k}_{+a}
  \bar{k}_{+b} \lambda^{2} + \bar{k}_{+a} \bar{k}_{-b} + \delta \lambda \left(\bar{k}_{+a} +
  \bar{k}_{+b}\right)\right)} \right)} \\
  &+ \log{\left(\frac{\bar{k}_{+b} \left(\bar{k}_{+a} \bar{k}_{-b} \lambda^{2} + \bar{k}_{-a} \bar{k}_{-b} + \delta \lambda \left(\bar{k}_{-a} + \bar{k}_{-b}\right)\right)}{\bar{k}_{-b} \left(\bar{k}_{+a} \bar{k}_{+b} \lambda^{2} + \bar{k}_{+b} \bar{k}_{-a} + \delta \lambda \left(\bar{k}_{+a} + \bar{k}_{+b}\right)\right)} \right)}
\end{align*}

    \hypertarget{switch-to-betagammarho-description}{%
\subsection{\texorpdfstring{Switch to \(\beta\)/\(\gamma\)/\(\rho\)
description}{Switch to \textbackslash beta/\textbackslash gamma/\textbackslash rho description}}\label{switch-to-betagammarho-description}}

    Previous description misses information. Switch to better parameters
\(\kappa\), \(\beta\), \(\gamma\), \(\rho\) \(\theta\). New parameters
suffixed by `x'. Suppose from here \(\kappa=1\)

    \begin{tcolorbox}[breakable, size=fbox, boxrule=1pt, pad at break*=1mm,colback=cellbackground, colframe=cellborder]
\prompt{In}{incolor}{45}{\boxspacing}
\begin{Verbatim}[commandchars=\\\{\}]
\PY{n}{kappa}\PY{p}{,} \PY{n}{rho}\PY{p}{,} \PY{n}{gamma}\PY{p}{,} \PY{n}{beta}\PY{p}{,} \PY{n}{epsilon} \PY{o}{=} \PY{n}{sp}\PY{o}{.}\PY{n}{symbols}\PY{p}{(}\PY{l+s+s2}{\PYZdq{}}\PY{l+s+s2}{kappa, rho, gamma, beta, }\PY{l+s+se}{\PYZbs{}\PYZbs{}}\PY{l+s+s2}{varepsilon}\PY{l+s+s2}{\PYZdq{}}\PY{p}{,} \PY{n}{real}\PY{o}{=}\PY{k+kc}{True}\PY{p}{)}
\PY{n}{epsilon}
\end{Verbatim}
\end{tcolorbox}

\prompt{Out}{outcolor}{45}{}
    
    \begin{equation*} \varepsilon\end{equation*}

    \begin{tcolorbox}[breakable, size=fbox, boxrule=1pt, pad at break*=1mm,colback=cellbackground, colframe=cellborder]
\prompt{In}{incolor}{46}{\boxspacing}
\begin{Verbatim}[commandchars=\\\{\}]
\PY{n}{kpax} \PY{o}{=} \PY{n}{kappa} \PY{o}{*} \PY{n}{rho}\PY{o}{*}\PY{n}{sp}\PY{o}{.}\PY{n}{exp}\PY{p}{(}\PY{o}{\PYZhy{}}\PY{n}{theta}\PY{o}{/}\PY{p}{(}\PY{l+m+mi}{1}\PY{o}{\PYZhy{}}\PY{n}{theta}\PY{p}{)}\PY{o}{*}\PY{p}{(}\PY{l+m+mi}{1}\PY{o}{+}\PY{n}{gamma}\PY{p}{)}\PY{o}{*}\PY{n}{beta}\PY{p}{)}
\end{Verbatim}
\end{tcolorbox}

    \begin{tcolorbox}[breakable, size=fbox, boxrule=1pt, pad at break*=1mm,colback=cellbackground, colframe=cellborder]
\prompt{In}{incolor}{47}{\boxspacing}
\begin{Verbatim}[commandchars=\\\{\}]
\PY{n}{kpbx} \PY{o}{=} \PY{n}{kappa} \PY{o}{*} \PY{n}{rho}\PY{o}{*}\PY{n}{sp}\PY{o}{.}\PY{n}{exp}\PY{p}{(}\PY{n}{theta}\PY{o}{/}\PY{p}{(}\PY{l+m+mi}{1}\PY{o}{+}\PY{n}{theta}\PY{p}{)}\PY{o}{*}\PY{p}{(}\PY{l+m+mi}{1}\PY{o}{+}\PY{n}{gamma}\PY{p}{)}\PY{o}{*}\PY{n}{beta}\PY{p}{)}
\end{Verbatim}
\end{tcolorbox}

    \begin{tcolorbox}[breakable, size=fbox, boxrule=1pt, pad at break*=1mm,colback=cellbackground, colframe=cellborder]
\prompt{In}{incolor}{48}{\boxspacing}
\begin{Verbatim}[commandchars=\\\{\}]
\PY{n}{kmax} \PY{o}{=} \PY{n}{kappa}\PY{o}{*}\PY{o}{*}\PY{o}{\PYZhy{}}\PY{l+m+mi}{1} \PY{o}{*} \PY{n}{rho}\PY{o}{*}\PY{n}{sp}\PY{o}{.}\PY{n}{exp}\PY{p}{(}\PY{o}{\PYZhy{}}\PY{n}{theta}\PY{o}{/}\PY{p}{(}\PY{l+m+mi}{1}\PY{o}{\PYZhy{}}\PY{n}{theta}\PY{p}{)}\PY{o}{*}\PY{p}{(}\PY{l+m+mi}{1}\PY{o}{\PYZhy{}}\PY{n}{gamma}\PY{p}{)}\PY{o}{*}\PY{n}{beta}\PY{p}{)}
\end{Verbatim}
\end{tcolorbox}

    \begin{tcolorbox}[breakable, size=fbox, boxrule=1pt, pad at break*=1mm,colback=cellbackground, colframe=cellborder]
\prompt{In}{incolor}{49}{\boxspacing}
\begin{Verbatim}[commandchars=\\\{\}]
\PY{n}{kmbx} \PY{o}{=} \PY{n}{kappa}\PY{o}{*}\PY{o}{*}\PY{o}{\PYZhy{}}\PY{l+m+mi}{1} \PY{o}{*} \PY{n}{rho}\PY{o}{*}\PY{n}{sp}\PY{o}{.}\PY{n}{exp}\PY{p}{(}\PY{n}{theta}\PY{o}{/}\PY{p}{(}\PY{l+m+mi}{1}\PY{o}{+}\PY{n}{theta}\PY{p}{)}\PY{o}{*}\PY{p}{(}\PY{l+m+mi}{1}\PY{o}{\PYZhy{}}\PY{n}{gamma}\PY{p}{)}\PY{o}{*}\PY{n}{beta}\PY{p}{)}
\end{Verbatim}
\end{tcolorbox}

    \begin{tcolorbox}[breakable, size=fbox, boxrule=1pt, pad at break*=1mm,colback=cellbackground, colframe=cellborder]
\prompt{In}{incolor}{50}{\boxspacing}
\begin{Verbatim}[commandchars=\\\{\}]
\PY{k}{def} \PY{n+nf}{replacek}\PY{p}{(}\PY{n}{x}\PY{p}{)}\PY{p}{:}
\PY{+w}{    }\PY{l+s+sd}{\PYZdq{}\PYZdq{}\PYZdq{} Substitute old k parameters with the new kx parameters\PYZdq{}\PYZdq{}\PYZdq{}}
    \PY{n}{res} \PY{o}{=} \PY{n}{x}\PY{o}{.}\PY{n}{replace}\PY{p}{(}\PY{n}{kpa}\PY{p}{,}\PY{n}{kpax}\PY{p}{)}\PY{o}{.}\PY{n}{simplify}\PY{p}{(}\PY{p}{)}
    \PY{n}{res} \PY{o}{=} \PY{n}{res}\PY{o}{.}\PY{n}{replace}\PY{p}{(}\PY{n}{kpb}\PY{p}{,}\PY{n}{kpbx}\PY{p}{)}\PY{o}{.}\PY{n}{simplify}\PY{p}{(}\PY{p}{)}
    \PY{n}{res} \PY{o}{=} \PY{n}{res}\PY{o}{.}\PY{n}{replace}\PY{p}{(}\PY{n}{kma}\PY{p}{,}\PY{n}{kmax}\PY{p}{)}\PY{o}{.}\PY{n}{simplify}\PY{p}{(}\PY{p}{)}
    \PY{n}{res} \PY{o}{=} \PY{n}{res}\PY{o}{.}\PY{n}{replace}\PY{p}{(}\PY{n}{kmb}\PY{p}{,}\PY{n}{kmbx}\PY{p}{)}\PY{o}{.}\PY{n}{simplify}\PY{p}{(}\PY{p}{)}
    \PY{k}{return} \PY{n}{res}
\end{Verbatim}
\end{tcolorbox}

    \begin{tcolorbox}[breakable, size=fbox, boxrule=1pt, pad at break*=1mm,colback=cellbackground, colframe=cellborder]
\prompt{In}{incolor}{51}{\boxspacing}
\begin{Verbatim}[commandchars=\\\{\}]
\PY{n}{alphax} \PY{o}{=} \PY{n}{replacek}\PY{p}{(}\PY{n}{sp}\PY{o}{.}\PY{n}{log}\PY{p}{(}\PY{n}{Kb}\PY{o}{/}\PY{n}{Ka}\PY{p}{)}\PY{p}{)}\PY{o}{.}\PY{n}{expand}\PY{p}{(}\PY{n}{force}\PY{o}{=}\PY{k+kc}{True}\PY{p}{)}\PY{o}{.}\PY{n}{simplify}\PY{p}{(}\PY{p}{)}
\PY{n}{alphax}
\end{Verbatim}
\end{tcolorbox}

\prompt{Out}{outcolor}{51}{}
    
    \begin{equation*} - \frac{4 \beta \gamma \theta}{\theta^{2} - 1}\end{equation*}

    \begin{tcolorbox}[breakable, size=fbox, boxrule=1pt, pad at break*=1mm,colback=cellbackground, colframe=cellborder]
\prompt{In}{incolor}{52}{\boxspacing}
\begin{Verbatim}[commandchars=\\\{\}]
\PY{n}{phix} \PY{o}{=} \PY{n}{replacek}\PY{p}{(}\PY{n}{phi}\PY{p}{)}\PY{o}{.}\PY{n}{expand}\PY{p}{(}\PY{n}{force}\PY{o}{=}\PY{k+kc}{True}\PY{p}{)}\PY{o}{.}\PY{n}{simplify}\PY{p}{(}\PY{p}{)}
\PY{n}{phix}
\end{Verbatim}
\end{tcolorbox}

\prompt{Out}{outcolor}{52}{}
\begin{equation*}
  \frac{2 \delta \kappa^{2} \lambda \rho \left(e^{\frac{2 \beta \theta \left(\gamma \left(\theta - 1\right) + 2
              \theta\right)}{\left(\theta - 1\right) \left(\theta + 1\right)}} - e^{\frac{2 \beta \theta \left(\gamma \left(\theta +
                1\right) + 2 \theta\right)}{\left(\theta - 1\right) \left(\theta + 1\right)}}\right)
      \cdot e^{- \beta \theta \left(\frac{1}{\theta + 1} + \frac{1}{\theta - 1}\right)}}{
      \begin{lgathered}
        2 \delta \kappa \lambda \left(\kappa^{2} e^{\frac{2 \beta \gamma \theta}{\theta - 1}} + 1\right) e^{\beta \theta \left(\frac{\gamma}{\theta + 1} +
            \frac{1}{\theta - 1}\right)} + 2 \delta \kappa \lambda \left(\kappa^{2} e^{\frac{2 \beta \gamma \theta}{\theta + 1}} + 1\right) e^{\beta
          \theta \left(\frac{\gamma}{\theta - 1} + \frac{1}{\theta + 1}\right)} + 2 \kappa^{4} \lambda^{2} \rho e^{\beta \theta \left(\frac{2 \gamma
              \theta}{\theta^{2} - 1} +
            \frac{\gamma}{\theta + 1} + \frac{\gamma}{\theta - 1} + \frac{2 \theta}{\theta^{2} - 1}\right)} \\
        \quad + \rho \left(\kappa^{2} \lambda^{2} e^{\frac{2 \beta \gamma \theta}{\theta - 1}} + \kappa^{2} \lambda^{2} e^{\frac{2 \beta \gamma \theta}{\theta + 1}} +
          \kappa^{2} e^{\frac{2 \beta \gamma \theta}{\theta - 1}} + \kappa^{2} e^{\frac{2 \beta \gamma \theta}{\theta + 1}} + 2\right) e^{\beta \theta
          \left(\frac{1}{\theta + 1} + \frac{1}{\theta - 1}\right)}
      \end{lgathered}
}
\end{equation*}

    \hypertarget{taylor-expansion-for-low-values-of-theta}{%
\subsection{\texorpdfstring{Taylor expansion for low values of
\(\theta\)}{Taylor expansion for low values of \textbackslash theta}}\label{taylor-expansion-for-low-values-of-theta}}

    This still remains complex\ldots{} and needs simplification. Taylor
expansion with \(\theta\) (order 3 is sufficient). Corresponding
parameters are suffixed by `d', and with `d2' with +O(\ldots) removed

    \begin{tcolorbox}[breakable, size=fbox, boxrule=1pt, pad at break*=1mm,colback=cellbackground, colframe=cellborder]
\prompt{In}{incolor}{53}{\boxspacing}
\begin{Verbatim}[commandchars=\\\{\}]
\PY{k}{def} \PY{n+nf}{develop}\PY{p}{(}\PY{n}{x}\PY{p}{,} \PY{n}{order}\PY{o}{=}\PY{l+m+mi}{3}\PY{p}{)}\PY{p}{:}
\PY{+w}{    }\PY{l+s+sd}{\PYZdq{}\PYZdq{}\PYZdq{}\PYZdq{}Shortcut for Taylor expension for θ\PYZdq{}\PYZdq{}\PYZdq{}}
    \PY{k}{return} \PY{n}{sp}\PY{o}{.}\PY{n}{series}\PY{p}{(}\PY{n}{x}\PY{p}{,}\PY{n}{theta}\PY{p}{,} \PY{n}{n}\PY{o}{=}\PY{n}{order}\PY{p}{)}

\PY{k}{def} \PY{n+nf}{full\PYZus{}removeO}\PY{p}{(}\PY{n}{x}\PY{p}{,} \PY{n}{order}\PY{o}{=}\PY{l+m+mi}{3}\PY{p}{)}\PY{p}{:}
\PY{+w}{    }\PY{l+s+sd}{\PYZdq{}\PYZdq{}\PYZdq{}Sympy sometimes need to be tricked for removing the O(...) }
\PY{l+s+sd}{    Sometimes usefull for better simplification after simplify() steps\PYZdq{}\PYZdq{}\PYZdq{}}
    \PY{k}{return} \PY{n}{develop}\PY{p}{(}\PY{n}{x}\PY{p}{,} \PY{n}{order}\PY{p}{)}\PY{o}{.}\PY{n}{removeO}\PY{p}{(}\PY{p}{)}\PY{o}{.}\PY{n}{simplify}\PY{p}{(}\PY{p}{)}\PY{o}{.}\PY{n}{factor}\PY{p}{(}\PY{p}{)}
\end{Verbatim}
\end{tcolorbox}

    \hypertarget{thermokinetic-constant}{%
\subsubsection{Thermokinetic constant}\label{thermokinetic-constant}}

    \begin{tcolorbox}[breakable, size=fbox, boxrule=1pt, pad at break*=1mm,colback=cellbackground, colframe=cellborder]
\prompt{In}{incolor}{54}{\boxspacing}
\begin{Verbatim}[commandchars=\\\{\}]
\PY{n}{kpad} \PY{o}{=} \PY{n}{develop}\PY{p}{(}\PY{n}{kpax}\PY{p}{)}\PY{o}{.}\PY{n}{simplify}\PY{p}{(}\PY{p}{)}
\end{Verbatim}
\end{tcolorbox}

    \begin{tcolorbox}[breakable, size=fbox, boxrule=1pt, pad at break*=1mm,colback=cellbackground, colframe=cellborder]
\prompt{In}{incolor}{55}{\boxspacing}
\begin{Verbatim}[commandchars=\\\{\}]
\PY{n}{kpad2} \PY{o}{=} \PY{n}{full\PYZus{}removeO}\PY{p}{(}\PY{n}{kpad}\PY{p}{)}\PY{o}{.}\PY{n}{factor}\PY{p}{(}\PY{p}{[}\PY{n}{theta}\PY{p}{,} \PY{n}{beta}\PY{p}{]}\PY{p}{)}
\PY{n}{kpad2}
\end{Verbatim}
\end{tcolorbox}

\prompt{Out}{outcolor}{55}{}
    
    \begin{equation*} \frac{\kappa \rho \left(\beta^{2} \theta^{2} \left(\gamma^{2} + 2 \gamma + 1\right) + \beta \theta^{2} \left(- 2 \gamma - 2\right) + \beta \theta \left(- 2 \gamma - 2\right) + 2\right)}{2}\end{equation*}

    \begin{tcolorbox}[breakable, size=fbox, boxrule=1pt, pad at break*=1mm,colback=cellbackground, colframe=cellborder]
\prompt{In}{incolor}{56}{\boxspacing}
\begin{Verbatim}[commandchars=\\\{\}]
\PY{n}{kpbd} \PY{o}{=} \PY{n}{develop}\PY{p}{(}\PY{n}{kpbx}\PY{p}{)}\PY{o}{.}\PY{n}{simplify}\PY{p}{(}\PY{p}{)}
\end{Verbatim}
\end{tcolorbox}

    \begin{tcolorbox}[breakable, size=fbox, boxrule=1pt, pad at break*=1mm,colback=cellbackground, colframe=cellborder]
\prompt{In}{incolor}{57}{\boxspacing}
\begin{Verbatim}[commandchars=\\\{\}]
\PY{n}{kpbd2} \PY{o}{=} \PY{n}{full\PYZus{}removeO}\PY{p}{(}\PY{n}{kpbd}\PY{p}{)}\PY{o}{.}\PY{n}{factor}\PY{p}{(}\PY{p}{[}\PY{n}{theta}\PY{p}{,} \PY{n}{beta}\PY{p}{]}\PY{p}{)}
\PY{n}{kpbd2}
\end{Verbatim}
\end{tcolorbox}

\prompt{Out}{outcolor}{57}{}
    
    \begin{equation*} \frac{\kappa \rho \left(\beta^{2} \theta^{2} \left(\gamma^{2} + 2 \gamma + 1\right) + \beta \theta^{2} \left(- 2 \gamma - 2\right) + \beta \theta \left(2 \gamma + 2\right) + 2\right)}{2}\end{equation*}

    \begin{tcolorbox}[breakable, size=fbox, boxrule=1pt, pad at break*=1mm,colback=cellbackground, colframe=cellborder]
\prompt{In}{incolor}{58}{\boxspacing}
\begin{Verbatim}[commandchars=\\\{\}]
\PY{n}{kmad} \PY{o}{=} \PY{n}{develop}\PY{p}{(}\PY{n}{kmax}\PY{p}{)}\PY{o}{.}\PY{n}{simplify}\PY{p}{(}\PY{p}{)}
\end{Verbatim}
\end{tcolorbox}

    \begin{tcolorbox}[breakable, size=fbox, boxrule=1pt, pad at break*=1mm,colback=cellbackground, colframe=cellborder]
\prompt{In}{incolor}{59}{\boxspacing}
\begin{Verbatim}[commandchars=\\\{\}]
\PY{n}{kmad2} \PY{o}{=} \PY{n}{full\PYZus{}removeO}\PY{p}{(}\PY{n}{kmad}\PY{p}{)}\PY{o}{.}\PY{n}{factor}\PY{p}{(}\PY{n}{theta}\PY{p}{)}
\PY{n}{kmad2}
\end{Verbatim}
\end{tcolorbox}

\prompt{Out}{outcolor}{59}{}
    
    \begin{equation*} \frac{\rho \left(\theta^{2} \left(\beta^{2} \gamma^{2} - 2 \beta^{2} \gamma + \beta^{2} + 2 \beta \gamma - 2 \beta\right) + \theta \left(2 \beta \gamma - 2 \beta\right) + 2\right)}{2 \kappa}\end{equation*}

    \begin{tcolorbox}[breakable, size=fbox, boxrule=1pt, pad at break*=1mm,colback=cellbackground, colframe=cellborder]
\prompt{In}{incolor}{60}{\boxspacing}
\begin{Verbatim}[commandchars=\\\{\}]
\PY{n}{kmbd} \PY{o}{=} \PY{n}{develop}\PY{p}{(}\PY{n}{kmbx}\PY{p}{)}\PY{o}{.}\PY{n}{simplify}\PY{p}{(}\PY{p}{)}
\end{Verbatim}
\end{tcolorbox}

    \begin{tcolorbox}[breakable, size=fbox, boxrule=1pt, pad at break*=1mm,colback=cellbackground, colframe=cellborder]
\prompt{In}{incolor}{61}{\boxspacing}
\begin{Verbatim}[commandchars=\\\{\}]
\PY{n}{kmbd2} \PY{o}{=} \PY{n}{full\PYZus{}removeO}\PY{p}{(}\PY{n}{kmbd}\PY{p}{)}\PY{o}{.}\PY{n}{factor}\PY{p}{(}\PY{n}{theta}\PY{p}{)}
\PY{n}{kmbd2}
\end{Verbatim}
\end{tcolorbox}

\prompt{Out}{outcolor}{61}{}
    
    \begin{equation*} \frac{\rho \left(\theta^{2} \left(\beta^{2} \gamma^{2} - 2 \beta^{2} \gamma + \beta^{2} + 2 \beta \gamma - 2 \beta\right) + \theta \left(- 2 \beta \gamma + 2 \beta\right) + 2\right)}{2 \kappa}\end{equation*}

    \begin{tcolorbox}[breakable, size=fbox, boxrule=1pt, pad at break*=1mm,colback=cellbackground, colframe=cellborder]
\prompt{In}{incolor}{62}{\boxspacing}
\begin{Verbatim}[commandchars=\\\{\}]
\PY{k}{def} \PY{n+nf}{replaced}\PY{p}{(}\PY{n}{x}\PY{p}{,} \PY{n}{order}\PY{o}{=}\PY{l+m+mi}{3}\PY{p}{)}\PY{p}{:}
\PY{+w}{    }\PY{l+s+sd}{\PYZdq{}\PYZdq{}\PYZdq{}Replace old parameters k with new DL\PYZsq{}d kd parameters}
\PY{l+s+sd}{    and computes corresponding DL\PYZdq{}\PYZdq{}\PYZdq{}}
    \PY{n}{res} \PY{o}{=} \PY{n}{x}\PY{o}{.}\PY{n}{replace}\PY{p}{(}\PY{n}{kpa}\PY{p}{,}\PY{n}{kpad}\PY{p}{)}
    \PY{n}{res} \PY{o}{=} \PY{n}{res}\PY{o}{.}\PY{n}{replace}\PY{p}{(}\PY{n}{kpb}\PY{p}{,}\PY{n}{kpbd}\PY{p}{)}
    \PY{n}{res} \PY{o}{=} \PY{n}{res}\PY{o}{.}\PY{n}{replace}\PY{p}{(}\PY{n}{kma}\PY{p}{,}\PY{n}{kmad}\PY{p}{)}
    \PY{n}{res} \PY{o}{=} \PY{n}{res}\PY{o}{.}\PY{n}{replace}\PY{p}{(}\PY{n}{kmb}\PY{p}{,}\PY{n}{kmbd}\PY{p}{)}
    \PY{k}{if} \PY{n}{order} \PY{o}{==} \PY{o}{\PYZhy{}}\PY{l+m+mi}{1}\PY{p}{:}
        \PY{k}{return} \PY{n}{res}\PY{o}{.}\PY{n}{simplify}\PY{p}{(}\PY{p}{)}\PY{o}{.}\PY{n}{factor}\PY{p}{(}\PY{p}{)}
    \PY{k}{return} \PY{n}{develop}\PY{p}{(}\PY{n}{res}\PY{p}{,} \PY{n}{order}\PY{p}{)}\PY{o}{.}\PY{n}{simplify}\PY{p}{(}\PY{p}{)}\PY{o}{.}\PY{n}{factor}\PY{p}{(}\PY{p}{)}
\end{Verbatim}
\end{tcolorbox}

    \hypertarget{ux3ba-simplification}{%
\subsubsection{κ Simplification}\label{ux3ba-simplification}}

Further, we will need identification of
\(\lambda' = \dfrac{2\lambda\kappa}{1+\lambda^2\kappa^2}\) and
\(\kappa' = \dfrac{2\kappa}{1+\kappa^2}\)

    \begin{tcolorbox}[breakable, size=fbox, boxrule=1pt, pad at break*=1mm,colback=cellbackground, colframe=cellborder]
\prompt{In}{incolor}{63}{\boxspacing}
\begin{Verbatim}[commandchars=\\\{\}]
\PY{n}{lp}\PY{p}{,} \PY{n}{kp}\PY{p}{,} \PY{n}{x}\PY{p}{,} \PY{n}{y}\PY{p}{,} \PY{n}{h} \PY{o}{=} \PY{n}{sp}\PY{o}{.}\PY{n}{symbols}\PY{p}{(}\PY{p}{[}\PY{l+s+s2}{\PYZdq{}}\PY{l+s+se}{\PYZbs{}\PYZbs{}}\PY{l+s+s2}{lambda}\PY{l+s+s2}{\PYZsq{}}\PY{l+s+s2}{\PYZdq{}}\PY{p}{,} \PY{l+s+s2}{\PYZdq{}}\PY{l+s+se}{\PYZbs{}\PYZbs{}}\PY{l+s+s2}{kappa}\PY{l+s+s2}{\PYZsq{}}\PY{l+s+s2}{\PYZdq{}}\PY{p}{,} \PY{l+s+s2}{\PYZdq{}}\PY{l+s+s2}{x}\PY{l+s+s2}{\PYZdq{}}\PY{p}{,} \PY{l+s+s2}{\PYZdq{}}\PY{l+s+s2}{y}\PY{l+s+s2}{\PYZdq{}}\PY{p}{,} \PY{l+s+s2}{\PYZdq{}}\PY{l+s+se}{\PYZbs{}\PYZbs{}}\PY{l+s+s2}{eta}\PY{l+s+s2}{\PYZdq{}}\PY{p}{]}\PY{p}{,} \PY{n}{real}\PY{o}{=}\PY{k+kc}{True}\PY{p}{)}
\PY{n}{lp}
\end{Verbatim}
\end{tcolorbox}

\prompt{Out}{outcolor}{63}{}
    
    \begin{equation*} \lambda'\end{equation*}

    \begin{tcolorbox}[breakable, size=fbox, boxrule=1pt, pad at break*=1mm,colback=cellbackground, colframe=cellborder]
\prompt{In}{incolor}{64}{\boxspacing}
\begin{Verbatim}[commandchars=\\\{\}]
\PY{k}{def} \PY{n+nf}{to\PYZus{}h\PYZus{}kp}\PY{p}{(}\PY{n}{res}\PY{p}{,} \PY{n}{debug}\PY{o}{=}\PY{k+kc}{False}\PY{p}{)}\PY{p}{:}
\PY{+w}{    }\PY{l+s+sd}{\PYZdq{}\PYZdq{}\PYZdq{} simplify the expression by attempting to identify }
\PY{l+s+sd}{    \PYZbs{}lambda\PYZsq{}, \PYZbs{}kappa\PYZsq{} and \PYZbs{}eta parameters}
\PY{l+s+sd}{    \PYZdq{}\PYZdq{}\PYZdq{}}
    \PY{k}{if} \PY{n}{debug}\PY{p}{:}
        \PY{n}{sp}\PY{o}{.}\PY{n}{pprint}\PY{p}{(}\PY{n}{res}\PY{p}{)}
    \PY{n}{res} \PY{o}{=} \PY{n}{res}\PY{o}{.}\PY{n}{subs}\PY{p}{(}\PY{n}{kappa}\PY{o}{*}\PY{o}{*}\PY{l+m+mi}{2}\PY{o}{*}\PY{n}{lamb}\PY{o}{*}\PY{o}{*}\PY{l+m+mi}{2}\PY{o}{+}\PY{l+m+mi}{1}\PY{p}{,} \PY{n}{x}\PY{p}{)}
    \PY{k}{if} \PY{n}{debug}\PY{p}{:}
        \PY{n}{sp}\PY{o}{.}\PY{n}{pprint}\PY{p}{(}\PY{n}{res}\PY{p}{)}
    \PY{n}{res} \PY{o}{=} \PY{n}{res}\PY{o}{.}\PY{n}{subs}\PY{p}{(}\PY{l+m+mi}{2}\PY{o}{*}\PY{n}{lamb}\PY{o}{*}\PY{n}{kappa}\PY{p}{,} \PY{n}{y}\PY{p}{)}
    \PY{k}{if} \PY{n}{debug}\PY{p}{:}
        \PY{n}{sp}\PY{o}{.}\PY{n}{pprint}\PY{p}{(}\PY{n}{res}\PY{p}{)}
    \PY{n}{res} \PY{o}{=} \PY{n}{res}\PY{o}{.}\PY{n}{factor}\PY{p}{(}\PY{n}{x}\PY{o}{/}\PY{n}{y}\PY{p}{)}
    \PY{k}{if} \PY{n}{debug}\PY{p}{:}
        \PY{n}{sp}\PY{o}{.}\PY{n}{pprint}\PY{p}{(}\PY{n}{res}\PY{p}{)}
    \PY{n}{res} \PY{o}{=} \PY{n}{res}\PY{o}{.}\PY{n}{subs}\PY{p}{(}\PY{n}{y}\PY{p}{,}\PY{n}{lp}\PY{o}{*}\PY{n}{x}\PY{p}{)}
    \PY{k}{if} \PY{n}{debug}\PY{p}{:}
        \PY{n}{sp}\PY{o}{.}\PY{n}{pprint}\PY{p}{(}\PY{n}{res}\PY{p}{)}
    \PY{n}{res} \PY{o}{=} \PY{n}{res}\PY{o}{.}\PY{n}{subs}\PY{p}{(}\PY{n}{beta}\PY{o}{*}\PY{n}{gamma}\PY{p}{,} \PY{n}{h}\PY{o}{/}\PY{l+m+mi}{2}\PY{p}{)}
    \PY{k}{if} \PY{n}{debug}\PY{p}{:}
        \PY{n}{sp}\PY{o}{.}\PY{n}{pprint}\PY{p}{(}\PY{n}{res}\PY{p}{)}
    \PY{n}{res} \PY{o}{=} \PY{n}{res}\PY{o}{.}\PY{n}{simplify}\PY{p}{(}\PY{p}{)}
    \PY{n}{res} \PY{o}{=} \PY{n}{res}\PY{o}{.}\PY{n}{subs}\PY{p}{(}\PY{n}{x}\PY{p}{,}\PY{n}{kappa}\PY{o}{*}\PY{o}{*}\PY{l+m+mi}{2}\PY{o}{*}\PY{n}{lamb}\PY{o}{*}\PY{o}{*}\PY{l+m+mi}{2}\PY{o}{+}\PY{l+m+mi}{1}\PY{p}{)}
    \PY{n}{res} \PY{o}{=} \PY{n}{res}\PY{o}{.}\PY{n}{subs}\PY{p}{(}\PY{n}{y}\PY{p}{,} \PY{l+m+mi}{2}\PY{o}{*}\PY{n}{lamb}\PY{o}{*}\PY{n}{kappa}\PY{p}{)}
    \PY{k}{if} \PY{n}{debug}\PY{p}{:}
        \PY{n}{sp}\PY{o}{.}\PY{n}{pprint}\PY{p}{(}\PY{n}{res}\PY{p}{)}
    \PY{n}{res} \PY{o}{=} \PY{n}{res}\PY{o}{.}\PY{n}{subs}\PY{p}{(}\PY{n}{kappa}\PY{o}{*}\PY{o}{*}\PY{l+m+mi}{2}\PY{o}{+}\PY{l+m+mi}{1}\PY{p}{,} \PY{n}{x}\PY{p}{)}
    \PY{k}{if} \PY{n}{debug}\PY{p}{:}
        \PY{n}{sp}\PY{o}{.}\PY{n}{pprint}\PY{p}{(}\PY{n}{res}\PY{p}{)}
    \PY{n}{res} \PY{o}{=} \PY{n}{res}\PY{o}{.}\PY{n}{subs}\PY{p}{(}\PY{l+m+mi}{2}\PY{o}{*}\PY{n}{kappa}\PY{p}{,} \PY{n}{y}\PY{p}{)}
    \PY{k}{if} \PY{n}{debug}\PY{p}{:}
        \PY{n}{sp}\PY{o}{.}\PY{n}{pprint}\PY{p}{(}\PY{n}{res}\PY{p}{)}
    \PY{n}{res} \PY{o}{=} \PY{n}{res}\PY{o}{.}\PY{n}{factor}\PY{p}{(}\PY{n}{x}\PY{o}{/}\PY{n}{y}\PY{p}{)}
    \PY{k}{if} \PY{n}{debug}\PY{p}{:}
        \PY{n}{sp}\PY{o}{.}\PY{n}{pprint}\PY{p}{(}\PY{n}{res}\PY{p}{)}
    \PY{n}{res} \PY{o}{=} \PY{n}{res}\PY{o}{.}\PY{n}{subs}\PY{p}{(}\PY{n}{y}\PY{p}{,}\PY{n}{kp}\PY{o}{*}\PY{n}{x}\PY{p}{)}
    \PY{k}{if} \PY{n}{debug}\PY{p}{:}
        \PY{n}{sp}\PY{o}{.}\PY{n}{pprint}\PY{p}{(}\PY{n}{res}\PY{p}{)}
    \PY{n}{res} \PY{o}{=} \PY{n}{res}\PY{o}{.}\PY{n}{simplify}\PY{p}{(}\PY{p}{)}
    \PY{k}{if} \PY{n}{debug}\PY{p}{:}
        \PY{n}{sp}\PY{o}{.}\PY{n}{pprint}\PY{p}{(}\PY{n}{res}\PY{p}{)}
    \PY{n}{res} \PY{o}{=} \PY{n}{res}\PY{o}{.}\PY{n}{subs}\PY{p}{(}\PY{n}{x}\PY{p}{,}\PY{n}{kappa}\PY{o}{*}\PY{o}{*}\PY{l+m+mi}{2}\PY{o}{+}\PY{l+m+mi}{1}\PY{p}{)}
    \PY{n}{res} \PY{o}{=} \PY{n}{res}\PY{o}{.}\PY{n}{subs}\PY{p}{(}\PY{n}{y}\PY{p}{,} \PY{l+m+mi}{2}\PY{o}{*}\PY{n}{kappa}\PY{p}{)}
    \PY{k}{return} \PY{n}{res}\PY{o}{.}\PY{n}{simplify}\PY{p}{(}\PY{p}{)}
\end{Verbatim}
\end{tcolorbox}

    \begin{tcolorbox}[breakable, size=fbox, boxrule=1pt, pad at break*=1mm,colback=cellbackground, colframe=cellborder]
\prompt{In}{incolor}{65}{\boxspacing}
\begin{Verbatim}[commandchars=\\\{\}]
\PY{n}{Kad} \PY{o}{=} \PY{n}{replaced}\PY{p}{(}\PY{n}{Ka}\PY{p}{)}
\end{Verbatim}
\end{tcolorbox}

    \begin{tcolorbox}[breakable, size=fbox, boxrule=1pt, pad at break*=1mm,colback=cellbackground, colframe=cellborder]
\prompt{In}{incolor}{66}{\boxspacing}
\begin{Verbatim}[commandchars=\\\{\}]
\PY{n}{Kad2} \PY{o}{=} \PY{n}{full\PYZus{}removeO}\PY{p}{(}\PY{n}{Kad}\PY{p}{)}\PY{o}{.}\PY{n}{factor}\PY{p}{(}\PY{n}{theta}\PY{p}{)}
\end{Verbatim}
\end{tcolorbox}

    \begin{tcolorbox}[breakable, size=fbox, boxrule=1pt, pad at break*=1mm,colback=cellbackground, colframe=cellborder]
\prompt{In}{incolor}{67}{\boxspacing}
\begin{Verbatim}[commandchars=\\\{\}]
\PY{n}{Ka\PYZus{}ok} \PY{o}{=} \PY{n}{to\PYZus{}h\PYZus{}kp}\PY{p}{(}\PY{n}{Kad2}\PY{p}{)}\PY{o}{.}\PY{n}{factor}\PY{p}{(}\PY{n}{theta}\PY{p}{)}
\PY{n}{Ka\PYZus{}ok}
\end{Verbatim}
\end{tcolorbox}

\prompt{Out}{outcolor}{67}{}
    
    \begin{equation*} \frac{\kappa^{2} \left(- 2 \eta \theta + \theta^{2} \left(\eta^{2} - 2 \eta\right) + 2\right)}{2}\end{equation*}

    \begin{tcolorbox}[breakable, size=fbox, boxrule=1pt, pad at break*=1mm,colback=cellbackground, colframe=cellborder]
\prompt{In}{incolor}{68}{\boxspacing}
\begin{Verbatim}[commandchars=\\\{\}]
\PY{n}{Kbd} \PY{o}{=} \PY{n}{replaced}\PY{p}{(}\PY{n}{Kb}\PY{p}{)}
\end{Verbatim}
\end{tcolorbox}

    \begin{tcolorbox}[breakable, size=fbox, boxrule=1pt, pad at break*=1mm,colback=cellbackground, colframe=cellborder]
\prompt{In}{incolor}{69}{\boxspacing}
\begin{Verbatim}[commandchars=\\\{\}]
\PY{n}{Kbd2} \PY{o}{=} \PY{n}{full\PYZus{}removeO}\PY{p}{(}\PY{n}{Kbd}\PY{p}{)}\PY{o}{.}\PY{n}{factor}\PY{p}{(}\PY{n}{theta}\PY{p}{)}
\end{Verbatim}
\end{tcolorbox}

    \begin{tcolorbox}[breakable, size=fbox, boxrule=1pt, pad at break*=1mm,colback=cellbackground, colframe=cellborder]
\prompt{In}{incolor}{70}{\boxspacing}
\begin{Verbatim}[commandchars=\\\{\}]
\PY{n}{Kb\PYZus{}ok} \PY{o}{=} \PY{n}{to\PYZus{}h\PYZus{}kp}\PY{p}{(}\PY{n}{Kbd2}\PY{p}{)}\PY{o}{.}\PY{n}{factor}\PY{p}{(}\PY{n}{theta}\PY{p}{)}
\PY{n}{Kb\PYZus{}ok}
\end{Verbatim}
\end{tcolorbox}

\prompt{Out}{outcolor}{70}{}
    
    \begin{equation*} \frac{\kappa^{2} \cdot \left(2 \eta \theta + \theta^{2} \left(\eta^{2} - 2 \eta\right) + 2\right)}{2}\end{equation*}

    \hypertarget{varphi}{%
\subsubsection{\texorpdfstring{\(\varphi\)}{\textbackslash varphi}}\label{varphi}}

    \begin{tcolorbox}[breakable, size=fbox, boxrule=1pt, pad at break*=1mm,colback=cellbackground, colframe=cellborder]
\prompt{In}{incolor}{71}{\boxspacing}
\begin{Verbatim}[commandchars=\\\{\}]
\PY{n}{phid}\PY{o}{=}\PY{n}{replaced}\PY{p}{(}\PY{n}{phi}\PY{p}{)}
\PY{n}{phid}
\end{Verbatim}
\end{tcolorbox}

\prompt{Out}{outcolor}{71}{}
    
    \begin{equation*} \frac{4 \beta \delta \gamma \kappa^{2} \lambda \rho \theta + O\left(\theta^{3}\right)}{\left(\kappa^{2} + 1\right) \left(2 \delta \kappa \lambda + \kappa^{2} \lambda^{2} \rho + \rho\right)}\end{equation*}

    \begin{tcolorbox}[breakable, size=fbox, boxrule=1pt, pad at break*=1mm,colback=cellbackground, colframe=cellborder]
\prompt{In}{incolor}{72}{\boxspacing}
\begin{Verbatim}[commandchars=\\\{\}]
\PY{n}{phid2} \PY{o}{=} \PY{n}{full\PYZus{}removeO}\PY{p}{(}\PY{n}{phid}\PY{p}{)}
\PY{n}{phid2}\PY{o}{.}\PY{n}{factor}\PY{p}{(}\PY{n}{rho}\PY{p}{)}
\end{Verbatim}
\end{tcolorbox}

\prompt{Out}{outcolor}{72}{}
    
    \begin{equation*} \frac{4 \beta \delta \gamma \kappa^{2} \lambda \rho \theta}{\left(\kappa^{2} + 1\right) \left(2 \delta \kappa \lambda + \rho \left(\kappa^{2} \lambda^{2} + 1\right)\right)}\end{equation*}

    \begin{tcolorbox}[breakable, size=fbox, boxrule=1pt, pad at break*=1mm,colback=cellbackground, colframe=cellborder]
\prompt{In}{incolor}{73}{\boxspacing}
\begin{Verbatim}[commandchars=\\\{\}]
\PY{n}{phi\PYZus{}ok} \PY{o}{=} \PY{n}{to\PYZus{}h\PYZus{}kp}\PY{p}{(}\PY{n}{phid2}\PY{o}{*}\PY{l+m+mi}{2}\PY{p}{)}\PY{o}{.}\PY{n}{factor}\PY{p}{(}\PY{n}{rho}\PY{p}{)}\PY{o}{/}\PY{l+m+mi}{2}
\PY{n}{phi\PYZus{}ok}
\end{Verbatim}
\end{tcolorbox}

\prompt{Out}{outcolor}{73}{}
    
    \begin{equation*} \frac{\eta \kappa' \lambda' \delta \rho \theta}{2 \left(\lambda' \delta + \rho\right)}\end{equation*}

    \hypertarget{alpha}{%
\subsubsection{\texorpdfstring{\(\alpha\)}{\textbackslash alpha}}\label{alpha}}

    \hypertarget{reaction-in-compartment-a}{%
\paragraph{Reaction in compartment A}\label{reaction-in-compartment-a}}

    \begin{tcolorbox}[breakable, size=fbox, boxrule=1pt, pad at break*=1mm,colback=cellbackground, colframe=cellborder]
\prompt{In}{incolor}{74}{\boxspacing}
\begin{Verbatim}[commandchars=\\\{\}]
\PY{n}{KQad} \PY{o}{=} \PY{n}{replaced}\PY{p}{(}\PY{n}{KQa}\PY{p}{)}
\PY{n}{KQad}
\end{Verbatim}
\end{tcolorbox}

\prompt{Out}{outcolor}{74}{}
    
\begin{equation*}
    \frac{
      \begin{lgathered}
        \rho^{2} + 2 \kappa^{2} \lambda^{2} \rho^{2} + \kappa^{4} \lambda^{4} \rho^{2} + 4 \delta \kappa \lambda \rho + 4 \delta \kappa^{3} \lambda^{3} \rho + 4 \delta^{2}
        \kappa^{2} \lambda^{2} \\
        \quad - 4 \beta \delta \gamma \kappa \lambda \rho \theta - 4 \beta \delta \gamma \kappa^{3} \lambda^{3} \rho \theta - 8 \beta \delta^{2} \gamma \kappa^{2} \lambda^{2} \theta - 4 \beta^{2} \delta \gamma \kappa \lambda \rho
        \theta^{2} - 4 \beta^{2} \delta
        \gamma \kappa^{3} \lambda^{3} \rho \theta^{2} \\
        \quad - 4 \beta^{2} \delta \gamma^{2} \kappa \lambda \rho \theta^{2} + 4 \beta^{2} \delta \gamma^{2} \kappa^{3} \lambda^{3} \rho \theta^{2} - 8 \beta^{2} \delta^{2} \gamma \kappa^{2}
        \lambda^{2} \theta^{2} + 8 \beta^{2} \delta^{2} \gamma^{2} \kappa^{2} \lambda^{2} \theta^{2} + O\left(\theta^{3}\right)
      \end{lgathered}
}{\left(2 \delta \kappa \lambda + \kappa^{2} \lambda^{2} \rho + \rho\right)^{2}}
  \end{equation*}

    \begin{tcolorbox}[breakable, size=fbox, boxrule=1pt, pad at break*=1mm,colback=cellbackground, colframe=cellborder]
\prompt{In}{incolor}{75}{\boxspacing}
\begin{Verbatim}[commandchars=\\\{\}]
\PY{n}{alpha_ad} \PY{o}{=} \PY{n}{develop}\PY{p}{(}\PY{n}{sp}\PY{o}{.}\PY{n}{log}\PY{p}{(}\PY{n}{KQad}\PY{p}{)}\PY{p}{)}\PY{o}{.}\PY{n}{simplify}\PY{p}{(}\PY{p}{)}
\PY{n}{alpha_ad}
\end{Verbatim}
\end{tcolorbox}

\prompt{Out}{outcolor}{75}{}
\begin{equation*}
  \frac{
    \begin{lgathered}
      - 4 \beta \delta \gamma \kappa \lambda \theta \left(2 \delta \kappa \lambda + \kappa^{2} \lambda^{2} \rho + \rho\right) \left(4 \delta^{2} \kappa^{2} \lambda^{2} + 4 \delta \kappa^{3}
        \lambda^{3} \rho + 4 \delta \kappa \lambda \rho + \kappa^{4} \lambda^{4} \rho^{2} + 2 \kappa^{2} \lambda^{2} \rho^{2} +
        \rho^{2}\right)^{3}  \\
      \quad - 4 \beta^{2} \delta \gamma \kappa \lambda \theta^{2} \left(4 \delta^{2} \kappa^{2} \lambda^{2} + 4 \delta \kappa^{3} \lambda^{3} \rho + 4 \delta \kappa \lambda \rho + \kappa^{4}
        \lambda^{4} \rho^{2} + 2 \kappa^{2}
        \lambda^{2} \rho^{2} + \rho^{2}\right)^{2} \\
      \qquad \cdot \Big(2 \delta \gamma \kappa \lambda \left(2 \delta \kappa \lambda + \kappa^{2} \lambda^{2} \rho + \rho\right)^{2} \\
      \quad \qquad + \left(4 \delta^{2} \kappa^{2} \lambda^{2} + 4 \delta \kappa^{3} \lambda^{3} \rho + 4 \delta \kappa \lambda \rho + \kappa^{4} \lambda^{4} \rho^{2} + 2 \kappa^{2}
        \lambda^{2} \rho^{2} + \rho^{2}\right) \\
      \qquad\qquad\cdot\left(- 2 \delta \gamma \kappa \lambda + 2 \delta \kappa \lambda - \gamma \kappa^{2} \lambda^{2} \rho +
        \gamma \rho + \kappa^{2} \lambda^{2} \rho + \rho\right)\Big) \\
      \quad + O\left(\theta^{3}\right)
    \end{lgathered}
}{\left(4 \delta^{2} \kappa^{2} \lambda^{2} + 4 \delta \kappa^{3} \lambda^{3} \rho + 4 \delta \kappa \lambda \rho + \kappa^{4} \lambda^{4} \rho^{2} + 2 \kappa^{2} \lambda^{2} \rho^{2} + \rho^{2}\right)^{4}}
  \end{equation*}

    \begin{tcolorbox}[breakable, size=fbox, boxrule=1pt, pad at break*=1mm,colback=cellbackground, colframe=cellborder]
\prompt{In}{incolor}{76}{\boxspacing}
\begin{Verbatim}[commandchars=\\\{\}]
\PY{n}{alpha_ad2}\PY{o}{=}\PY{n}{full\PYZus{}removeO}\PY{p}{(}\PY{n}{alpha_ad}\PY{p}{)}
\PY{n}{alpha_ad2}
\end{Verbatim}
\end{tcolorbox}

\prompt{Out}{outcolor}{76}{}
    
    \begin{equation*} - \frac{4 \beta \delta \gamma \kappa \lambda \theta \left(2 \beta \delta \kappa \lambda \theta - \beta \gamma \kappa^{2} \lambda^{2} \rho \theta + \beta \gamma \rho \theta + \beta \kappa^{2} \lambda^{2} \rho \theta + \beta \rho \theta + 2 \delta \kappa \lambda + \kappa^{2} \lambda^{2} \rho + \rho\right)}{\left(2 \delta \kappa \lambda + \kappa^{2} \lambda^{2} \rho + \rho\right)^{2}}\end{equation*}

    \begin{tcolorbox}[breakable, size=fbox, boxrule=1pt, pad at break*=1mm,colback=cellbackground, colframe=cellborder]
\prompt{In}{incolor}{77}{\boxspacing}
\begin{Verbatim}[commandchars=\\\{\}]
\PY{n}{alpha_a\PYZus{}ok} \PY{o}{=} \PY{n}{to\PYZus{}h\PYZus{}kp}\PY{p}{(}\PY{n}{alpha_ad2}\PY{p}{)}
\PY{n}{alpha_a\PYZus{}ok}
\end{Verbatim}
\end{tcolorbox}

\prompt{Out}{outcolor}{77}{}
\begin{equation*}
- \dfrac{
  \begin{lgathered}
    \eta \lambda' \delta \theta \left(\kappa^{2} \lambda^{2} + 1\right) \\
    \quad \cdot\Big(- \eta \kappa^{2} \lambda^{2} \rho \theta + \eta \rho \theta + \kappa' \beta \kappa \lambda^{2} \rho
    \theta \left(\kappa^{2} + 1\right) + \kappa' \kappa \lambda^{2} \rho \left(\kappa^{2} + 1\right) \\
    \qquad + 2 \lambda' \beta \delta \kappa^{2} \lambda^{2} \theta + 2 \lambda' \beta \delta \theta + 2 \lambda' \delta \kappa^{2} \lambda^{2} + 2 \lambda' \delta + 2 \beta \rho \theta + 2 \rho\Big)
  \end{lgathered}
}{2 \left(\lambda' \delta \kappa^{2} \lambda^{2} + \lambda' \delta + \kappa^{2} \lambda^{2} \rho + \rho\right)^{2}}
\end{equation*}

    \hypertarget{reaction-in-compartment-b}{%
\paragraph{Reaction in compartment B}\label{reaction-in-compartment-b}}

    \begin{tcolorbox}[breakable, size=fbox, boxrule=1pt, pad at break*=1mm,colback=cellbackground, colframe=cellborder]
\prompt{In}{incolor}{78}{\boxspacing}
\begin{Verbatim}[commandchars=\\\{\}]
\PY{n}{KQbd} \PY{o}{=} \PY{n}{replaced}\PY{p}{(}\PY{n}{KQb}\PY{p}{)}
\PY{n}{KQbd}
\end{Verbatim}
\end{tcolorbox}

\prompt{Out}{outcolor}{78}{}
\begin{equation*}
  \frac{
    \begin{lgathered}
      \rho^{2} + 2 \kappa^{2} \lambda^{2} \rho^{2} + \kappa^{4} \lambda^{4} \rho^{2} + 4 \delta \kappa \lambda \rho + 4 \delta \kappa^{3} \lambda^{3} \rho + 4 \delta^{2}
      \kappa^{2} \lambda^{2} + 4 \beta \delta \gamma \kappa \lambda \rho \theta + 4 \beta \delta \gamma \kappa^{3} \lambda^{3} \rho \theta + 8 \beta \delta^{2} \gamma \kappa^{2} \lambda^{2} \theta \\
      \quad - 4 \beta^{2} \delta \gamma \kappa \lambda \rho \theta^{2} - 4 \beta^{2} \delta \gamma \kappa^{3} \lambda^{3} \rho \theta^{2} - 4 \beta^{2} \delta \gamma^{2} \kappa \lambda \rho \theta^{2} +
      4 \beta^{2} \delta \gamma^{2} \kappa^{3} \lambda^{3} \rho \theta^{2} - 8 \beta^{2} \delta^{2} \gamma \kappa^{2} \lambda^{2} \theta^{2} + 8 \beta^{2} \delta^{2} \gamma^{2}
      \kappa^{2} \lambda^{2} \theta^{2} + O\left(\theta^{3}\right)
    \end{lgathered}
}{\left(2 \delta \kappa \lambda + \kappa^{2} \lambda^{2} \rho + \rho\right)^{2}}
\end{equation*}

    \begin{tcolorbox}[breakable, size=fbox, boxrule=1pt, pad at break*=1mm,colback=cellbackground, colframe=cellborder]
\prompt{In}{incolor}{79}{\boxspacing}
\begin{Verbatim}[commandchars=\\\{\}]
\PY{n}{alpha_bd} \PY{o}{=} \PY{n}{develop}\PY{p}{(}\PY{n}{sp}\PY{o}{.}\PY{n}{log}\PY{p}{(}\PY{n}{KQbd}\PY{p}{)}\PY{p}{)}\PY{o}{.}\PY{n}{simplify}\PY{p}{(}\PY{p}{)}
\PY{n}{alpha_bd}
\end{Verbatim}
\end{tcolorbox}

\prompt{Out}{outcolor}{79}{}
    
\begin{equation*}
  \frac{
    \begin{lgathered}
      4 \beta \delta \gamma \kappa \lambda \theta \left(2 \delta \kappa \lambda + \kappa^{2} \lambda^{2} \rho + \rho\right) \left(4 \delta^{2} \kappa^{2} \lambda^{2} + 4 \delta \kappa^{3}
        \lambda^{3} \rho + 4 \delta \kappa \lambda \rho + \kappa^{4} \lambda^{4} \rho^{2} + 2 \kappa^{2} \lambda^{2} \rho^{2} + \rho^{2}\right)^{3} \\
      \quad - 4 \beta^{2} \delta \gamma \kappa \lambda \theta^{2} \left(4 \delta^{2} \kappa^{2} \lambda^{2} + 4 \delta \kappa^{3} \lambda^{3} \rho + 4 \delta \kappa \lambda \rho + \kappa^{4} \lambda^{4} \rho^{2} + 2 \kappa^{2}
        \lambda^{2} \rho^{2} + \rho^{2}\right)^{2}\\
      \qquad \cdot \Big(2 \delta \gamma \kappa \lambda \left(2 \delta \kappa \lambda + \kappa^{2} \lambda^{2} \rho + \rho\right)^{2} \\
      \quad \qquad + \left(4 \delta^{2} \kappa^{2} \lambda^{2} + 4 \delta \kappa^{3} \lambda^{3} \rho + 4 \delta \kappa \lambda \rho + \kappa^{4} \lambda^{4} \rho^{2} + 2 \kappa^{2}
        \lambda^{2} \rho^{2} + \rho^{2}\right) \\
      \qquad\qquad\cdot\left(- 2 \delta \gamma \kappa \lambda + 2 \delta \kappa \lambda - \gamma \kappa^{2} \lambda^{2} \rho + \gamma \rho + \kappa^{2}
        \lambda^{2} \rho + \rho\right)\Big) \\
      \quad + O\left(\theta^{3}\right)
    \end{lgathered}
}{\left(4 \delta^{2} \kappa^{2} \lambda^{2} + 4 \delta \kappa^{3} \lambda^{3} \rho + 4 \delta \kappa \lambda \rho + \kappa^{4} \lambda^{4} \rho^{2} + 2 \kappa^{2} \lambda^{2} \rho^{2} + \rho^{2}\right)^{4}}
\end{equation*}

    \begin{tcolorbox}[breakable, size=fbox, boxrule=1pt, pad at break*=1mm,colback=cellbackground, colframe=cellborder]
\prompt{In}{incolor}{80}{\boxspacing}
\begin{Verbatim}[commandchars=\\\{\}]
\PY{n}{alpha_bd2}\PY{o}{=}\PY{n}{full\PYZus{}removeO}\PY{p}{(}\PY{n}{alpha_bd}\PY{p}{)}
\PY{n}{alpha_bd2}
\end{Verbatim}
\end{tcolorbox}

\prompt{Out}{outcolor}{80}{}
    
    \begin{equation*} - \frac{4 \beta \delta \gamma \kappa \lambda \theta \left(2 \beta \delta \kappa \lambda \theta - \beta \gamma \kappa^{2} \lambda^{2} \rho \theta + \beta \gamma \rho \theta + \beta \kappa^{2} \lambda^{2} \rho \theta + \beta \rho \theta - 2 \delta \kappa \lambda - \kappa^{2} \lambda^{2} \rho - \rho\right)}{\left(2 \delta \kappa \lambda + \kappa^{2} \lambda^{2} \rho + \rho\right)^{2}}\end{equation*}

    \hypertarget{total-reaction}{%
\paragraph{Total reaction}\label{total-reaction}}

    \begin{tcolorbox}[breakable, size=fbox, boxrule=1pt, pad at break*=1mm,colback=cellbackground, colframe=cellborder]
\prompt{In}{incolor}{81}{\boxspacing}
\begin{Verbatim}[commandchars=\\\{\}]
\PY{n}{alpha\PYZus{}rd} \PY{o}{=} \PY{p}{(}\PY{n}{alpha_bd}\PY{o}{\PYZhy{}}\PY{n}{alpha_ad}\PY{p}{)}\PY{o}{.}\PY{n}{simplify}\PY{p}{(}\PY{p}{)}
\PY{n}{alpha\PYZus{}rd}
\end{Verbatim}
\end{tcolorbox}

\prompt{Out}{outcolor}{81}{}
    
    \begin{equation*} \frac{8 \beta \delta \gamma \kappa \lambda \theta \left(2 \delta \kappa \lambda + \kappa^{2} \lambda^{2} \rho + \rho\right) \left(4 \delta^{2} \kappa^{2} \lambda^{2} + 4 \delta \kappa^{3} \lambda^{3} \rho + 4 \delta \kappa \lambda \rho + \kappa^{4} \lambda^{4} \rho^{2} + 2 \kappa^{2} \lambda^{2} \rho^{2} + \rho^{2}\right)^{3} + O\left(\theta^{3}\right)}{\left(4 \delta^{2} \kappa^{2} \lambda^{2} + 4 \delta \kappa^{3} \lambda^{3} \rho + 4 \delta \kappa \lambda \rho + \kappa^{4} \lambda^{4} \rho^{2} + 2 \kappa^{2} \lambda^{2} \rho^{2} + \rho^{2}\right)^{4}}\end{equation*}

    \begin{tcolorbox}[breakable, size=fbox, boxrule=1pt, pad at break*=1mm,colback=cellbackground, colframe=cellborder]
\prompt{In}{incolor}{82}{\boxspacing}
\begin{Verbatim}[commandchars=\\\{\}]
\PY{n}{alpha\PYZus{}rd2} \PY{o}{=} \PY{p}{(}\PY{n}{alpha_bd2}\PY{o}{\PYZhy{}}\PY{n}{alpha_ad2}\PY{p}{)}\PY{o}{.}\PY{n}{simplify}\PY{p}{(}\PY{p}{)}
\PY{n}{alpha\PYZus{}rd2}
\end{Verbatim}
\end{tcolorbox}

\prompt{Out}{outcolor}{82}{}
    
    \begin{equation*} \frac{8 \beta \delta \gamma \kappa \lambda \theta}{2 \delta \kappa \lambda + \kappa^{2} \lambda^{2} \rho + \rho}\end{equation*}

    \begin{tcolorbox}[breakable, size=fbox, boxrule=1pt, pad at break*=1mm,colback=cellbackground, colframe=cellborder]
\prompt{In}{incolor}{83}{\boxspacing}
\begin{Verbatim}[commandchars=\\\{\}]
\PY{n}{alpha\PYZus{}r\PYZus{}ok} \PY{o}{=} \PY{n}{to\PYZus{}h\PYZus{}kp}\PY{p}{(}\PY{n}{alpha\PYZus{}rd2}\PY{p}{)}
\PY{n}{alpha\PYZus{}r\PYZus{}ok}
\end{Verbatim}
\end{tcolorbox}

\prompt{Out}{outcolor}{83}{}
    
    \begin{equation*} \frac{2 \eta \lambda' \delta \theta}{\lambda' \delta + \rho}\end{equation*}

    \hypertarget{further-simplification-of-independent-reactions}{%
\paragraph{Further simplification of independent
reactions:}\label{further-simplification-of-independent-reactions}}

    \hypertarget{identification-of-varepsilon}{%
\paragraph{\texorpdfstring{identification of
\(\varepsilon\)}{identification of \textbackslash varepsilon}}\label{identification-of-varepsilon}}

    \begin{tcolorbox}[breakable, size=fbox, boxrule=1pt, pad at break*=1mm,colback=cellbackground, colframe=cellborder]
\prompt{In}{incolor}{84}{\boxspacing}
\begin{Verbatim}[commandchars=\\\{\}]
\PY{n}{f\PYZus{}d2} \PY{o}{=} \PY{p}{(}\PY{p}{(}\PY{p}{(}\PY{o}{\PYZhy{}}\PY{n}{alpha_ad2}\PY{o}{\PYZhy{}}\PY{n}{alpha_bd2}\PY{p}{)}\PY{o}{.}\PY{n}{simplify}\PY{p}{(}\PY{p}{)}\PY{o}{/}\PY{n}{alpha\PYZus{}rd2}\PY{o}{/}\PY{n}{theta}\PY{p}{)}\PY{o}{\PYZhy{}}\PY{n}{beta}\PY{p}{)}\PY{o}{.}\PY{n}{simplify}\PY{p}{(}\PY{p}{)}
\PY{n}{f\PYZus{}d2}
\end{Verbatim}
\end{tcolorbox}

\prompt{Out}{outcolor}{84}{}
    
    \begin{equation*} \frac{\beta \gamma \rho \left(- \kappa^{2} \lambda^{2} + 1\right)}{2 \delta \kappa \lambda + \kappa^{2} \lambda^{2} \rho + \rho}\end{equation*}

    \begin{tcolorbox}[breakable, size=fbox, boxrule=1pt, pad at break*=1mm,colback=cellbackground, colframe=cellborder]
\prompt{In}{incolor}{85}{\boxspacing}
\begin{Verbatim}[commandchars=\\\{\}]
\PY{n}{epsilon\PYZus{}d2} \PY{o}{=} \PY{p}{(}\PY{n}{beta}\PY{o}{+}\PY{n}{f\PYZus{}d2}\PY{p}{)}
\PY{n}{epsilon\PYZus{}d2}
\end{Verbatim}
\end{tcolorbox}

\prompt{Out}{outcolor}{85}{}
    
    \begin{equation*} \frac{\beta \gamma \rho \left(- \kappa^{2} \lambda^{2} + 1\right)}{2 \delta \kappa \lambda + \kappa^{2} \lambda^{2} \rho + \rho} + \beta\end{equation*}

    \begin{tcolorbox}[breakable, size=fbox, boxrule=1pt, pad at break*=1mm,colback=cellbackground, colframe=cellborder]
\prompt{In}{incolor}{86}{\boxspacing}
\begin{Verbatim}[commandchars=\\\{\}]
\PY{n}{epsilon\PYZus{}ok} \PY{o}{=} \PY{n}{to\PYZus{}h\PYZus{}kp}\PY{p}{(}\PY{n}{f\PYZus{}d2}\PY{o}{.}\PY{n}{factor}\PY{p}{(}\PY{n}{rho}\PY{p}{)}\PY{p}{)}\PY{o}{+}\PY{n}{beta}
\PY{n}{epsilon\PYZus{}ok}
\end{Verbatim}
\end{tcolorbox}

\prompt{Out}{outcolor}{86}{}
    
    \begin{equation*} - \frac{\eta \rho \left(\kappa^{2} \lambda^{2} - 1\right)}{2 \left(\lambda' \delta + \rho\right) \left(\kappa^{2} \lambda^{2} + 1\right)} + \beta\end{equation*}

    \hypertarget{varepsilon-properties}{%
\paragraph{\texorpdfstring{\(\varepsilon\)
properties}{\textbackslash varepsilon properties}}\label{varepsilon-properties}}

    \begin{tcolorbox}[breakable, size=fbox, boxrule=1pt, pad at break*=1mm,colback=cellbackground, colframe=cellborder]
\prompt{In}{incolor}{87}{\boxspacing}
\begin{Verbatim}[commandchars=\\\{\}]
\PY{n}{epsilon\PYZus{}d2}\PY{o}{.}\PY{n}{replace}\PY{p}{(}\PY{n}{kappa}\PY{p}{,}\PY{l+m+mi}{0}\PY{p}{)}\PY{o}{.}\PY{n}{simplify}\PY{p}{(}\PY{p}{)}
\end{Verbatim}
\end{tcolorbox}

\prompt{Out}{outcolor}{87}{}
    
    \begin{equation*} \beta \left(\gamma + 1\right)\end{equation*}

    \begin{tcolorbox}[breakable, size=fbox, boxrule=1pt, pad at break*=1mm,colback=cellbackground, colframe=cellborder]
\prompt{In}{incolor}{88}{\boxspacing}
\begin{Verbatim}[commandchars=\\\{\}]
\PY{n}{epsilon\PYZus{}d2}\PY{o}{.}\PY{n}{replace}\PY{p}{(}\PY{n}{lamb}\PY{p}{,}\PY{l+m+mi}{0}\PY{p}{)}\PY{o}{.}\PY{n}{simplify}\PY{p}{(}\PY{p}{)}
\end{Verbatim}
\end{tcolorbox}

\prompt{Out}{outcolor}{88}{}
    
    \begin{equation*} \beta \left(\gamma + 1\right)\end{equation*}

    \begin{tcolorbox}[breakable, size=fbox, boxrule=1pt, pad at break*=1mm,colback=cellbackground, colframe=cellborder]
\prompt{In}{incolor}{89}{\boxspacing}
\begin{Verbatim}[commandchars=\\\{\}]
\PY{n}{epsilon\PYZus{}d2}\PY{o}{.}\PY{n}{replace}\PY{p}{(}\PY{n}{kappa}\PY{p}{,}\PY{l+m+mi}{1}\PY{o}{/}\PY{n}{lamb}\PY{p}{)}\PY{o}{.}\PY{n}{replace}\PY{p}{(}\PY{n}{lp}\PY{p}{,}\PY{l+m+mi}{1}\PY{p}{)}\PY{o}{.}\PY{n}{simplify}\PY{p}{(}\PY{p}{)}
\end{Verbatim}
\end{tcolorbox}

\prompt{Out}{outcolor}{89}{}
    
    \begin{equation*} \beta\end{equation*}

    \begin{tcolorbox}[breakable, size=fbox, boxrule=1pt, pad at break*=1mm,colback=cellbackground, colframe=cellborder]
\prompt{In}{incolor}{90}{\boxspacing}
\begin{Verbatim}[commandchars=\\\{\}]
\PY{n}{sp}\PY{o}{.}\PY{n}{limit}\PY{p}{(}\PY{n}{epsilon\PYZus{}d2}\PY{p}{,} \PY{n}{kappa}\PY{p}{,}\PY{n}{sp}\PY{o}{.}\PY{n}{oo}\PY{p}{)}\PY{o}{.}\PY{n}{simplify}\PY{p}{(}\PY{p}{)}
\end{Verbatim}
\end{tcolorbox}

\prompt{Out}{outcolor}{90}{}
    
    \begin{equation*} \beta \left(1 - \gamma\right)\end{equation*}

    \begin{tcolorbox}[breakable, size=fbox, boxrule=1pt, pad at break*=1mm,colback=cellbackground, colframe=cellborder]
\prompt{In}{incolor}{91}{\boxspacing}
\begin{Verbatim}[commandchars=\\\{\}]
\PY{n}{sp}\PY{o}{.}\PY{n}{limit}\PY{p}{(}\PY{n}{epsilon\PYZus{}d2}\PY{p}{,} \PY{n}{lamb}\PY{p}{,}\PY{n}{sp}\PY{o}{.}\PY{n}{oo}\PY{p}{)}\PY{o}{.}\PY{n}{simplify}\PY{p}{(}\PY{p}{)}
\end{Verbatim}
\end{tcolorbox}

\prompt{Out}{outcolor}{91}{}
    
    \begin{equation*} \beta \left(1 - \gamma\right)\end{equation*}

    \begin{tcolorbox}[breakable, size=fbox, boxrule=1pt, pad at break*=1mm,colback=cellbackground, colframe=cellborder]
\prompt{In}{incolor}{92}{\boxspacing}
\begin{Verbatim}[commandchars=\\\{\}]
\PY{n}{sp}\PY{o}{.}\PY{n}{limit}\PY{p}{(}\PY{n}{epsilon\PYZus{}ok}\PY{o}{\PYZhy{}}\PY{n}{beta}\PY{p}{,} \PY{n}{rho}\PY{p}{,}\PY{n}{sp}\PY{o}{.}\PY{n}{oo}\PY{p}{)}\PY{o}{+}\PY{n}{beta}
\end{Verbatim}
\end{tcolorbox}

\prompt{Out}{outcolor}{92}{}
    
    \begin{equation*} - \frac{\eta \left(\kappa^{2} \lambda^{2} - 1\right)}{2 \left(\kappa^{2} \lambda^{2} + 1\right)} + \beta\end{equation*}

    \begin{tcolorbox}[breakable, size=fbox, boxrule=1pt, pad at break*=1mm,colback=cellbackground, colframe=cellborder]
\prompt{In}{incolor}{93}{\boxspacing}
\begin{Verbatim}[commandchars=\\\{\}]
\PY{n}{sp}\PY{o}{.}\PY{n}{limit}\PY{p}{(}\PY{n}{epsilon\PYZus{}d2}\PY{p}{,} \PY{n}{rho}\PY{p}{,}\PY{l+m+mi}{0}\PY{p}{)}\PY{o}{.}\PY{n}{simplify}\PY{p}{(}\PY{p}{)}
\end{Verbatim}
\end{tcolorbox}

\prompt{Out}{outcolor}{93}{}
    
    \begin{equation*} \beta\end{equation*}

    \hypertarget{simplified-expressions}{%
\paragraph{Simplified expressions}\label{simplified-expressions}}

    \begin{tcolorbox}[breakable, size=fbox, boxrule=1pt, pad at break*=1mm,colback=cellbackground, colframe=cellborder]
\prompt{In}{incolor}{94}{\boxspacing}
\begin{Verbatim}[commandchars=\\\{\}]
\PY{p}{(}\PY{n}{alpha_ad2} \PY{o}{+} \PY{n}{alpha\PYZus{}rd2}\PY{o}{/}\PY{l+m+mi}{2}\PY{o}{*}\PY{p}{(}\PY{n}{epsilon\PYZus{}d2}\PY{o}{*}\PY{n}{theta}\PY{o}{+}\PY{l+m+mi}{1}\PY{p}{)}\PY{p}{)}\PY{o}{.}\PY{n}{simplify}\PY{p}{(}\PY{p}{)}
\end{Verbatim}
\end{tcolorbox}

\prompt{Out}{outcolor}{94}{}
    
    \begin{equation*} 0\end{equation*}

    \begin{tcolorbox}[breakable, size=fbox, boxrule=1pt, pad at break*=1mm,colback=cellbackground, colframe=cellborder]
\prompt{In}{incolor}{95}{\boxspacing}
\begin{Verbatim}[commandchars=\\\{\}]
\PY{n}{alpha_a\PYZus{}s} \PY{o}{=} \PY{o}{\PYZhy{}}\PY{n}{alpha\PYZus{}rd2}\PY{o}{/}\PY{l+m+mi}{2}\PY{o}{*}\PY{p}{(}\PY{n}{epsilon}\PY{o}{*}\PY{n}{theta}\PY{o}{+}\PY{l+m+mi}{1}\PY{p}{)}
\PY{n}{alpha_a\PYZus{}s}
\end{Verbatim}
\end{tcolorbox}

\prompt{Out}{outcolor}{95}{}
    
    \begin{equation*} - \frac{4 \beta \delta \gamma \kappa \lambda \theta \left(\varepsilon \theta + 1\right)}{2 \delta \kappa \lambda + \kappa^{2} \lambda^{2} \rho + \rho}\end{equation*}

    \begin{tcolorbox}[breakable, size=fbox, boxrule=1pt, pad at break*=1mm,colback=cellbackground, colframe=cellborder]
\prompt{In}{incolor}{96}{\boxspacing}
\begin{Verbatim}[commandchars=\\\{\}]
\PY{n}{alpha_a\PYZus{}ok} \PY{o}{=} \PY{n}{to\PYZus{}h\PYZus{}kp}\PY{p}{(}\PY{n}{alpha_a\PYZus{}s}\PY{p}{)}
\PY{n}{alpha_a\PYZus{}ok}
\end{Verbatim}
\end{tcolorbox}

\prompt{Out}{outcolor}{96}{}
    
    \begin{equation*} - \frac{\eta \lambda' \delta \theta \left(\varepsilon \theta + 1\right)}{\lambda' \delta + \rho}\end{equation*}

    \begin{tcolorbox}[breakable, size=fbox, boxrule=1pt, pad at break*=1mm,colback=cellbackground, colframe=cellborder]
\prompt{In}{incolor}{97}{\boxspacing}
\begin{Verbatim}[commandchars=\\\{\}]
\PY{n}{alpha_a\PYZus{}ok}\PY{o}{/}\PY{n}{alpha\PYZus{}r\PYZus{}ok}
\end{Verbatim}
\end{tcolorbox}

\prompt{Out}{outcolor}{97}{}
    
    \begin{equation*} - \frac{\varepsilon \theta}{2} - \frac{1}{2}\end{equation*}

    \begin{tcolorbox}[breakable, size=fbox, boxrule=1pt, pad at break*=1mm,colback=cellbackground, colframe=cellborder]
\prompt{In}{incolor}{98}{\boxspacing}
\begin{Verbatim}[commandchars=\\\{\}]
\PY{p}{(}\PY{n}{alpha_bd2} \PY{o}{\PYZhy{}}  \PY{n}{alpha\PYZus{}rd2}\PY{o}{/}\PY{l+m+mi}{2}\PY{o}{*}\PY{p}{(}\PY{l+m+mi}{1}\PY{o}{\PYZhy{}}\PY{n}{epsilon\PYZus{}d2}\PY{o}{*}\PY{n}{theta}\PY{p}{)}\PY{p}{)}\PY{o}{.}\PY{n}{simplify}\PY{p}{(}\PY{p}{)}
\end{Verbatim}
\end{tcolorbox}

\prompt{Out}{outcolor}{98}{}
    
    \begin{equation*} 0\end{equation*}

    \begin{tcolorbox}[breakable, size=fbox, boxrule=1pt, pad at break*=1mm,colback=cellbackground, colframe=cellborder]
\prompt{In}{incolor}{99}{\boxspacing}
\begin{Verbatim}[commandchars=\\\{\}]
\PY{n}{alpha_b\PYZus{}s} \PY{o}{=} \PY{n}{alpha\PYZus{}rd2}\PY{o}{/}\PY{l+m+mi}{2}\PY{o}{*}\PY{p}{(}\PY{l+m+mi}{1}\PY{o}{\PYZhy{}}\PY{n}{epsilon}\PY{o}{*}\PY{n}{theta}\PY{p}{)}
\PY{n}{alpha_b\PYZus{}s}
\end{Verbatim}
\end{tcolorbox}

\prompt{Out}{outcolor}{99}{}
    
    \begin{equation*} \frac{4 \beta \delta \gamma \kappa \lambda \theta \left(- \varepsilon \theta + 1\right)}{2 \delta \kappa \lambda + \kappa^{2} \lambda^{2} \rho + \rho}\end{equation*}

    \begin{tcolorbox}[breakable, size=fbox, boxrule=1pt, pad at break*=1mm,colback=cellbackground, colframe=cellborder]
\prompt{In}{incolor}{100}{\boxspacing}
\begin{Verbatim}[commandchars=\\\{\}]
\PY{n}{alpha_b\PYZus{}ok} \PY{o}{=} \PY{n}{to\PYZus{}h\PYZus{}kp}\PY{p}{(}\PY{n}{alpha_b\PYZus{}s}\PY{p}{)}
\PY{n}{alpha_b\PYZus{}ok}
\end{Verbatim}
\end{tcolorbox}

\prompt{Out}{outcolor}{100}{}
    
    \begin{equation*} \frac{\eta \lambda' \delta \theta \left(- \varepsilon \theta + 1\right)}{\lambda' \delta + \rho}\end{equation*}

    \begin{tcolorbox}[breakable, size=fbox, boxrule=1pt, pad at break*=1mm,colback=cellbackground, colframe=cellborder]
\prompt{In}{incolor}{101}{\boxspacing}
\begin{Verbatim}[commandchars=\\\{\}]
\PY{n}{alpha_b\PYZus{}ok}\PY{o}{/}\PY{n}{alpha\PYZus{}r\PYZus{}ok}
\end{Verbatim}
\end{tcolorbox}

\prompt{Out}{outcolor}{101}{}
    
    \begin{equation*} - \frac{\varepsilon \theta}{2} + \frac{1}{2}\end{equation*}

    \hypertarget{diffusion-of-u_1}{%
\paragraph{\texorpdfstring{Diffusion of
\(U_1\)}{Diffusion of U\_1}}\label{diffusion-of-u_1}}

    \begin{tcolorbox}[breakable, size=fbox, boxrule=1pt, pad at break*=1mm,colback=cellbackground, colframe=cellborder]
\prompt{In}{incolor}{102}{\boxspacing}
\begin{Verbatim}[commandchars=\\\{\}]
\PY{n}{KQ\PYZus{}diff1d} \PY{o}{=} \PY{n}{replaced}\PY{p}{(}\PY{n}{KQ\PYZus{}diff1}\PY{p}{)}
\PY{n}{KQ\PYZus{}diff1d}
\end{Verbatim}
\end{tcolorbox}

\prompt{Out}{outcolor}{102}{}
    
\begin{equation*}
  \frac{
    \begin{lgathered}
      \rho^{2} + 2 \kappa^{2} \lambda^{2} \rho^{2} + \kappa^{4} \lambda^{4} \rho^{2} + 4 \delta \kappa \lambda \rho + 4 \delta \kappa^{3} \lambda^{3} \rho + 4 \delta^{2}
      \kappa^{2} \lambda^{2}\\
      \quad  + 4 \beta \gamma \kappa^{2} \lambda^{2} \rho^{2} \theta + 4 \beta \gamma \kappa^{4} \lambda^{4} \rho^{2} \theta + 8 \beta \delta \gamma \kappa^{3} \lambda^{3} \rho \theta + 8 \beta^{2} \gamma^{2} \kappa^{4} \lambda^{4} \rho^{2}
      \theta^{2} + O\left(\theta^{3}\right)
    \end{lgathered}
  }{\left(2 \delta \kappa \lambda + \kappa^{2} \lambda^{2} \rho + \rho\right)^{2}}
\end{equation*}

    \begin{tcolorbox}[breakable, size=fbox, boxrule=1pt, pad at break*=1mm,colback=cellbackground, colframe=cellborder]
\prompt{In}{incolor}{103}{\boxspacing}
\begin{Verbatim}[commandchars=\\\{\}]
\PY{n}{alpha\PYZus{}diff1d} \PY{o}{=} \PY{n}{develop}\PY{p}{(}\PY{n}{sp}\PY{o}{.}\PY{n}{log}\PY{p}{(}\PY{n}{KQ\PYZus{}diff1d}\PY{p}{)}\PY{p}{)}\PY{o}{.}\PY{n}{simplify}\PY{p}{(}\PY{p}{)}
\PY{n}{alpha\PYZus{}diff1d}
\end{Verbatim}
\end{tcolorbox}

\prompt{Out}{outcolor}{103}{}
    
\begin{equation*}
  \frac{
    \begin{lgathered}
      4 \beta \gamma \kappa^{2} \lambda^{2} \rho \theta \left(2 \delta \kappa \lambda + \kappa^{2} \lambda^{2} \rho + \rho\right) \left(4 \delta^{2} \kappa^{2} \lambda^{2} + 4 \delta
        \kappa^{3} \lambda^{3} \rho + 4 \delta \kappa \lambda \rho + \kappa^{4} \lambda^{4} \rho^{2} + 2 \kappa^{2} \lambda^{2} \rho^{2} +
        \rho^{2}\right)^{3} \\
      \quad + 8 \beta^{2} \gamma^{2} \kappa^{4} \lambda^{4} \rho^{2} \theta^{2} \left(4 \delta^{2} \kappa^{2} \lambda^{2} + 4 \delta \kappa^{3} \lambda^{3} \rho + 4 \delta
        \kappa
        \lambda \rho + \kappa^{4} \lambda^{4} \rho^{2} + 2 \kappa^{2} \lambda^{2} \rho^{2} + \rho^{2}\right)^{2} \\
      \qquad \cdot \left(4 \delta^{2} \kappa^{2} \lambda^{2} + 4 \delta \kappa^{3} \lambda^{3} \rho + 4 \delta \kappa \lambda \rho + \kappa^{4} \lambda^{4} \rho^{2} + 2 \kappa^{2}
        \lambda^{2} \rho^{2} + \rho^{2} - \left(2 \delta \kappa \lambda + \kappa^{2} \lambda^{2} \rho + \rho\right)^{2}\right) +
      O\left(\theta^{3}\right)
    \end{lgathered}
  }{\left(4 \delta^{2} \kappa^{2} \lambda^{2} + 4 \delta \kappa^{3} \lambda^{3} \rho + 4 \delta \kappa \lambda \rho + \kappa^{4} \lambda^{4} \rho^{2} + 2 \kappa^{2} \lambda^{2} \rho^{2} + \rho^{2}\right)^{4}}
\end{equation*}

    \begin{tcolorbox}[breakable, size=fbox, boxrule=1pt, pad at break*=1mm,colback=cellbackground, colframe=cellborder]
\prompt{In}{incolor}{104}{\boxspacing}
\begin{Verbatim}[commandchars=\\\{\}]
\PY{n}{alpha\PYZus{}diff1d2}\PY{o}{=}\PY{n}{full\PYZus{}removeO}\PY{p}{(}\PY{n}{alpha\PYZus{}diff1d}\PY{p}{)}
\PY{n}{alpha\PYZus{}diff1d2}
\end{Verbatim}
\end{tcolorbox}

\prompt{Out}{outcolor}{104}{}
    
    \begin{equation*} \frac{4 \beta \gamma \kappa^{2} \lambda^{2} \rho \theta}{2 \delta \kappa \lambda + \kappa^{2} \lambda^{2} \rho + \rho}\end{equation*}

    \begin{tcolorbox}[breakable, size=fbox, boxrule=1pt, pad at break*=1mm,colback=cellbackground, colframe=cellborder]
\prompt{In}{incolor}{105}{\boxspacing}
\begin{Verbatim}[commandchars=\\\{\}]
\PY{n}{alpha\PYZus{}diff1\PYZus{}ok} \PY{o}{=} \PY{n}{to\PYZus{}h\PYZus{}kp}\PY{p}{(}\PY{n}{alpha\PYZus{}diff1d2}\PY{o}{/}\PY{n}{kappa}\PY{p}{)}\PY{o}{*}\PY{n}{kappa}
\PY{n}{alpha\PYZus{}diff1\PYZus{}ok}
\end{Verbatim}
\end{tcolorbox}

\prompt{Out}{outcolor}{105}{}
    
    \begin{equation*} \frac{\eta \lambda' \kappa \lambda \rho \theta}{\lambda' \delta + \rho}\end{equation*}

    \hypertarget{diffusion-of-u_2}{%
\paragraph{\texorpdfstring{Diffusion of
\(U_2\)}{Diffusion of U\_2}}\label{diffusion-of-u_2}}

    \begin{tcolorbox}[breakable, size=fbox, boxrule=1pt, pad at break*=1mm,colback=cellbackground, colframe=cellborder]
\prompt{In}{incolor}{106}{\boxspacing}
\begin{Verbatim}[commandchars=\\\{\}]
\PY{n}{KQ\PYZus{}diff2d} \PY{o}{=} \PY{n}{replaced}\PY{p}{(}\PY{n}{KQ\PYZus{}diff2}\PY{p}{)}
\PY{n}{KQ\PYZus{}diff2d}
\end{Verbatim}
\end{tcolorbox}

\prompt{Out}{outcolor}{106}{}
    
\begin{equation*}
  \frac{
    \begin{lgathered}
      \rho^{2} + 2 \kappa^{2} \lambda^{2} \rho^{2} + \kappa^{4} \lambda^{4} \rho^{2} + 4 \delta \kappa \lambda \rho + 4 \delta \kappa^{3} \lambda^{3} \rho + 4 \delta^{2}
      \kappa^{2} \lambda^{2} \\
      \quad- 4 \beta \gamma \rho^{2} \theta - 4 \beta \gamma \kappa^{2} \lambda^{2} \rho^{2} \theta - 8 \beta \delta \gamma \kappa \lambda \rho \theta + 8 \beta^{2} \gamma^{2} \rho^{2} \theta^{2} +
      O\left(\theta^{3}\right)
    \end{lgathered}
  }{\left(2 \delta \kappa \lambda + \kappa^{2} \lambda^{2} \rho + \rho\right)^{2}}
\end{equation*}

    \begin{tcolorbox}[breakable, size=fbox, boxrule=1pt, pad at break*=1mm,colback=cellbackground, colframe=cellborder]
\prompt{In}{incolor}{107}{\boxspacing}
\begin{Verbatim}[commandchars=\\\{\}]
\PY{n}{alpha\PYZus{}diff2d} \PY{o}{=} \PY{n}{develop}\PY{p}{(}\PY{n}{sp}\PY{o}{.}\PY{n}{log}\PY{p}{(}\PY{n}{KQ\PYZus{}diff2d}\PY{p}{)}\PY{p}{)}\PY{o}{.}\PY{n}{simplify}\PY{p}{(}\PY{p}{)}
\PY{n}{alpha\PYZus{}diff2d}
\end{Verbatim}
\end{tcolorbox}

\prompt{Out}{outcolor}{107}{}
    
\begin{equation*}
  \frac{
    \begin{lgathered}
      - 4 \beta \gamma \rho \theta \left(2 \delta \kappa \lambda + \kappa^{2} \lambda^{2} \rho + \rho\right) \left(4 \delta^{2} \kappa^{2} \lambda^{2} + 4 \delta \kappa^{3}
        \lambda^{3} \rho + 4 \delta \kappa \lambda \rho + \kappa^{4} \lambda^{4} \rho^{2} + 2 \kappa^{2} \lambda^{2} \rho^{2} +
        \rho^{2}\right)^{3} \\
      \quad + 8 \beta^{2} \gamma^{2} \rho^{2} \theta^{2} \left(4 \delta^{2} \kappa^{2} \lambda^{2} + 4 \delta \kappa^{3} \lambda^{3} \rho + 4 \delta \kappa \lambda \rho +
        \kappa^{4} \lambda^{4} \rho^{2} + 2 \kappa^{2} \lambda^{2} \rho^{2} +
        \rho^{2}\right)^{2} \\
      \qquad \cdot \left(4 \delta^{2} \kappa^{2} \lambda^{2} + 4 \delta \kappa^{3} \lambda^{3} \rho + 4 \delta \kappa \lambda \rho + \kappa^{4} \lambda^{4} \rho^{2} + 2 \kappa^{2}
        \lambda^{2} \rho^{2} + \rho^{2} - \left(2 \delta \kappa \lambda + \kappa^{2} \lambda^{2} \rho + \rho\right)^{2}\right) +
      O\left(\theta^{3}\right)
    \end{lgathered}
  }{\left(4 \delta^{2} \kappa^{2} \lambda^{2} + 4 \delta \kappa^{3} \lambda^{3} \rho + 4 \delta \kappa \lambda \rho + \kappa^{4} \lambda^{4} \rho^{2} + 2 \kappa^{2} \lambda^{2} \rho^{2} + \rho^{2}\right)^{4}}
\end{equation*}

    \begin{tcolorbox}[breakable, size=fbox, boxrule=1pt, pad at break*=1mm,colback=cellbackground, colframe=cellborder]
\prompt{In}{incolor}{108}{\boxspacing}
\begin{Verbatim}[commandchars=\\\{\}]
\PY{n}{alpha\PYZus{}diff2d2}\PY{o}{=}\PY{n}{full\PYZus{}removeO}\PY{p}{(}\PY{n}{alpha\PYZus{}diff2d}\PY{p}{)}
\PY{n}{alpha\PYZus{}diff2d2}
\end{Verbatim}
\end{tcolorbox}

\prompt{Out}{outcolor}{108}{}
    
    \begin{equation*} - \frac{4 \beta \gamma \rho \theta}{2 \delta \kappa \lambda + \kappa^{2} \lambda^{2} \rho + \rho}\end{equation*}

    \begin{tcolorbox}[breakable, size=fbox, boxrule=1pt, pad at break*=1mm,colback=cellbackground, colframe=cellborder]
\prompt{In}{incolor}{109}{\boxspacing}
\begin{Verbatim}[commandchars=\\\{\}]
\PY{n}{alpha\PYZus{}diff2\PYZus{}ok} \PY{o}{=} \PY{n}{to\PYZus{}h\PYZus{}kp}\PY{p}{(}\PY{n}{alpha\PYZus{}diff2d2}\PY{o}{*}\PY{n}{kappa}\PY{o}{*}\PY{n}{lamb}\PY{o}{*}\PY{l+m+mi}{4}\PY{p}{)}\PY{o}{/}\PY{n}{kappa}\PY{o}{/}\PY{n}{lamb}\PY{o}{/}\PY{l+m+mi}{4}
\PY{n}{alpha\PYZus{}diff2\PYZus{}ok}
\end{Verbatim}
\end{tcolorbox}

\prompt{Out}{outcolor}{109}{}
    
    \begin{equation*} - \frac{\eta \lambda' \rho \theta}{\kappa \lambda \left(\lambda' \delta + \rho\right)}\end{equation*}

    \hypertarget{total-diffusion}{%
\paragraph{Total diffusion}\label{total-diffusion}}

    \begin{tcolorbox}[breakable, size=fbox, boxrule=1pt, pad at break*=1mm,colback=cellbackground, colframe=cellborder]
\prompt{In}{incolor}{110}{\boxspacing}
\begin{Verbatim}[commandchars=\\\{\}]
\PY{n}{alpha\PYZus{}diffd} \PY{o}{=} \PY{p}{(}\PY{n}{alpha\PYZus{}diff1d}\PY{o}{\PYZhy{}}\PY{n}{alpha\PYZus{}diff2d}\PY{p}{)}\PY{o}{.}\PY{n}{simplify}\PY{p}{(}\PY{p}{)}
\PY{n}{alpha\PYZus{}diffd}
\end{Verbatim}
\end{tcolorbox}

\prompt{Out}{outcolor}{110}{}
    
\begin{equation*}
  \frac{
    \begin{lgathered}
      4 \beta \gamma \rho \theta \left(2 \delta \kappa \lambda + \kappa^{2} \lambda^{2} \rho + \rho\right) \left(4 \delta^{2} \kappa^{2} \lambda^{2} + 4 \delta \kappa^{3} \lambda^{3}
        \rho + 4 \delta \kappa \lambda \rho + \kappa^{4} \lambda^{4} \rho^{2} + 2 \kappa^{2} \lambda^{2} \rho^{2} +   \rho^{2}\right)^{3}  \\
      \quad + 4 \beta \gamma \kappa^{2} \lambda^{2} \rho \theta \left(2 \delta \kappa \lambda + \kappa^{2} \lambda^{2} \rho + \rho\right) \left(4 \delta^{2} \kappa^{2} \lambda^{2} +
        4 \delta \kappa^{3} \lambda^{3} \rho + 4 \delta \kappa \lambda \rho + \kappa^{4} \lambda^{4} \rho^{2} + 2 \kappa^{2} \lambda^{2} \rho^{2} + \rho^{2}\right)^{3}
      \\
      \quad + 8 \beta^{2} \gamma^{2} \rho^{2} \theta^{2} \left(4 \delta^{2} \kappa^{2} \lambda^{2} + 4 \delta \kappa^{3} \lambda^{3} \rho + 4 \delta \kappa \lambda \rho + \kappa^{4}
        \lambda^{4} \rho^{2} + 2 \kappa^{2} \lambda^{2} \rho^{2} + \rho^{2}\right)^{2} \\
      \qquad\cdot\left(- 4 \delta^{2} \kappa^{2} \lambda^{2} - 4 \delta \kappa^{3}
        \lambda^{3} \rho - 4 \delta \kappa \lambda \rho - \kappa^{4} \lambda^{4} \rho^{2} - 2 \kappa^{2} \lambda^{2} \rho^{2} - \rho^{2} + \left(2 \delta \kappa \lambda +
          \kappa^{2} \lambda^{2} \rho + \rho\right)^{2}\right) \\
      \quad + 8 \beta^{2} \gamma^{2} \kappa^{4} \lambda^{4} \rho^{2} \theta^{2} \left(4 \delta^{2}
        \kappa^{2} \lambda^{2} + 4 \delta \kappa^{3} \lambda^{3} \rho + 4 \delta \kappa \lambda \rho + \kappa^{4} \lambda^{4} \rho^{2} + 2 \kappa^{2} \lambda^{2} \rho^{2} +
        \rho^{2}\right)^{2} \\
      \qquad\cdot \left(4 \delta^{2} \kappa^{2} \lambda^{2} + 4 \delta \kappa^{3} \lambda^{3} \rho + 4 \delta \kappa \lambda \rho + \kappa^{4} \lambda^{4}
        \rho^{2} + 2 \kappa^{2} \lambda^{2} \rho^{2} + \rho^{2} - \left(2 \delta \kappa \lambda + \kappa^{2} \lambda^{2} \rho + \rho\right)^{2}\right) +
      O\left(\theta^{3}\right)
    \end{lgathered}
}{\left(4 \delta^{2} \kappa^{2} \lambda^{2} + 4 \delta \kappa^{3} \lambda^{3} \rho + 4 \delta \kappa \lambda \rho + \kappa^{4} \lambda^{4} \rho^{2} + 2 \kappa^{2} \lambda^{2} \rho^{2} + \rho^{2}\right)^{4}}
  \end{equation*}

    \begin{tcolorbox}[breakable, size=fbox, boxrule=1pt, pad at break*=1mm,colback=cellbackground, colframe=cellborder]
\prompt{In}{incolor}{111}{\boxspacing}
\begin{Verbatim}[commandchars=\\\{\}]
\PY{n}{alpha\PYZus{}diffd2} \PY{o}{=} \PY{p}{(}\PY{n}{alpha\PYZus{}diff1d2}\PY{o}{\PYZhy{}}\PY{n}{alpha\PYZus{}diff2d2}\PY{p}{)}\PY{o}{.}\PY{n}{simplify}\PY{p}{(}\PY{p}{)}
\PY{n}{alpha\PYZus{}diffd2}
\end{Verbatim}
\end{tcolorbox}

\prompt{Out}{outcolor}{111}{}
    
    \begin{equation*} \frac{4 \beta \gamma \rho \theta \left(\kappa^{2} \lambda^{2} + 1\right)}{2 \delta \kappa \lambda + \kappa^{2} \lambda^{2} \rho + \rho}\end{equation*}

    \begin{tcolorbox}[breakable, size=fbox, boxrule=1pt, pad at break*=1mm,colback=cellbackground, colframe=cellborder]
\prompt{In}{incolor}{112}{\boxspacing}
\begin{Verbatim}[commandchars=\\\{\}]
\PY{n}{alpha\PYZus{}diff\PYZus{}ok} \PY{o}{=} \PY{n}{to\PYZus{}h\PYZus{}kp}\PY{p}{(}\PY{n}{alpha\PYZus{}diffd2}\PY{p}{)}
\PY{n}{alpha\PYZus{}diff\PYZus{}ok}
\end{Verbatim}
\end{tcolorbox}

\prompt{Out}{outcolor}{112}{}
    
    \begin{equation*} \frac{2 \eta \rho \theta}{\lambda' \delta + \rho}\end{equation*}

    \begin{tcolorbox}[breakable, size=fbox, boxrule=1pt, pad at break*=1mm,colback=cellbackground, colframe=cellborder]
\prompt{In}{incolor}{113}{\boxspacing}
\begin{Verbatim}[commandchars=\\\{\}]
\PY{n}{alpha\PYZus{}diff1\PYZus{}ok}\PY{o}{/}\PY{n}{alpha\PYZus{}diff\PYZus{}ok}
\end{Verbatim}
\end{tcolorbox}

\prompt{Out}{outcolor}{113}{}
    
    \begin{equation*} \frac{\lambda' \kappa \lambda}{2}\end{equation*}

    \begin{tcolorbox}[breakable, size=fbox, boxrule=1pt, pad at break*=1mm,colback=cellbackground, colframe=cellborder]
\prompt{In}{incolor}{114}{\boxspacing}
\begin{Verbatim}[commandchars=\\\{\}]
\PY{n}{alpha\PYZus{}diff2\PYZus{}ok}\PY{o}{/}\PY{n}{alpha\PYZus{}diff\PYZus{}ok}
\end{Verbatim}
\end{tcolorbox}

\prompt{Out}{outcolor}{114}{}
    
    \begin{equation*} - \frac{\lambda'}{2 \kappa \lambda}\end{equation*}

    \hypertarget{total-alpha}{%
\paragraph{\texorpdfstring{Total
\(\alpha\)}{Total \textbackslash alpha}}\label{total-alpha}}

    \begin{tcolorbox}[breakable, size=fbox, boxrule=1pt, pad at break*=1mm,colback=cellbackground, colframe=cellborder]
\prompt{In}{incolor}{115}{\boxspacing}
\begin{Verbatim}[commandchars=\\\{\}]
\PY{n}{alpha\PYZus{}d} \PY{o}{=} \PY{p}{(}\PY{n}{alpha\PYZus{}rd}\PY{o}{+}\PY{n}{alpha\PYZus{}diffd}\PY{p}{)}\PY{o}{.}\PY{n}{simplify}\PY{p}{(}\PY{p}{)}
\PY{n}{alpha\PYZus{}d}
\end{Verbatim}
\end{tcolorbox}

\prompt{Out}{outcolor}{115}{}
    
\begin{equation*}
  \frac{
    \begin{lgathered}
      4 \beta \gamma \rho \theta \left(2 \delta \kappa \lambda + \kappa^{2} \lambda^{2} \rho + \rho\right) \left(4 \delta^{2} \kappa^{2} \lambda^{2} + 4 \delta \kappa^{3}
        \lambda^{3} \rho + 4 \delta \kappa \lambda \rho + \kappa^{4} \lambda^{4} \rho^{2} + 2 \kappa^{2} \lambda^{2} \rho^{2} + \rho^{2}\right)^{3} \\
      \quad + 4 \beta \gamma \kappa^{2} \lambda^{2} \rho \theta \left(2 \delta \kappa \lambda + \kappa^{2} \lambda^{2} \rho + \rho\right) \left(4 \delta^{2} \kappa^{2} \lambda^{2} +
        4 \delta
        \kappa^{3} \lambda^{3} \rho + 4 \delta \kappa \lambda \rho + \kappa^{4} \lambda^{4} \rho^{2} + 2 \kappa^{2} \lambda^{2} \rho^{2} + \rho^{2}\right)^{3} \\
      \quad + 8 \beta \delta \gamma \kappa \lambda \theta \left(2 \delta \kappa \lambda + \kappa^{2} \lambda^{2} \rho + \rho\right) \left(4 \delta^{2} \kappa^{2} \lambda^{2} + 4 \delta
        \kappa^{3} \lambda^{3} \rho + 4 \delta \kappa \lambda \rho + \kappa^{4} \lambda^{4} \rho^{2} + 2 \kappa^{2} \lambda^{2} \rho^{2} + \rho^{2}\right)^{3} \\
      \quad + 8 \beta^{2} \gamma^{2} \rho^{2} \theta^{2} \left(4 \delta^{2} \kappa^{2} \lambda^{2} + 4 \delta \kappa^{3} \lambda^{3} \rho + 4 \delta \kappa \lambda \rho +
        \kappa^{4} \lambda^{4} \rho^{2} + 2 \kappa^{2} \lambda^{2} \rho^{2} + \rho^{2}\right)^{2} \\
      \qquad \cdot \left(- 4 \delta^{2} \kappa^{2} \lambda^{2} - 4
        \delta \kappa^{3} \lambda^{3} \rho - 4 \delta \kappa \lambda \rho - \kappa^{4} \lambda^{4} \rho^{2} - 2 \kappa^{2} \lambda^{2} \rho^{2} - \rho^{2} + \left(2 \delta
          \kappa \lambda + \kappa^{2} \lambda^{2} \rho + \rho\right)^{2}\right) \\
      \quad + 8 \beta^{2} \gamma^{2} \kappa^{4} \lambda^{4} \rho^{2} \theta^{2} \left(4 \delta^{2} \kappa^{2} \lambda^{2} + 4 \delta \kappa^{3} \lambda^{3} \rho + 4 \delta
        \kappa \lambda \rho + \kappa^{4} \lambda^{4} \rho^{2} + 2 \kappa^{2} \lambda^{2} \rho^{2} + \rho^{2}\right)^{2} \\
      \qquad \cdot \left(4 \delta^{2} \kappa^{2}
        \lambda^{2} + 4 \delta \kappa^{3} \lambda^{3} \rho + 4 \delta \kappa \lambda \rho + \kappa^{4} \lambda^{4} \rho^{2} + 2 \kappa^{2} \lambda^{2} \rho^{2} + \rho^{2} -
        \left(2 \delta \kappa \lambda + \kappa^{2} \lambda^{2} \rho + \rho\right)^{2}\right) \\
      + O\left(\theta^{3}\right)
    \end{lgathered}
  }{\left(4 \delta^{2} \kappa^{2} \lambda^{2} + 4 \delta \kappa^{3} \lambda^{3} \rho + 4 \delta \kappa \lambda \rho + \kappa^{4} \lambda^{4} \rho^{2} + 2 \kappa^{2} \lambda^{2} \rho^{2} + \rho^{2}\right)^{4}}
\end{equation*}

    \begin{tcolorbox}[breakable, size=fbox, boxrule=1pt, pad at break*=1mm,colback=cellbackground, colframe=cellborder]
\prompt{In}{incolor}{116}{\boxspacing}
\begin{Verbatim}[commandchars=\\\{\}]
\PY{n}{alpha\PYZus{}d2} \PY{o}{=} \PY{p}{(}\PY{n}{alpha\PYZus{}rd2}\PY{o}{+}\PY{n}{alpha\PYZus{}diffd2}\PY{p}{)}\PY{o}{.}\PY{n}{simplify}\PY{p}{(}\PY{p}{)}
\PY{n}{alpha\PYZus{}d2}
\end{Verbatim}
\end{tcolorbox}

\prompt{Out}{outcolor}{116}{}
    
    \begin{equation*} 4 \beta \gamma \theta\end{equation*}

    \begin{tcolorbox}[breakable, size=fbox, boxrule=1pt, pad at break*=1mm,colback=cellbackground, colframe=cellborder]
\prompt{In}{incolor}{117}{\boxspacing}
\begin{Verbatim}[commandchars=\\\{\}]
\PY{n}{alpha\PYZus{}ok} \PY{o}{=} \PY{n}{to\PYZus{}h\PYZus{}kp}\PY{p}{(}\PY{n}{alpha\PYZus{}d2}\PY{p}{)}
\PY{n}{alpha\PYZus{}ok}
\end{Verbatim}
\end{tcolorbox}

\prompt{Out}{outcolor}{117}{}
    
    \begin{equation*} 2 \eta \theta\end{equation*}

    \hypertarget{ratios}{%
\paragraph{Ratios}\label{ratios}}

    \begin{tcolorbox}[breakable, size=fbox, boxrule=1pt, pad at break*=1mm,colback=cellbackground, colframe=cellborder]
\prompt{In}{incolor}{118}{\boxspacing}
\begin{Verbatim}[commandchars=\\\{\}]
\PY{n}{alpha\PYZus{}diff\PYZus{}ok}\PY{o}{/}\PY{n}{alpha\PYZus{}ok}
\end{Verbatim}
\end{tcolorbox}

\prompt{Out}{outcolor}{118}{}
    
    \begin{equation*} \frac{\rho}{\lambda' \delta + \rho}\end{equation*}

    \begin{tcolorbox}[breakable, size=fbox, boxrule=1pt, pad at break*=1mm,colback=cellbackground, colframe=cellborder]
\prompt{In}{incolor}{119}{\boxspacing}
\begin{Verbatim}[commandchars=\\\{\}]
\PY{n}{alpha\PYZus{}r\PYZus{}ok}\PY{o}{/}\PY{n}{alpha\PYZus{}ok}
\end{Verbatim}
\end{tcolorbox}

\prompt{Out}{outcolor}{119}{}
    
    \begin{equation*} \frac{\lambda' \delta}{\lambda' \delta + \rho}\end{equation*}

    \begin{tcolorbox}[breakable, size=fbox, boxrule=1pt, pad at break*=1mm,colback=cellbackground, colframe=cellborder]
\prompt{In}{incolor}{120}{\boxspacing}
\begin{Verbatim}[commandchars=\\\{\}]
\PY{p}{(}\PY{n}{alpha_a\PYZus{}ok}\PY{o}{/}\PY{n}{alpha\PYZus{}r\PYZus{}ok}\PY{p}{)}\PY{o}{.}\PY{n}{factor}\PY{p}{(}\PY{p}{)}
\end{Verbatim}
\end{tcolorbox}

\prompt{Out}{outcolor}{120}{}
    
    \begin{equation*} - \frac{\varepsilon \theta + 1}{2}\end{equation*}

    \begin{tcolorbox}[breakable, size=fbox, boxrule=1pt, pad at break*=1mm,colback=cellbackground, colframe=cellborder]
\prompt{In}{incolor}{121}{\boxspacing}
\begin{Verbatim}[commandchars=\\\{\}]
\PY{p}{(}\PY{n}{alpha_b\PYZus{}ok}\PY{o}{/}\PY{n}{alpha\PYZus{}r\PYZus{}ok}\PY{p}{)}\PY{o}{.}\PY{n}{factor}\PY{p}{(}\PY{p}{)}
\end{Verbatim}
\end{tcolorbox}

\prompt{Out}{outcolor}{121}{}
    
    \begin{equation*} - \frac{\varepsilon \theta - 1}{2}\end{equation*}

    \begin{tcolorbox}[breakable, size=fbox, boxrule=1pt, pad at break*=1mm,colback=cellbackground, colframe=cellborder]
\prompt{In}{incolor}{122}{\boxspacing}
\begin{Verbatim}[commandchars=\\\{\}]
\PY{p}{(}\PY{n}{alpha\PYZus{}diff1\PYZus{}ok}\PY{o}{/}\PY{n}{alpha\PYZus{}diff\PYZus{}ok}\PY{p}{)}\PY{o}{.}\PY{n}{simplify}\PY{p}{(}\PY{p}{)}
\end{Verbatim}
\end{tcolorbox}

\prompt{Out}{outcolor}{122}{}
    
    \begin{equation*} \frac{\lambda' \kappa \lambda}{2}\end{equation*}

    \begin{tcolorbox}[breakable, size=fbox, boxrule=1pt, pad at break*=1mm,colback=cellbackground, colframe=cellborder]
\prompt{In}{incolor}{123}{\boxspacing}
\begin{Verbatim}[commandchars=\\\{\}]
\PY{p}{(}\PY{o}{\PYZhy{}}\PY{n}{alpha\PYZus{}diff2\PYZus{}ok}\PY{o}{/}\PY{n}{alpha\PYZus{}diff\PYZus{}ok}\PY{p}{)}\PY{o}{.}\PY{n}{simplify}\PY{p}{(}\PY{p}{)}
\end{Verbatim}
\end{tcolorbox}

\prompt{Out}{outcolor}{123}{}
    
    \begin{equation*} \frac{\lambda'}{2 \kappa \lambda}\end{equation*}

    \begin{tcolorbox}[breakable, size=fbox, boxrule=1pt, pad at break*=1mm,colback=cellbackground, colframe=cellborder]
\prompt{In}{incolor}{124}{\boxspacing}
\begin{Verbatim}[commandchars=\\\{\}]
\PY{n}{alpha\PYZus{}r\PYZus{}ok}\PY{o}{/}\PY{n}{alpha\PYZus{}diff\PYZus{}ok}
\end{Verbatim}
\end{tcolorbox}

\prompt{Out}{outcolor}{124}{}
    
    \begin{equation*} \frac{\lambda' \delta}{\rho}\end{equation*}

    \hypertarget{chemical-resistance-ralphavarphi}{%
\subsubsection{\texorpdfstring{Chemical resistance
\(R=\alpha/\varphi\)}{Chemical resistance R=\textbackslash alpha/\textbackslash varphi}}\label{chemical-resistance-ralphavarphi}}

    \begin{tcolorbox}[breakable, size=fbox, boxrule=1pt, pad at break*=1mm,colback=cellbackground, colframe=cellborder]
\prompt{In}{incolor}{125}{\boxspacing}
\begin{Verbatim}[commandchars=\\\{\}]
\PY{n}{r\PYZus{}tot} \PY{o}{=} \PY{n}{alpha\PYZus{}ok}\PY{o}{/}\PY{n}{phi\PYZus{}ok}
\PY{n}{r\PYZus{}tot}
\end{Verbatim}
\end{tcolorbox}

\prompt{Out}{outcolor}{125}{}
    
    \begin{equation*} \frac{4 \left(\lambda' \delta + \rho\right)}{\kappa' \lambda' \delta \rho}\end{equation*}

    \begin{tcolorbox}[breakable, size=fbox, boxrule=1pt, pad at break*=1mm,colback=cellbackground, colframe=cellborder]
\prompt{In}{incolor}{126}{\boxspacing}
\begin{Verbatim}[commandchars=\\\{\}]
\PY{n}{r\PYZus{}r} \PY{o}{=} \PY{n}{alpha\PYZus{}r\PYZus{}ok}\PY{o}{/}\PY{n}{phi\PYZus{}ok}
\PY{n}{r\PYZus{}r}
\end{Verbatim}
\end{tcolorbox}

\prompt{Out}{outcolor}{126}{}
    
    \begin{equation*} \frac{4}{\kappa' \rho}\end{equation*}

    \begin{tcolorbox}[breakable, size=fbox, boxrule=1pt, pad at break*=1mm,colback=cellbackground, colframe=cellborder]
\prompt{In}{incolor}{127}{\boxspacing}
\begin{Verbatim}[commandchars=\\\{\}]
\PY{n}{r\PYZus{}diff} \PY{o}{=} \PY{n}{alpha\PYZus{}diff\PYZus{}ok}\PY{o}{/}\PY{n}{phi\PYZus{}ok}
\PY{n}{r\PYZus{}diff}
\end{Verbatim}
\end{tcolorbox}

\prompt{Out}{outcolor}{127}{}
    
    \begin{equation*} \frac{4}{\kappa' \lambda' \delta}\end{equation*}

    \begin{tcolorbox}[breakable, size=fbox, boxrule=1pt, pad at break*=1mm,colback=cellbackground, colframe=cellborder]
\prompt{In}{incolor}{128}{\boxspacing}
\begin{Verbatim}[commandchars=\\\{\}]
\PY{n}{r_a} \PY{o}{=} \PY{o}{\PYZhy{}}\PY{n}{alpha_a\PYZus{}ok}\PY{o}{/}\PY{n}{phi\PYZus{}ok}
\PY{n}{r_a}
\end{Verbatim}
\end{tcolorbox}

\prompt{Out}{outcolor}{128}{}
    
    \begin{equation*} \frac{2 \left(\varepsilon \theta + 1\right)}{\kappa' \rho}\end{equation*}

    \begin{tcolorbox}[breakable, size=fbox, boxrule=1pt, pad at break*=1mm,colback=cellbackground, colframe=cellborder]
\prompt{In}{incolor}{129}{\boxspacing}
\begin{Verbatim}[commandchars=\\\{\}]
\PY{n}{r_b} \PY{o}{=} \PY{n}{alpha_b\PYZus{}ok}\PY{o}{/}\PY{n}{phi\PYZus{}ok}
\PY{n}{r_b}
\end{Verbatim}
\end{tcolorbox}

\prompt{Out}{outcolor}{129}{}
    
    \begin{equation*} \frac{2 \left(- \varepsilon \theta + 1\right)}{\kappa' \rho}\end{equation*}

    \begin{tcolorbox}[breakable, size=fbox, boxrule=1pt, pad at break*=1mm,colback=cellbackground, colframe=cellborder]
\prompt{In}{incolor}{130}{\boxspacing}
\begin{Verbatim}[commandchars=\\\{\}]
\PY{n}{r\PYZus{}diff1} \PY{o}{=} \PY{n}{alpha\PYZus{}diff1\PYZus{}ok}\PY{o}{/}\PY{n}{phi\PYZus{}ok}
\PY{n}{r\PYZus{}diff1}
\end{Verbatim}
\end{tcolorbox}

\prompt{Out}{outcolor}{130}{}
    
    \begin{equation*} \frac{2 \kappa \lambda}{\kappa' \delta}\end{equation*}

    \begin{tcolorbox}[breakable, size=fbox, boxrule=1pt, pad at break*=1mm,colback=cellbackground, colframe=cellborder]
\prompt{In}{incolor}{131}{\boxspacing}
\begin{Verbatim}[commandchars=\\\{\}]
\PY{n}{r\PYZus{}diff2} \PY{o}{=} \PY{o}{\PYZhy{}}\PY{n}{alpha\PYZus{}diff2\PYZus{}ok}\PY{o}{/}\PY{n}{phi\PYZus{}ok}
\PY{n}{r\PYZus{}diff2}
\end{Verbatim}
\end{tcolorbox}

\prompt{Out}{outcolor}{131}{}
    
    \begin{equation*} \frac{2}{\kappa' \delta \kappa \lambda}\end{equation*}

    \hypertarget{sigma}{%
\subsubsection{\texorpdfstring{\(\sigma\)}{\textbackslash sigma}}\label{sigma}}

    \begin{tcolorbox}[breakable, size=fbox, boxrule=1pt, pad at break*=1mm,colback=cellbackground, colframe=cellborder]
\prompt{In}{incolor}{132}{\boxspacing}
\begin{Verbatim}[commandchars=\\\{\}]
\PY{n}{sigmad} \PY{o}{=} \PY{p}{(}\PY{n}{alpha\PYZus{}d}\PY{o}{*}\PY{n}{phid}\PY{p}{)}\PY{o}{.}\PY{n}{simplify}\PY{p}{(}\PY{p}{)}
\PY{n}{sigmad}
\end{Verbatim}
\end{tcolorbox}

\prompt{Out}{outcolor}{132}{}
    
\begin{equation*}
  \frac{
    \begin{lgathered}
      \left(4 \beta \delta \gamma \kappa^{2} \lambda \rho \theta + O\left(\theta^{3}\right)\right) \\
      \quad \cdot \bigg(4 \beta \gamma \rho \theta \left(2 \delta \kappa \lambda + \kappa^{2}
          \lambda^{2} \rho + \rho\right) \left(4 \delta^{2} \kappa^{2} \lambda^{2} 
          + 4 \delta \kappa^{3} \lambda^{3} \rho + 4 \delta \kappa \lambda \rho + \kappa^{4} \lambda^{4}
          \rho^{2} + 2 \kappa^{2} \lambda^{2} \rho^{2} + \rho^{2}\right)^{3} \\
        \qquad + 4 \beta \gamma \kappa^{2} \lambda^{2} \rho \theta \left(2 \delta \kappa \lambda +
          \kappa^{2} \lambda^{2} \rho + \rho\right) \left(4 \delta^{2} \kappa^{2} \lambda^{2} + 4 \delta \kappa^{3} \lambda^{3} \rho + 4 \delta \kappa \lambda \rho + \kappa^{4}
          \lambda^{4} \rho^{2} + 2 \kappa^{2} \lambda^{2} \rho^{2} + \rho^{2}\right)^{3} \\
        \qquad + 8 \beta \delta \gamma \kappa \lambda \theta \left(2 \delta \kappa \lambda + \kappa^{2}
          \lambda^{2} \rho + \rho\right) \left(4 \delta^{2} \kappa^{2} \lambda^{2} + 4 \delta \kappa^{3} \lambda^{3} \rho + 4 \delta \kappa \lambda \rho + \kappa^{4} \lambda^{4}
          \rho^{2} + 2 \kappa^{2} \lambda^{2} \rho^{2} + \rho^{2}\right)^{3} \\
        \qquad + 8 \beta^{2} \gamma^{2} \rho^{2} \theta^{2} \left(4 \delta^{2}
          \kappa^{2} \lambda^{2} + 4 \delta \kappa^{3} \lambda^{3} \rho + 4 \delta \kappa \lambda \rho + \kappa^{4} \lambda^{4} \rho^{2} + 2 \kappa^{2} \lambda^{2} \rho^{2} +
          \rho^{2}\right)^{2} \\
        \quad\qquad\cdot\left(- 4 \delta^{2} \kappa^{2} \lambda^{2} - 4 \delta \kappa^{3} \lambda^{3} \rho - 4 \delta \kappa \lambda \rho - \kappa^{4} \lambda^{4}
          \rho^{2} - 2 \kappa^{2} \lambda^{2} \rho^{2} - \rho^{2} + \left(2 \delta \kappa \lambda + \kappa^{2} \lambda^{2} \rho +
            \rho\right)^{2}\right)\\
        \qquad + 8 \beta^{2} \gamma^{2} \kappa^{4} \lambda^{4} \rho^{2} \theta^{2} \left(4 \delta^{2} \kappa^{2} \lambda^{2} + 4 \delta \kappa^{3} \lambda^{3} \rho + 4 \delta
          \kappa \lambda \rho + \kappa^{4} \lambda^{4} \rho^{2} + 2 \kappa^{2} \lambda^{2} \rho^{2} + \rho^{2}\right)^{2} \\
        \quad\qquad \cdot \left(4 \delta^{2} \kappa^{2}
          \lambda^{2} + 4 \delta \kappa^{3} \lambda^{3} \rho + 4 \delta \kappa \lambda \rho + \kappa^{4} \lambda^{4} \rho^{2} + 2 \kappa^{2} \lambda^{2} \rho^{2} + \rho^{2} -
          \left(2 \delta \kappa \lambda + \kappa^{2} \lambda^{2} \rho + \rho\right)^{2}\right) \\
        \qquad + O\left(\theta^{3}\right)\bigg)
    \end{lgathered}
}{\left(\kappa^{2} + 1\right)
    \left(2 \delta \kappa \lambda + \kappa^{2} \lambda^{2} \rho + \rho\right) \left(4 \delta^{2} \kappa^{2} \lambda^{2} + 4 \delta \kappa^{3} \lambda^{3} \rho + 4 \delta \kappa \lambda
      \rho + \kappa^{4} \lambda^{4} \rho^{2} + 2 \kappa^{2} \lambda^{2} \rho^{2} + \rho^{2}\right)^{4}}
\end{equation*}

    \begin{tcolorbox}[breakable, size=fbox, boxrule=1pt, pad at break*=1mm,colback=cellbackground, colframe=cellborder]
\prompt{In}{incolor}{133}{\boxspacing}
\begin{Verbatim}[commandchars=\\\{\}]
\PY{n}{sigmad2} \PY{o}{=} \PY{n}{full\PYZus{}removeO}\PY{p}{(}\PY{n}{alpha\PYZus{}d2}\PY{o}{*}\PY{n}{phid2}\PY{p}{)}\PY{o}{.}\PY{n}{simplify}\PY{p}{(}\PY{p}{)}
\PY{n}{sigmad2}
\end{Verbatim}
\end{tcolorbox}

\prompt{Out}{outcolor}{133}{}
    
    \begin{equation*} \frac{16 \beta^{2} \delta \gamma^{2} \kappa^{2} \lambda \rho \theta^{2}}{\left(\kappa^{2} + 1\right) \left(2 \delta \kappa \lambda + \kappa^{2} \lambda^{2} \rho + \rho\right)}\end{equation*}

    \begin{tcolorbox}[breakable, size=fbox, boxrule=1pt, pad at break*=1mm,colback=cellbackground, colframe=cellborder]
\prompt{In}{incolor}{134}{\boxspacing}
\begin{Verbatim}[commandchars=\\\{\}]
\PY{n}{sigma\PYZus{}ok} \PY{o}{=} \PY{n}{to\PYZus{}h\PYZus{}kp}\PY{p}{(}\PY{n}{sigmad2}\PY{p}{)}
\PY{n}{sigma\PYZus{}ok}
\end{Verbatim}
\end{tcolorbox}

\prompt{Out}{outcolor}{134}{}
    
    \begin{equation*} \frac{\eta^{2} \kappa' \lambda' \delta \rho \theta^{2}}{\lambda' \delta + \rho}\end{equation*}

    \begin{tcolorbox}[breakable, size=fbox, boxrule=1pt, pad at break*=1mm,colback=cellbackground, colframe=cellborder]
\prompt{In}{incolor}{135}{\boxspacing}
\begin{Verbatim}[commandchars=\\\{\}]
\PY{n}{sigma\PYZus{}diffd} \PY{o}{=} \PY{p}{(}\PY{n}{alpha\PYZus{}diffd}\PY{o}{*}\PY{n}{phid}\PY{p}{)}\PY{o}{.}\PY{n}{simplify}\PY{p}{(}\PY{p}{)}
\PY{n}{sigma\PYZus{}diffd}
\end{Verbatim}
\end{tcolorbox}

\prompt{Out}{outcolor}{135}{}
    
\begin{equation*}
\frac{
  \begin{lgathered}
    \left(4 \beta \delta \gamma \kappa^{2} \lambda \rho \theta + O\left(\theta^{3}\right)\right) \\
    \quad \cdot \Big(4 \beta \gamma \rho \theta \left(2 \delta \kappa \lambda + \kappa^{2} \lambda^{2} \rho + \rho\right) \left(4 \delta^{2} \kappa^{2} \lambda^{2} + 4 \delta
      \kappa^{3} \lambda^{3} \rho + 4 \delta \kappa \lambda \rho + \kappa^{4} \lambda^{4} \rho^{2} + 2 \kappa^{2} \lambda^{2} \rho^{2} + \rho^{2}\right)^{3} \\
    \qquad + 4 \beta \gamma \kappa^{2} \lambda^{2} \rho \theta \left(2 \delta \kappa \lambda + \kappa^{2} \lambda^{2} \rho + \rho\right) \left(4 \delta^{2} \kappa^{2} \lambda^{2} + 4
      \delta \kappa^{3} \lambda^{3} \rho + 4 \delta \kappa \lambda \rho + \kappa^{4} \lambda^{4} \rho^{2} + 2 \kappa^{2} \lambda^{2} \rho^{2} + \rho^{2}\right)^{3} \\
    \qquad + 8 \beta^{2} \gamma^{2} \rho^{2} \theta^{2} \left(4 \delta^{2} \kappa^{2} \lambda^{2} + 4 \delta \kappa^{3} \lambda^{3} \rho + 4 \delta \kappa \lambda \rho + \kappa^{4}
      \lambda^{4} \rho^{2} + 2 \kappa^{2} \lambda^{2} \rho^{2} + \rho^{2}\right)^{2} \\
    \quad\qquad\cdot\left(- 4 \delta^{2} \kappa^{2} \lambda^{2} - 4 \delta \kappa^{3}
      \lambda^{3} \rho - 4 \delta \kappa \lambda \rho - \kappa^{4} \lambda^{4}
      \rho^{2} - 2 \kappa^{2} \lambda^{2} \rho^{2} - \rho^{2} + \left(2 \delta \kappa \lambda + \kappa^{2} \lambda^{2} \rho + \rho\right)^{2}\right) \\
    \qquad + 8 \beta^{2} \gamma^{2} \kappa^{4} \lambda^{4} \rho^{2} \theta^{2} \left(4 \delta^{2} \kappa^{2} \lambda^{2} + 4 \delta \kappa^{3} \lambda^{3} \rho + 4 \delta \kappa
      \lambda \rho + \kappa^{4} \lambda^{4} \rho^{2} + 2 \kappa^{2} \lambda^{2} \rho^{2} + \rho^{2}\right)^{2} \\
    \quad\qquad\cdot \left(4 \delta^{2} \kappa^{2} \lambda^{2} +
      4 \delta \kappa^{3} \lambda^{3} \rho + 4 \delta \kappa \lambda \rho + \kappa^{4} \lambda^{4} \rho^{2} + 2 \kappa^{2} \lambda^{2} \rho^{2} + \rho^{2} - \left(2 \delta \kappa
        \lambda + \kappa^{2} \lambda^{2} \rho + \rho\right)^{2}\right) + O\left(\theta^{3}\right)\Big)
  \end{lgathered}
}{\left(\kappa^{2} + 1\right) \left(2 \delta \kappa \lambda + \kappa^{2} \lambda^{2} \rho + \rho\right) \left(4 \delta^{2} \kappa^{2} \lambda^{2} + 4 \delta \kappa^{3} \lambda^{3} \rho + 4 \delta \kappa \lambda \rho + \kappa^{4} \lambda^{4} \rho^{2} + 2 \kappa^{2} \lambda^{2} \rho^{2} + \rho^{2}\right)^{4}}
  \end{equation*}

    \begin{tcolorbox}[breakable, size=fbox, boxrule=1pt, pad at break*=1mm,colback=cellbackground, colframe=cellborder]
\prompt{In}{incolor}{136}{\boxspacing}
\begin{Verbatim}[commandchars=\\\{\}]
\PY{n}{sigma\PYZus{}diffd2} \PY{o}{=} \PY{n}{full\PYZus{}removeO}\PY{p}{(}\PY{n}{sigma\PYZus{}diffd}\PY{p}{)}
\PY{n}{sigma\PYZus{}diffd2}
\end{Verbatim}
\end{tcolorbox}

\prompt{Out}{outcolor}{136}{}
    
    \begin{equation*} \frac{16 \beta^{2} \delta \gamma^{2} \kappa^{2} \lambda \rho^{2} \theta^{2} \left(\kappa^{2} \lambda^{2} + 1\right)}{\left(\kappa^{2} + 1\right) \left(2 \delta \kappa \lambda + \kappa^{2} \lambda^{2} \rho + \rho\right)^{2}}\end{equation*}

    \begin{tcolorbox}[breakable, size=fbox, boxrule=1pt, pad at break*=1mm,colback=cellbackground, colframe=cellborder]
\prompt{In}{incolor}{137}{\boxspacing}
\begin{Verbatim}[commandchars=\\\{\}]
\PY{n}{sigma\PYZus{}diff\PYZus{}ok} \PY{o}{=} \PY{n}{to\PYZus{}h\PYZus{}kp}\PY{p}{(}\PY{n}{sigma\PYZus{}diffd2}\PY{p}{)}\PY{o}{.}\PY{n}{factor}\PY{p}{(}\PY{p}{)}
\PY{n}{sigma\PYZus{}diff\PYZus{}ok}
\end{Verbatim}
\end{tcolorbox}

\prompt{Out}{outcolor}{137}{}
    
    \begin{equation*} \frac{\eta^{2} \kappa' \lambda' \delta \rho^{2} \theta^{2}}{\left(\lambda' \delta + \rho\right)^{2}}\end{equation*}

    \begin{tcolorbox}[breakable, size=fbox, boxrule=1pt, pad at break*=1mm,colback=cellbackground, colframe=cellborder]
\prompt{In}{incolor}{138}{\boxspacing}
\begin{Verbatim}[commandchars=\\\{\}]
\PY{n}{sigma\PYZus{}rd} \PY{o}{=} \PY{p}{(}\PY{n}{alpha\PYZus{}rd}\PY{o}{*}\PY{n}{phid}\PY{p}{)}\PY{o}{.}\PY{n}{simplify}\PY{p}{(}\PY{p}{)}
\PY{n}{sigma\PYZus{}rd}
\end{Verbatim}
\end{tcolorbox}

\prompt{Out}{outcolor}{138}{}
    
\begin{equation*}
  \frac{
    \begin{lgathered}
      \left(4 \beta \delta \gamma \kappa^{2} \lambda \rho \theta + O\left(\theta^{3}\right)\right) \\
      \quad \cdot \left(8 \beta \delta \gamma \kappa \lambda \theta \left(2 \delta \kappa \lambda +
          \kappa^{2} \lambda^{2} \rho + \rho\right) \left(4 \delta^{2} \kappa^{2} \lambda^{2} + 4 \delta \kappa^{3} \lambda^{3} \rho + 4 \delta \kappa \lambda \rho + \kappa^{4}
          \lambda^{4} \rho^{2} + 2 \kappa^{2} \lambda^{2} \rho^{2} + \rho^{2}\right)^{3} + O\left(\theta^{3}\right)\right)
    \end{lgathered}
}{\left(\kappa^{2} + 1\right) \left(2 \delta \kappa \lambda + \kappa^{2} \lambda^{2} \rho + \rho\right) \left(4 \delta^{2} \kappa^{2} \lambda^{2} + 4 \delta \kappa^{3} \lambda^{3} \rho + 4 \delta \kappa \lambda \rho + \kappa^{4} \lambda^{4} \rho^{2} + 2 \kappa^{2} \lambda^{2} \rho^{2} + \rho^{2}\right)^{4}}
  \end{equation*}

    \begin{tcolorbox}[breakable, size=fbox, boxrule=1pt, pad at break*=1mm,colback=cellbackground, colframe=cellborder]
\prompt{In}{incolor}{139}{\boxspacing}
\begin{Verbatim}[commandchars=\\\{\}]
\PY{n}{sigma\PYZus{}rd2} \PY{o}{=} \PY{n}{full\PYZus{}removeO}\PY{p}{(}\PY{n}{sigma\PYZus{}rd}\PY{p}{)}
\PY{n}{sigma\PYZus{}rd2}
\end{Verbatim}
\end{tcolorbox}

\prompt{Out}{outcolor}{139}{}
    
    \begin{equation*} \frac{32 \beta^{2} \delta^{2} \gamma^{2} \kappa^{3} \lambda^{2} \rho \theta^{2}}{\left(\kappa^{2} + 1\right) \left(2 \delta \kappa \lambda + \kappa^{2} \lambda^{2} \rho + \rho\right)^{2}}\end{equation*}

    \begin{tcolorbox}[breakable, size=fbox, boxrule=1pt, pad at break*=1mm,colback=cellbackground, colframe=cellborder]
\prompt{In}{incolor}{140}{\boxspacing}
\begin{Verbatim}[commandchars=\\\{\}]
\PY{n}{sigma\PYZus{}r\PYZus{}ok} \PY{o}{=} \PY{n}{to\PYZus{}h\PYZus{}kp}\PY{p}{(}\PY{n}{sigma\PYZus{}rd2}\PY{p}{)}\PY{o}{.}\PY{n}{factor}\PY{p}{(}\PY{p}{)}
\PY{n}{sigma\PYZus{}r\PYZus{}ok}
\end{Verbatim}
\end{tcolorbox}

\prompt{Out}{outcolor}{140}{}
    
    \begin{equation*} \frac{\eta^{2} \kappa' \lambda'^{2} \delta^{2} \rho \theta^{2}}{\left(\lambda' \delta + \rho\right)^{2}}\end{equation*}

    \begin{tcolorbox}[breakable, size=fbox, boxrule=1pt, pad at break*=1mm,colback=cellbackground, colframe=cellborder]
\prompt{In}{incolor}{141}{\boxspacing}
\begin{Verbatim}[commandchars=\\\{\}]
\PY{p}{(}\PY{n}{sigma\PYZus{}diff\PYZus{}ok}\PY{o}{+}\PY{n}{sigma\PYZus{}r\PYZus{}ok}\PY{p}{)}\PY{o}{.}\PY{n}{simplify}\PY{p}{(}\PY{p}{)}
\end{Verbatim}
\end{tcolorbox}

\prompt{Out}{outcolor}{141}{}
    
    \begin{equation*} \frac{\eta^{2} \kappa' \lambda' \delta \rho \theta^{2}}{\lambda' \delta + \rho}\end{equation*}

    \hypertarget{ratiosb}{%
\paragraph{Ratios}\label{ratiosb}}

    \begin{tcolorbox}[breakable, size=fbox, boxrule=1pt, pad at break*=1mm,colback=cellbackground, colframe=cellborder]
\prompt{In}{incolor}{142}{\boxspacing}
\begin{Verbatim}[commandchars=\\\{\}]
\PY{n}{sigma\PYZus{}r\PYZus{}ok}\PY{o}{/}\PY{n}{sigma\PYZus{}diff\PYZus{}ok}
\end{Verbatim}
\end{tcolorbox}

\prompt{Out}{outcolor}{142}{}
    
    \begin{equation*} \frac{\lambda' \delta}{\rho}\end{equation*}

    \begin{tcolorbox}[breakable, size=fbox, boxrule=1pt, pad at break*=1mm,colback=cellbackground, colframe=cellborder]
\prompt{In}{incolor}{143}{\boxspacing}
\begin{Verbatim}[commandchars=\\\{\}]
\PY{n}{sigma\PYZus{}diff\PYZus{}ok}\PY{o}{/}\PY{n}{sigma\PYZus{}ok}
\end{Verbatim}
\end{tcolorbox}

\prompt{Out}{outcolor}{143}{}
    
    \begin{equation*} \frac{\rho}{\lambda' \delta + \rho}\end{equation*}

    \hypertarget{barmathcala}{%
\subsubsection{\texorpdfstring{\(\bar{\mathcal{A}}\)}{\textbackslash bar\{\textbackslash mathcal\{A\}\}}}\label{barmathcala}}

    \begin{tcolorbox}[breakable, size=fbox, boxrule=1pt, pad at break*=1mm,colback=cellbackground, colframe=cellborder]
\prompt{In}{incolor}{144}{\boxspacing}
\begin{Verbatim}[commandchars=\\\{\}]
\PY{n}{a_a\PYZus{}d} \PY{o}{=} \PY{n}{develop}\PY{p}{(}\PY{p}{(}\PY{l+m+mi}{1}\PY{o}{\PYZhy{}}\PY{n}{theta}\PY{p}{)}\PY{o}{*}\PY{n}{alpha_ad}\PY{p}{)}
\PY{n}{a_a\PYZus{}d}
\end{Verbatim}
\end{tcolorbox}

\prompt{Out}{outcolor}{144}{}
\begin{align*}
  \theta \Bigg(&- \frac{8 \beta \delta^{2} \gamma \kappa^{2} \lambda^{2}}{4 \delta^{2} \kappa^{2} \lambda^{2} + 4 \delta \kappa^{3} \lambda^{3} \rho + 4 \delta \kappa \lambda \rho +
      \kappa^{4} \lambda^{4} \rho^{2} + 2 \kappa^{2} \lambda^{2} \rho^{2} + \rho^{2}} \\
    &- \frac{4 \beta \delta \gamma \kappa^{3} \lambda^{3} \rho}{4 \delta^{2} \kappa^{2}
      \lambda^{2} + 4 \delta \kappa^{3} \lambda^{3} \rho + 4 \delta \kappa \lambda \rho + \kappa^{4} \lambda^{4} \rho^{2} + 2 \kappa^{2} \lambda^{2} \rho^{2} + \rho^{2}} \\
    &- \frac{4 \beta \delta \gamma \kappa \lambda \rho}{4 \delta^{2} \kappa^{2} \lambda^{2} + 4 \delta \kappa^{3} \lambda^{3} \rho + 4 \delta \kappa \lambda \rho + \kappa^{4} \lambda^{4} \rho^{2} + 2
      \kappa^{2} \lambda^{2} \rho^{2} + \rho^{2}}\Bigg) \\
  + \theta^{2} \Bigg(&- \frac{8 \beta^{2} \delta^{2} \gamma^{2} \kappa^{2} \lambda^{2}
      \left(2 \delta \kappa \lambda + \kappa^{2} \lambda^{2} \rho + \rho\right)^{2}}{\left(4 \delta^{2} \kappa^{2} \lambda^{2} + 4 \delta \kappa^{3} \lambda^{3} \rho +
        4 \delta \kappa \lambda \rho + \kappa^{4} \lambda^{4} \rho^{2} + 2 \kappa^{2} \lambda^{2} \rho^{2} + \rho^{2}\right)^{2}} \\
    &- \frac{4 \beta^{2} \delta \gamma
      \kappa \lambda \left(- 2 \delta \gamma \kappa \lambda + 2 \delta \kappa \lambda - \gamma \kappa^{2} \lambda^{2} \rho + \gamma \rho + \kappa^{2} \lambda^{2} \rho + \rho\right)}{4 \delta^{2}
      \kappa^{2} \lambda^{2} + 4 \delta \kappa^{3} \lambda^{3} \rho + 4 \delta \kappa \lambda \rho + \kappa^{4} \lambda^{4} \rho^{2} + 2 \kappa^{2} \lambda^{2} \rho^{2} + \rho^{2}}\\
    &+ \frac{8 \beta \delta^{2} \gamma \kappa^{2} \lambda^{2}}{4 \delta^{2} \kappa^{2} \lambda^{2} + 4 \delta \kappa^{3} \lambda^{3} \rho + 4 \delta \kappa \lambda \rho + \kappa^{4}
      \lambda^{4} \rho^{2} + 2 \kappa^{2} \lambda^{2} \rho^{2} + \rho^{2}} \\
    &+ \frac{4 \beta \delta \gamma \kappa^{3} \lambda^{3} \rho}{4 \delta^{2} \kappa^{2} \lambda^{2}
      + 4 \delta \kappa^{3} \lambda^{3} \rho + 4 \delta \kappa \lambda \rho + \kappa^{4} \lambda^{4} \rho^{2} + 2 \kappa^{2} \lambda^{2} \rho^{2} + \rho^{2}} \\
    &+ \frac{4 \beta
      \delta \gamma \kappa \lambda \rho}{4 \delta^{2} \kappa^{2} \lambda^{2} + 4 \delta \kappa^{3} \lambda^{3} \rho + 4 \delta \kappa \lambda \rho + \kappa^{4} \lambda^{4} \rho^{2} + 2 \kappa^{2}
      \lambda^{2} \rho^{2} + \rho^{2}}\Bigg) \\
  + O\left(\theta^{3}\right)
\end{align*}

    \begin{tcolorbox}[breakable, size=fbox, boxrule=1pt, pad at break*=1mm,colback=cellbackground, colframe=cellborder]
\prompt{In}{incolor}{145}{\boxspacing}
\begin{Verbatim}[commandchars=\\\{\}]
\PY{n}{a_a\PYZus{}d2} \PY{o}{=} \PY{n}{full\PYZus{}removeO}\PY{p}{(}\PY{n}{a_a\PYZus{}d}\PY{p}{)}
\PY{n}{a_a\PYZus{}d2}
\end{Verbatim}
\end{tcolorbox}

\prompt{Out}{outcolor}{145}{}
    
    \begin{equation*} - \frac{4 \beta \delta \gamma \kappa \lambda \theta \left(2 \beta \delta \kappa \lambda \theta - \beta \gamma \kappa^{2} \lambda^{2} \rho \theta + \beta \gamma \rho \theta + \beta \kappa^{2} \lambda^{2} \rho \theta + \beta \rho \theta - 2 \delta \kappa \lambda \theta + 2 \delta \kappa \lambda - \kappa^{2} \lambda^{2} \rho \theta + \kappa^{2} \lambda^{2} \rho - \rho \theta + \rho\right)}{\left(2 \delta \kappa \lambda + \kappa^{2} \lambda^{2} \rho + \rho\right)^{2}}\end{equation*}

    \begin{tcolorbox}[breakable, size=fbox, boxrule=1pt, pad at break*=1mm,colback=cellbackground, colframe=cellborder]
\prompt{In}{incolor}{146}{\boxspacing}
\begin{Verbatim}[commandchars=\\\{\}]
\PY{n}{a_a\PYZus{}s} \PY{o}{=} \PY{p}{(}\PY{n}{alpha_a\PYZus{}s}\PY{o}{+}\PY{n}{a_a\PYZus{}d2}\PY{o}{\PYZhy{}}\PY{n}{alpha_ad2}\PY{p}{)}\PY{o}{.}\PY{n}{simplify}\PY{p}{(}\PY{p}{)}
\PY{n}{a_a\PYZus{}s}
\end{Verbatim}
\end{tcolorbox}

\prompt{Out}{outcolor}{146}{}
    
    \begin{equation*} \frac{4 \beta \delta \gamma \kappa \lambda \theta \left(- \varepsilon \theta + \theta - 1\right)}{2 \delta \kappa \lambda + \kappa^{2} \lambda^{2} \rho + \rho}\end{equation*}

    \begin{tcolorbox}[breakable, size=fbox, boxrule=1pt, pad at break*=1mm,colback=cellbackground, colframe=cellborder]
\prompt{In}{incolor}{147}{\boxspacing}
\begin{Verbatim}[commandchars=\\\{\}]
\PY{n}{a_a\PYZus{}ok} \PY{o}{=} \PY{n}{to\PYZus{}h\PYZus{}kp}\PY{p}{(}\PY{n}{a_a\PYZus{}s}\PY{p}{)}
\PY{n}{a_a\PYZus{}ok}
\end{Verbatim}
\end{tcolorbox}

\prompt{Out}{outcolor}{147}{}
    
    \begin{equation*} \frac{\eta \lambda' \delta \theta \left(- \varepsilon \theta + \theta - 1\right)}{\lambda' \delta + \rho}\end{equation*}

    \begin{tcolorbox}[breakable, size=fbox, boxrule=1pt, pad at break*=1mm,colback=cellbackground, colframe=cellborder]
\prompt{In}{incolor}{148}{\boxspacing}
\begin{Verbatim}[commandchars=\\\{\}]
\PY{n}{a_b\PYZus{}d} \PY{o}{=} \PY{n}{develop}\PY{p}{(}\PY{p}{(}\PY{l+m+mi}{1}\PY{o}{+}\PY{n}{theta}\PY{p}{)}\PY{o}{*}\PY{n}{alpha_bd}\PY{p}{)}
\PY{n}{a_b\PYZus{}d2} \PY{o}{=} \PY{n}{full\PYZus{}removeO}\PY{p}{(}\PY{n}{a_b\PYZus{}d}\PY{p}{)}
\PY{n}{a_b\PYZus{}d2}
\end{Verbatim}
\end{tcolorbox}

\prompt{Out}{outcolor}{148}{}
    
    \begin{equation*} - \frac{4 \beta \delta \gamma \kappa \lambda \theta \left(2 \beta \delta \kappa \lambda \theta - \beta \gamma \kappa^{2} \lambda^{2} \rho \theta + \beta \gamma \rho \theta + \beta \kappa^{2} \lambda^{2} \rho \theta + \beta \rho \theta - 2 \delta \kappa \lambda \theta - 2 \delta \kappa \lambda - \kappa^{2} \lambda^{2} \rho \theta - \kappa^{2} \lambda^{2} \rho - \rho \theta - \rho\right)}{\left(2 \delta \kappa \lambda + \kappa^{2} \lambda^{2} \rho + \rho\right)^{2}}\end{equation*}

    \begin{tcolorbox}[breakable, size=fbox, boxrule=1pt, pad at break*=1mm,colback=cellbackground, colframe=cellborder]
\prompt{In}{incolor}{149}{\boxspacing}
\begin{Verbatim}[commandchars=\\\{\}]
\PY{n}{a_b\PYZus{}s} \PY{o}{=} \PY{p}{(}\PY{n}{alpha_b\PYZus{}s}\PY{o}{+}\PY{n}{a_b\PYZus{}d2}\PY{o}{\PYZhy{}}\PY{n}{alpha_bd2}\PY{p}{)}\PY{o}{.}\PY{n}{simplify}\PY{p}{(}\PY{p}{)}
\PY{n}{a_b\PYZus{}s}
\end{Verbatim}
\end{tcolorbox}

\prompt{Out}{outcolor}{149}{}
    
    \begin{equation*} \frac{4 \beta \delta \gamma \kappa \lambda \theta \left(- \varepsilon \theta + \theta + 1\right)}{2 \delta \kappa \lambda + \kappa^{2} \lambda^{2} \rho + \rho}\end{equation*}

    \begin{tcolorbox}[breakable, size=fbox, boxrule=1pt, pad at break*=1mm,colback=cellbackground, colframe=cellborder]
\prompt{In}{incolor}{150}{\boxspacing}
\begin{Verbatim}[commandchars=\\\{\}]
\PY{n}{a_b\PYZus{}ok} \PY{o}{=} \PY{n}{to\PYZus{}h\PYZus{}kp}\PY{p}{(}\PY{n}{a_b\PYZus{}s}\PY{p}{)}
\PY{n}{a_b\PYZus{}ok}
\end{Verbatim}
\end{tcolorbox}

\prompt{Out}{outcolor}{150}{}
    
    \begin{equation*} \frac{\eta \lambda' \delta \theta \left(- \varepsilon \theta + \theta + 1\right)}{\lambda' \delta + \rho}\end{equation*}

    \begin{tcolorbox}[breakable, size=fbox, boxrule=1pt, pad at break*=1mm,colback=cellbackground, colframe=cellborder]
\prompt{In}{incolor}{151}{\boxspacing}
\begin{Verbatim}[commandchars=\\\{\}]
\PY{n}{E} \PY{o}{=} \PY{p}{(}\PY{n}{a_b\PYZus{}s} \PY{o}{\PYZhy{}}\PY{n}{a_a\PYZus{}s}\PY{p}{)}\PY{o}{.}\PY{n}{simplify}\PY{p}{(}\PY{p}{)}
\PY{n}{E}
\end{Verbatim}
\end{tcolorbox}

\prompt{Out}{outcolor}{151}{}
    
    \begin{equation*} \frac{8 \beta \delta \gamma \kappa \lambda \theta}{2 \delta \kappa \lambda + \kappa^{2} \lambda^{2} \rho + \rho}\end{equation*}

    Same values are thus obtained for \(\alpha\) and \(\bar{\mathcal{A}}\)
considering the Taylor expansion.

    \begin{tcolorbox}[breakable, size=fbox, boxrule=1pt, pad at break*=1mm,colback=cellbackground, colframe=cellborder]
\prompt{In}{incolor}{152}{\boxspacing}
\begin{Verbatim}[commandchars=\\\{\}]
\PY{n}{E\PYZus{}ok} \PY{o}{=} \PY{n}{to\PYZus{}h\PYZus{}kp}\PY{p}{(}\PY{n}{E}\PY{p}{)}
\PY{n}{E\PYZus{}ok}
\end{Verbatim}
\end{tcolorbox}

\prompt{Out}{outcolor}{152}{}
    
    \begin{equation*} \frac{2 \eta \lambda' \delta \theta}{\lambda' \delta + \rho}\end{equation*}

    \begin{tcolorbox}[breakable, size=fbox, boxrule=1pt, pad at break*=1mm,colback=cellbackground, colframe=cellborder]
\prompt{In}{incolor}{153}{\boxspacing}
\begin{Verbatim}[commandchars=\\\{\}]
\PY{n}{alpha\PYZus{}r\PYZus{}ok}
\end{Verbatim}
\end{tcolorbox}

\prompt{Out}{outcolor}{153}{}
    
\begin{equation*} \frac{2 \eta \lambda' \delta \theta}{\lambda' \delta + \rho}\end{equation*}


\newpage{}

\listoffigures

\newpage{}

\begin{figure}[p]
  \begin{center}
    \includegraphics[width=17cm]{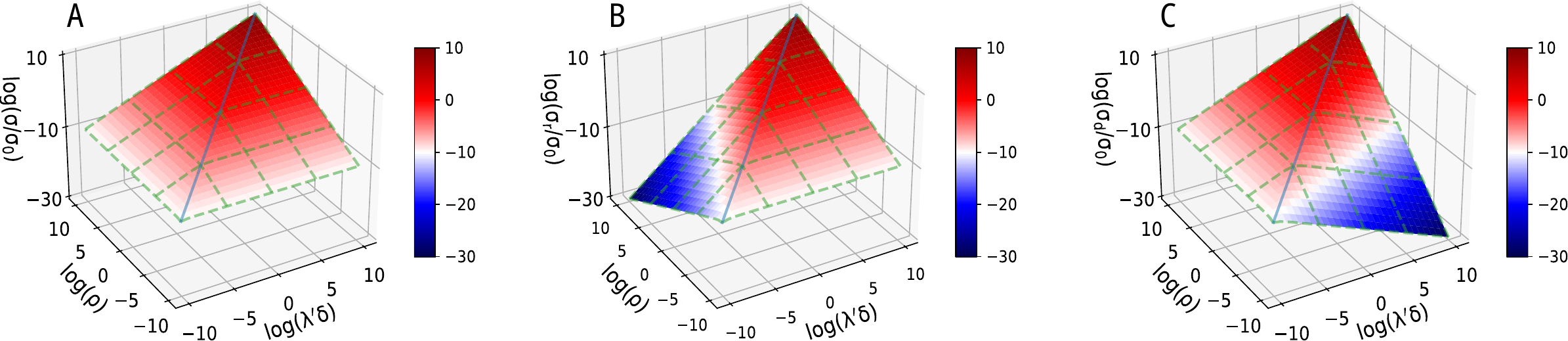}    
  \end{center}
  \caption{Normalized entropy production $\sfrac{\sigma_{i}}{\sigma_{0}}$ in a two-compartment system, as a
    function of $\rho \in [10^{-10}, 10^{10}]$ and $\lambda'\delta \in [10^{-10}, 10^{10}]$ (defined in Eq. (C24) in
    MT). \textbf{A}, total entropy production $\sigma$; \textbf{B}, entropy production by chemical
    reaction $\sigma_{r}$; \textbf{C}, entropy production by chemical diffusion $\sigma_{d}$.}
  \label{fig:entro-opt}
\end{figure}

\begin{figure}[p]
  \begin{center}
    \includegraphics[width=12cm]{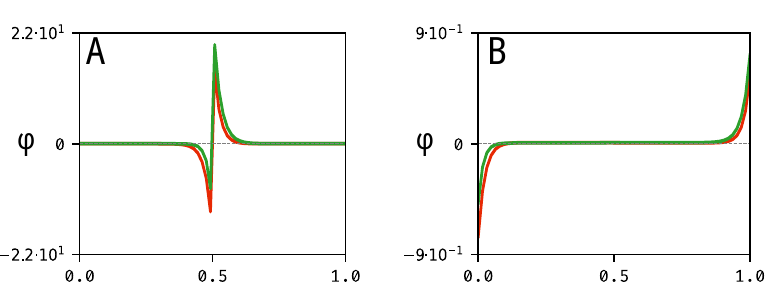}
  \end{center}
  \caption{Effect of the dependence on temperature on the diffusion coefficient. A first order
    correction on the diffusion coefficient was introduced as
    $d = d_{0} (1 + \delta d \cdot \theta)$. $\beta=1$, $\gamma=0.2$, $\theta_{g=0.1}$,
    $\rho=1000$, $\kappa=1$, $\lambda=1$, $d_{0}=1$, $\delta d=0$ (red) or $\delta d=5$ (green). A: Heaviside temperature
    profile; B: Linear temperature profile.}
\end{figure}

\begin{figure}[p]
  \centering
  \includegraphics[width=14cm]{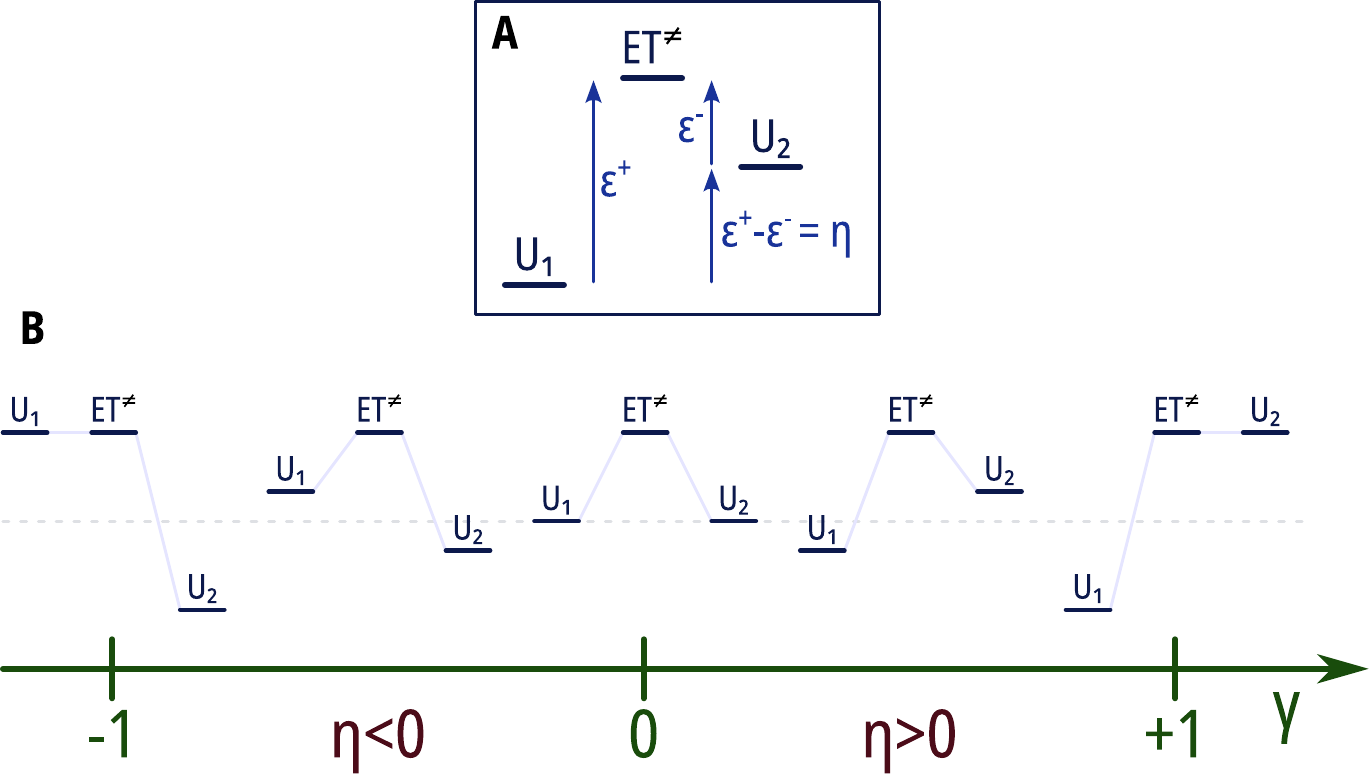}
  \caption{Energetic profiles. A: activation energies $\varepsilon_{+}$ and $\varepsilon_{-}$, and reaction enthalpy
    $\eta$. B: Evolution of energetic profiles as a function of $\gamma$ for a constant $\beta$ value.
    }
  \label{fig:S3}
\end{figure}

\begin{figure}[p]
  \includegraphics[width=17cm]{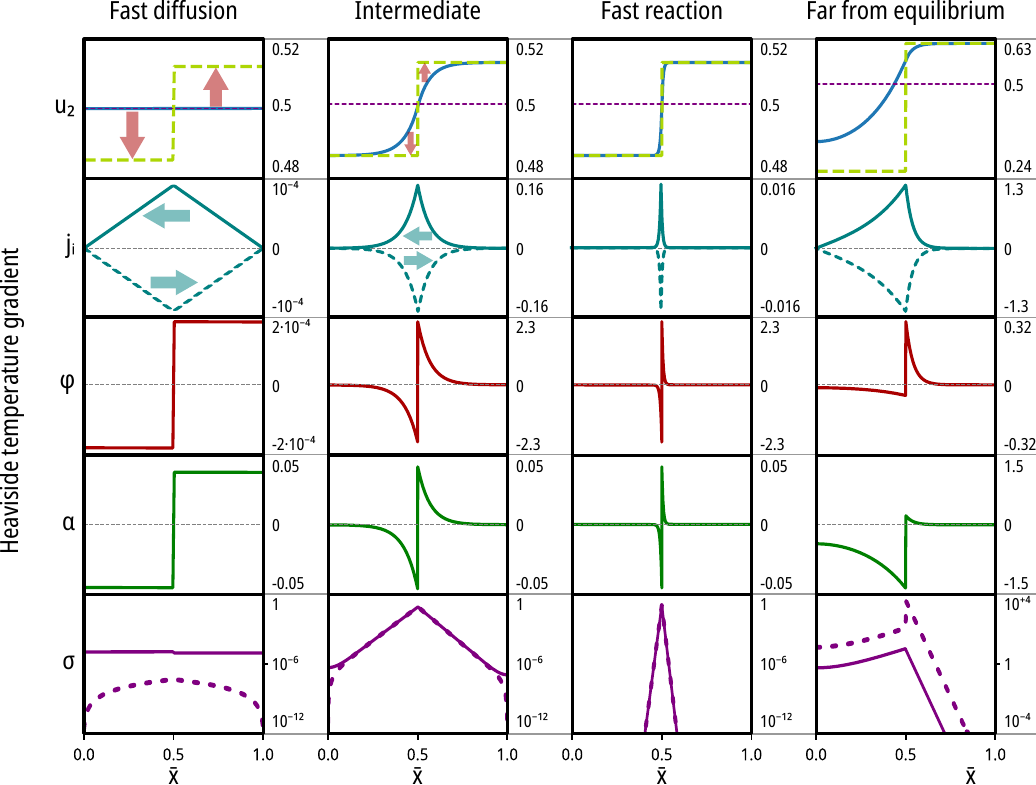}
  \caption{Steady state spatial profiles of key parameters in unstirred systems driven by Heaviside
    temperature profiles.  Solid blue, steady $u_{2}$ concentration profile; dotted green, local
    equilibrium $u_{2}$ concentration profile; solid light blue, $j_{2}$ concentration gradient;
    dotted light blue, $j_{1}$ concentration gradient; red: steady chemical flux of reaction
    $\varphi$; dark green, steady chemical force $\alpha$, purple: steady dissipation of entropy
    $\sigma_{r}$ (plain line) and $\sigma_{d}$ (dotted line). The horizontal wide arrows represent the
    direction of \ch{U1} and \ch{U2} diffusion; vertical wide arrows represent the global reaction
    fluxes \ch{U1 -> U2} (up arrows) and \ch{U2 -> U1} (down arrows). Model parameters: $\kappa=1$,
    $\lambda=1$, $\gamma=0.2$, and \textbf{Fast diffusion}: $d=1$, $\rho=10^{-2}$, $\beta=1$,
    $\theta_{g}=10^{-1}$, \textbf{Intermediate}: $d=1$, $\rho=10^{2}$, $\beta=1$,
    $\theta_{g}=10^{-1}$, \textbf{Fast reaction}: $d=10^{-2}$, $\rho=10^{2}$, $\beta=1$,
    $\theta_{g}=10^{-1}$, \textbf{Far from equilibrium}: $d=1$, $\rho=10^{2}$, $\beta=4$, $\theta_{g}=0.4$.}
  \label{fig:FigS4}
\end{figure}

\begin{figure}[p]
  \includegraphics[width=17cm]{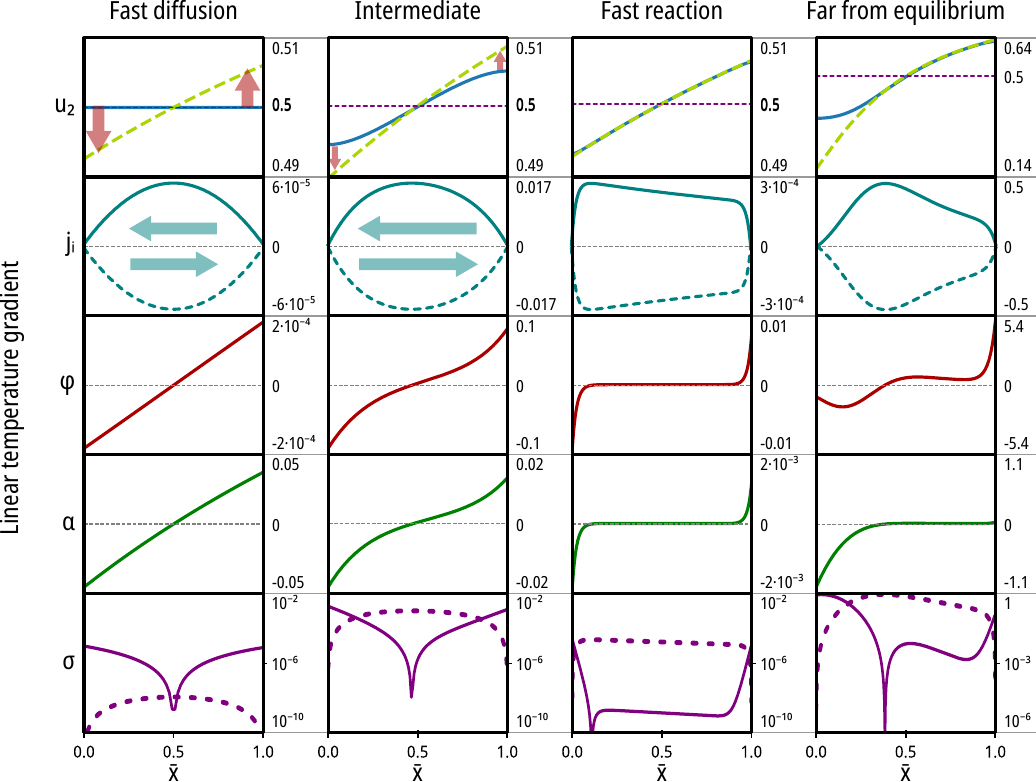}
  \caption{Steady state spatial profiles of key parameters in unstirred systems driven by linear
    temperature profiles.  Solid blue, steady $u_{2}$ concentration profile; dotted green, local
    equilibrium $u_{2}$ concentration profile; solid light blue, $j_{2}$ concentration gradient;
    dotted light blue, $j_{1}$ concentration gradient; red: steady chemical flux of reaction
    $\varphi$; dark green, steady chemical force $\alpha$, purple: steady dissipation of entropy
    $\sigma_{r}$ (plain line) and $\sigma_{d}$ (dotted line). The horizontal wide arrows represent the
    direction of \ch{U1} and \ch{U2} diffusion; vertical wide arrows represent the global reaction
    fluxes \ch{U1 -> U2} (up arrows) and \ch{U2 -> U1} (down arrows). Model parameters: $\kappa=1$,
    $\lambda=1$, $\gamma=0.2$, and \textbf{Fast diffusion}: $d=1$, $\rho=10^{-2}$, $\beta=1$,
    $\theta_{g}=10^{-1}$, \textbf{Intermediate}: $d=1$, $\rho=10$, $\beta=1$, $\theta_{g}=10^{-1}$, \textbf{Fast
      reaction}: $d=10^{-2}$, $\rho=10$, $\beta=1$, $\theta_{g}=10^{-1}$, \textbf{Far from equilibrium}:
    $d=1$, $\rho=10^{2}$, $\beta=4$, $\theta_{g}=0.5$.}
  \label{fig:FigS5}
\end{figure}

\begin{figure}[p]
  \includegraphics[width=17cm]{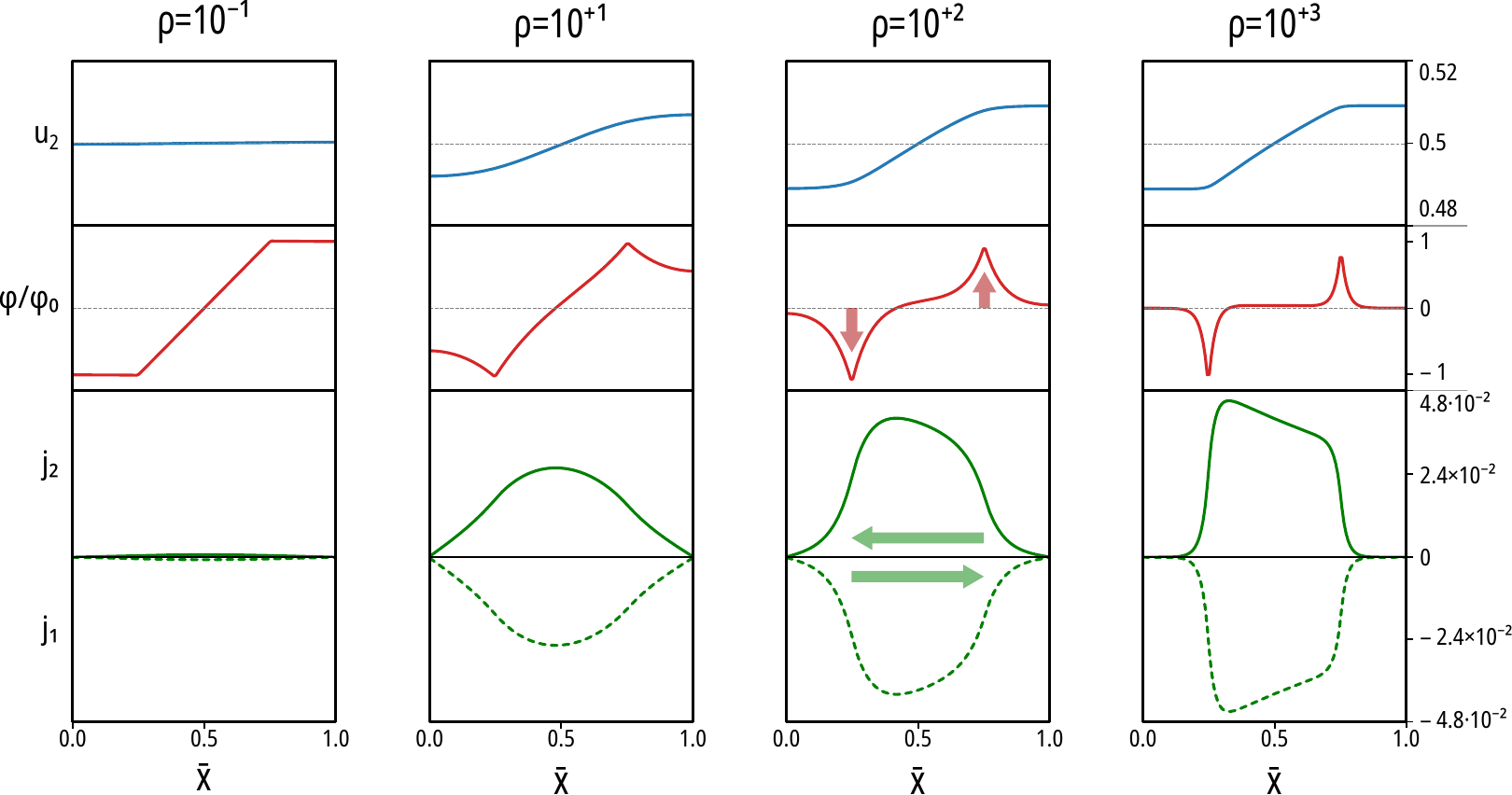}
  \caption{Mixed temperature profile. A precise localization of the reaction diffusion cycle is
    achieved by a tailored temperature gradient profile. Diffusion and reaction processes are
    physically separated. \textbf{A}: temperature, with constant temperature $-\theta_{g}$ until
    \textcircled{1}, linear variation of $\theta$ from \textcircled{1} to \textcircled{2}, and constant
    temperature $+\theta_{g}$ from \textcircled{2}; \textbf{B, G}, diffusion flux $j_{2}$; \textbf{C, F},
    normalized reaction flux $\varphi/\varphi_{0}$, with $\varphi_{0}=2\beta\gamma\theta_{g}\rho$; \textbf{D, H}, diffusion flux
    $j_{1}$; \textbf{E}, concentration $u_{2}$. Model parameters: $\kappa=1$, $\lambda=1$,
    $\gamma=0.2$, $d=1$, $\beta=1$, $\theta_{g}=10^{-1}$, \textbf{A-D}: $\rho = 10^{3}$, \textbf{E-H}:
    $\rho \in [10^{-1}, 10^{3}]$.  \label{fig:FigS6}}
\end{figure}

\begin{figure}[p]
  \centering
  \includegraphics[width=17cm]{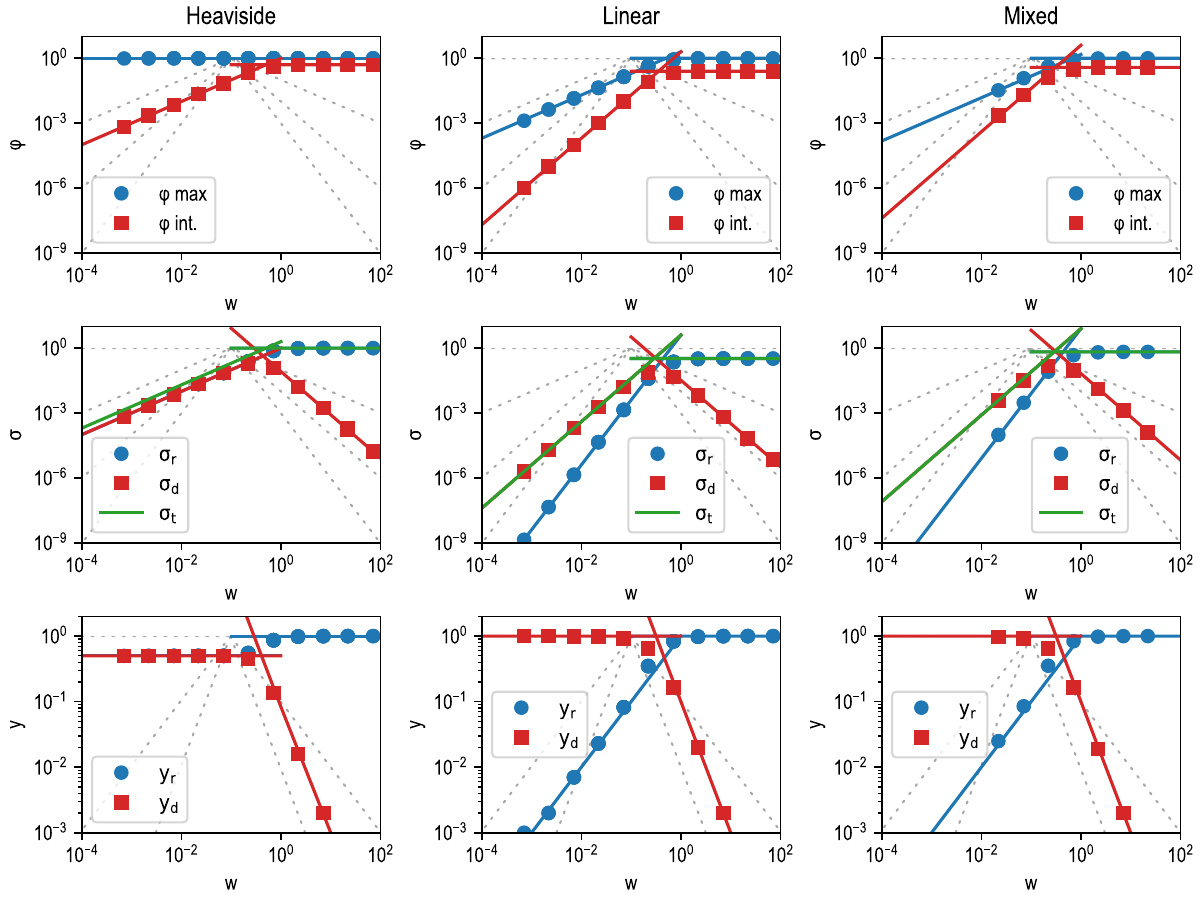}
  \caption{Profile scaling of maximal and integral values on $\varphi$, $\sigma$ and $y$ (see Table 2 in Main
    Text). The dotted lines represent, for convenience of reading, slopes corresponding on scaling
    in $w^{n}$. Parameters: $\beta=1$, $\kappa=1$, $\gamma=0.2$, $\theta_{g}=0.1$. Heaviside and linear profiles:
    $d \in [10^{-3}, 10^{3}]$ and $\rho \in [10^{-3}, 10^{3}]$. Mixed profile: $d=1$ and
    $\rho \in [10^{-3}, 10^{3}]$.}
  \label{fig:S7}
\end{figure}

\begin{figure}[p]
  \begin{center}
    \includegraphics[width=17cm]{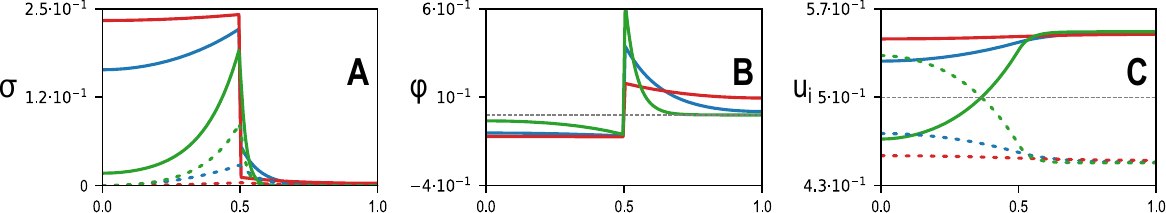}
  \end{center}
  \caption{Peptide exchange reaction in Heaviside temperature gradient. \textbf{A}, entropy
    production ($\sigma_{r}$, solid lines; $\sigma_{d}$, dotted line); \textbf{B}, reaction flux
    $\varphi$, \textbf{C} concentrations ($u_{1}$, dotted line; $u_{2}$, solid line).  $\beta=18.8$,
    $\gamma=0.122$, $\theta_{g}=0.1$, $d=1$, $\kappa=1$ and $\lambda=1$. \textbf{Red lines},
    $\rho=0.28$, $l_{1}=1$~cm; \textbf{blue lines}, $\rho=2.5$, $l_{0}=3$~cm; \textbf{green lines},
    $\rho=28$, $l_{2}=10$~cm.}
  \label{fig:S8}
\end{figure}

\clearpage{}

\section*{Reaction diffusion profiles}

In the next pages are available the automatically generated figures of the spatial profiles for all
simulations reported in this work.



\maketitle

The following figures give the profile of $\alpha, \varphi, R_{\textrm{tot}}, u_{2}, \sigma_{r}, \sigma_{d},
\sigma_{\textrm{tot}}$ as a function of $[d, \rho]$ values or $[\beta, \theta_{g}]$ values:
\begin{center}
  \begin{tabular}{lllllllll}
    \toprule
                                &                        & $\alpha$                    & $\varphi$                    & $R_{\textrm{tot}}$     & $u_{2}$                & $\sigma_{r}$                 & $\sigma_{d}$                 & $\sigma_{\textrm{tot}}$      \\
    \midrule
    \multirow[c]{3}*{Heaviside} & $[d,\rho]$ (low $\theta_{g}$)  & Fig.~\ref{fig:HSl-a}   & Fig.~\ref{fig:HSl-p}   & Fig.~\ref{fig:HSl-r}   & Fig.~\ref{fig:HSl-u}   & Fig.~\ref{fig:HSl-sr}   & Fig.~\ref{fig:HSl-sd}   & Fig.~\ref{fig:HSl-st}   \\
                                & $[d,\rho]$ (high $\theta_{g}$) & Fig.~\ref{fig:HSh-a}   & Fig.~\ref{fig:HSh-p}   & Fig.~\ref{fig:HSh-r}   & Fig.~\ref{fig:HSh-u}   & Fig.~\ref{fig:HSh-sr}   & Fig.~\ref{fig:HSh-sd}   & Fig.~\ref{fig:HSh-st}   \\
                                & $[\beta,\theta_{g}]$            & Fig.~\ref{fig:HSbt-a}  & Fig.~\ref{fig:HSbt-p}  & Fig.~\ref{fig:HSbt-r}  & Fig.~\ref{fig:HSbt-u}  & Fig.~\ref{fig:HSbt-sr}  & Fig.~\ref{fig:HSbt-sd}  & Fig.~\ref{fig:HSbt-st}  \\
    \midrule                                                                                                                                                                  
    \multirow[c]{3}*{Linear}    & $[d,\rho]$ (low $\theta_{g}$)  & Fig.~\ref{fig:Linl-a}  & Fig.~\ref{fig:Linl-p}  & Fig.~\ref{fig:Linl-r}  & Fig.~\ref{fig:Linl-u}  & Fig.~\ref{fig:Linl-sr}  & Fig.~\ref{fig:Linl-sd}  & Fig.~\ref{fig:Linl-st}  \\
                                & $[d,\rho]$ (high $\theta_{g}$) & Fig.~\ref{fig:Linh-a}  & Fig.~\ref{fig:Linh-p}  & Fig.~\ref{fig:Linh-r}  & Fig.~\ref{fig:Linh-u}  & Fig.~\ref{fig:Linh-sr}  & Fig.~\ref{fig:Linh-sd}  & Fig.~\ref{fig:Linh-st}  \\
                                & $[\beta/\theta_{g}]$            & Fig.~\ref{fig:Linbt-a} & Fig.~\ref{fig:Linbt-p} & Fig.~\ref{fig:Linbt-r} & Fig.~\ref{fig:Linbt-u} & Fig.~\ref{fig:Linbt-sr} & Fig.~\ref{fig:Linbt-sd} & Fig.~\ref{fig:Linbt-st} \\
    \bottomrule
  \end{tabular}
\end{center}
Dotted line in $R_{\textrm{tot}}$ figures represent the R value in the linear regime. Dotted line in
$u_{2}$ figures represent the equilibrium value.

Common parameter values:
\begin{itemize}
\item $\kappa=1$
\item $\lambda=1$
\item $\gamma=0.2$
\end{itemize}


\begin{figure}[p]
  \centering
  \includegraphics[width=0.87\linewidth]{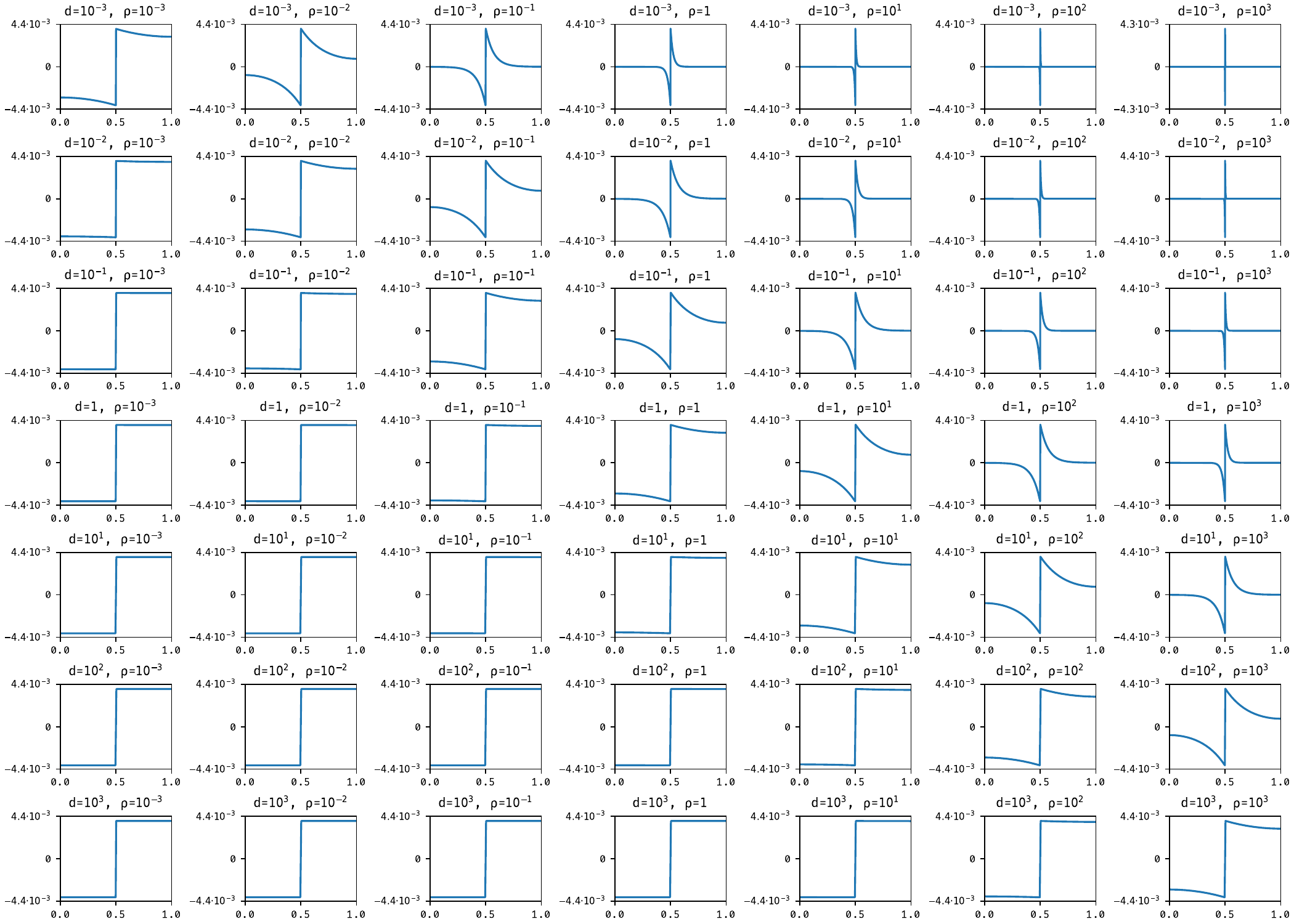}
  \caption{$\alpha$ for Heaviside profile, low $\theta_{g}$ value ($\beta=1$, $\theta_g=0.01$).}
  \label{fig:HSl-a}
\end{figure}

\begin{figure}[p]
  \centering
  \includegraphics[width=0.87\linewidth]{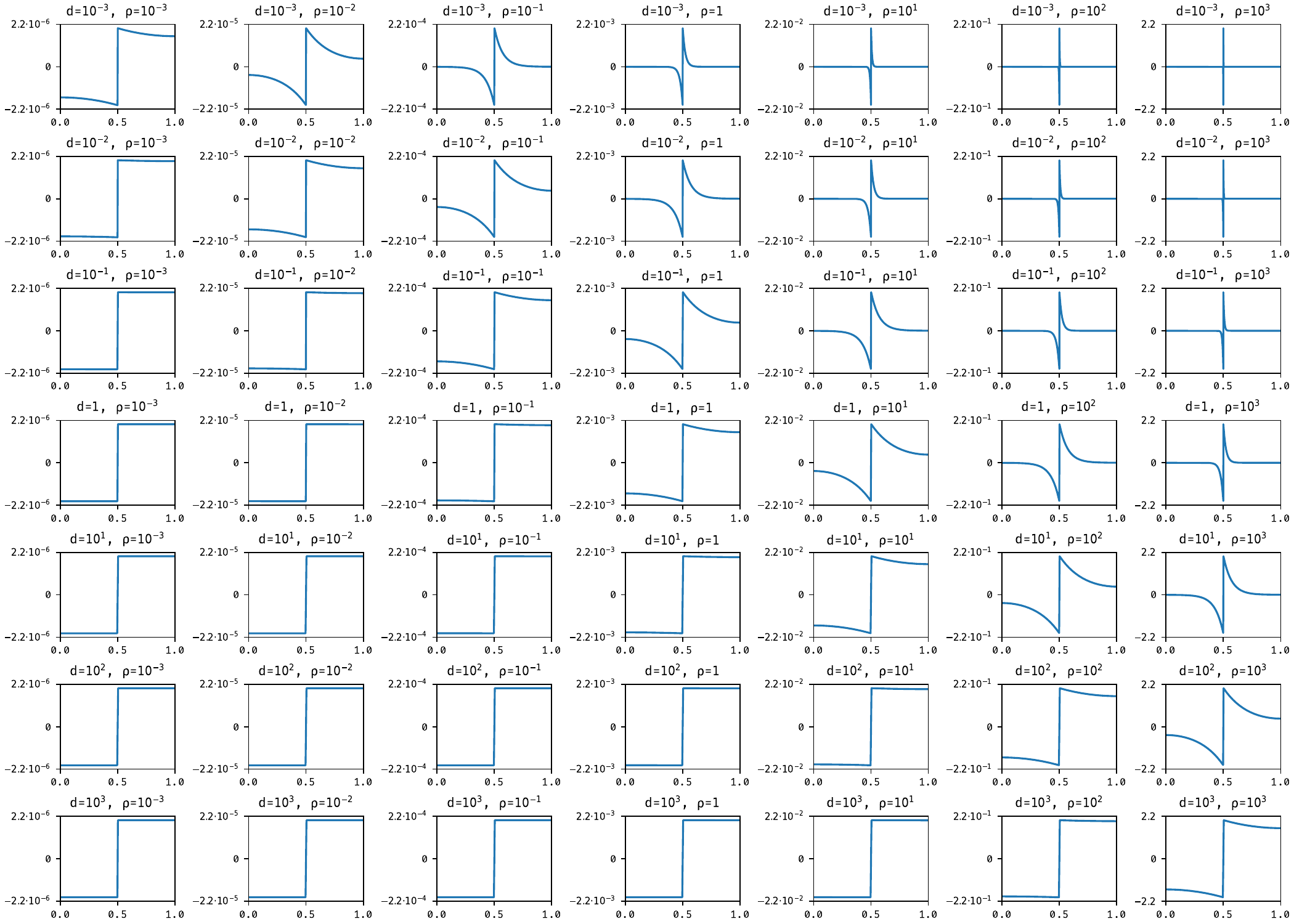}
  \caption{$\varphi$ for Heaviside profile, low $\theta_{g}$ value ($\beta=1$, $\theta_g=0.01$).}
  \label{fig:HSl-p}
\end{figure}

\begin{figure}[p]
  \centering
  \includegraphics[width=0.87\linewidth]{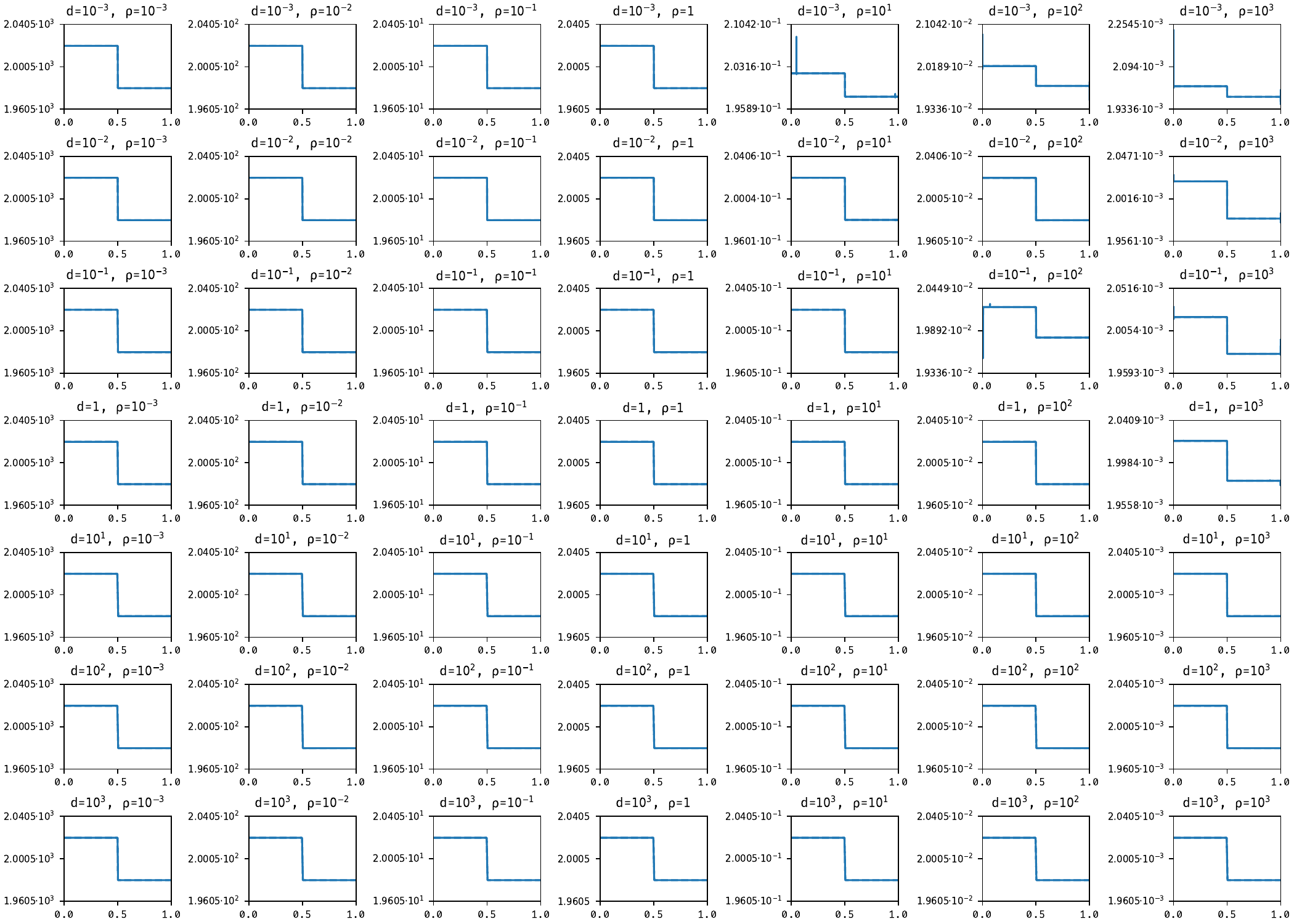}
  \caption{$R_{\textrm{tot}}$ for Heaviside profile, low $\theta_{g}$ value ($\beta=1$, $\theta_g=0.01$).}
  \label{fig:HSl-r}
\end{figure}

\begin{figure}[p]
  \centering
  \includegraphics[width=0.87\linewidth]{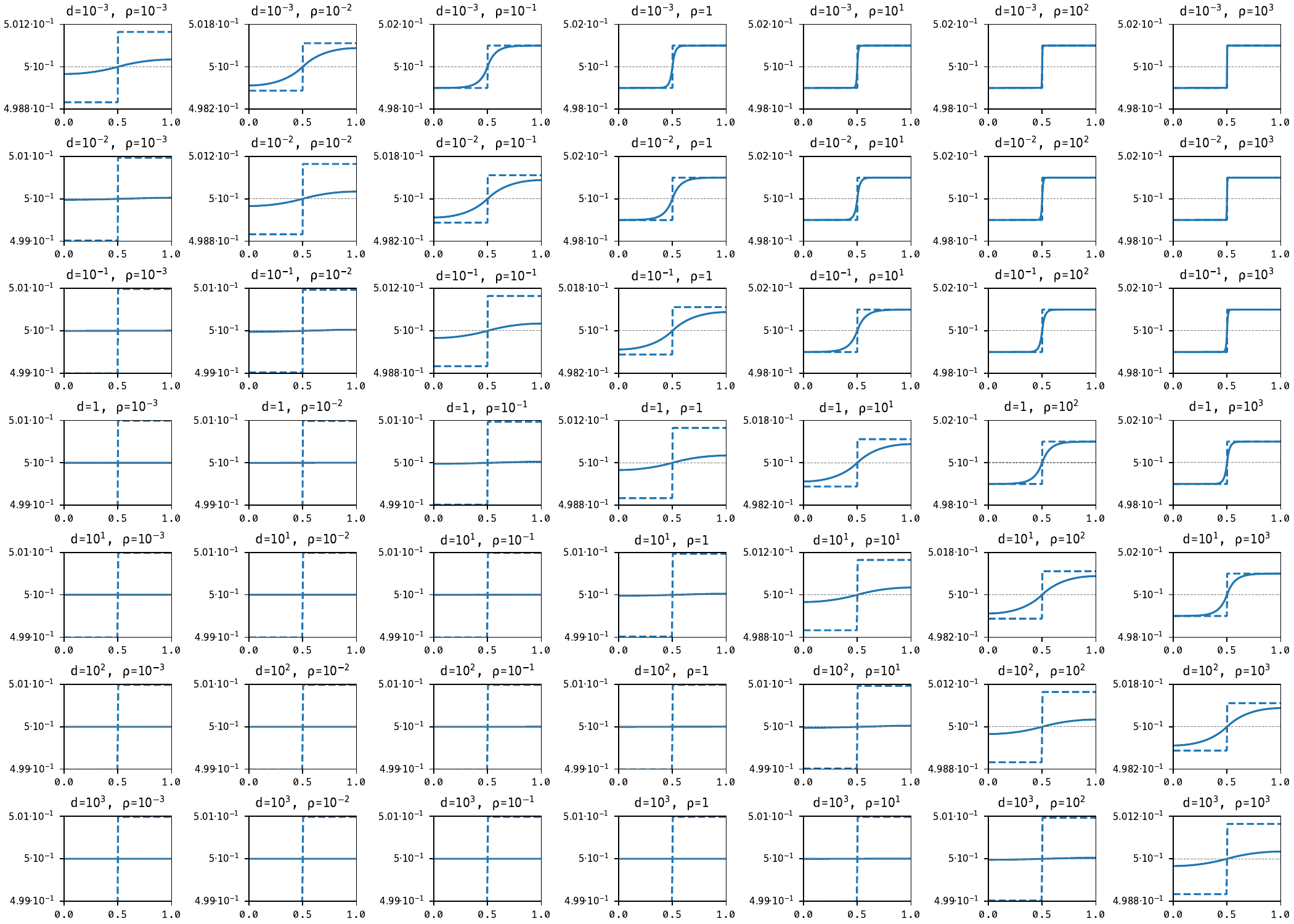}
  \caption{$u_{2}$ for Heaviside profile, low $\theta_{g}$ value ($\beta=1$, $\theta_g=0.01$).}
  \label{fig:HSl-u}
\end{figure}

\begin{figure}[p]
  \centering
  \includegraphics[width=0.87\linewidth]{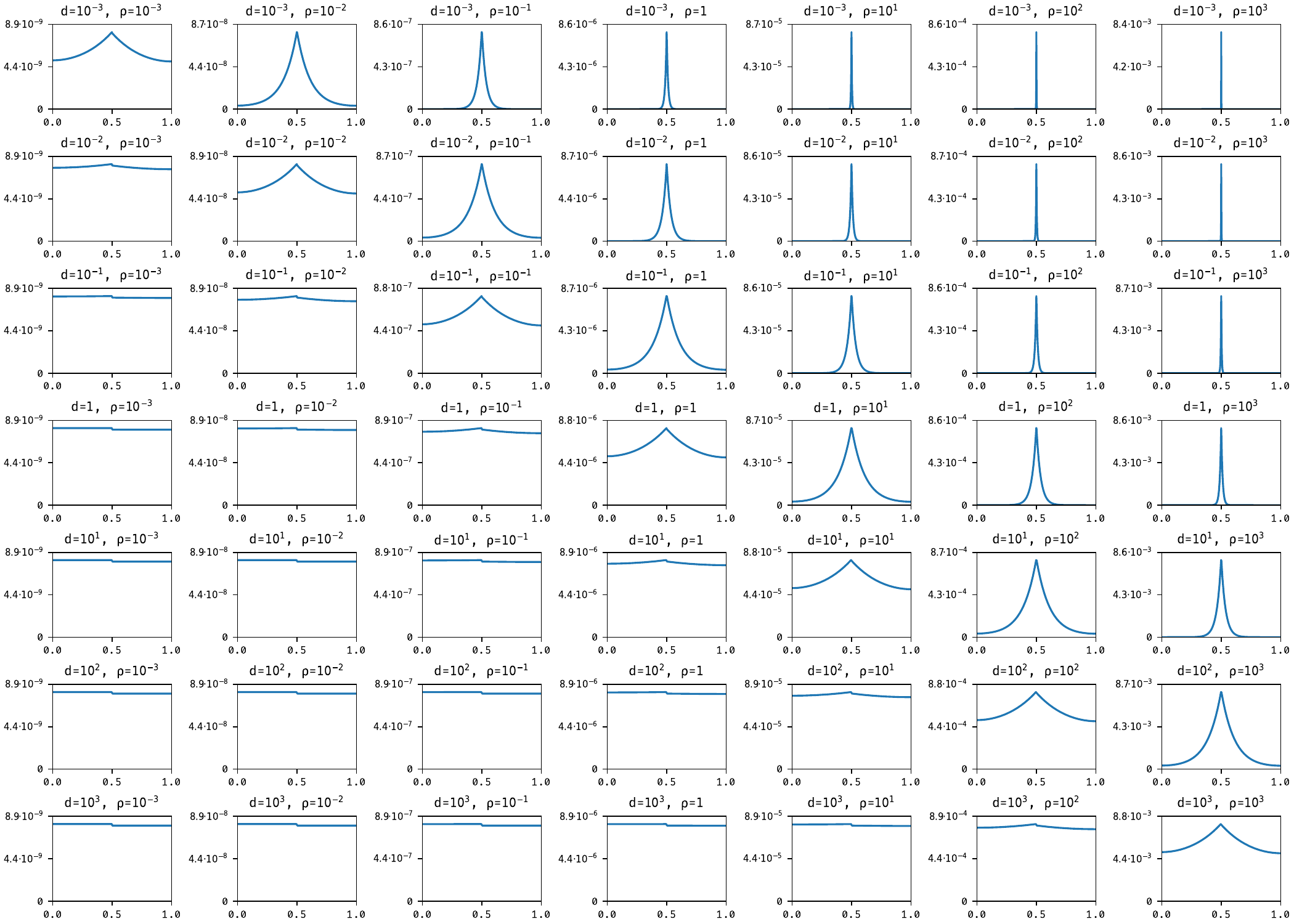}
  \caption{$\sigma_{r}$ for Heaviside profile, low $\theta_{g}$ value ($\beta=1$, $\theta_g=0.01$).}
  \label{fig:HSl-sr}
\end{figure}

\begin{figure}[p]
  \centering
  \includegraphics[width=0.87\linewidth]{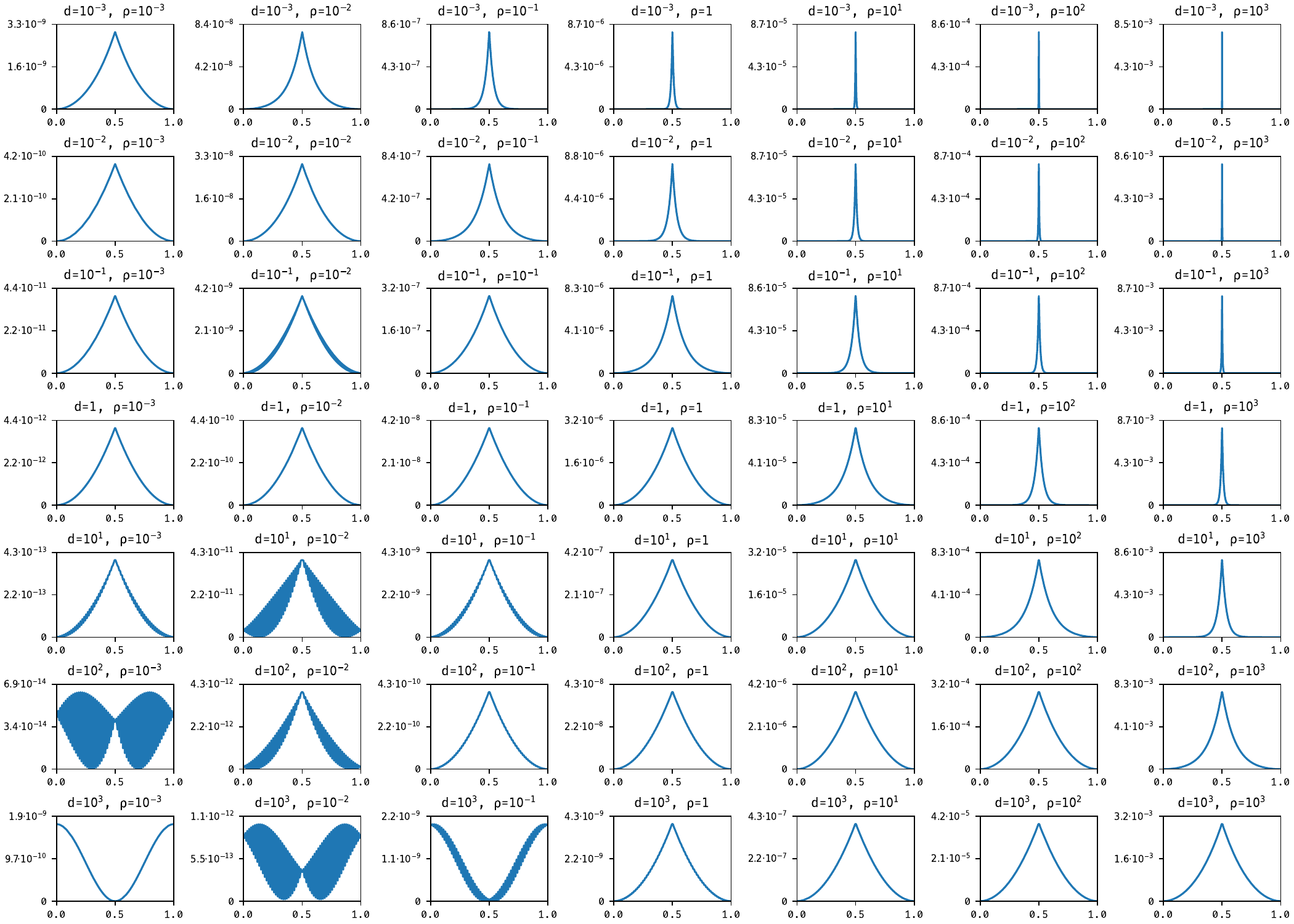}
  \caption{$\sigma_{d}$ for Heaviside profile, low $\theta_{g}$ value ($\beta=1$, $\theta_g=0.01$).}
  \label{fig:HSl-sd}
\end{figure}

\begin{figure}[p]
  \centering
  \includegraphics[width=0.87\linewidth]{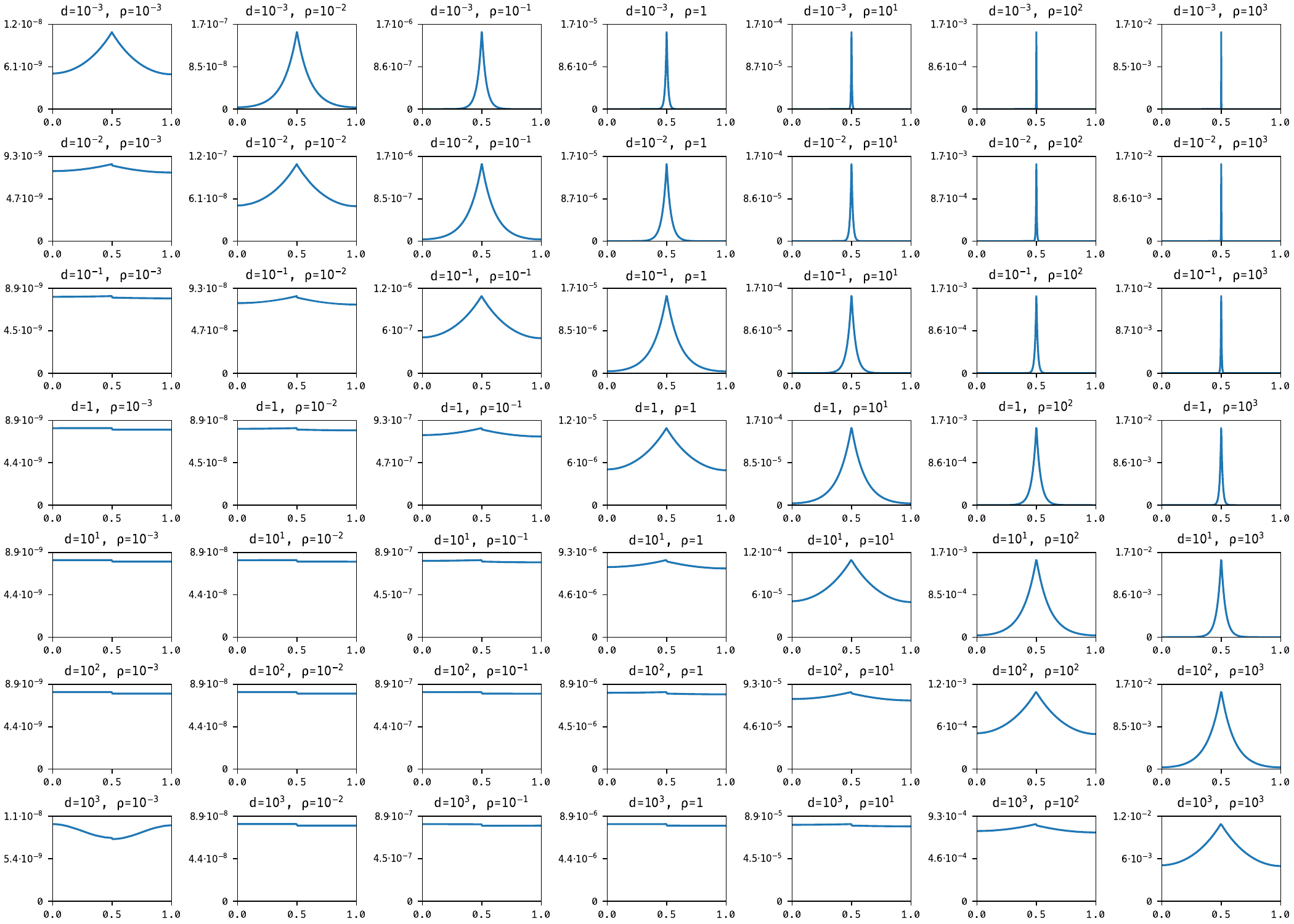}
  \caption{$\sigma_{\textrm{tot}}$ for Heaviside profile, low $\theta_{g}$ value ($\beta=1$, $\theta_g=0.01$).}
  \label{fig:HSl-st}
\end{figure}


\begin{figure}[p]
  \centering
  \includegraphics[width=0.87\linewidth]{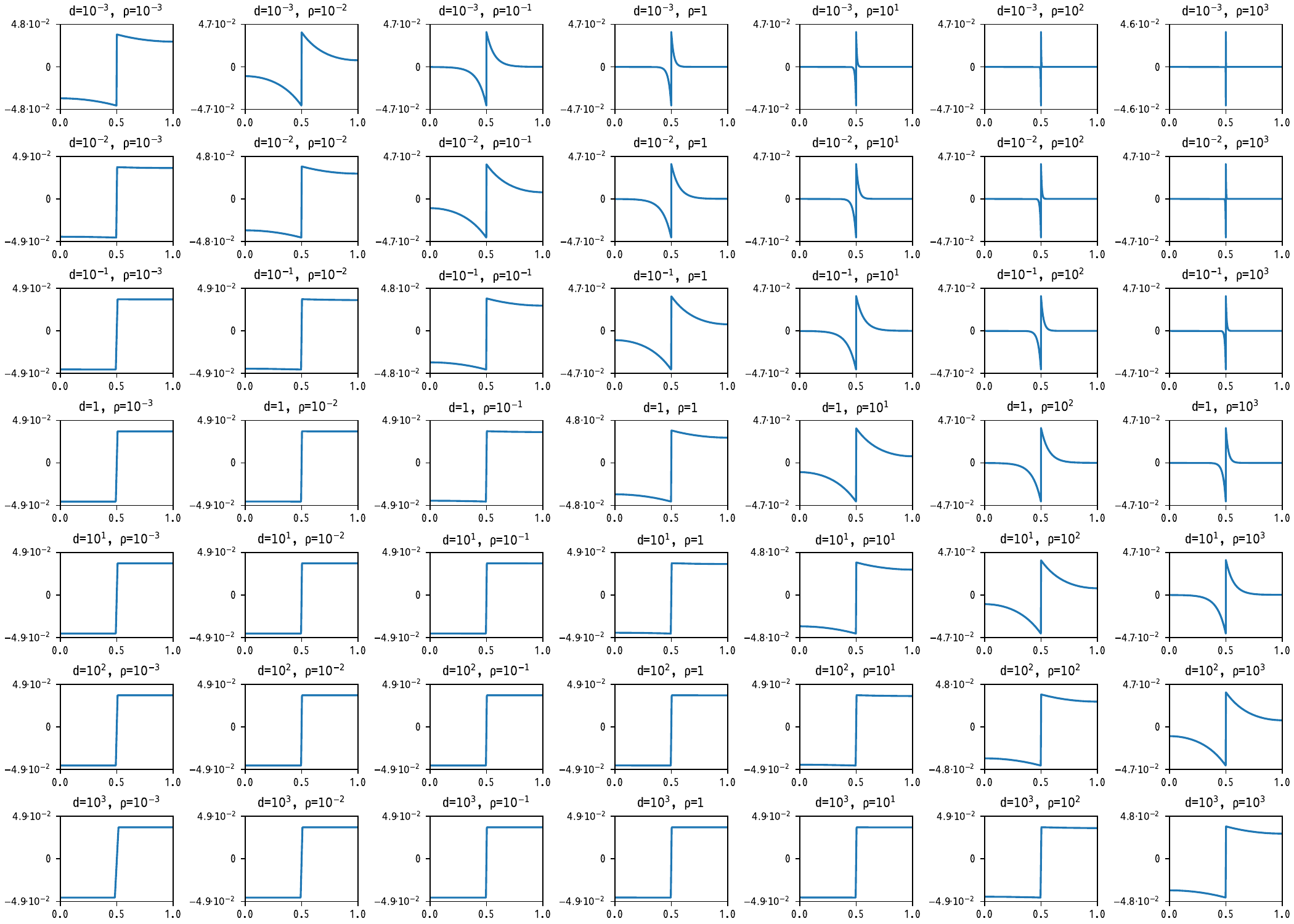}
  \caption{$\alpha$ for Heaviside profile, high $\theta_{g}$ value ($\beta=1$, $\theta_g=0.1$).}
  \label{fig:HSh-a}
\end{figure}

\begin{figure}[p]
  \centering
  \includegraphics[width=0.87\linewidth]{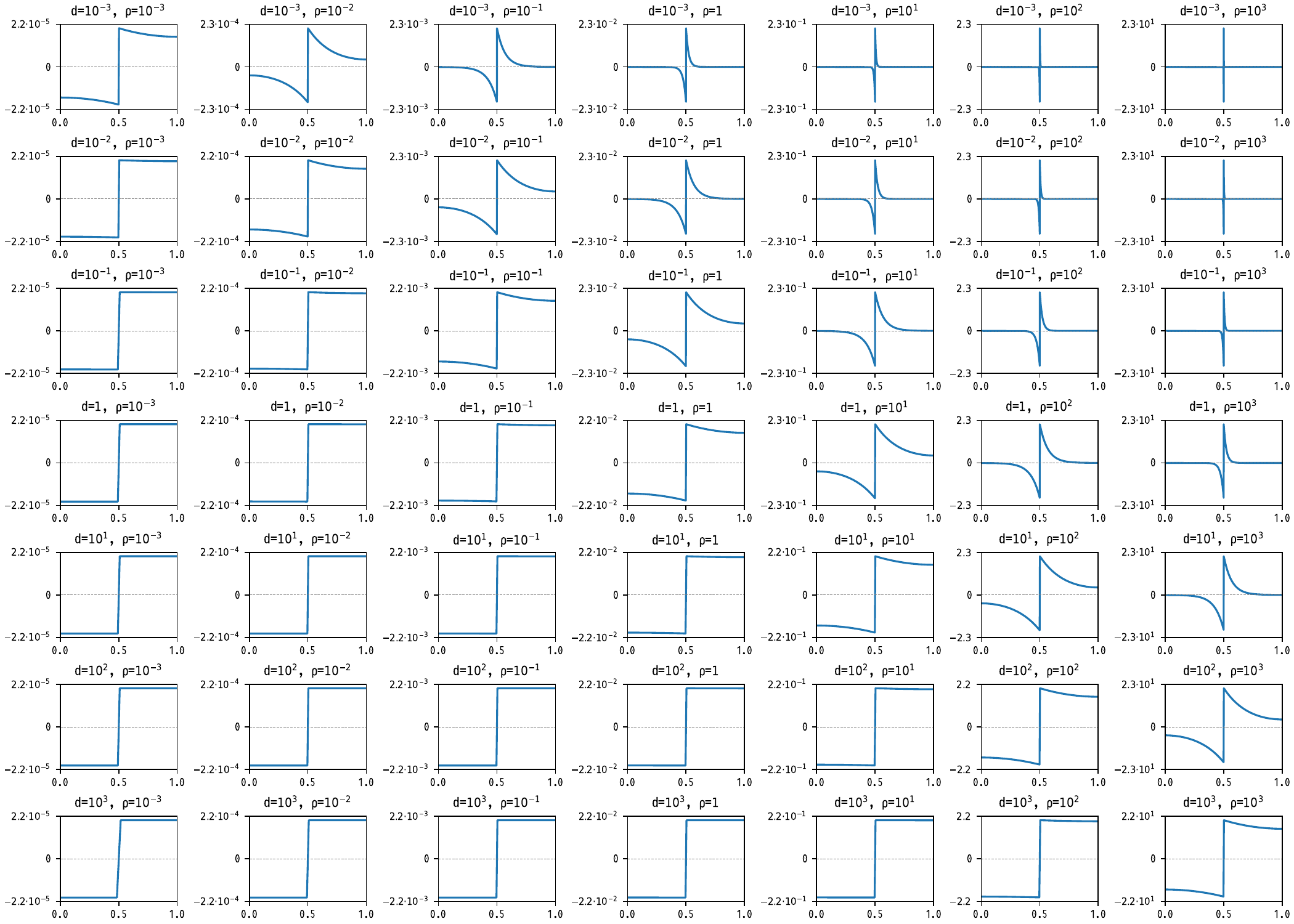}
  \caption{$\varphi$ for Heaviside profile, high $\theta_{g}$ value ($\beta=1$, $\theta_g=0.1$).}
  \label{fig:HSh-p}
\end{figure}

\begin{figure}[p]
  \centering
  \includegraphics[width=0.87\linewidth]{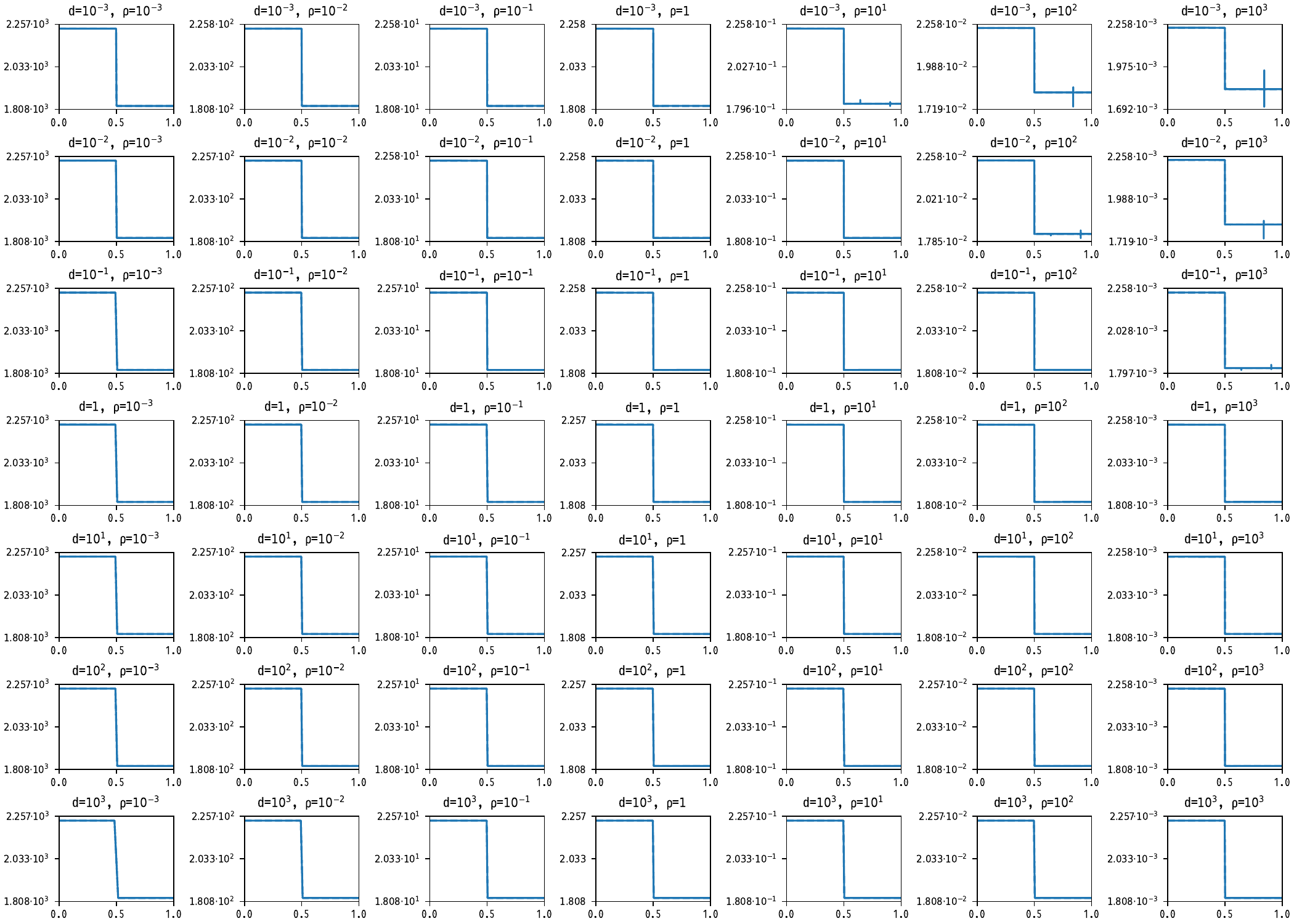}
  \caption{$R_{\textrm{tot}}$ for Heaviside profile, high $\theta_{g}$ value ($\beta=1$, $\theta_g=0.1$).}
  \label{fig:HSh-r}
\end{figure}

\begin{figure}[p]
  \centering
  \includegraphics[width=0.87\linewidth]{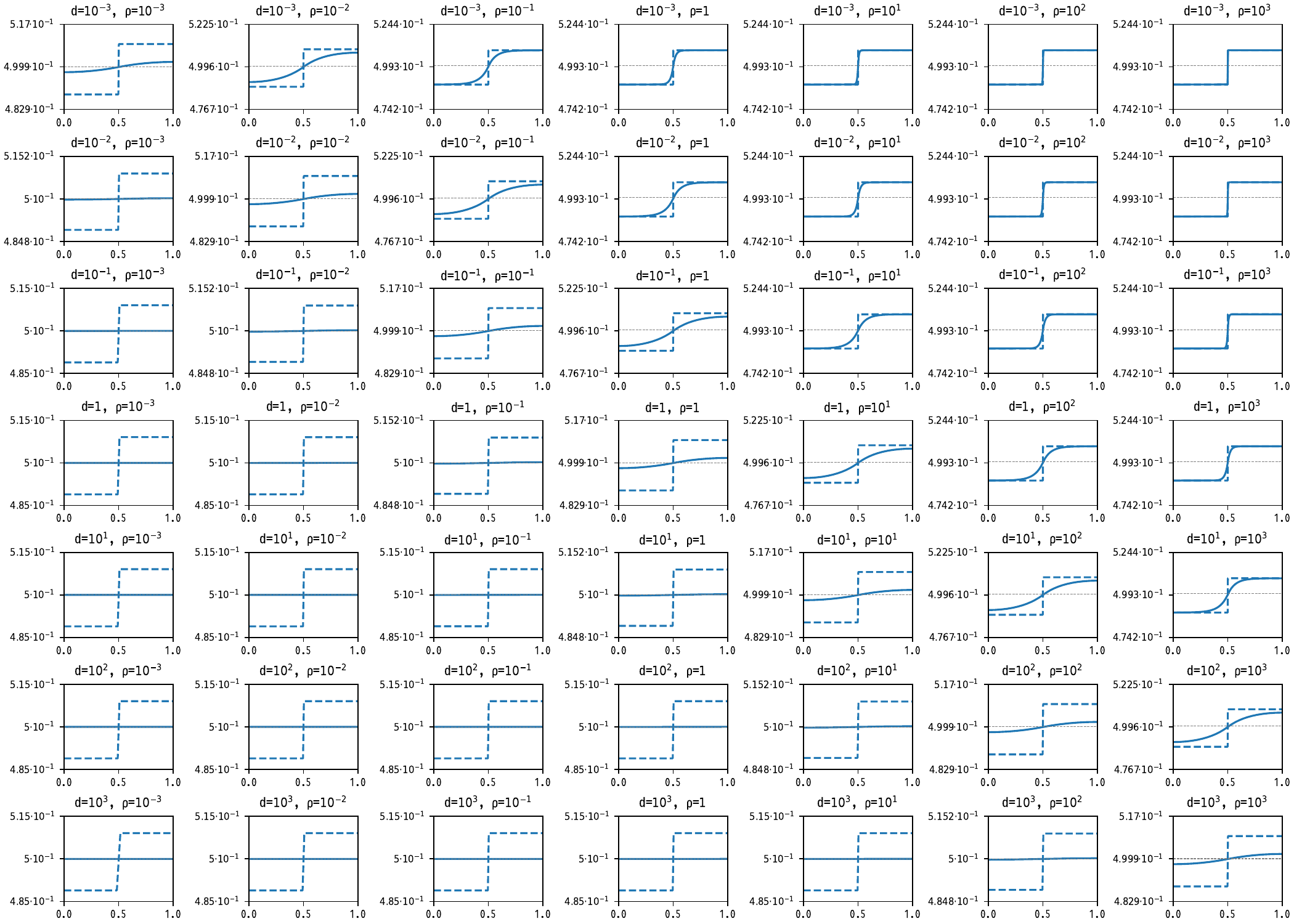}
  \caption{$u_{2}$ for Heaviside profile, high $\theta_{g}$ value ($\beta=1$, $\theta_g=0.1$).}
  \label{fig:HSh-u}
\end{figure}

\begin{figure}[p]
  \centering
  \includegraphics[width=0.87\linewidth]{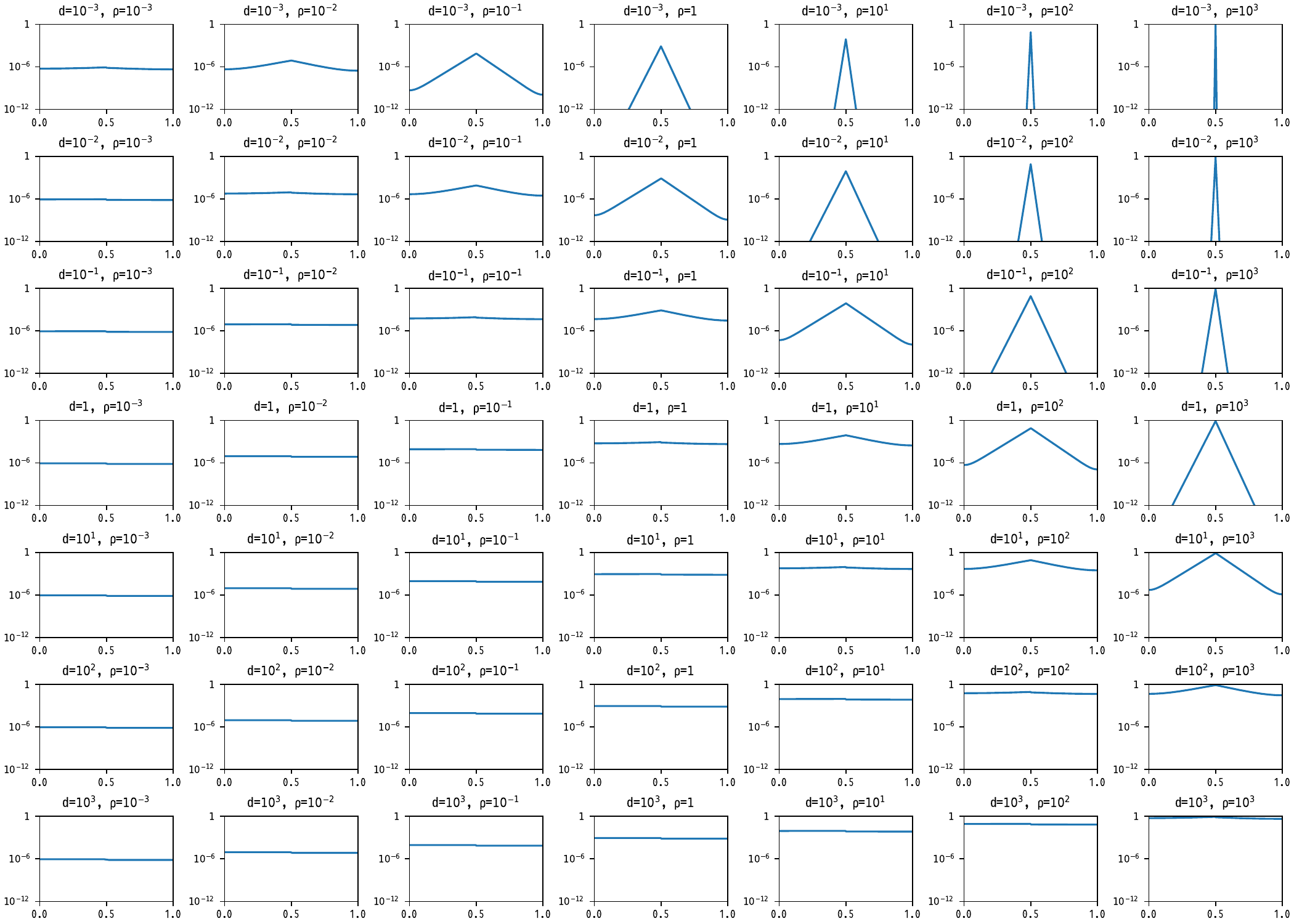}
  \caption{$\sigma_{r}$ for Heaviside profile, high $\theta_{g}$ value ($\beta=1$, $\theta_g=0.1$).}
  \label{fig:HSh-sr}
\end{figure}

\begin{figure}[p]
  \centering
  \includegraphics[width=0.87\linewidth]{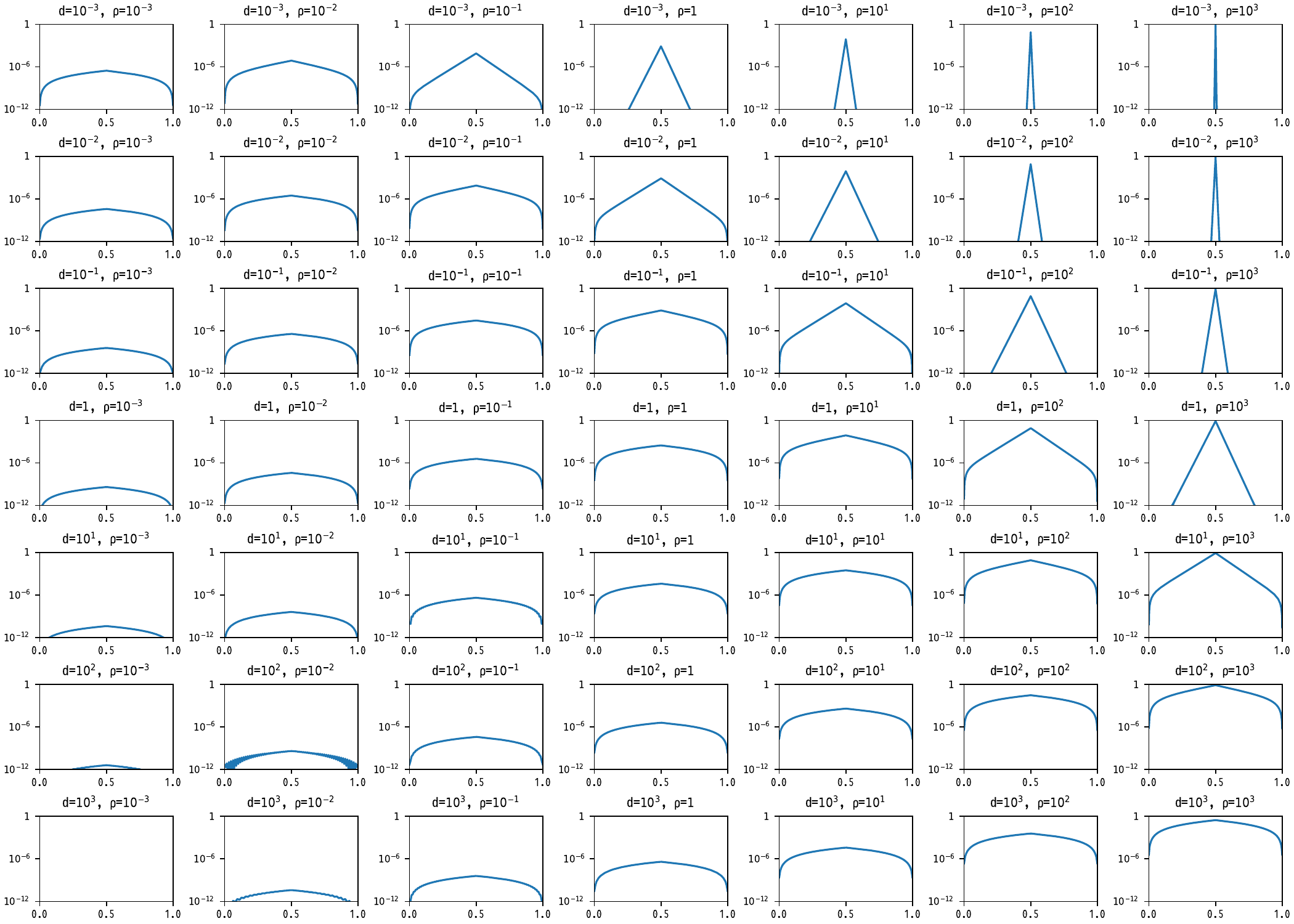}
  \caption{$\sigma_{d}$ for Heaviside profile, high $\theta_{g}$ value ($\beta=1$, $\theta_g=0.1$).}
  \label{fig:HSh-sd}
\end{figure}

\begin{figure}[p]
  \centering
  \includegraphics[width=0.87\linewidth]{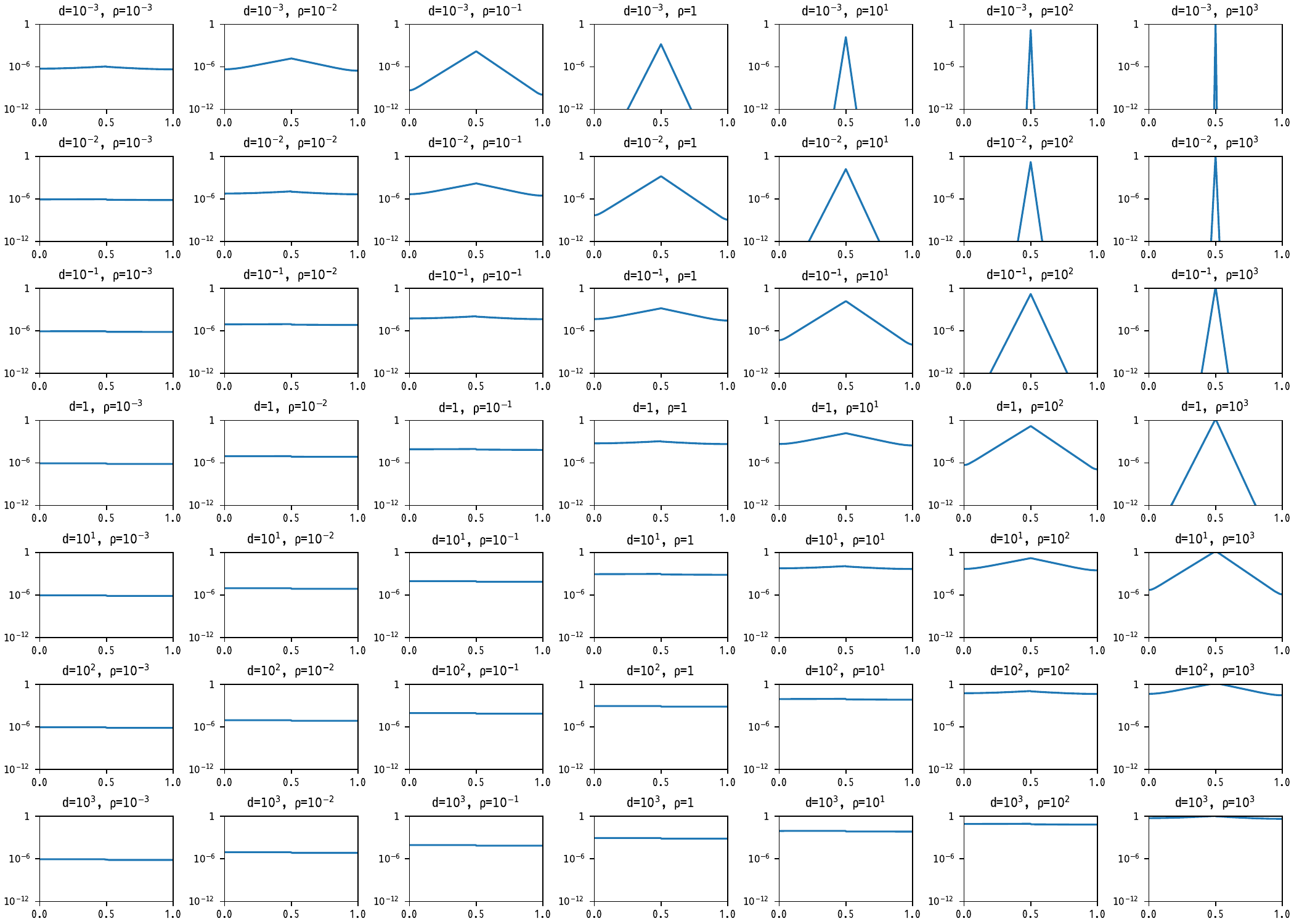}
  \caption{$\sigma_{\textrm{tot}}$ for Heaviside profile, high $\theta_{g}$ value ($\beta=1$, $\theta_g=0.1$).}
  \label{fig:HSh-st}
\end{figure}


\begin{figure}[p]
  \centering
  \includegraphics[width=0.87\linewidth]{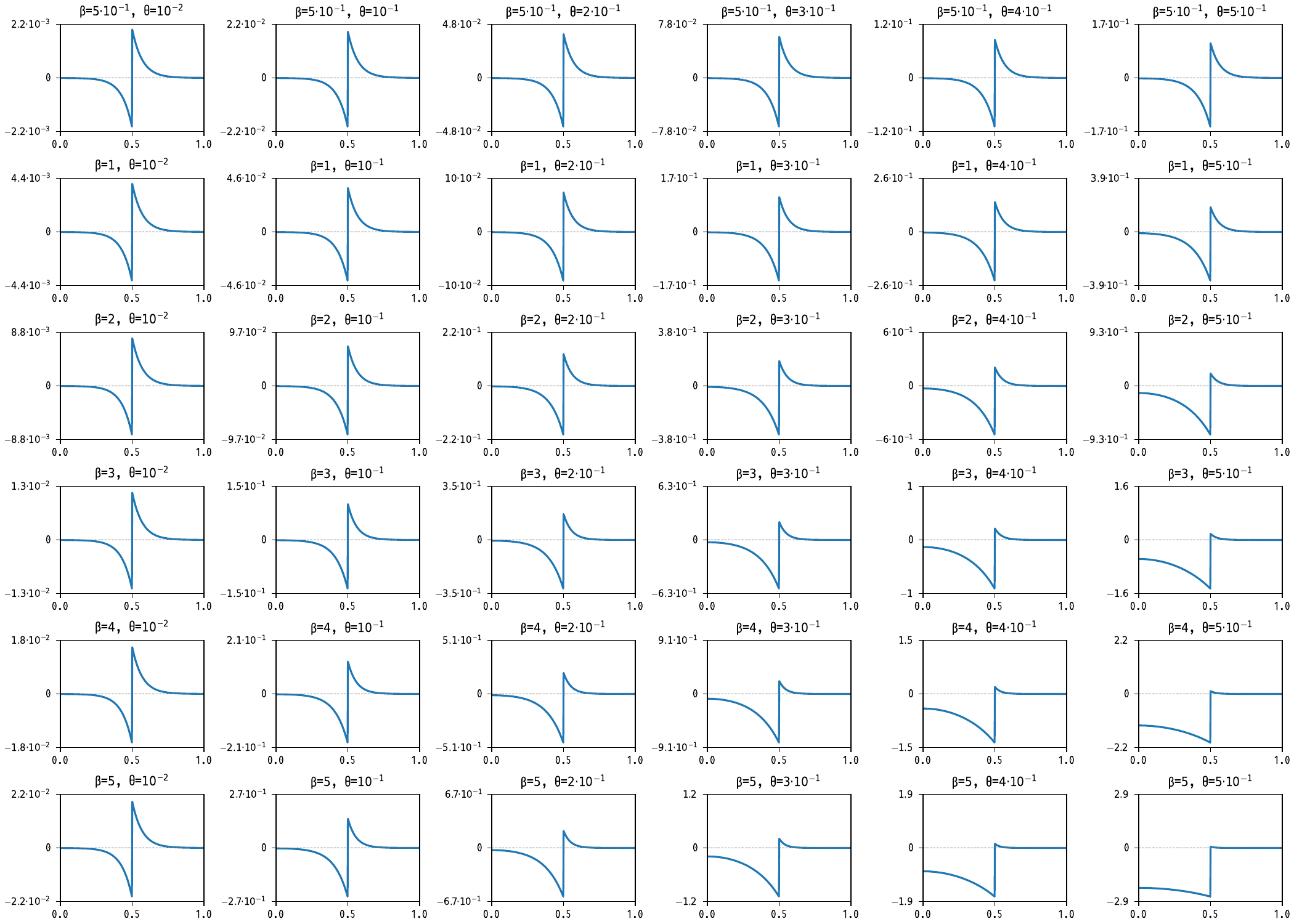}
  \caption{$\alpha$ for Heaviside profile, as a function of $\beta$ and $\theta_g$ ($d=1$, $\rho=100$).}
  \label{fig:HSbt-a}
\end{figure}

\begin{figure}[p]
  \centering
  \includegraphics[width=0.87\linewidth]{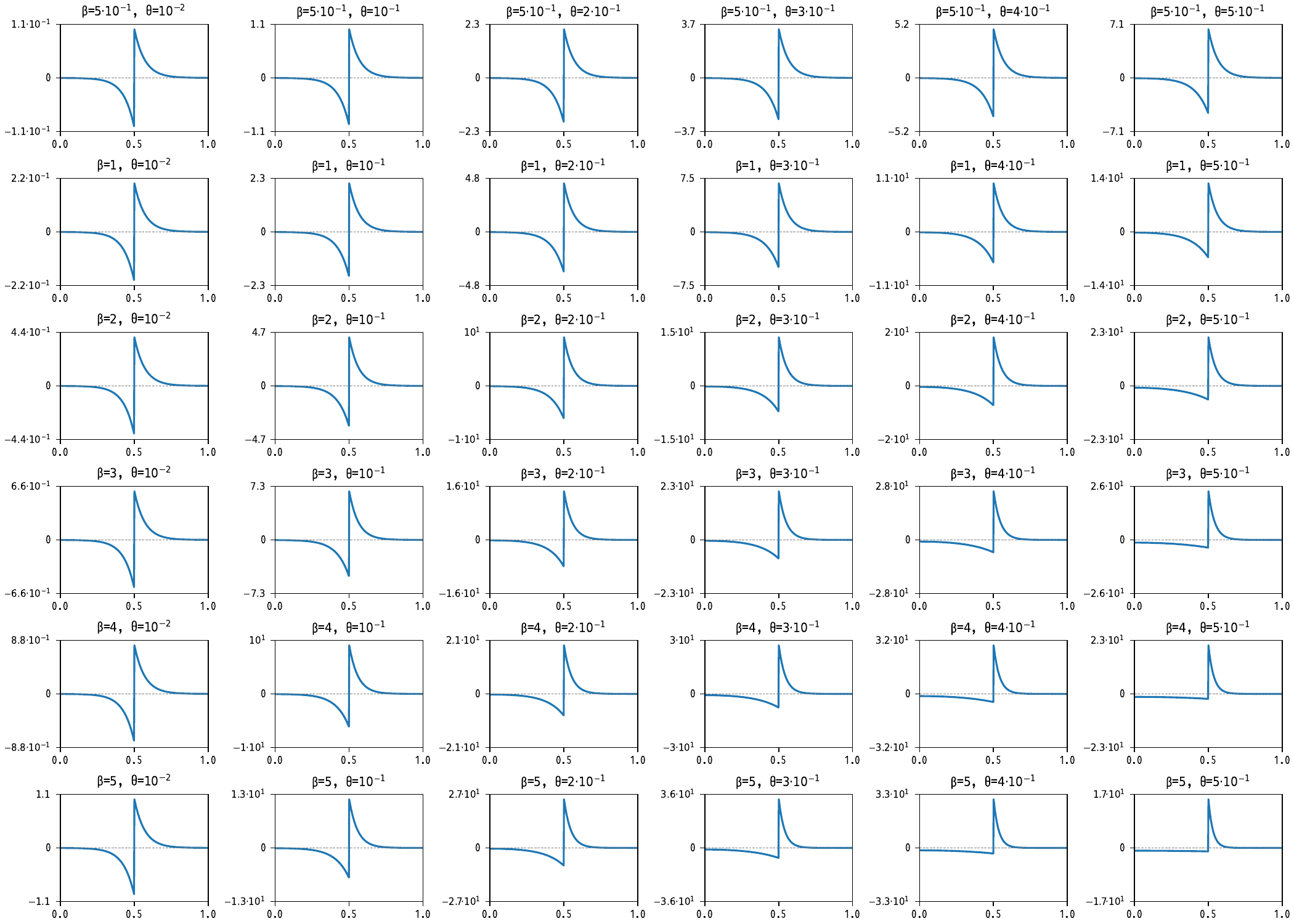}
  \caption{$\varphi$ for Heaviside profile, as a function of $\beta$ and $\theta_g$ ($d=1$, $\rho=100$).}
  \label{fig:HSbt-p}
\end{figure}

\begin{figure}[p]
  \centering
  \includegraphics[width=0.87\linewidth]{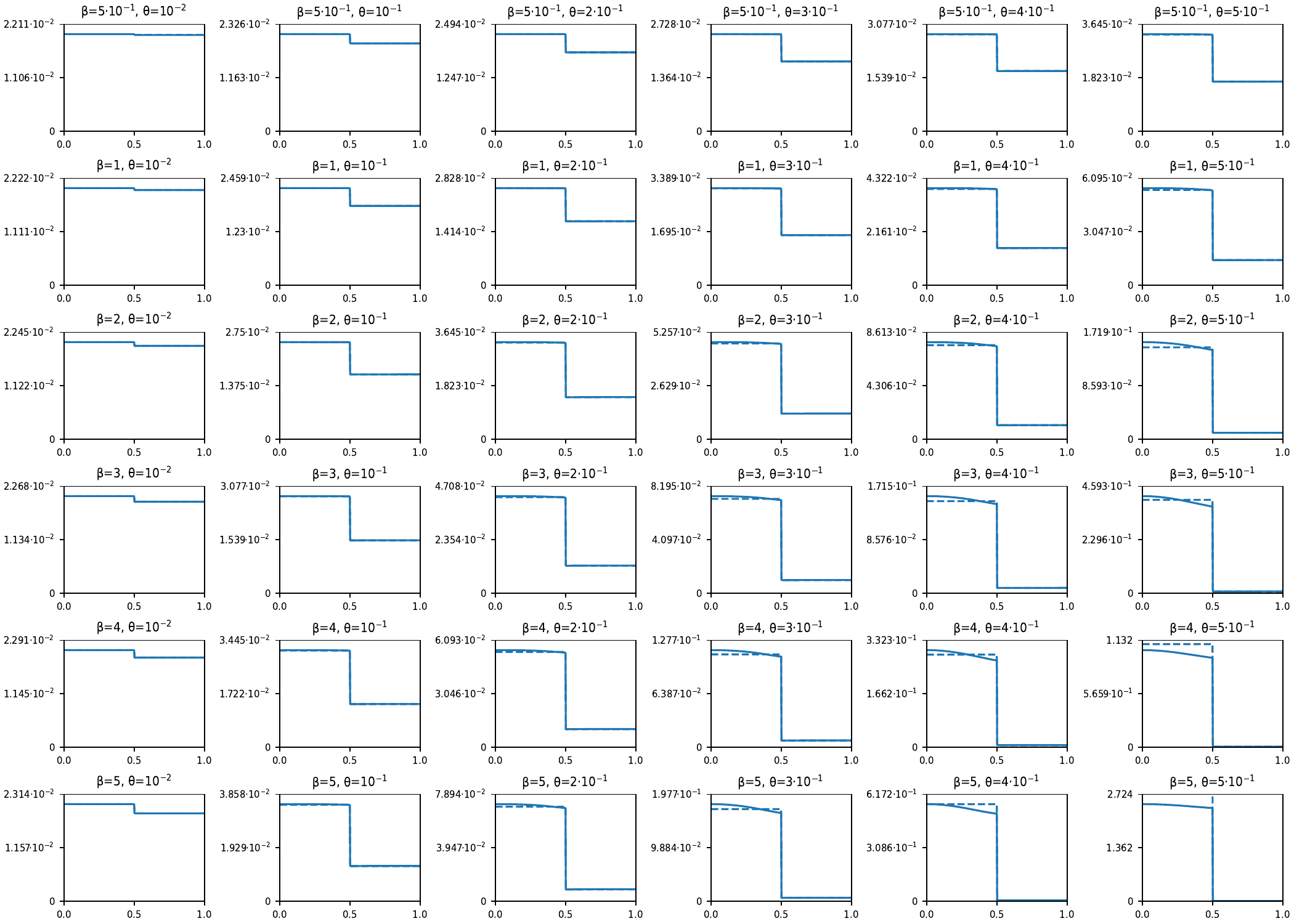}
  \caption{$R_{\textrm{tot}}$ for Heaviside profile, as a function of $\beta$ and $\theta_g$ ($d=1$, $\rho=100$).}
  \label{fig:HSbt-r}
\end{figure}

\begin{figure}[p]
  \centering
  \includegraphics[width=0.87\linewidth]{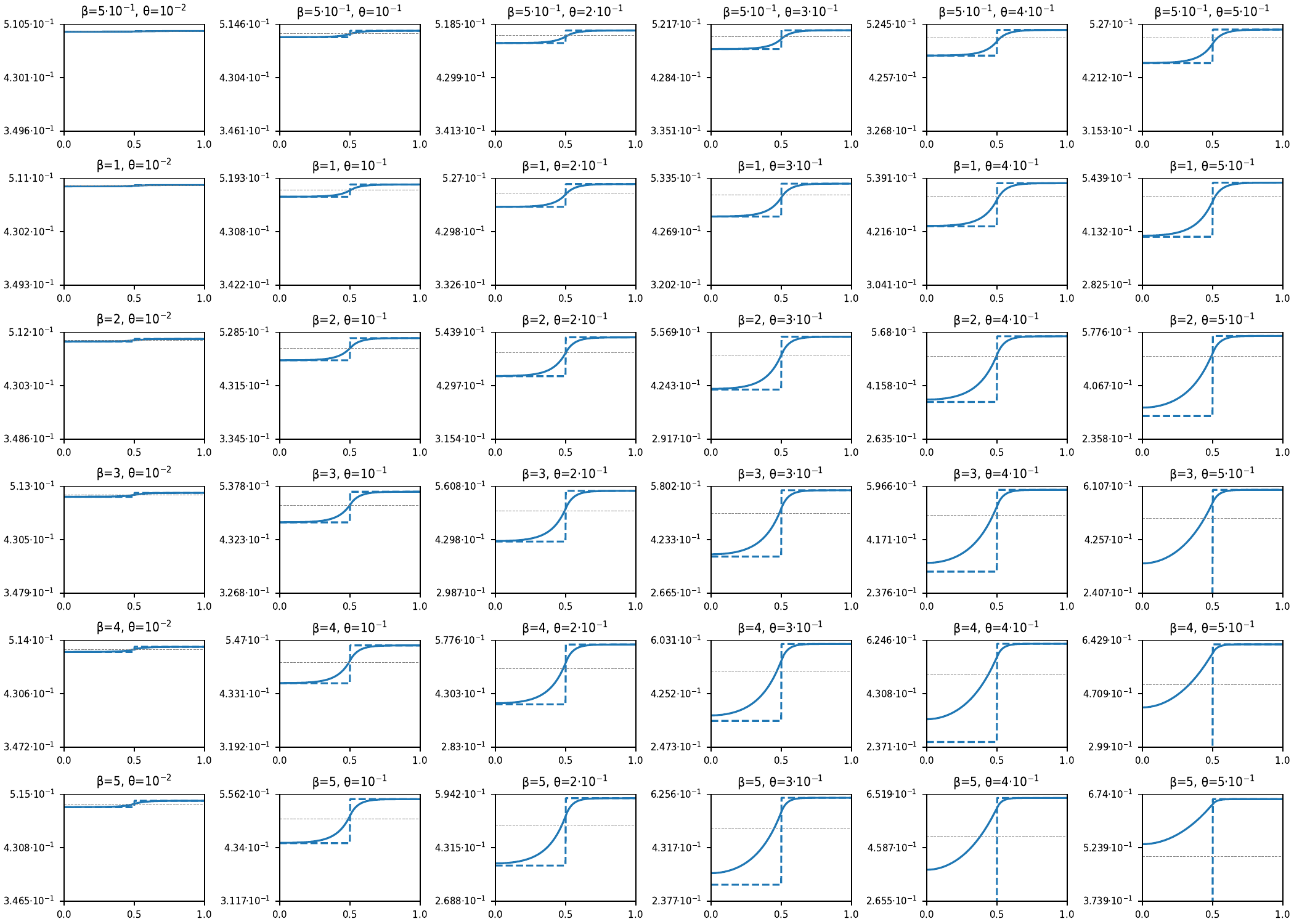}
  \caption{$u_{2}$ for Heaviside profile, as a function of $\beta$ and $\theta_g$ ($d=1$, $\rho=100$).}
  \label{fig:HSbt-u}
\end{figure}

\begin{figure}[p]
  \centering
  \includegraphics[width=0.87\linewidth]{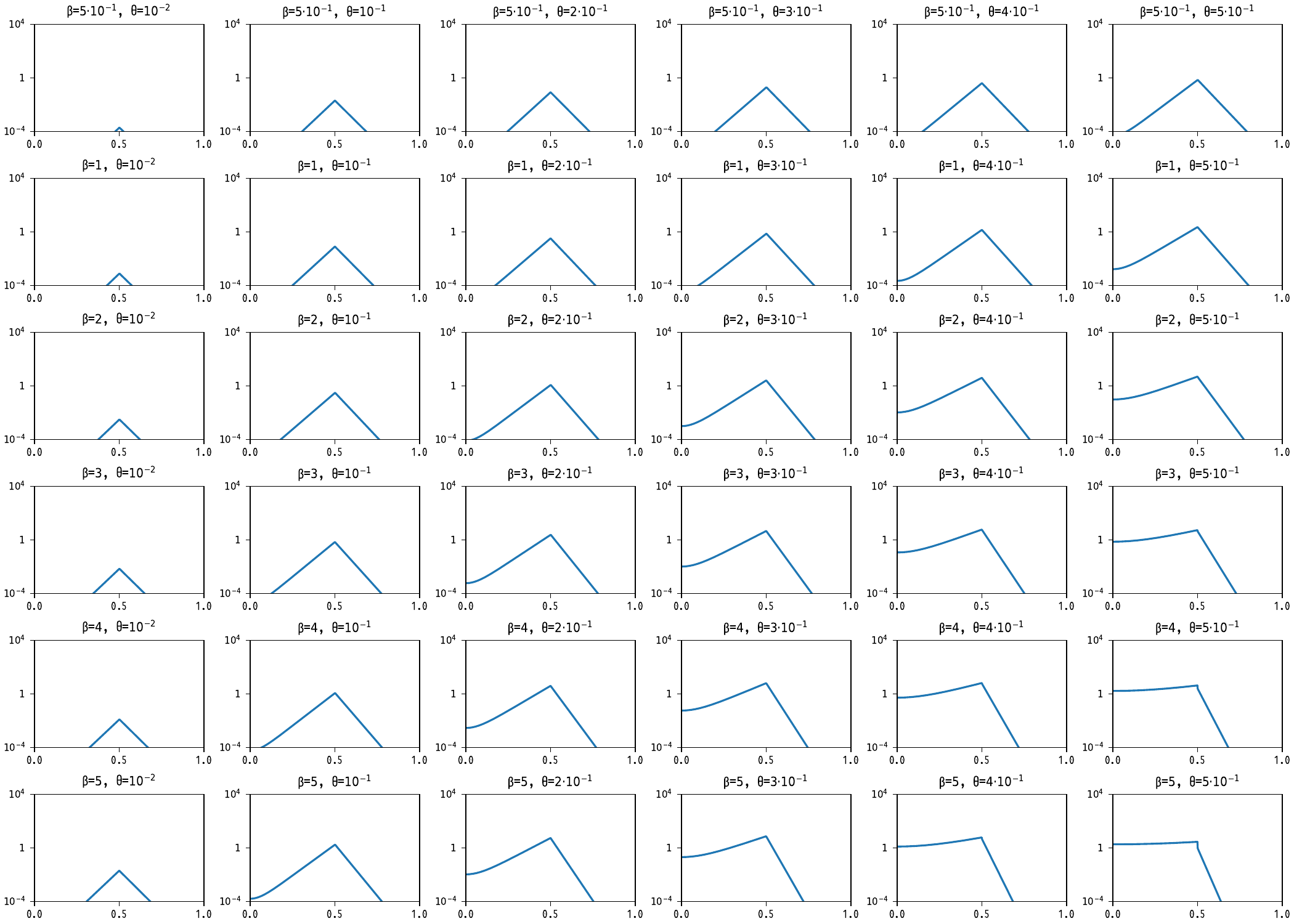}
  \caption{$\sigma_{r}$ for Heaviside profile, as a function of $\beta$ and $\theta_g$ ($d=1$, $\rho=100$).}
  \label{fig:HSbt-sr}
\end{figure}

\begin{figure}[p]
  \centering
  \includegraphics[width=0.87\linewidth]{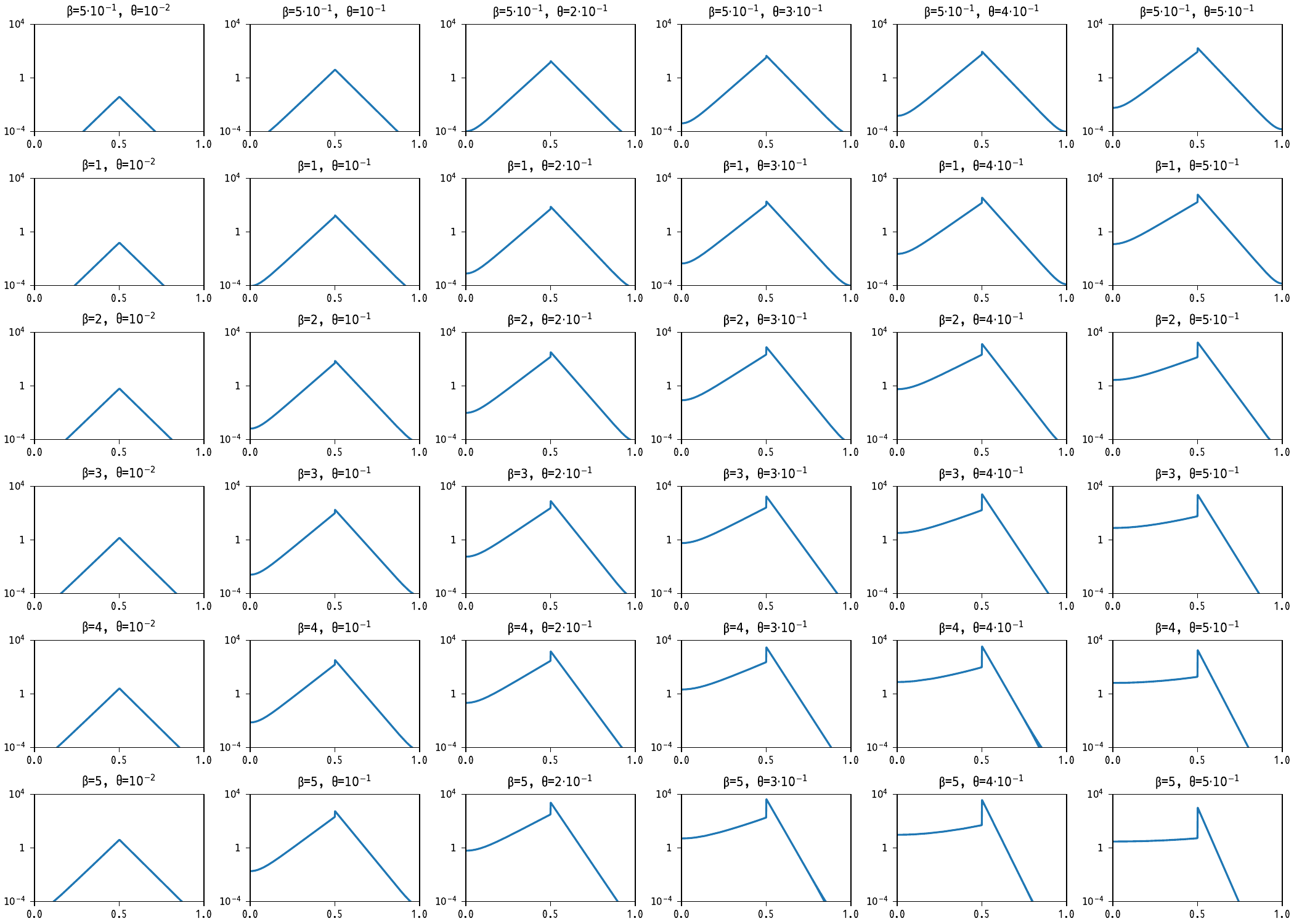}
  \caption{$\sigma_{d}$ for Heaviside profile, as a function of $\beta$ and $\theta_g$ ($d=1$, $\rho=100$).}
  \label{fig:HSbt-sd}
\end{figure}

\begin{figure}[p]
  \centering
  \includegraphics[width=0.87\linewidth]{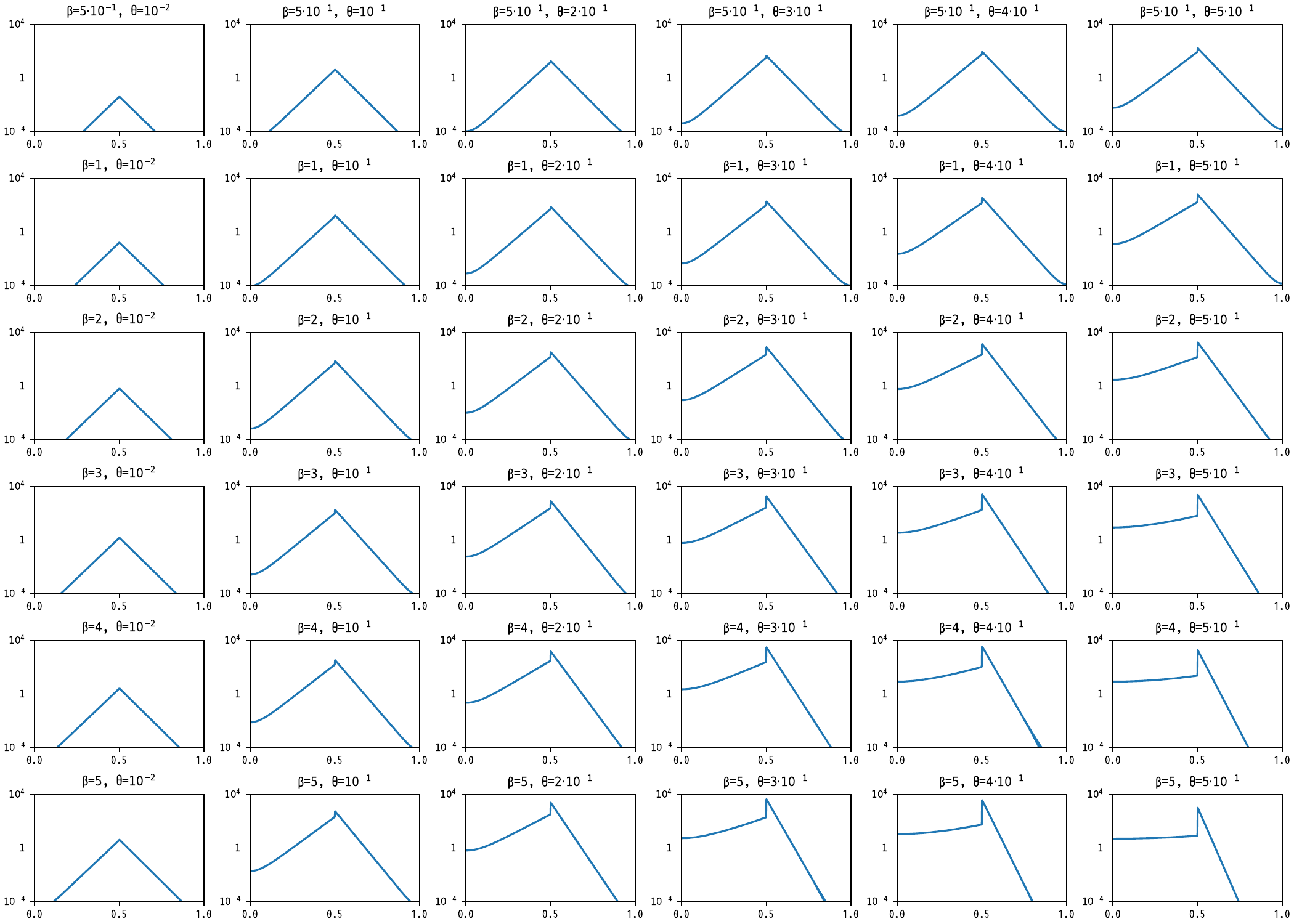}
  \caption{$\sigma_{\textrm{tot}}$ for Heaviside profile, as a function of $\beta$ and $\theta_g$ ($d=1$, $\rho=100$).}
  \label{fig:HSbt-st}
\end{figure}


\begin{figure}[p]
  \centering
  \includegraphics[width=0.87\linewidth]{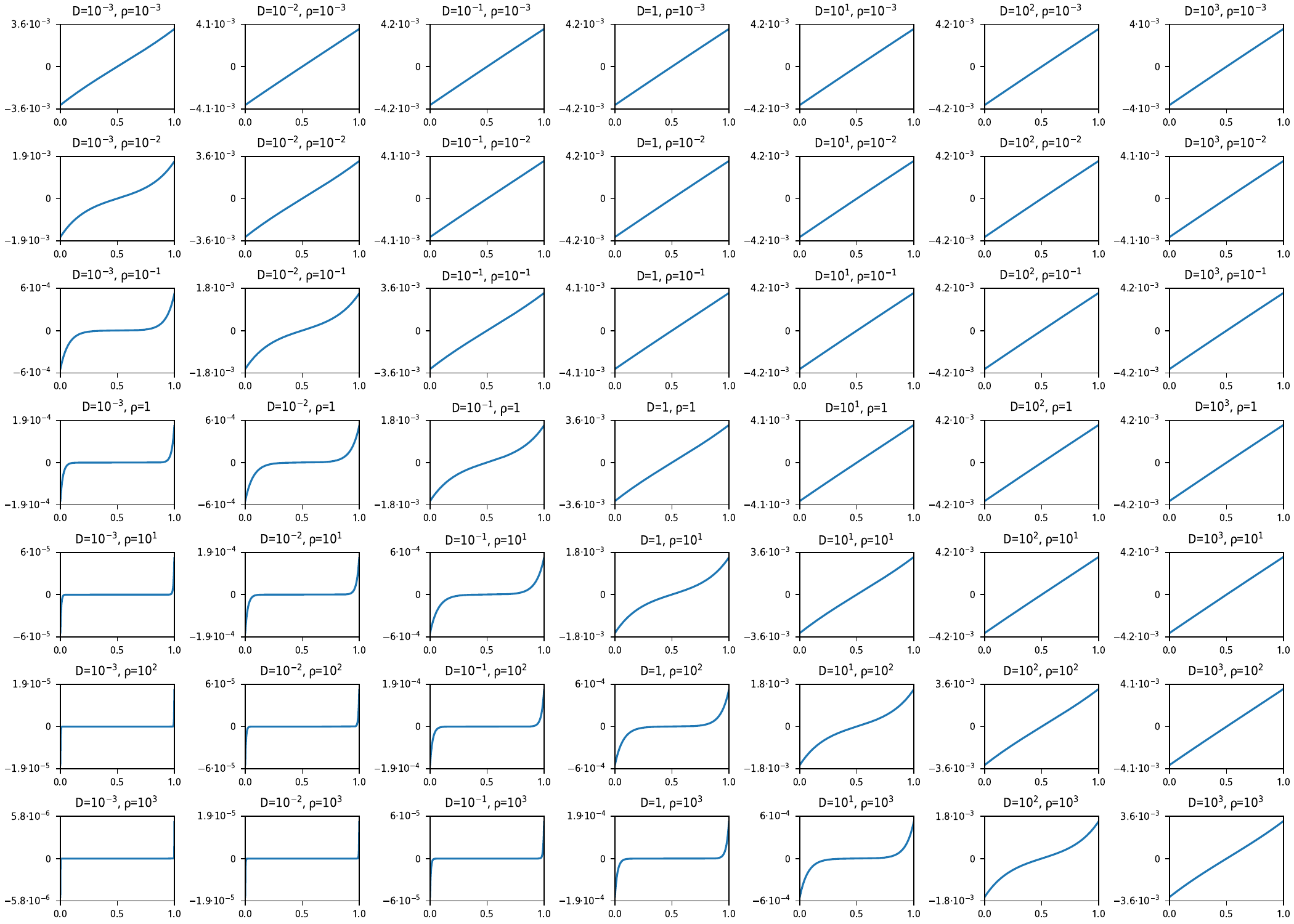}
  \caption{$\alpha$ for Linear profile, low $\theta_{g}$ value ($\beta=1$, $\theta_g=0.01$).}
  \label{fig:Linl-a}
\end{figure}

\begin{figure}[p]
  \centering
  \includegraphics[width=0.87\linewidth]{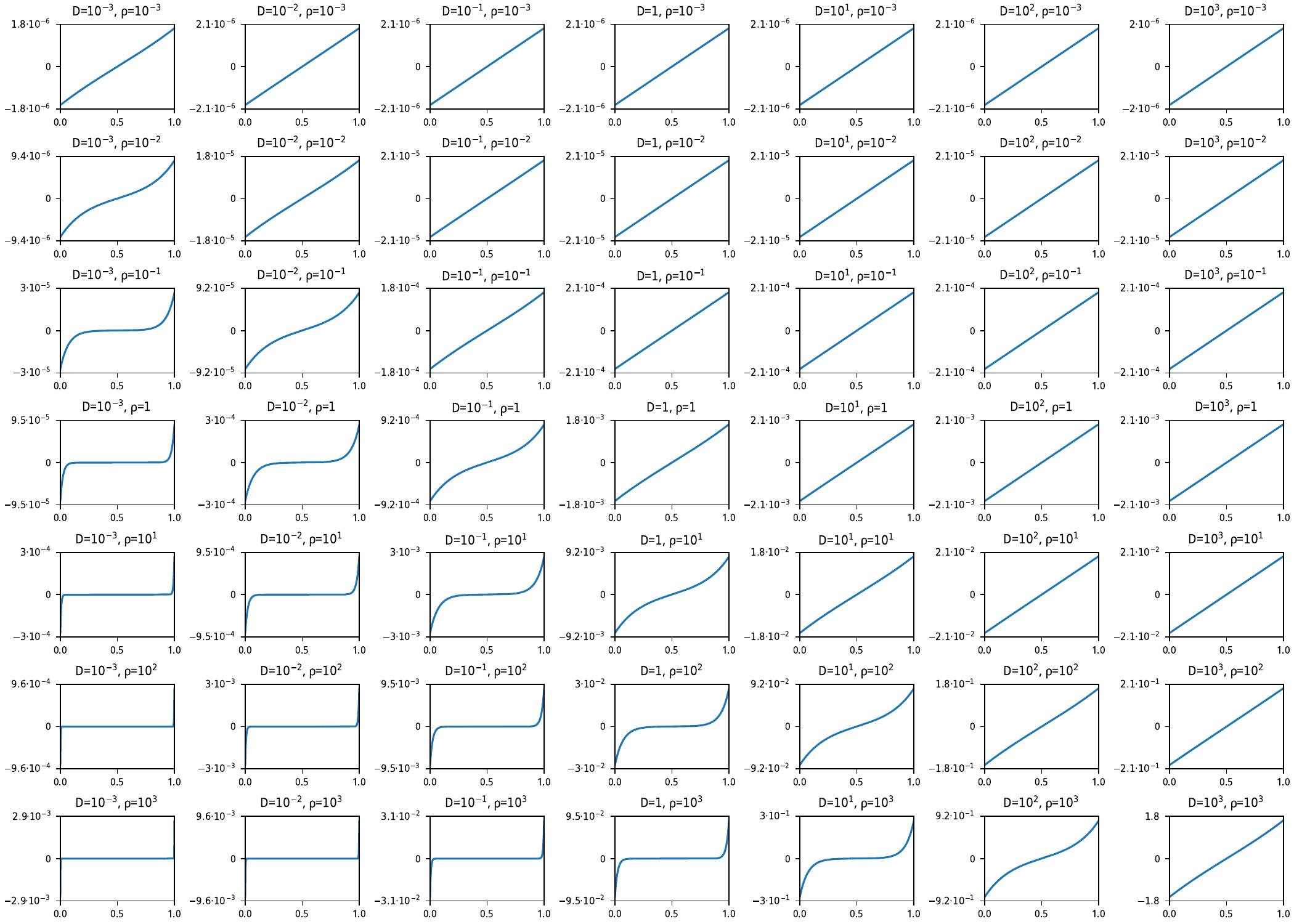}
  \caption{$\varphi$ for Linear profile, low $\theta_{g}$ value ($\beta=1$, $\theta_g=0.01$).}
  \label{fig:Linl-p}
\end{figure}

\begin{figure}[p]
  \centering
  \includegraphics[width=0.87\linewidth]{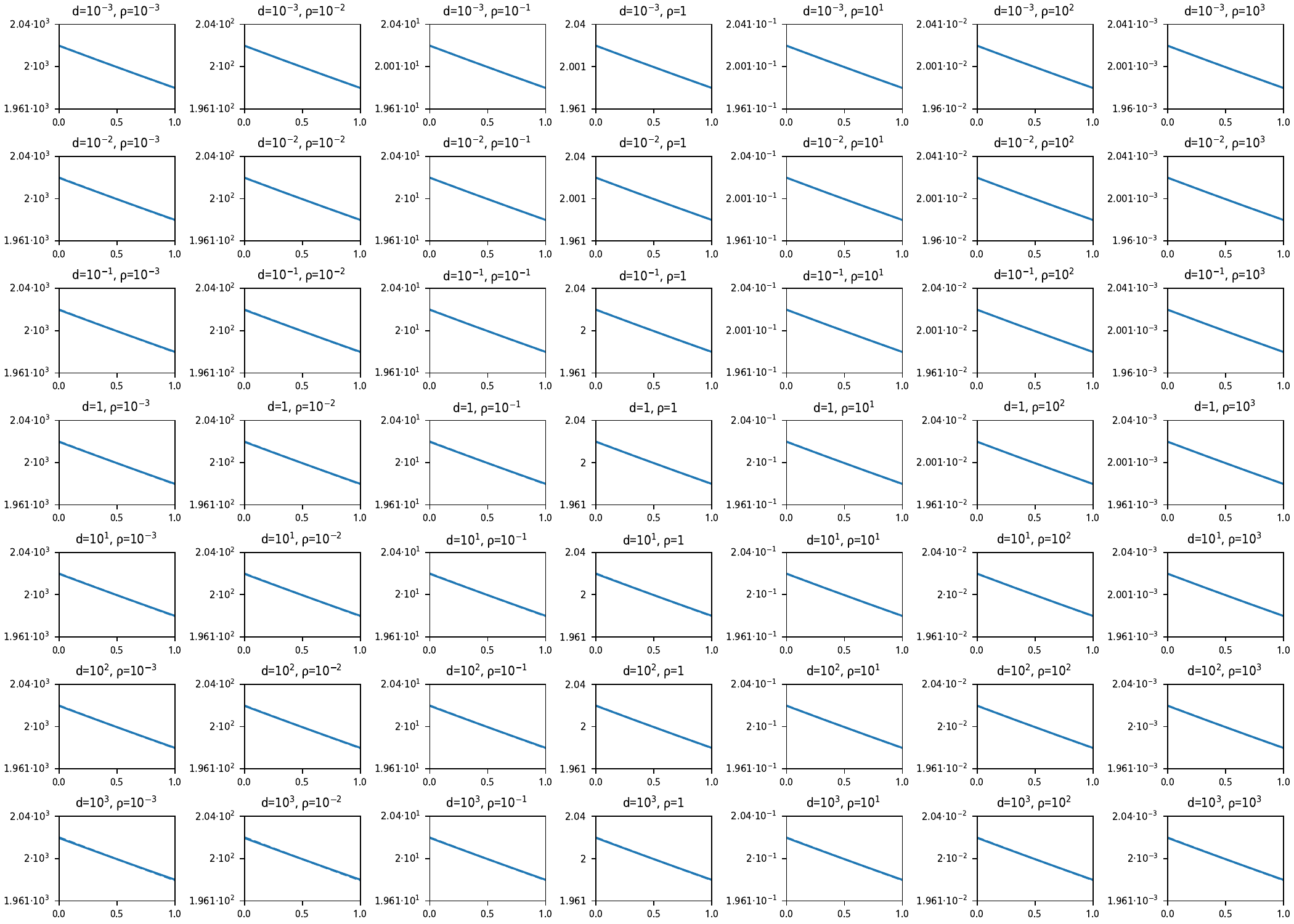}
  \caption{$R_{\textrm{tot}}$ for Linear profile, low $\theta_{g}$ value ($\beta=1$, $\theta_g=0.01$).}
  \label{fig:Linl-r}
\end{figure}

\begin{figure}[p]
  \centering
  \includegraphics[width=0.87\linewidth]{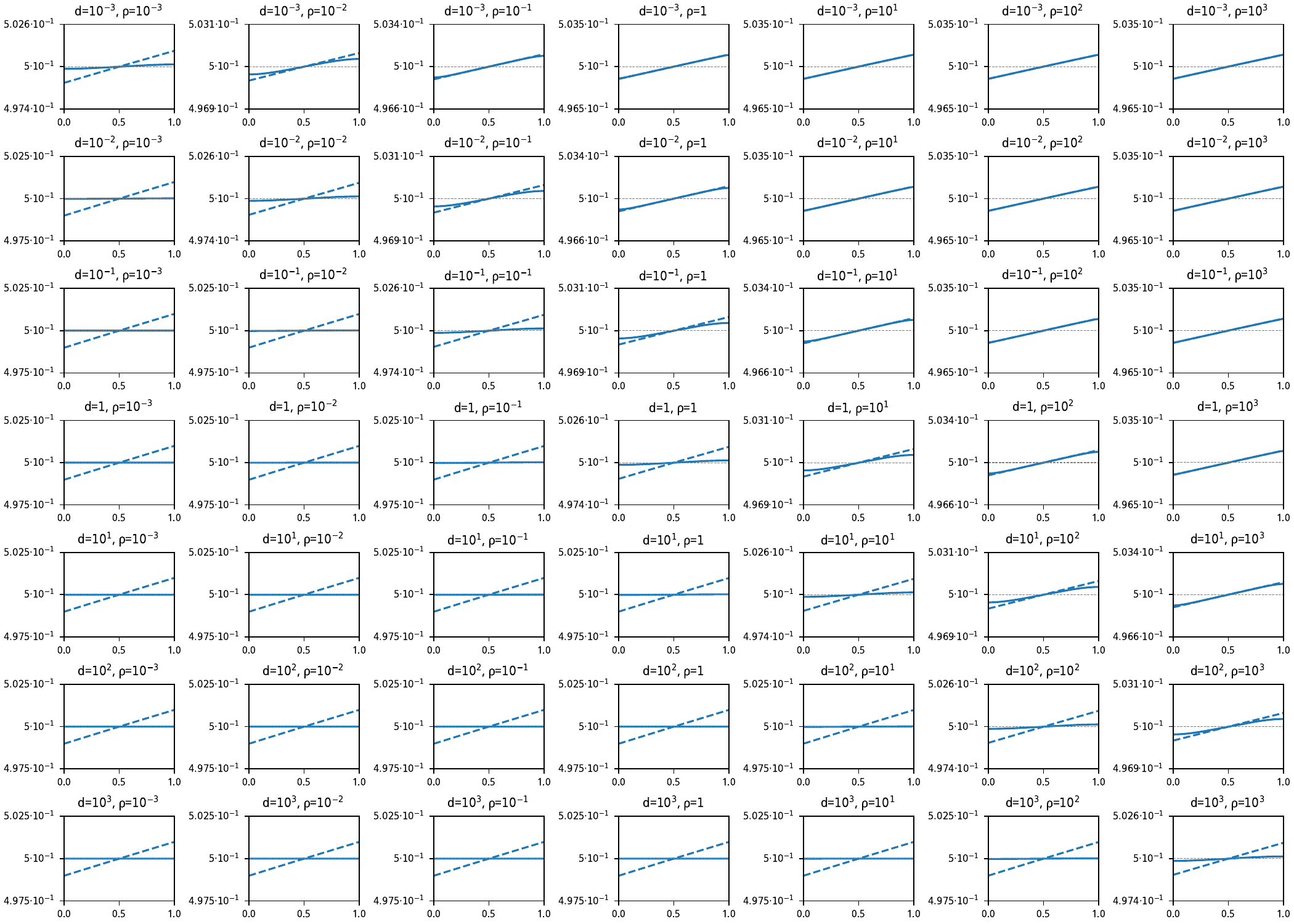}
  \caption{$u_{2}$ for Linear profile, low $\theta_{g}$ value ($\beta=1$, $\theta_g=0.01$).}
  \label{fig:Linl-u}
\end{figure}

\begin{figure}[p]
  \centering
  \includegraphics[width=0.87\linewidth]{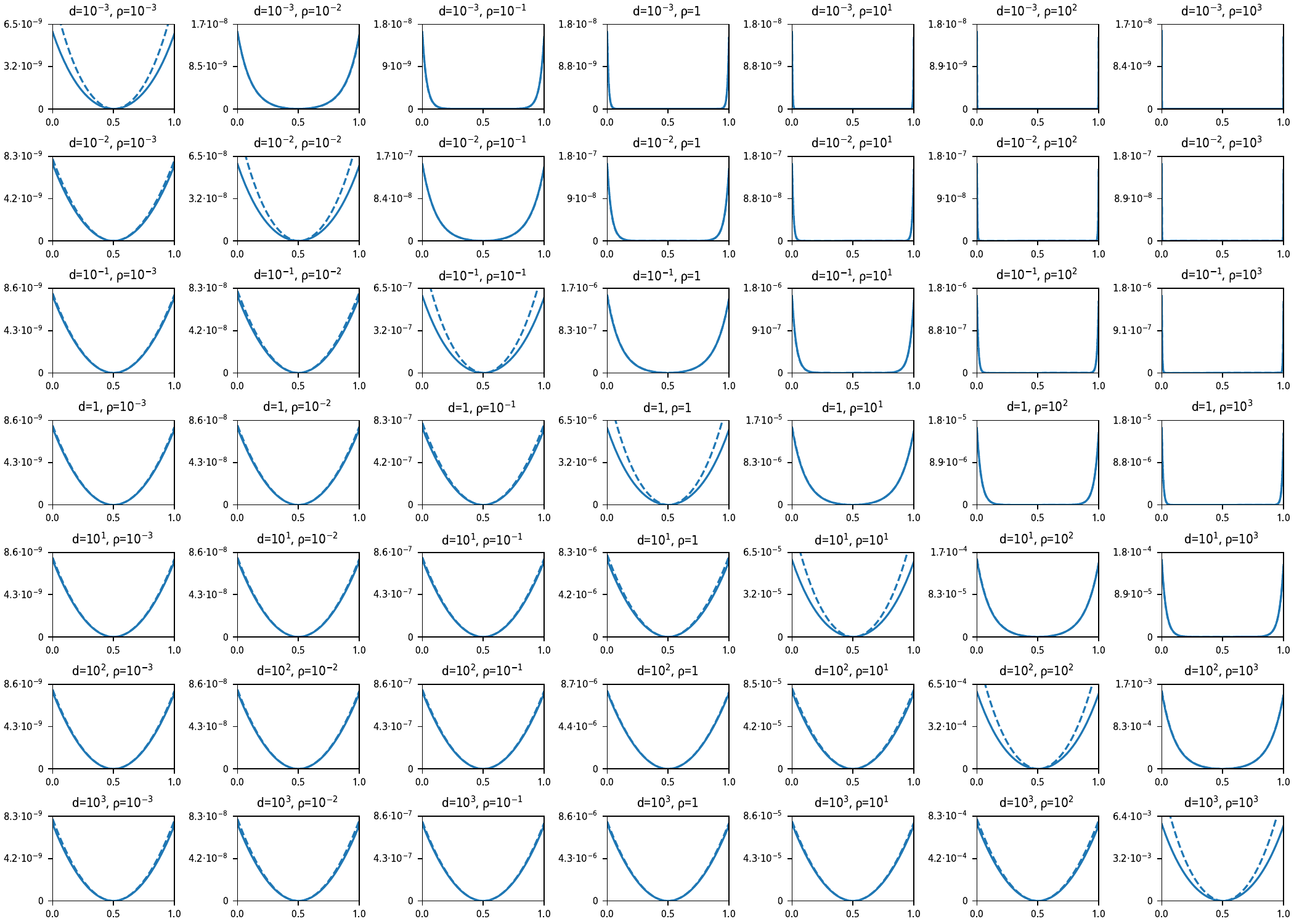}
  \caption{$\sigma_{r}$ for Linear profile, low $\theta_{g}$ value ($\beta=1$, $\theta_g=0.01$).}
  \label{fig:Linl-sr}
\end{figure}

\begin{figure}[p]
  \centering
  \includegraphics[width=0.87\linewidth]{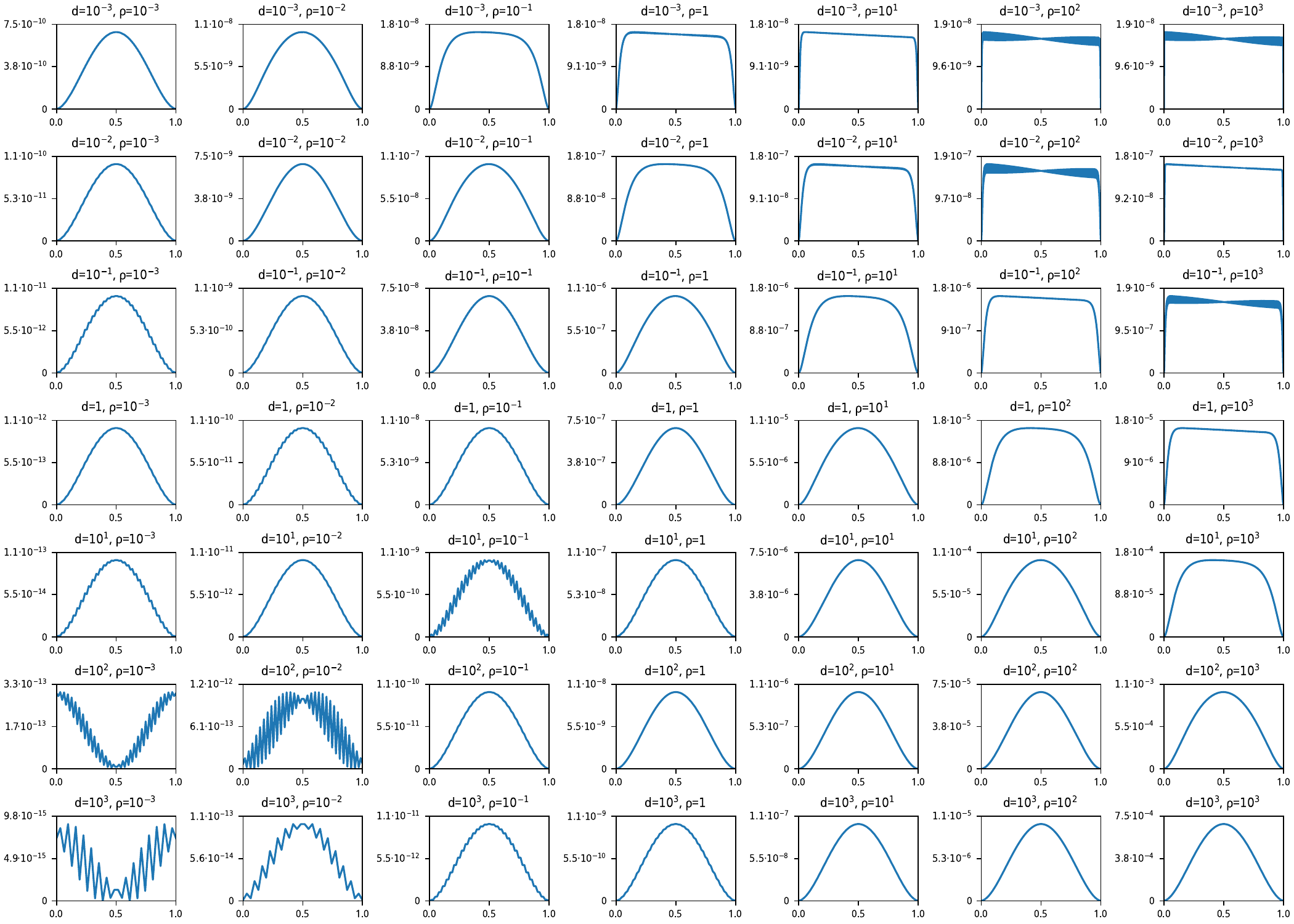}
  \caption{$\sigma_{d}$ for Linear profile, low $\theta_{g}$ value ($\beta=1$, $\theta_g=0.01$).}
  \label{fig:Linl-sd}
\end{figure}

\begin{figure}[p]
  \centering
  \includegraphics[width=0.87\linewidth]{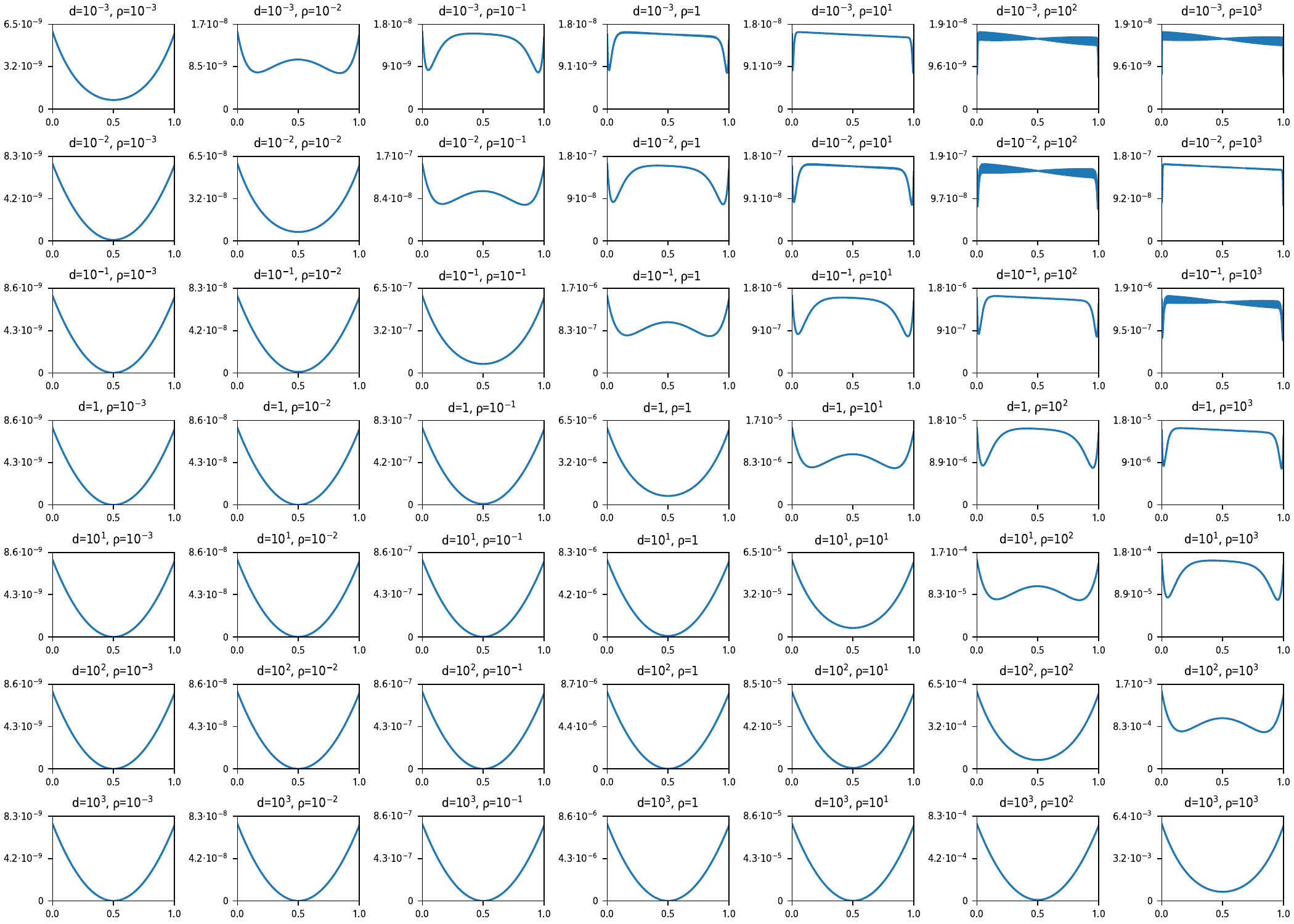}
  \caption{$\sigma_{\textrm{tot}}$ for Linear profile, low $\theta_{g}$ value ($\beta=1$, $\theta_g=0.01$).}
  \label{fig:Linl-st}
\end{figure}


\begin{figure}[p]
  \centering
  \includegraphics[width=0.87\linewidth]{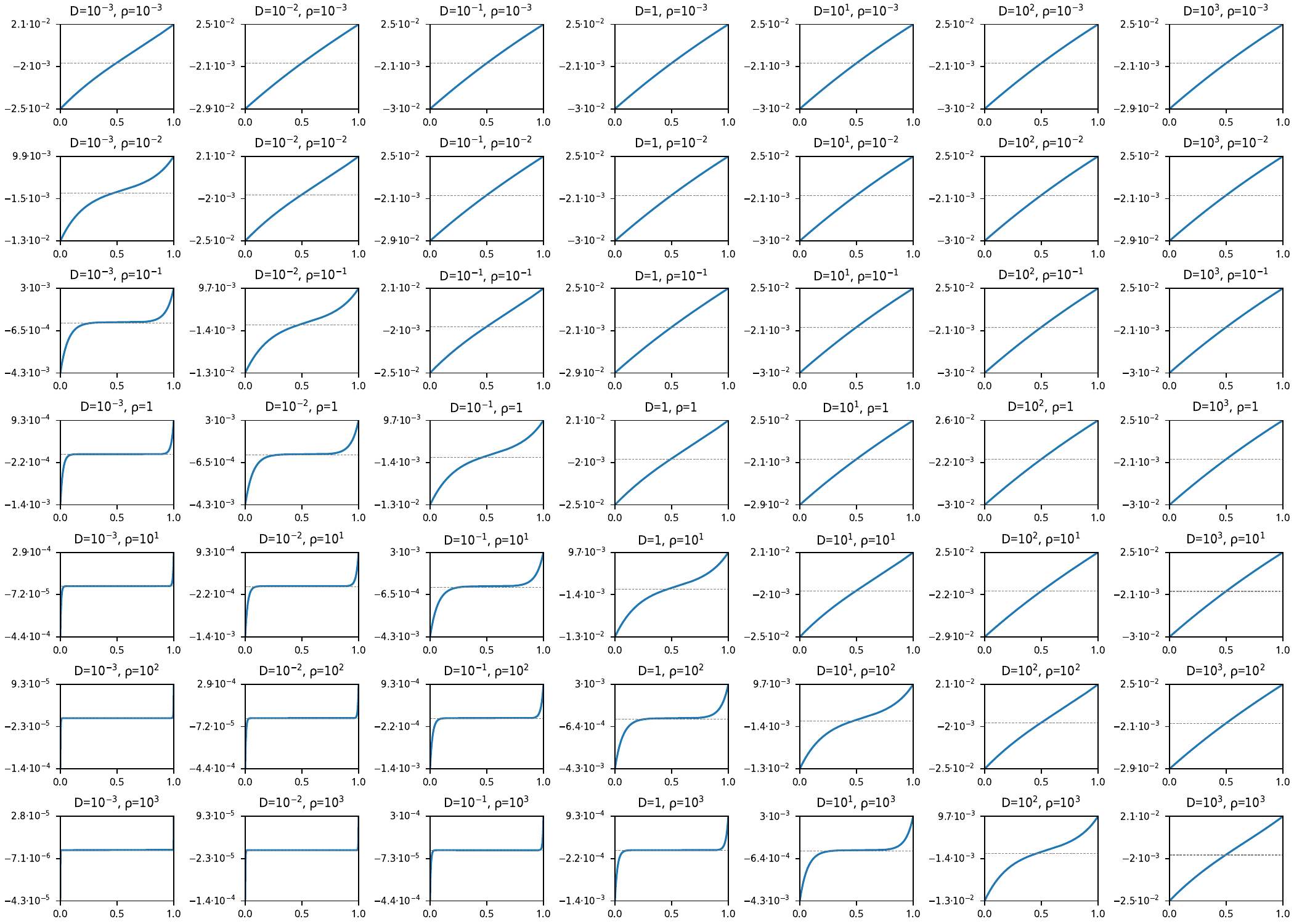}
  \caption{$\alpha$ for Linear profile, high $\theta_{g}$ value ($\beta=1$, $\theta_g=0.1$).}
  \label{fig:Linh-a}
\end{figure}

\begin{figure}[p]
  \centering
  \includegraphics[width=0.87\linewidth]{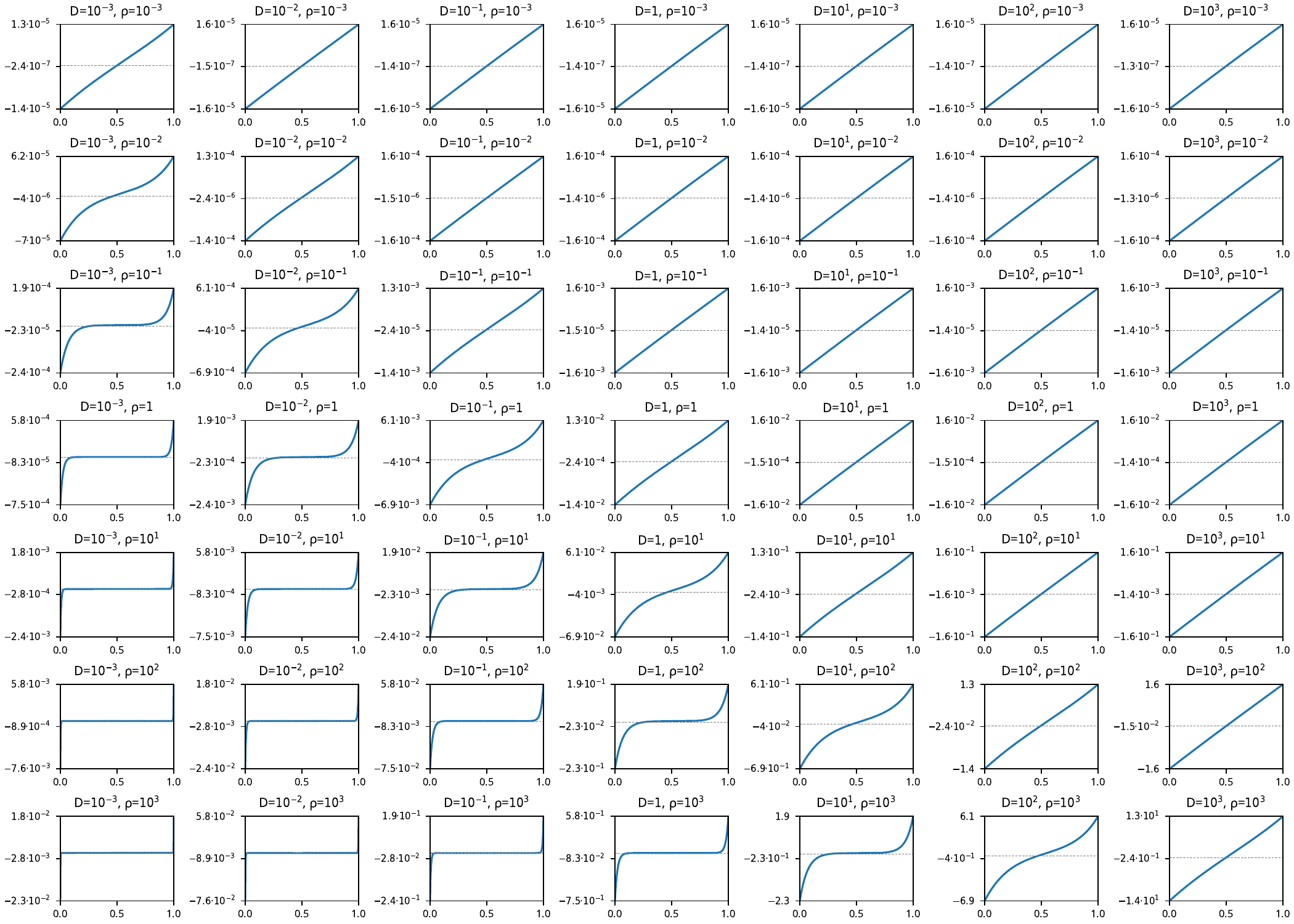}
  \caption{$\varphi$ for Linear profile, high $\theta_{g}$ value ($\beta=1$, $\theta_g=0.1$).}
  \label{fig:Linh-p}
\end{figure}

\begin{figure}[p]
  \centering
  \includegraphics[width=0.87\linewidth]{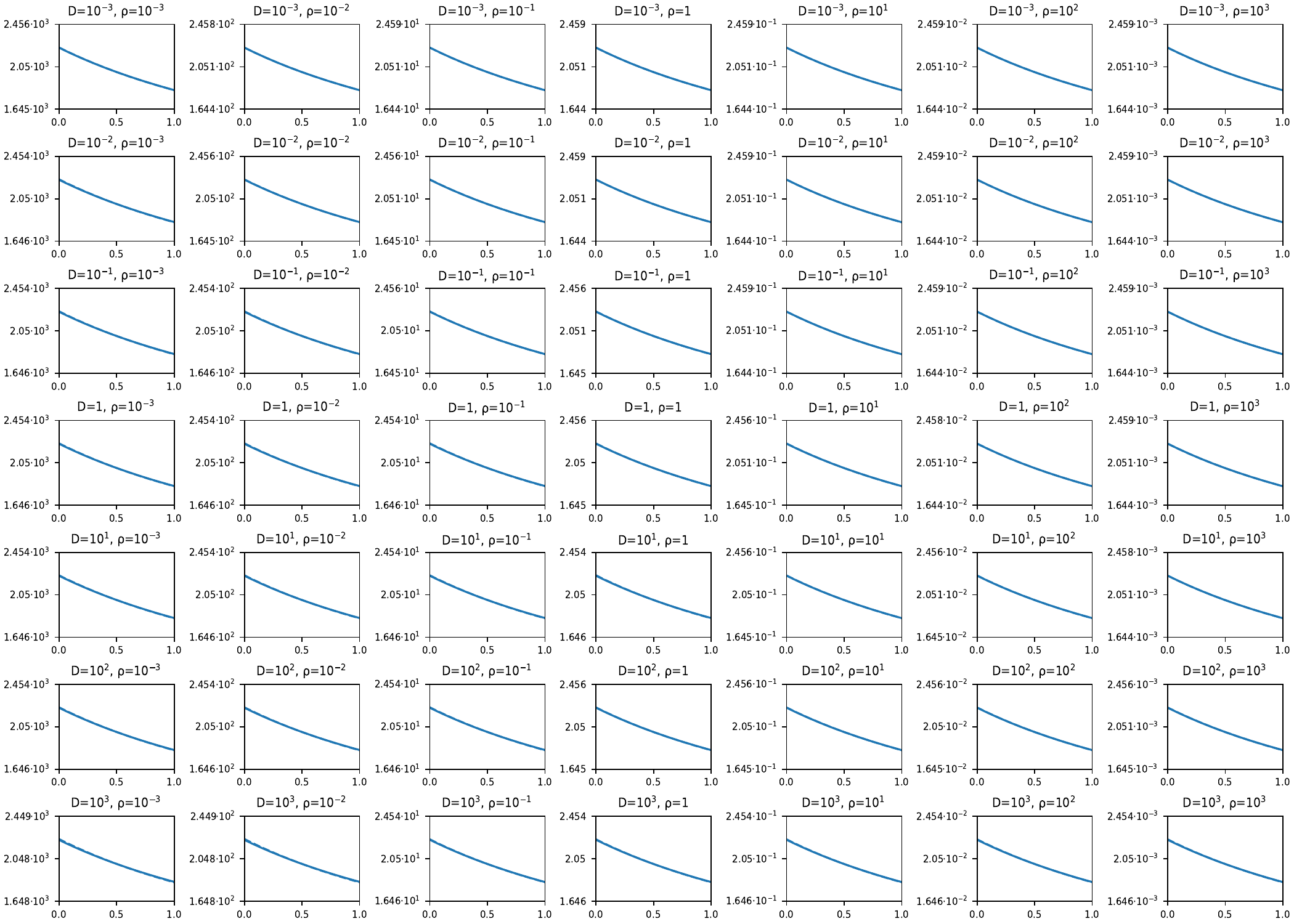}
  \caption{$R_{\textrm{tot}}$ for Linear profile, high $\theta_{g}$ value ($\beta=1$, $\theta_g=0.1$).}
  \label{fig:Linh-r}
\end{figure}

\begin{figure}[p]
  \centering
  \includegraphics[width=0.87\linewidth]{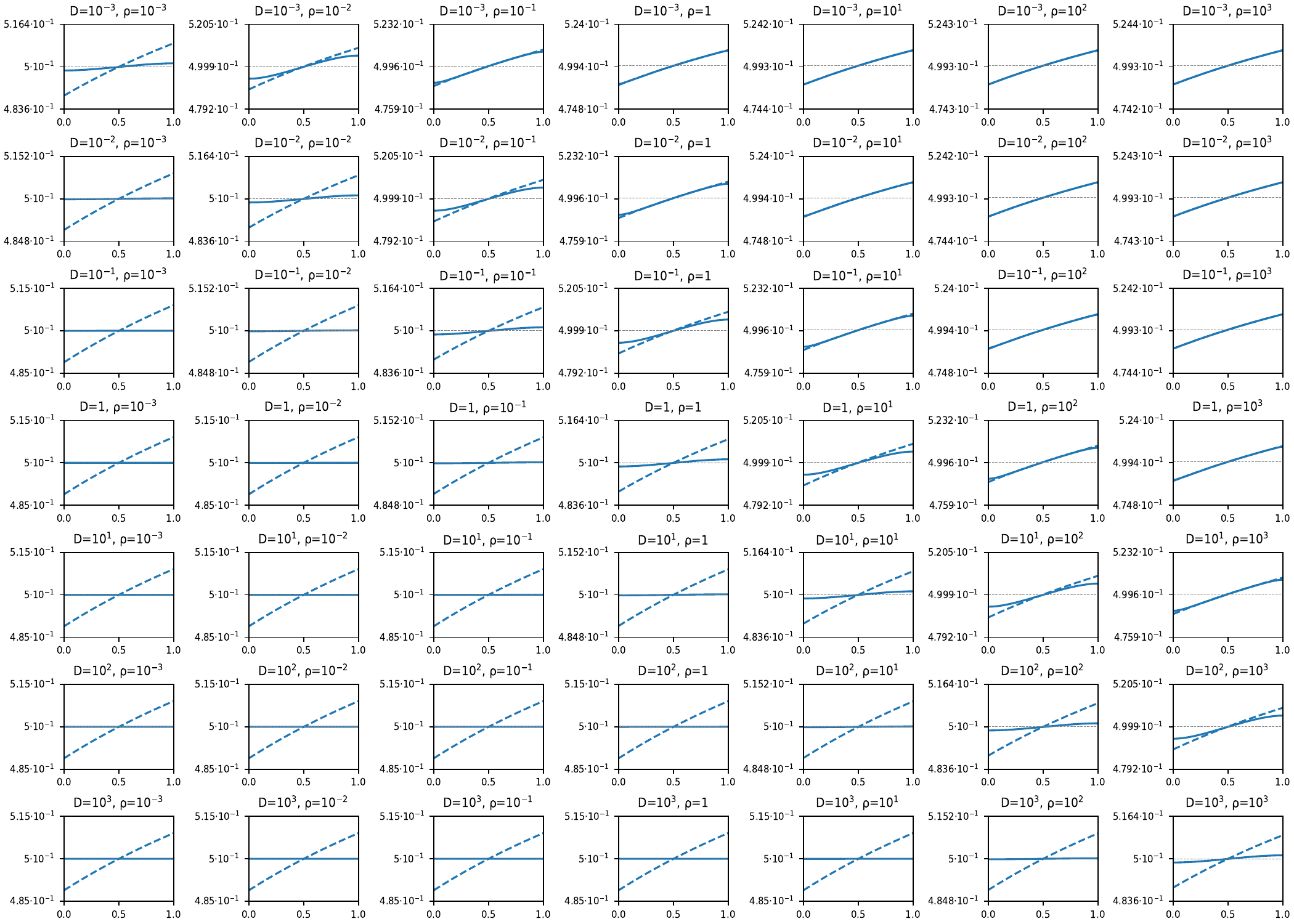}
  \caption{$u_{2}$ for Linear profile, high $\theta_{g}$ value ($\beta=1$, $\theta_g=0.1$).}
  \label{fig:Linh-u}
\end{figure}

\begin{figure}[p]
  \centering
  \includegraphics[width=0.87\linewidth]{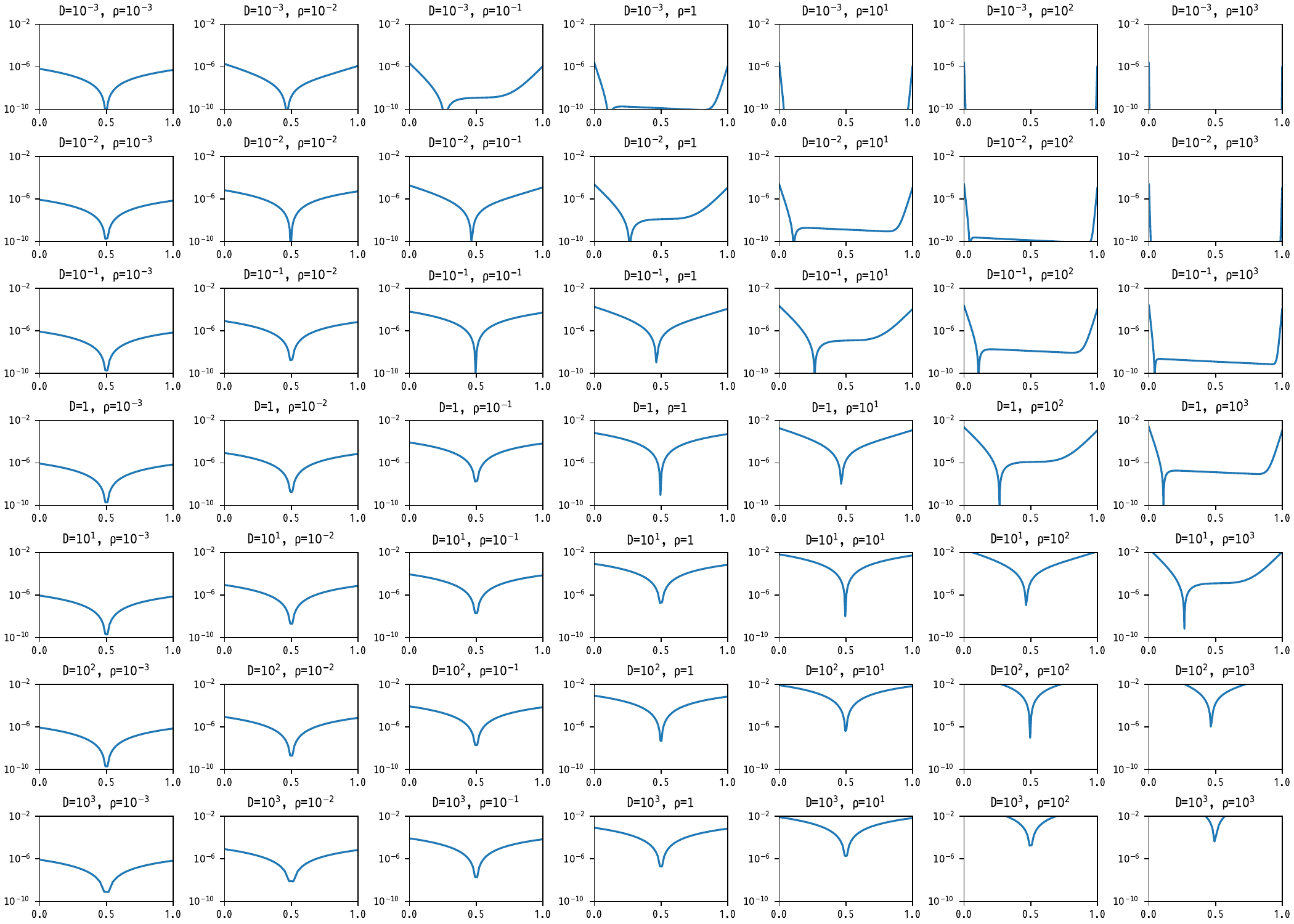}
  \caption{$\sigma_{r}$ for Linear profile, high $\theta_{g}$ value ($\beta=1$, $\theta_g=0.1$).}
  \label{fig:Linh-sr}
\end{figure}

\begin{figure}[p]
  \centering
  \includegraphics[width=0.87\linewidth]{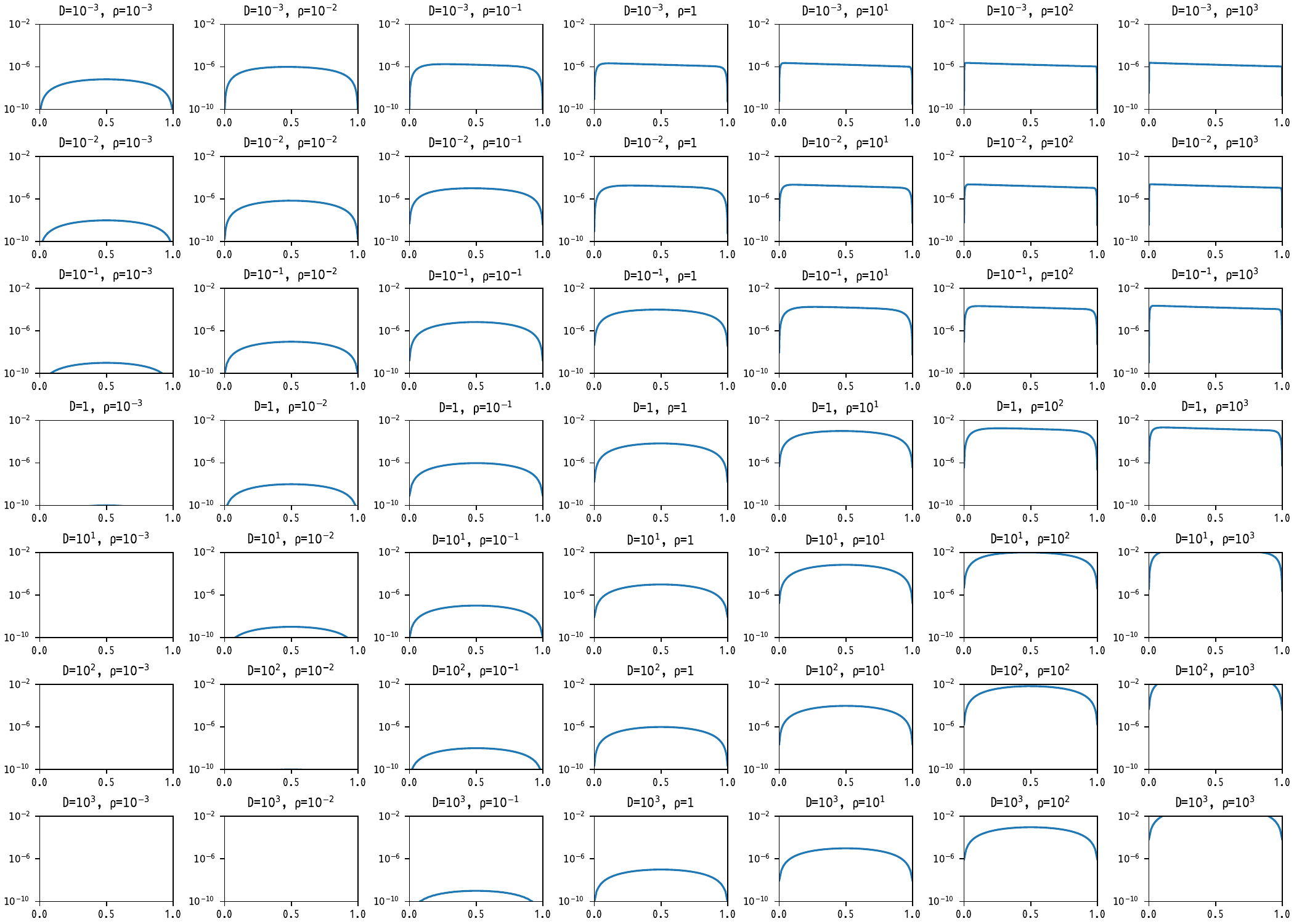}
  \caption{$\sigma_{d}$ for Linear profile, high $\theta_{g}$ value ($\beta=1$, $\theta_g=0.1$).}
  \label{fig:Linh-sd}
\end{figure}

\begin{figure}[p]
  \centering
  \includegraphics[width=0.87\linewidth]{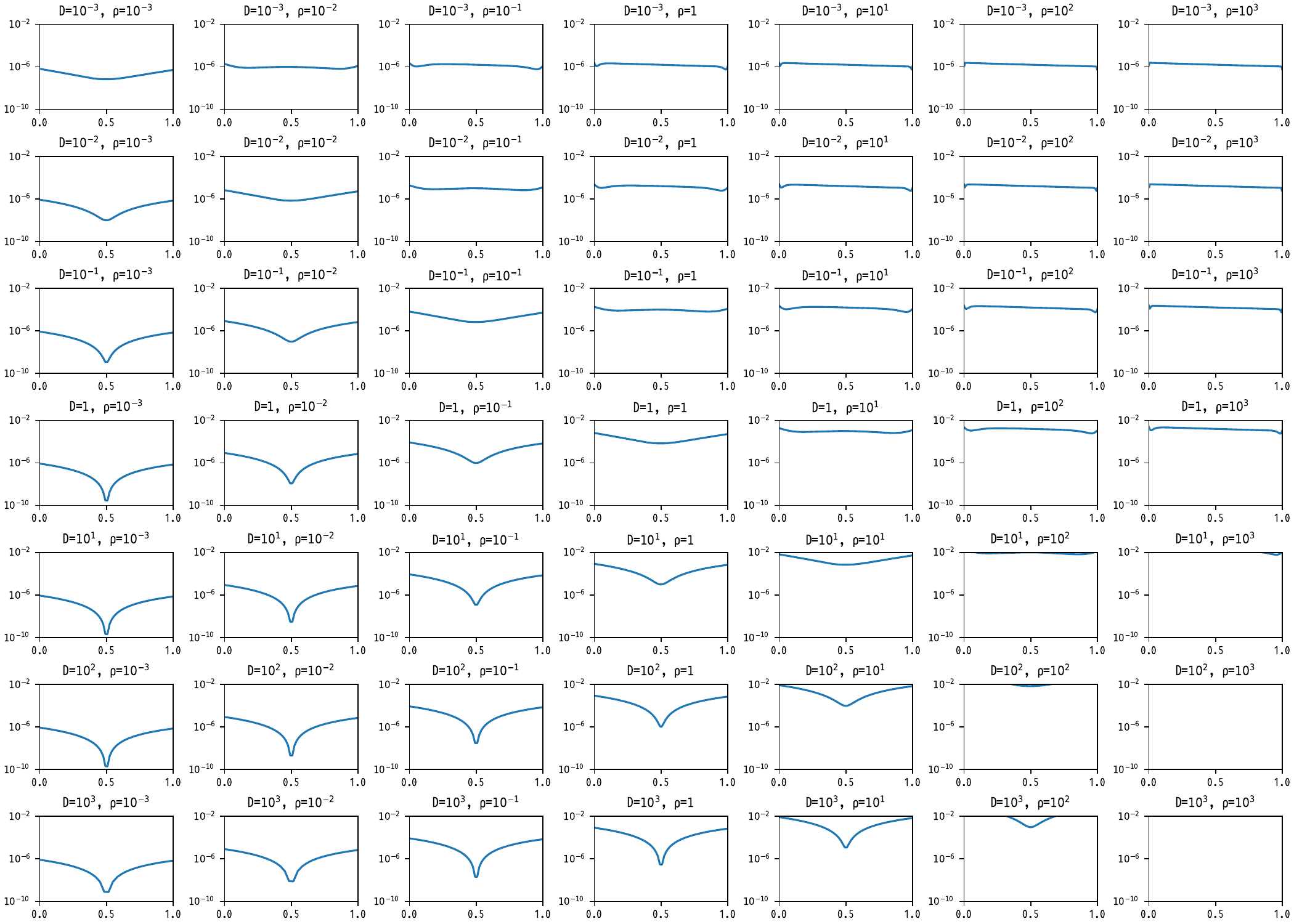}
  \caption{$\sigma_{\textrm{tot}}$ for Linear profile, high $\theta_{g}$ value ($\beta=1$, $\theta_g=0.1$).}
  \label{fig:Linh-st}
\end{figure}


\begin{figure}[p]
  \centering
  \includegraphics[width=0.87\linewidth]{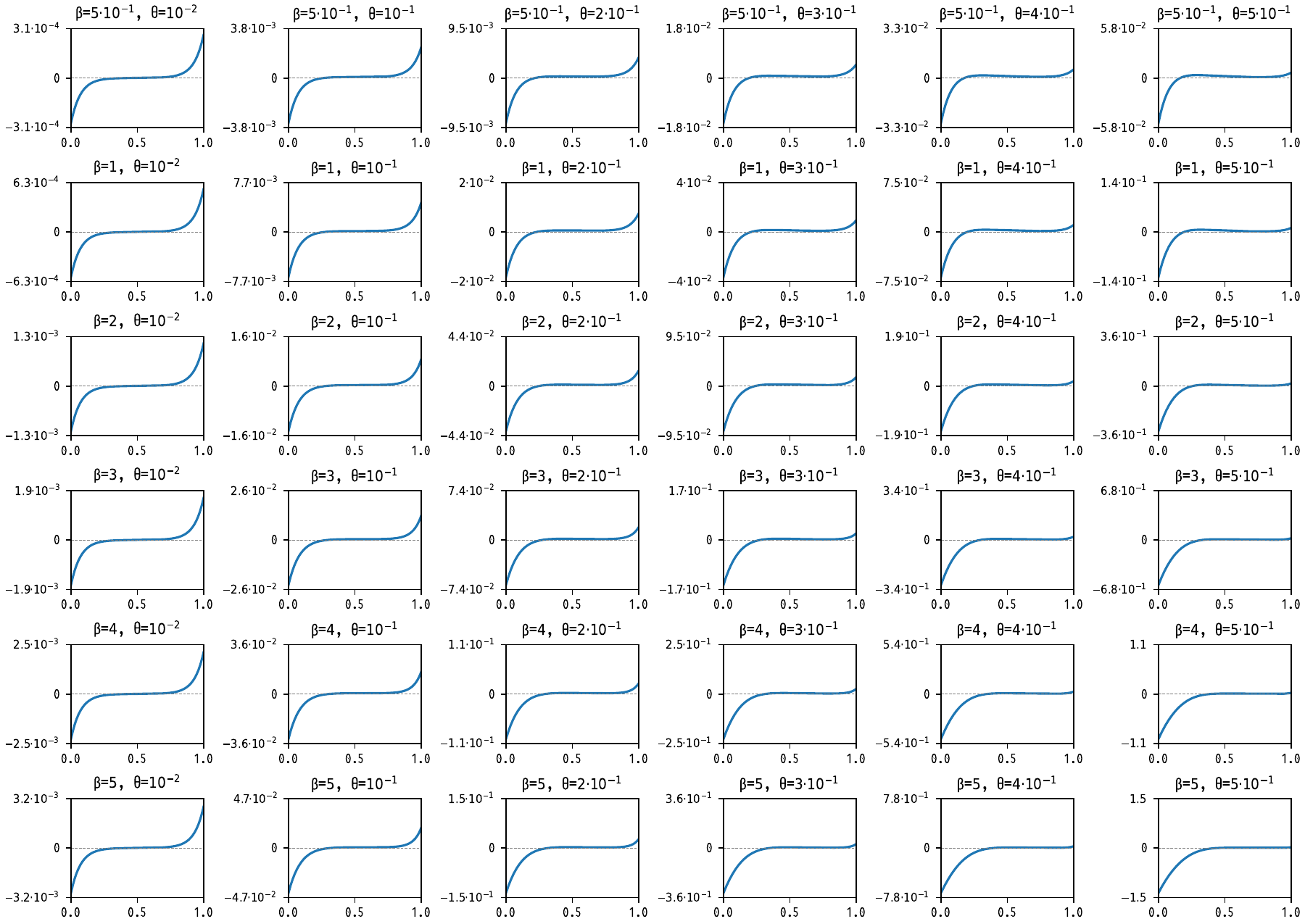}
  \caption{$\alpha$ for Linear profile, as a function of $\beta$ and $\theta_g$ ($d=1$, $\rho=100$).}
  \label{fig:Linbt-a}
\end{figure}

\begin{figure}[p]
  \centering
  \includegraphics[width=0.87\linewidth]{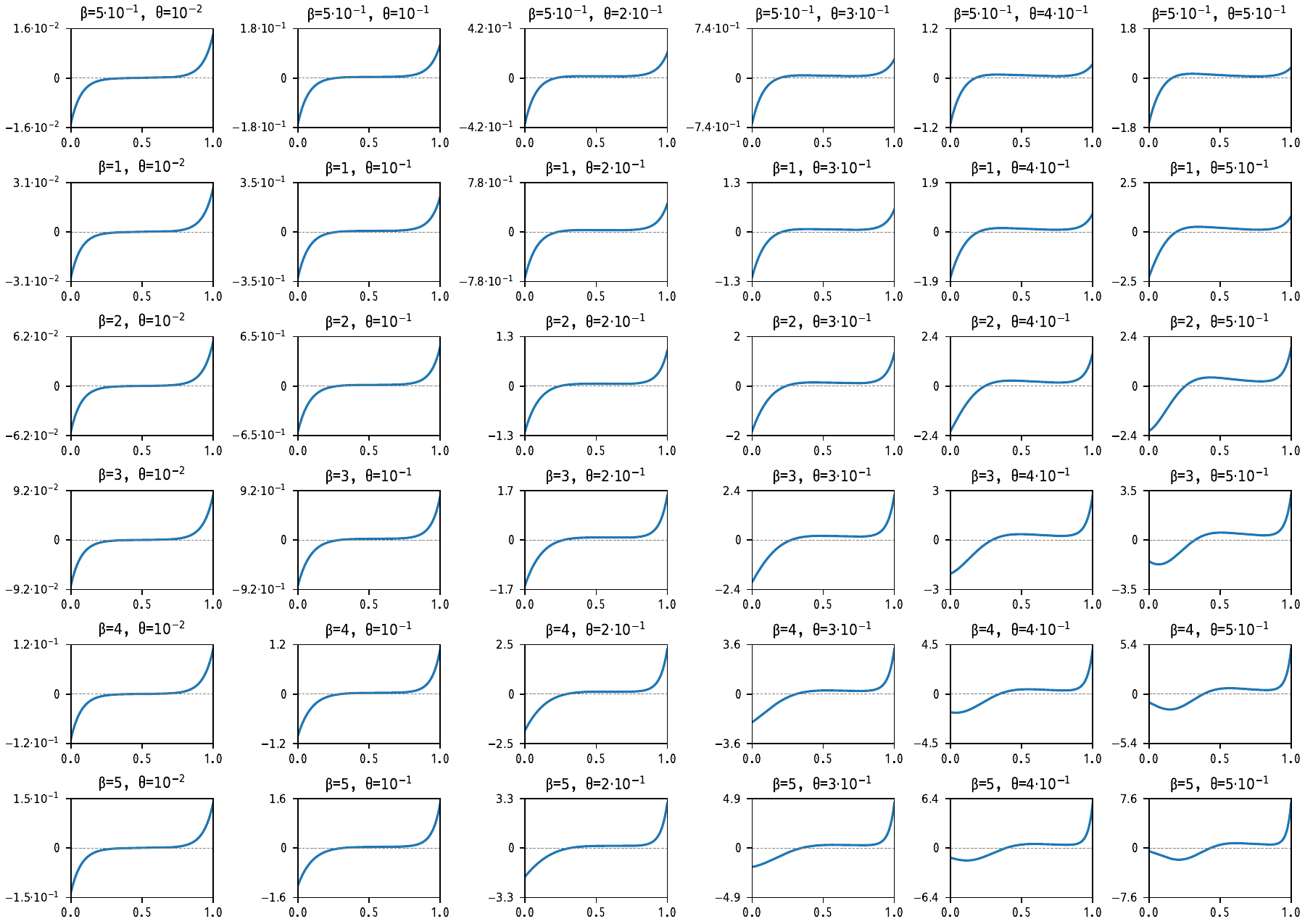}
  \caption{$\varphi$ for Linear profile, as a function of $\beta$ and $\theta_g$ ($d=1$, $\rho=100$).}
  \label{fig:Linbt-p}
\end{figure}

\begin{figure}[p]
  \centering
  \includegraphics[width=0.87\linewidth]{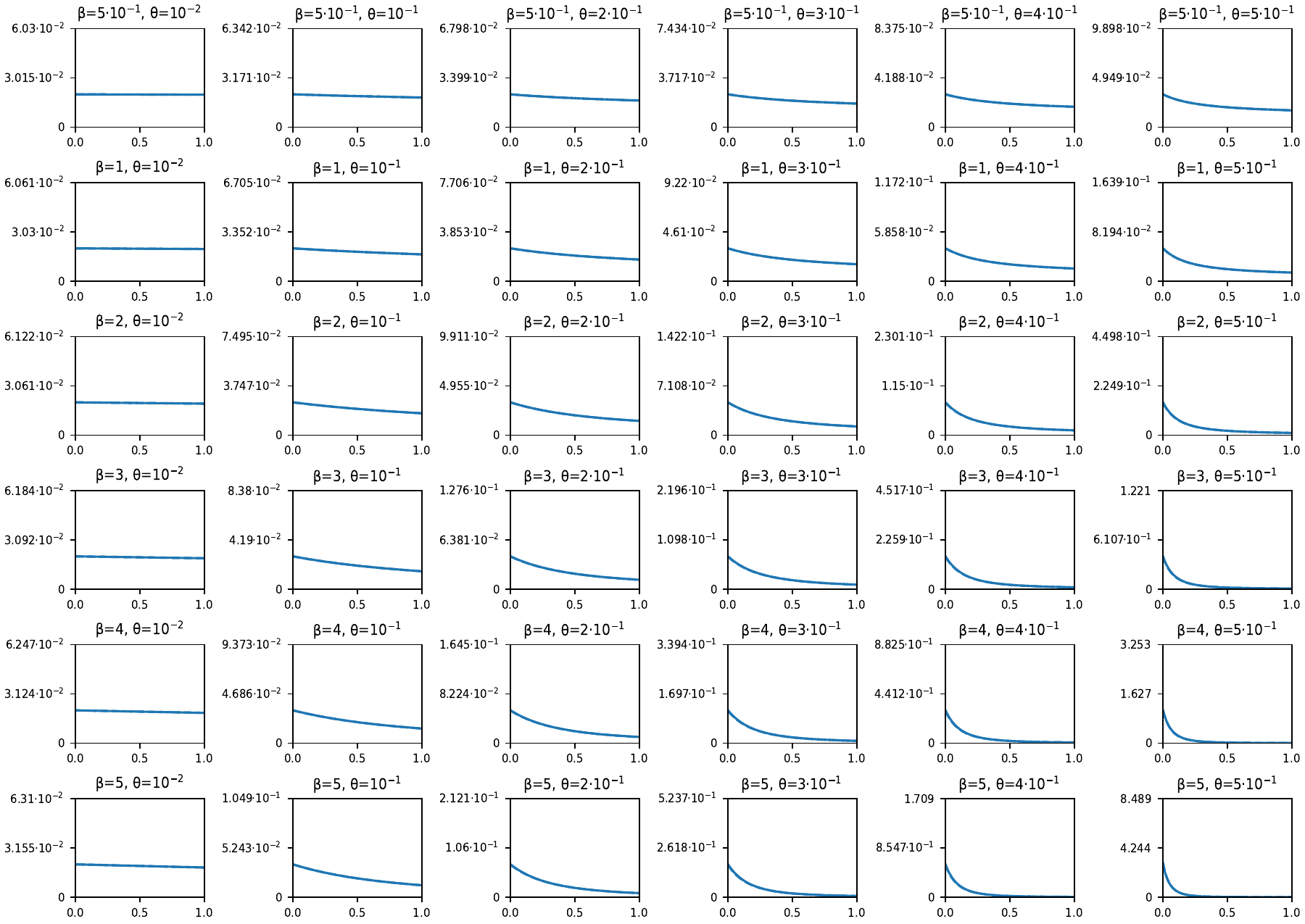}
  \caption{$R_{\textrm{tot}}$ for Linear profile, as a function of $\beta$ and $\theta_g$ ($d=1$, $\rho=100$).}
  \label{fig:Linbt-r}
\end{figure}

\begin{figure}[p]
  \centering
  \includegraphics[width=0.87\linewidth]{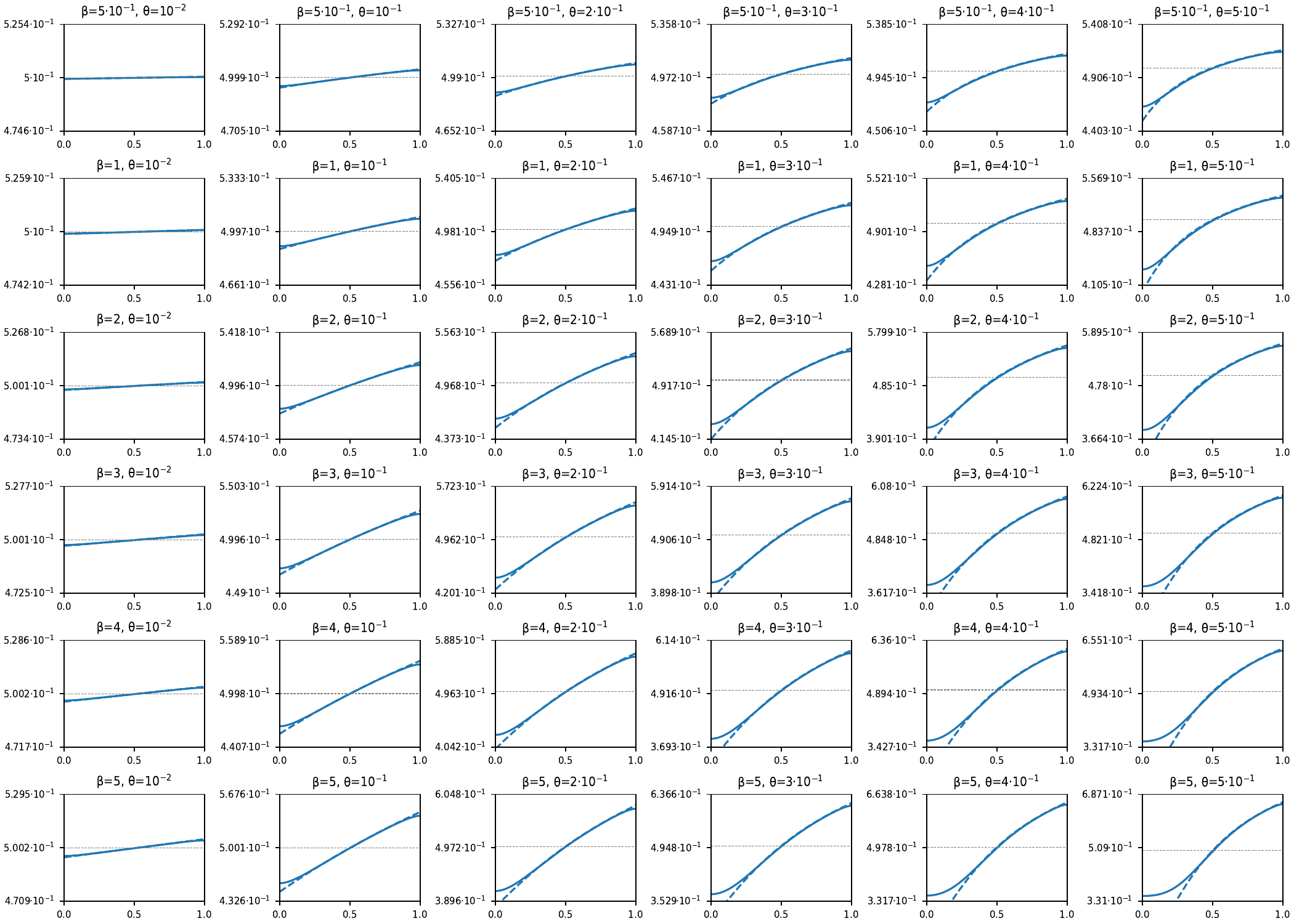}
  \caption{$u_{2}$ for Linear profile, as a function of $\beta$ and $\theta_g$ ($d=1$, $\rho=100$).}
  \label{fig:Linbt-u}
\end{figure}

\begin{figure}[p]
  \centering
  \includegraphics[width=0.87\linewidth]{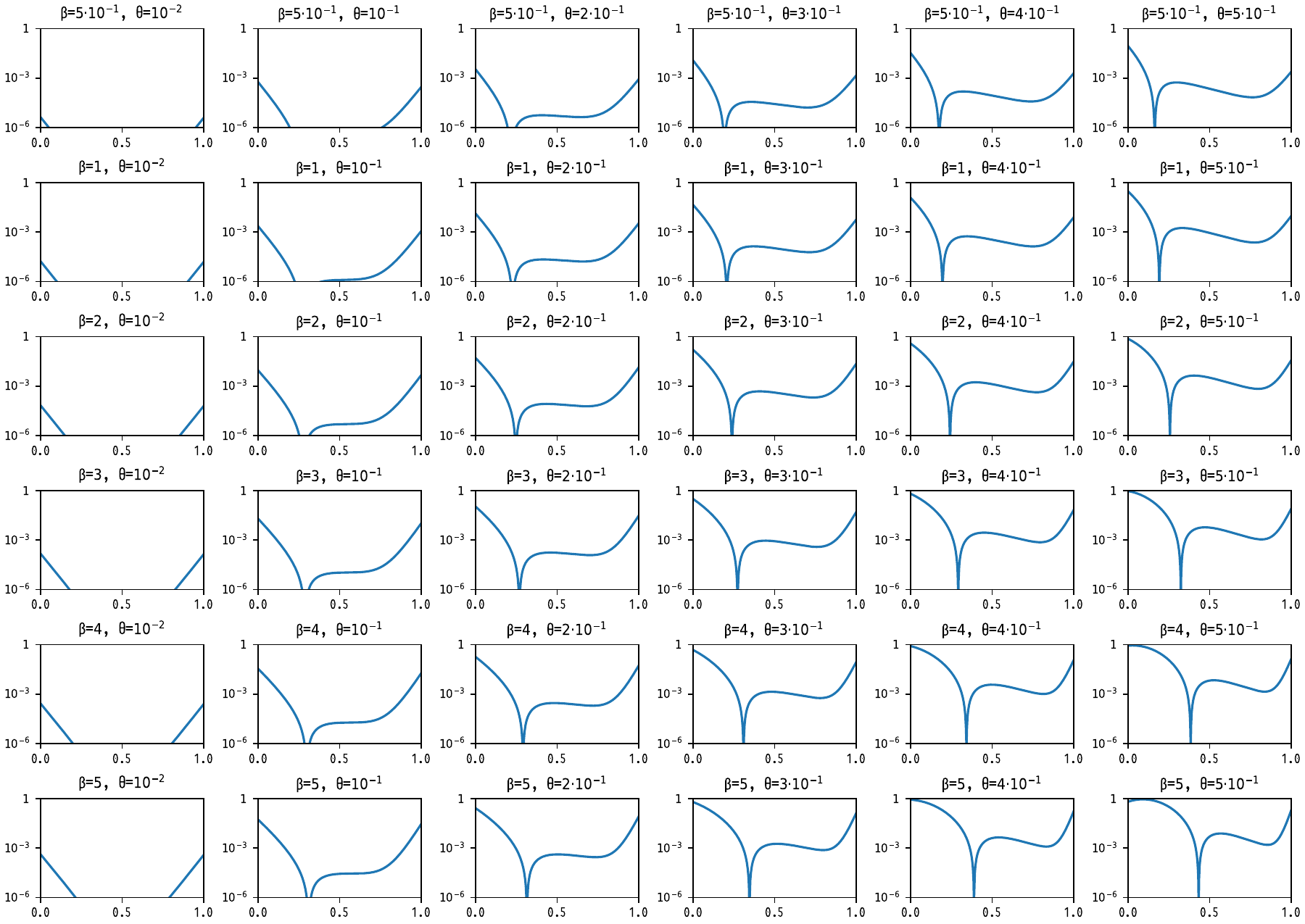}
  \caption{$\sigma_{r}$ for Linear profile, as a function of $\beta$ and $\theta_g$ ($d=1$, $\rho=100$).}
  \label{fig:Linbt-sr}
\end{figure}

\begin{figure}[p]
  \centering
  \includegraphics[width=0.87\linewidth]{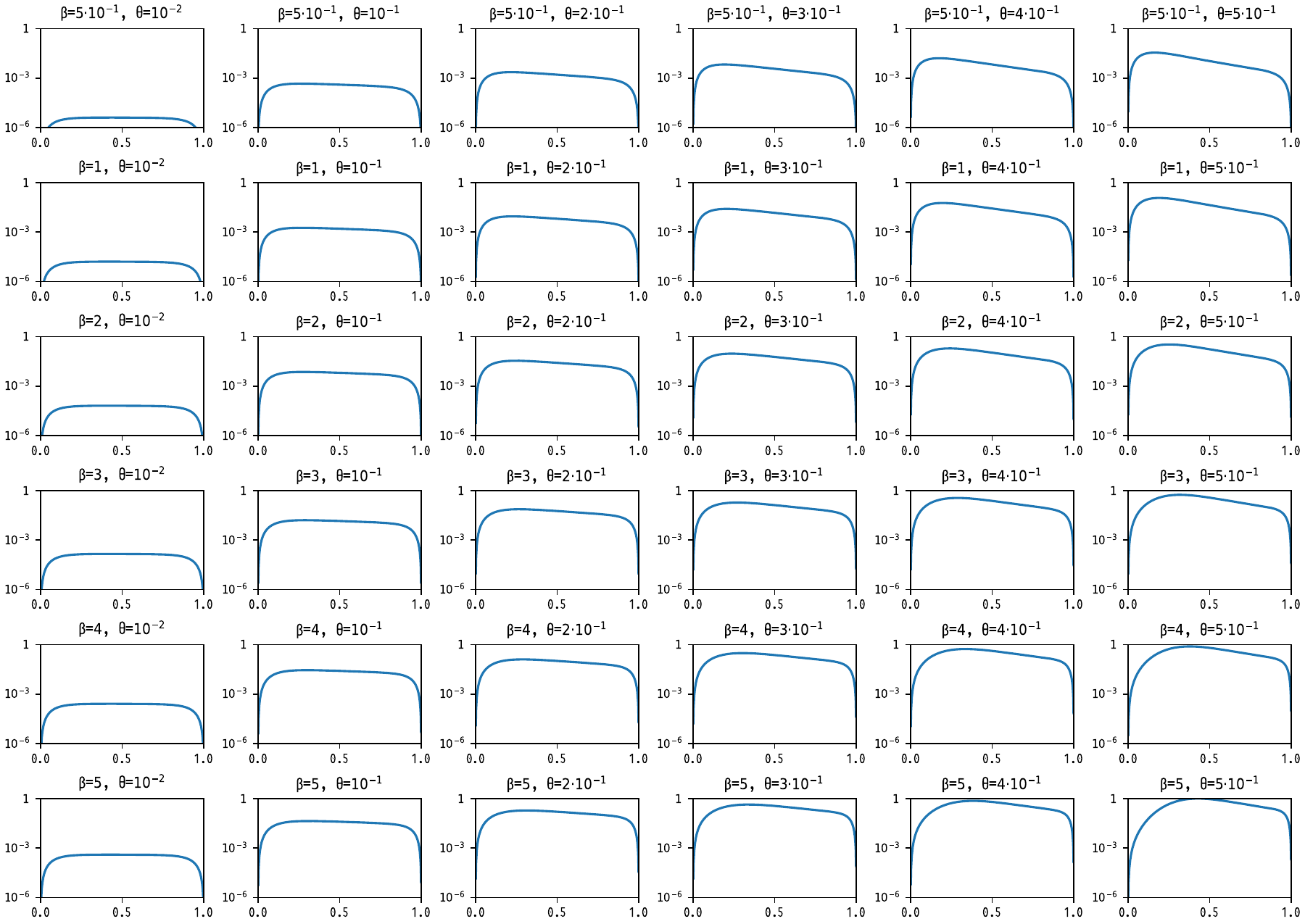}
  \caption{$\sigma_{d}$ for Linear profile, as a function of $\beta$ and $\theta_g$ ($d=1$, $\rho=100$).}
  \label{fig:Linbt-sd}
\end{figure}

\begin{figure}[p]
  \centering
  \includegraphics[width=0.87\linewidth]{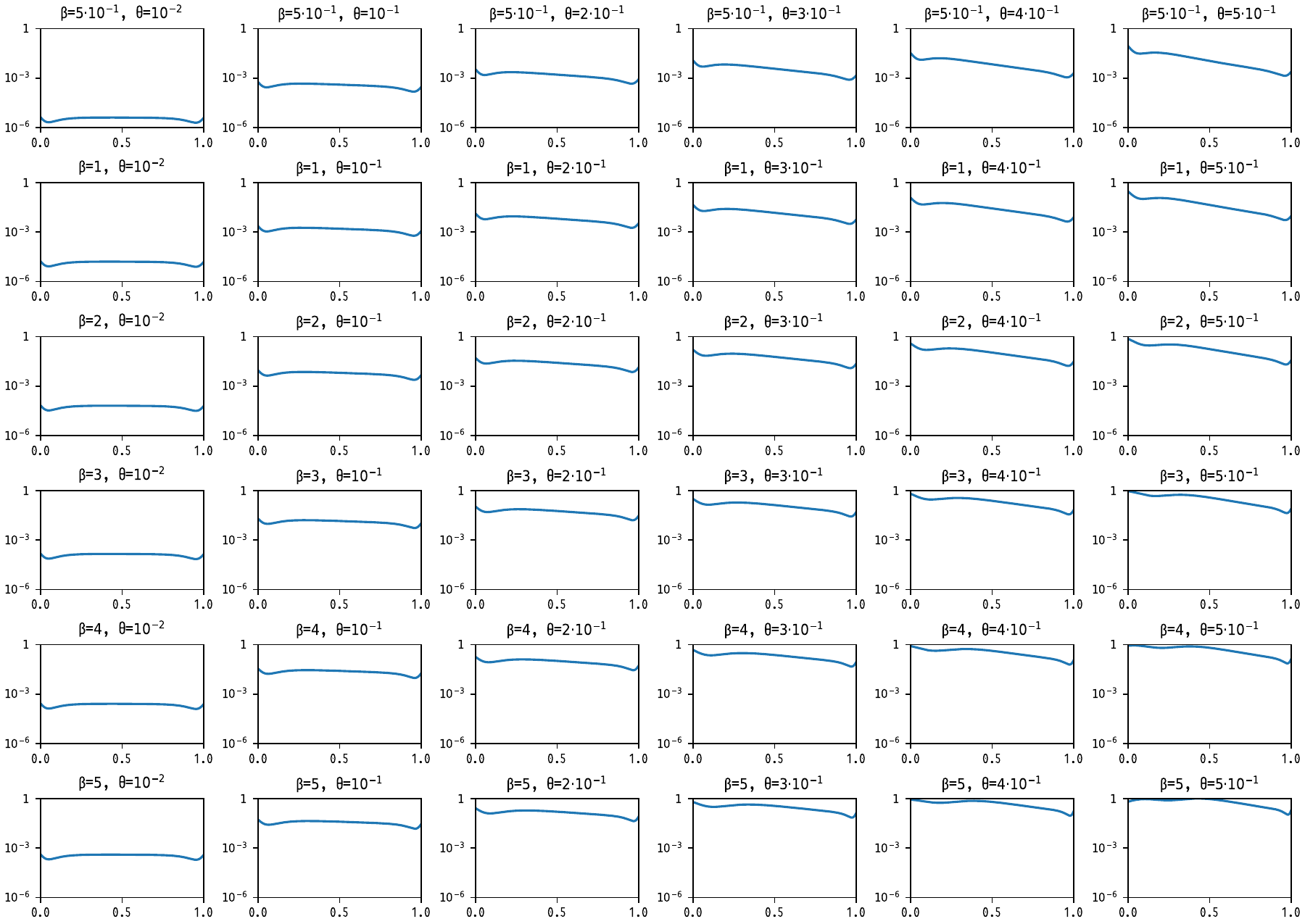}
  \caption{$\sigma_{\textrm{tot}}$ for Linear profile, as a function of $\beta$ and $\theta_g$ ($d=1$, $\rho=100$).}
  \label{fig:Linbt-st}
\end{figure}